\documentclass{report}

\usepackage{macros}
\usepackage{macros-gen-en}
\usepackage{macros-math}
\usepackage{macros-logic}
\usepackage{macros-ML}
\usepackage{macros-ML-subst}

\begin{document}

\title{Matching logic -  a new axiomatization}
\author{Lauren{\c t}iu Leu{\c s}tean${}^{a,b,c}$, Dafina Trufa\c{s}${}^{a,b}$,\\[2mm]
	\footnotesize ${}^{a}$ Institute for Logic and Data Science, Bucharest\\
	\footnotesize ${}^{b} $ LOS, Faculty of  Mathematics and Computer Science, University of Bucharest,\\
	\footnotesize ${}^{c}$ Simion Stoilow Institute of Mathematics of the Romanian Academy\\[1mm]
}

\date{}

\maketitle

\setlength{\parindent}{0em}

\tableofcontents

\chapter{Introduction}

Matching logic was introduced by Grigore Ro\c{s}u \cite{RosEllSch10,Ros17}.  Different versions 
of matching logic were recently studied  (see, for example, \cite{CheRos19a,CheRos19,CheRos20,CheLucRos21}). \\

In Chapter \ref{AML-Gc} we propose a new proof system for $\appML$, first-order matching logic with application.
This proof system, denoted  $\appMLGc$,  is obtained by adapting to matching logic G\"odel's proof sytem for 
first-order intuitionistic logic (see \cite{God58,Tro73}). We prove  different useful theorems and derived rules, 
important properties of the proof system. For different proofs in the propositional 
fragment of $\appMLGc$, Kleene's textbook \cite{Kle52} was a source of inspiration.

Finally, we prove that $\appMLGc$ is equivalent with the proof system $\appMLnow$ for \appML,  obtained 
from the proof system defined in \cite{BerCheHorMizPenTus22} by considering only the axioms and deduction rules from the categories 
FOL reasoning and Frame reasoning. \\

In Chapter \ref{AML-Tdef} we give a new, simpler proof system $\appMLnewGc$  for first-order matching 
logic with application  and definedness ($\appMLdef$ \!). The proof system $\appMLnewGc$ is inspired by 
Tarski's  \cite{Tar65,Tar51} axiomatization for first order-logic with equality (simplified by Kalish and 
Montague \cite{KalMon65,MonKal56}), that  does not involve the notions of a free variable and free substitution. 
We prove  different useful theorems and derived rules for the propositional  and first-order connectives, 
as well as for definedness, equality, totality, membership and application (contexts). Some of the results 
are obtained by adapting to our setting proofs from \cite[Chapter 10]{Mon76}. The propositional axioms 
and deduction rules of 
$\appMLnewGc$  coincide with the ones of the proof system $\appMLGc$, hence we can use in this setting 
propositional theorems and derived deduction rules from Chapter \ref{AML-Gc}.

We point out the following important results:
\be
\item replacement theorems (Subsection  \ref{section-replacement-thms});
\item  substitution of equal variables (Subsection \ref{section-free-subst}); 
\item the bounded substitution theorem (Subsection \ref{section-bounded-subst-thm});
\item universal and existential specification (Subsection \ref{section-univ-exist-spec});
\item substitution of equal patterns (Subsection \ref{section-subst-equal-patterns}); 
\item the deduction theorem (Subsection \ref{section-deduction-thm});
\item the propagation and framing rules from the proof system for applicative matching logic used in
 \cite{CheRos19a,CheLucRos21,Che23} (Subsection \ref{section-app-contexts});
\item theorems and rules for definedness, totality, equality (Subsections  \ref{section-def},  \ref{section-def-tot-eq})
and  membership (Subsection \ref{section-membership}). 
\ee 

Theorem \ref{Gc-weaker-MGc}, the main result of Section \ref{section-Gc-weaker-MGc}, shows that $\appMLGc$  
is weaker than $\appMLnewGc$.  As $\appMLGc$ is equivalent with $\appMLnow$, we get that $\appMLnow$  is 
also weaker than $\appMLnewGc$. By using notions and results from \cite{Leu25}, we prove in Section \ref{MG-sound} a soundness 
theorem for $\appMLnewGc$.

\mbox{}

Throughout these notes, we use the terminology for abstract matching logic from \cite{Leu25a}. Thus, we refer 
to \cite{Leu25a} for definitions and properties of free and bound variables, 
free substitution $Subf$ and bounded substitution $Subb$, variable free for patterns, 
congruences, contexts, proof systems, as well as for all the undefined notions from these notes.

\section*{Acknowledgement}

This research was supported by a sponsorship from Pi Squared Inc.

\chapter{Applicative matching logic with application $\appML$}\label{AML-Gc}


The logic \textit{\appML} is defined as follows:
\be
\item $\lc_{\appML}$: 

$\PropConstants=\{\bot\}$, $\PropUnary=\{\neg\}$, $\PropBinary=\{\to, \si, \sau, \Appl\}$, 
$\FolQ=\{\forall, \exists\}$, 	$\Equal=\SolQ=\svar=\emptyset$.
\item Its proof system, $\appMLGc$, is given in Appendix \ref{appML-Gc}.
\ee

We write simply $\vdash$ instead of $\vdash_{\cG^c}$.\\

We shall  use the infix notation for patterns. Thus, we write $\vp\to\psi$ instead of $\to\vp\psi$, 
$\vp\si\psi$ instead of $\si\vp\psi$, $\vp\sau\psi$ instead of $\sau\vp\psi$, 
$\appln{\vp}{\psi}$ instead of $\Appl \vp\psi$. 
We also use parentheses when it is necessary to make clear the 
structure of a pattern. For example, we write $(\vp\to \psi)\to \chi$. \\

We introduce the derived  connective $\dnd$ by the following abbreviation
$$
\vp \dnd \psi := (\vp \to \psi) \si (\psi \to \vp).
$$
To reduce the use of parentheses, we assume that 
\be
\item quantifiers $\exists$, $\forall$, application, $\neg$ have higher precedence 
than $\ra$, $\si$, $\sau$, $\dnd$;
\item $\si, \sau$ have higher precedence than $\ra$, $\dnd$.
\ee

\newpage
\section{Theorems and derived rules in $\cG^c$}

In the following, $\Gamma$ is an arbitrary set of patterns and $\vp$, $\psi$, $\chi$, $\vp'$, $\psi'$ are arbitrary patterns.

\subsection{Propositional connectives}

\begin{lemma}
\begin{align}
\vdash & \, \psi \to \vp \sau \psi, \label{weak-sau-2} \\
\vdash & \, \vp \si \psi \to \psi, \label{weak-si-2} \\
\vdash & \, \vp\to(\psi\to\vp), \label{vp-to-psi-to-vp}\\
\vdash & \, \vp\to\vp, \label{vp-to-vp} \\
\vdash & \,\vp\to (\psi\to \vp\si \psi), \label{vp-to-psi-to-vpsipsi} \\
\vdash & \,(\vp\to \psi)\si\vp \to \psi, \label{MP-axiom-1}\\
\vdash & \,\vp\si (\vp\to \psi) \to \psi, \label{MP-axiom-2}\\
\vdash & \,\vp\to ((\vp\to\psi)\to \psi), \label{MP-axiom-3}\\
\vdash & \,(\vp\si\psi)\si\chi \to \vp, \label{weak-si-3-l} \\
\vdash & \,\vp\si(\psi\si\chi) \to \psi,\label{weak-si-3-l1} \\
\vdash & \,(\vp\si\psi)\si\chi \to \psi, \label{weak-si-3-r}\\
\vdash & \vp\si(\psi\si\chi)\to \chi, \label{weak-si-3-r1}\\
\vdash & \psi\to \vp\sau(\psi\sau \chi), \label{weak-sau-3-l1}\\
\vdash & \chi\to \vp\sau(\psi\sau \chi), \label{weak-sau-3-r1}\\
\vdash & \vp\to (\vp\sau\psi)\sau \chi, \label{weak-sau-3-l2}\\
\vdash & \psi\to (\vp\sau\psi)\sau \chi.  \label{weak-sau-3-r2}
\end{align}
\end{lemma}
\solution{\,\\[1mm]
\eqref{weak-sau-2}:\\[1mm]
\begin{tabular}{lll}
(1) & $\vdash \psi \to \psi \sau \vp $  & \prule{weakening}\\[1mm]
(2)  & $\vdash \psi \sau \vp \to \vp\sau\psi$ & \prule{permutation}\\[1mm]
(3)  & $\vdash \psi \to \vp \sau \psi$ & \prule{syllogism}: (1), (2)\\[2mm]
\end{tabular}

\eqref{weak-si-2}:\\[1mm]
\begin{tabular}{lll}
(1) & $\vdash \vp \si \psi \to \psi \si \vp $  & \prule{permutation}\\[1mm]
(2)  & $\vdash \psi \si \vp  \to \psi$ & \prule{weakening}\\[1mm]
(3)  & $\vdash \vp \si \psi \to \psi$ & \prule{syllogism}: (1), (2)\\[2mm]
\end{tabular}

\eqref{vp-to-psi-to-vp}:\\[1mm]
\begin{tabular}{lll}
(1) & $\vdash \vp\si \psi \to \vp$  & \prule{weakening}\\[1mm]
(2)  & $\vdash \vp\to(\psi\to\vp)$ & \prule{exportation}: (1)\\[2mm]
\end{tabular}

\eqref{vp-to-vp}:\\[1mm]
\begin{tabular}{lll}
(1) & $\vdash \vp\to\vp\si\vp$  & \prule{contraction}\\[1mm]
(2)  & $\vdash \vp\si\vp\to\vp$ & \prule{weakening}\\[1mm]
(3)  & $\vdash \vp\to\vp$ & \prule{syllogism}: (1), (2)\\[2mm]
\end{tabular}

\eqref{vp-to-psi-to-vpsipsi}:\\[1mm]
\begin{tabular}{lll}
(1) & $\vdash \vp\si \psi \to\vp\si\psi$  & \eqref{vp-to-vp}\\[1mm]
(2)  & $\vdash \vp\to (\psi\to \vp\si \psi) $ & \prule{exportation}: (1)\\[2mm]
\end{tabular}

\eqref{MP-axiom-1}:\\[1mm]
\begin{tabular}{lll}
(1) & $\vdash (\vp\to\psi) \to (\vp\to\psi)$  & \eqref{vp-to-vp}\\[1mm]
(2)  & $\vdash (\vp\to\psi)\si \vp\to\psi $ & \prule{importation}: (1)\\[2mm]
\end{tabular}

\eqref{MP-axiom-2}:\\[1mm]
\begin{tabular}{lll}
(1) & $\vdash (\vp\to\psi)\si \vp\to\psi $  & \eqref{MP-axiom-1}\\[1mm]
(2)  & $\vdash \vp\si(\vp\to\psi)\to (\vp\to\psi)\si \vp $ & \prule{permutation}\\[1mm]
(3)  & $\vdash \vp\si(\vp\to\psi)\to \psi$ & \prule{syllogism}: (2), (1)\\[2mm]
\end{tabular}

\eqref{MP-axiom-3}:\\[1mm]
\begin{tabular}{lll}
(1)  & $\vdash \vp\si(\vp\to\psi)\to \psi$ & \eqref{MP-axiom-2} \\[1mm]
(2)  & $\vdash \vp\to ((\vp\to\psi)\to \psi)$ & \prule{exportation}: (1)\\[2mm]
\end{tabular}

\eqref{weak-si-3-l}:\\[1mm]
\begin{tabular}{lll}
(1) & $\vdash (\vp\si\psi)\si\chi \to \vp\si \psi $  & \prule{weakening} \\[1mm]
(2)  & $\vdash \vp\si \psi \to \vp $ & \prule{weakening} \\[1mm]
(3)  & $\vdash (\vp\si\psi)\si\chi \to  \vp$ & \prule{syllogism}: (1), (2)\\[2mm]
\end{tabular}

\eqref{weak-si-3-l1}:\\[1mm]
\begin{tabular}{lll}
(1) & $\vdash \vp\si(\psi\si\chi) \to \psi\si\chi$  & \eqref{weak-si-2}\\[1mm]
(2)  & $\vdash \psi\si\chi \to \psi$ & \prule{weakening} \\[1mm]
(3)  & $\vdash \vp\si(\psi\si\chi) \to \psi$ & \prule{syllogism}: (1), (2)\\[2mm]
\end{tabular}

\eqref{weak-si-3-r}:\\[1mm]
\begin{tabular}{lll}
(1) & $\vdash (\vp\si\psi)\si\chi \to \vp\si \psi $  & \prule{weakening} \\[1mm]
(2)  & $\vdash \vp\si \psi \to \psi $ & \eqref{weak-si-2}\\[1mm]
(3)  & $\vdash (\vp\si\psi)\si\chi \to  \psi$ & \prule{syllogism}: (1), (2)\\[2mm]
\end{tabular}

\eqref{weak-si-3-r1}:\\[1mm]
\begin{tabular}{lll}
(1) & $\vdash \vp\si(\psi\si\chi) \to \psi\si\chi$  & \eqref{weak-si-2}\\[1mm]
(2)  & $\vdash \psi\si\chi \to \chi$ & \eqref{weak-si-2}\\[1mm]
(3)  & $\vdash \vp\si(\psi\si\chi) \to \chi$ & (SYLLOGISM): (1), (2)\\[2mm]
\end{tabular}

\eqref{weak-sau-3-l1}:\\[1mm]
\begin{tabular}{lll}
(1) & $\vdash \psi\to \psi\sau\chi$  & \prule{weakening}\\[1mm]
(2)  & $\vdash \psi\sau\chi \to \vp\sau(\psi\sau \chi)$ & \eqref{weak-sau-2}\\[1mm]
(3)  & $\vdash \psi\to \vp\sau(\psi\sau \chi)$ & \prule{syllogism}: (1), (2)\\[2mm]
\end{tabular}

\eqref{weak-sau-3-r1}:\\[1mm]
\begin{tabular}{lll}
(1) & $\vdash \chi\to \psi\sau\chi$  & \eqref{weak-sau-2}\\[1mm]
(2)  & $\vdash \psi\sau\chi \to \vp\sau(\psi\sau \chi)$ & \eqref{weak-sau-2}\\[1mm]
(3)  & $\vdash \chi\to \vp\sau(\psi\sau \chi)$ & \prule{syllogism}: (1), (2)\\[2mm]
\end{tabular}

\eqref{weak-sau-3-l2}:\\[1mm]
\begin{tabular}{lll}
(1) & $\vdash \vp\to \vp\sau\psi$  & \prule{weakening}\\[1mm]
(2)  & $\vdash \vp\sau\psi \to (\vp\sau\psi)\sau \chi$ & \prule{weakening}\\[1mm]
(3)  & $\vdash \vp\to (\vp\sau\psi)\sau \chi$ & \prule{syllogism}: (1), (2)\\[2mm]
\end{tabular}

\eqref{weak-sau-3-r2}:\\[1mm]
\begin{tabular}{lll}
(1) & $\vdash \psi\to \vp\sau\psi$  & \eqref{weak-sau-2}\\[1mm]
(2)  & $\vdash \vp\sau\psi \to (\vp\sau\psi)\sau \chi$ & \prule{weakening}\\[1mm]
(3)  & $\vdash \psi\to (\vp\sau\psi)\sau \chi$ & \prule{syllogism}: (1), (2)
\end{tabular}

\,
}

\begin{lemma}
\bea
\Gamma \vdash \vp  & \text{implies} &  \Gamma \vdash\psi\to\vp, 
\label{vp-implies-psi-to-vp}\\
\Gamma \vdash \vp \,\, \text{and}  \,\, \Gamma \vdash \psi  & \text{iff} &   \Gamma \vdash \vp \si \psi, 
\label{Gamma-vp-psi-vp-si-psi} \\
\Gamma\vdash\vp\to\psi   & \text{implies} &   \Gamma\vdash \vp\sau\chi \to \psi\sau\chi, 
\label{expansion-2}\\
\Gamma\vdash\vp\to\psi & \text{implies} & \Gamma\vdash \vp\sau\psi \to \psi  \text{~and~} \Gamma\vdash \psi\sau\vp \to \psi, 
\label{vp-psi-vp-to-psi-implies two}\\
\Gamma \vdash \vp \to \psi & \text{imply} &  \Gamma \vdash \vp\si\chi\to\psi \text{~and~} \Gamma \vdash \chi\si \vp\to\psi,
\label{vp-to-psi-imp-vp-si-chi-to-psi}\\
\Gamma\vdash \vp\to\psi \,\, \text{and}  \,\,\Gamma\vdash\vp\to\chi &  \text{iff}   & \Gamma\vdash\vp\to\psi\si\chi,
 \label{vp-to-psi-si-vp-to-chi-implies-vp-to-psi-si-chi}\\
\Gamma\vdash\vp\to\chi \,\, \text{and}  \,\,\Gamma\vdash \psi\to\chi & \text{imply} & \Gamma\vdash\vp\sau\psi\to\chi,
\label{vp-to-chi-and-psi-to-chi-implies-vp-sau-psi-to-chi}\\
\Gamma \vdash \vp\to \psi & \text{implies} &   \Gamma \vdash (\chi\to\vp)\to (\chi\to\psi),
\label{Gamma-vp-to-psi-imp-Gamma-chi-to-vp-chi-to-psi} \\
\Gamma\vdash \vp\to\psi \,\, \text{and}  \,\,\Gamma\vdash\chi\to\gamma & \text{imply} & 
\Gamma\vdash\vp\si\chi\to \psi\si\gamma, 
\label{psi-to-vp-chi-to-gamma-vp-si-chi-to-psi-si-gamma}\\
\Gamma\vdash \vp\to\psi \,\, \text{and}  \,\,\Gamma\vdash\chi\to\gamma & \text{imply} & \Gamma\vdash\vp\sau\chi\to \psi\sau\gamma,
 \label{vp-to-psi-and-chi-to-gamma-implies-sau}\\
\Gamma \vdash \vp \si \psi \to \chi  & \text{imply} &   \Gamma \vdash \psi \si \vp \to \chi,
\label{vp-si-psi-implies-chi-perm}\\
\Gamma \vdash \vp \to (\psi \to \chi) & \text{imply} &   \Gamma \vdash \psi \si \vp \to \chi,
\label{vp-implies-psi-implies-chi-perm}\\
\Gamma \vdash \vp \to (\psi \to \chi) & \text{imply} &   \Gamma \vdash \psi \to (\vp \to \chi),
\label{vp-to-psi-to-chi--psi-to-vp-to-chi}\\
\Gamma \vdash \vp\to \psi  & \text{implies} & \Gamma \vdash \vp \si \chi \to \psi \si\chi \\
&& \text{~and~} \Gamma \vdash  \chi \si \vp \to  \chi \si \psi,
\label{Gamma-vp-to-psi-imp-Gamma-vp-si-chi-to-psi-si-chi} \\
\Gamma \vdash \vp\to\psi & \text{implies} &    \Gamma \vdash (\psi\to\chi) \to (\vp\to\chi), 
\label{vp-to-psi-imp-psi-to-chi-tto-vp-to-chi}\\
\Gamma \vdash \vp \to \psi & \text{implies} &   \Gamma \vdash \vp \to \vp\si\psi \text{~and~} 
\Gamma \vdash \vp \to \psi\si\vp, 
\label{vp-vp-si-psi-implies-two}\\
\Gamma \vdash \vp\to\psi \,\, \text{and}  \,\,\Gamma \vdash\vp\to(\psi\to\chi) & \text{imply} &   
\Gamma \vdash\vp\to\chi,
\label{vp-to-psi-and-vp-to-psi-to-chi-implies-vp-to-chi}\\
\Gamma \vdash \vp\to(\psi\to\chi) \,\, \text{and}  \,\,\Gamma \vdash \chi\to\gamma & \text{implies} & 
\Gamma \vdash \vp\to(\psi\to\gamma).
\label{vp-tto-psi-to-chi-and-chi-togamma-imp-vp-tto-psi-to-gamma}
\eea
\end{lemma}
\solution{\,\\[1mm]
\eqref{vp-implies-psi-to-vp}:
Immediate from \eqref{vp-to-psi-to-vp} by \prule{modus ponens}.\\

\eqref{Gamma-vp-psi-vp-si-psi}: $(\Ra)$: \\[1mm]
\begin{tabular}{lll}
(1) & $\Gamma \vdash \vp$  & (Assumption)\\[1mm]
(2) & $\Gamma \vdash \psi$  & (Assumption)\\[1mm]
(3) & $\Gamma \vdash \vp \si \psi \to \vp \si \psi$  & \eqref{vp-to-vp}\\[1mm]
(4) & $\Gamma \vdash \vp \to (\psi \to \vp \si \psi)$ & \prule{exportation}: (3)\\[1mm]
(5) & $\Gamma \vdash \psi \to \vp \si \psi$ & \prule{modus ponens}: (1), (4)\\[1mm]
(6) & $\Gamma \vdash  \vp \si \psi$ & \prule{modus ponens}: (2), (5)\\[2mm]
\end{tabular}

$(\La)$: \\[1mm]
\begin{tabular}{lll}
(1) & $\Gamma \vdash  \vp \si \psi$  & (Assumption)\\[1mm]
(2) & $\Gamma \vdash  \vp \si \psi \to \vp$  & \prule{weakening}\\[1mm]
(3) & $\Gamma \vdash \vp$  & \prule{modus ponens}: (1), (2)\\[1mm]
(4) & $\Gamma \vdash  \vp \si \psi$  & (Assumption)\\[1mm]
(5) & $\Gamma \vdash  \vp \si \psi \to \psi$  & \eqref{weak-si-2}\\[1mm]
(6) & $\Gamma \vdash \psi$  & \prule{modus ponens}: (4), (5)\\[2mm]
\end{tabular}

\eqref{expansion-2}:\\[1mm]
\begin{tabular}{lll}
(1) & $\Gamma\vdash\vp\to\psi$ & (Assumption)\\[1mm]
(2) & $\Gamma\vdash\chi\sau\vp\to\chi\sau\psi$ & \prule{expansion}: (1) \\[1mm]
(3) & $\Gamma\vdash \vp\sau\chi \to \chi\sau\vp$ & \prule{permutation} \\[1mm]
(4) & $\Gamma\vdash \vp\sau\chi \to \chi\sau\psi$ & \prule{syllogism}: (3), (2) \\[1mm]
(5) & $\Gamma\vdash \chi\sau\psi \to \psi\sau\chi$ & \prule{permutation} \\[1mm]
(6) & $\Gamma\vdash \vp\sau\chi \to \psi\sau\chi$ & \prule{syllogism}: (4), (5) \\[2mm]
\end{tabular}

\eqref{vp-psi-vp-to-psi-implies two}:\\[1mm]
\begin{tabular}{lll}
(1) & $\Gamma\vdash\vp\to\psi$ & (Assumption)\\[1mm]
(2) & $\Gamma\vdash \vp\sau\psi \to\psi\sau\psi$ & \eqref{expansion-2}: (1) \\[1mm]
(3) & $\Gamma\vdash\psi\sau\psi\to\psi$ & \prule{contraction} \\[1mm]
(4) & $\Gamma\vdash\vp\sau\psi\to\psi$ & \prule{syllogism}: (2), (3)\\[2mm]
\end{tabular}

and \\[2mm]
\begin{tabular}{lll}
(1) & $\Gamma \vdash \vp \to \psi$  & (Assumption)\\[1mm]
(2) & $\Gamma\vdash \psi\sau\vp \to\psi\sau\psi$ & \prule{expansion}: (1) \\[1mm]
(3) & $\Gamma\vdash\psi\sau\psi\to\psi$ & \prule{contraction} \\[1mm]
(4) & $\Gamma\vdash\psi\sau\vp\to\psi$ & \prule{syllogism}: (2), (3)\\[2mm]
\end{tabular}

\eqref{vp-to-psi-imp-vp-si-chi-to-psi}:\\[1mm]
\begin{tabular}{lll}
(1) & $\Gamma \vdash \vp \to \psi$  & (Assumption)\\[1mm]
(2) & $\Gamma \vdash \vp\si\chi\to \vp$  & \prule{weakening}\\[1mm]
(3) & $\Gamma \vdash  \vp\si\chi \to \psi$ & \prule{syllogism}: (2), (1)\\[2mm]
\end{tabular}

and \\[2mm]
\begin{tabular}{lll}
(1) & $\Gamma \vdash \vp \to \psi$  & (Assumption)\\[1mm]
(2) & $\Gamma \vdash \chi\si\vp\to \vp$  & \eqref{weak-si-2}\\[1mm]
(3) & $\Gamma \vdash \chi\si\vp \to \psi$ & \prule{syllogism}: (2), (1)\\[2mm]
\end{tabular}

\eqref{vp-to-psi-si-vp-to-chi-implies-vp-to-psi-si-chi}: \\[1mm]
$(\Ra)$: \\[1mm]
\begin{tabular}{lll}
(1) & $\Gamma\vdash\psi\si\chi\to\psi\si\chi$ & \eqref{vp-to-vp}\\[1mm]
(2) & $\Gamma\vdash\psi\to(\chi\to\psi\si\chi)$ & \prule{exportation}: (1)\\[1mm]
(3) & $\Gamma\vdash\vp\to\psi$ & (Assumption)\\[1mm]
(4) & $\Gamma\vdash\vp\to(\chi\to\psi\si\chi)$ & \prule{syllogism}: (3), (2)\\[1mm]
(5) & $\Gamma\vdash\chi\si\vp\to\vp\si\chi$ & \prule{permutation}\\[1mm]
(6) & $\Gamma\vdash\vp\si\chi\to\psi\si\chi$ & \prule{importation}: (4)\\[1mm]
(7) & $\Gamma\vdash\chi\si\vp\to\psi\si\chi$ & \prule{syllogism}: (5), (6)\\[1mm]
(8) & $\Gamma\vdash\vp\to\chi$ & (Assumption)\\[1mm]
(9) & $\Gamma\vdash\chi\to(\vp\to\psi\si\chi)$ & \prule{exportation}: (7)\\[1mm]
(10) & $\Gamma\vdash\vp\to(\vp\to\psi\si\chi)$ & \prule{syllogism}: (8), (9)\\[1mm]
(11) & $\Gamma\vdash\vp\to\vp\si\vp$ & \prule{contraction}\\[1mm]
(12) & $\Gamma\vdash\vp\si\vp\to\psi\si\chi$ & \prule{importation}: (10)\\[1mm]
(13) & $\Gamma\vdash\vp\to\psi\si\chi$ & \prule{syllogism}: (11), (12)\\[2mm]
\end{tabular}

$(\La)$: \\[1mm]
\begin{tabular}{lll}
(1) & $\Gamma\vdash\vp\to\psi\si\chi$ & (Assumption)\\[1mm]
(2) & $\Gamma\vdash\psi\si\chi\to\psi$ & \prule{weakening}\\[1mm]
(3) & $\Gamma\vdash\vp\to\psi$ & \prule{syllogism}: (1), (2)\\[1mm]
(4) & $\Gamma\vdash\vp\to\psi\si\chi$ & (Assumption)\\[1mm]
(5) & $\Gamma\vdash\psi\si\chi\to\chi$ & \eqref{weak-si-2}\\[1mm]
(6) & $\Gamma\vdash\vp\to\chi$ & \prule{syllogism}: (4), (5)\\[2mm]
\end{tabular}

\eqref{vp-to-chi-and-psi-to-chi-implies-vp-sau-psi-to-chi}:\\[1mm]
\begin{tabular}{lll}
(1) & $\Gamma\vdash\vp\to\chi$ & (Assumption)\\[1mm]
(2) & $\Gamma\vdash\vp\sau\chi\to\chi$ & \eqref{vp-psi-vp-to-psi-implies two}: (1)\\[1mm]
(3) & $\Gamma\vdash\psi\to\chi$ & (Assumption)\\[1mm]
(4) & $\Gamma\vdash\vp\sau\psi\to\vp\sau\chi$ & \prule{expansion}: (3)\\[1mm]
(5) & $\Gamma\vdash\vp\sau\psi\to\chi$ & \prule{syllogism}: (4), (2)\\[2mm]
\end{tabular}

\eqref{Gamma-vp-to-psi-imp-Gamma-chi-to-vp-chi-to-psi}:\\[1mm]
\begin{tabular}{lll}
(1) & $\Gamma\vdash(\chi\to\vp)\si\chi\to\vp$ & \eqref{MP-axiom-1}\\[1mm]
(2) & $\Gamma\vdash\vp\to\psi$ & (Assumption)\\[1mm]
(3) & $\Gamma\vdash(\chi\to\vp)\si\chi\to\psi$ & \prule{syllogism}: (1), (2)\\[1mm]
(4) & $\Gamma\vdash(\chi\to\vp)\to(\chi\to\psi)$ & \prule{exportation}: (3)\\[2mm]
\end{tabular}

\eqref{psi-to-vp-chi-to-gamma-vp-si-chi-to-psi-si-gamma}:\\[1mm]
\begin{tabular}{lll}
(1) & $\Gamma\vdash \vp\to\psi$ & (Assumption)\\[1mm]
(2) & $\Gamma\vdash \chi\to\gamma$ & (Assumption)\\[1mm]
(3) & $\Gamma\vdash \vp\si\chi\to\psi$ & \eqref{vp-to-psi-imp-vp-si-chi-to-psi}: (1)\\[1mm]
(4) & $\Gamma\vdash \vp\si\chi\to\gamma$ & \eqref{vp-to-psi-imp-vp-si-chi-to-psi}: (2)\\[1mm]
(5) & $\Gamma\vdash\vp\si\chi \to \psi\si\gamma$ & \eqref{vp-to-psi-si-vp-to-chi-implies-vp-to-psi-si-chi}: (3), (4)\\[2mm]
\end{tabular}

\eqref{vp-to-psi-and-chi-to-gamma-implies-sau}:\\[1mm]
\begin{tabular}{lll}
(1) & $\Gamma\vdash \vp\to\psi$ & (Assumption)\\[1mm]
(2) & $\Gamma\vdash \chi\to\gamma$ & (Assumption)\\[1mm]
(3) & $\Gamma\vdash \psi\to \psi \sau\gamma$ & \prule{weakening}\\[1mm]
(4) & $\Gamma\vdash \vp\to \psi \sau\gamma$ & \prule{syllogism}: (1), (3)\\[1mm]
(5) & $\Gamma\vdash \gamma \to \psi \sau\gamma$ & \eqref{weak-sau-2}\\[1mm]
(6) & $\Gamma\vdash \chi\to \psi \sau\gamma$ & \prule{syllogism}: (2), (5)\\[1mm]
(7) & $\Gamma\vdash\vp\sau\chi \to \psi\sau\gamma$ & \eqref{vp-to-chi-and-psi-to-chi-implies-vp-sau-psi-to-chi}: (4), (6)\\[2mm]
\end{tabular}

\eqref{vp-si-psi-implies-chi-perm}:\\[1mm]
\begin{tabular}{lll}
(1) & $\Gamma \vdash \vp \si \psi \to \chi$  & (Assumption)\\[1mm]
(2) & $\Gamma \vdash \psi\si\vp \to \vp \si \psi$  & \prule{permutation}\\[1mm]
(3) & $\Gamma \vdash  \psi \si \vp \to \chi$ & \prule{syllogism}: (2), (1)\\[2mm]
\end{tabular}

\eqref{vp-implies-psi-implies-chi-perm}:\\[1mm]
\begin{tabular}{lll}
(1) & $\Gamma \vdash \vp \to (\psi \to \chi)$  & (Assumption)\\[1mm]
(2) & $\Gamma \vdash \vp\si\psi \to \chi$  & \prule{importation}: (1)\\[1mm]
(3) & $\Gamma \vdash  \psi \si \vp \to \chi$ & \eqref{vp-si-psi-implies-chi-perm}: (2)\\[2mm]
\end{tabular}

\eqref{vp-to-psi-to-chi--psi-to-vp-to-chi}:\\[1mm]
\begin{tabular}{lll}
(1) & $\Gamma \vdash \vp \to (\psi \to \chi)$  & (Assumption)\\[1mm]
(2) & $\Gamma \vdash  \psi \si \vp \to \chi$ & \eqref{vp-implies-psi-implies-chi-perm}: (2)\\[1mm]
(3) & $\Gamma \vdash \psi \to (\vp \to \chi)$ & \prule{exportation}: (2)\\[2mm]
\end{tabular}

\eqref{Gamma-vp-to-psi-imp-Gamma-vp-si-chi-to-psi-si-chi}:\\[1mm]
\begin{tabular}{lll}
(1) & $\Gamma\vdash\vp\to\psi$ & (Assumption)\\[1mm]
(2) & $\Gamma\vdash\vp\si\chi\to\vp$ & \prule{weakening} \\[1mm]
(3) & $\Gamma\vdash\vp\si\chi\to\psi$ & \prule{syllogism}: (2), (1) \\[1mm]
(4) & $\Gamma\vdash\vp\si\chi\to\chi$ & \eqref{weak-si-2} \\[1mm]
(5) & $\Gamma\vdash\vp\si\chi\to\psi\si\chi$ & \eqref{vp-to-psi-si-vp-to-chi-implies-vp-to-psi-si-chi}: (3), (4)  \\[2mm]
\end{tabular}

and\\[2mm]
\begin{tabular}{lll}
(1) & $\Gamma\vdash\vp\to\psi$ & (Assumption)\\[1mm]
(2) & $\Gamma\vdash\chi\si\vp\to\vp$ & \eqref{weak-si-2} \\[1mm]
(3) & $\Gamma\vdash\chi\si\vp\to\psi$ & \prule{syllogism}: (2), (1)  \\[1mm]
(4) & $\Gamma\vdash\chi\si\vp\to\chi$ & \prule{weakening} \\[1mm]
(5) & $\Gamma\vdash\chi\si\vp\to\chi\si\psi$ & \eqref{vp-to-psi-si-vp-to-chi-implies-vp-to-psi-si-chi}: (4), (3)  \\[2mm]
\end{tabular}

\eqref{vp-to-psi-imp-psi-to-chi-tto-vp-to-chi}:\\[1mm]
\begin{tabular}{lll}
(1) & $\Gamma\vdash\vp\to\psi$ & (Assumption)\\[1mm]
(2) & $\Gamma\vdash (\psi\to\chi) \to (\psi\to\chi)$ & \eqref{vp-to-vp}\\[1mm]
(3) & $\Gamma\vdash \vp\si(\psi\to\chi) \to \psi\si(\psi\to\chi)$ & 
\eqref{psi-to-vp-chi-to-gamma-vp-si-chi-to-psi-si-gamma}: (1), (2)\\[1mm]
(4) & $\Gamma\vdash \psi\si(\psi\to\chi)\to \chi$ & \eqref{MP-axiom-2} \\[1mm]
(5) & $\Gamma\vdash \vp\si(\psi\to\chi) \to \chi$  & \prule{syllogism}: (3), (4) \\[1mm]
(6) & $\Gamma\vdash (\psi\to\chi)\si \vp \to \chi$  & \eqref{vp-si-psi-implies-chi-perm}: (5) \\[1mm]
(7) & $\Gamma\vdash (\psi\to\chi) \to (\vp \to \chi)$  & \prule{exportation}: (6) \\[2mm]
\end{tabular}

\eqref{vp-vp-si-psi-implies-two}:\\[1mm]
\begin{tabular}{lll}
(1) & $\Gamma\vdash\vp\to\vp$ & \eqref{vp-to-vp}\\[1mm]
(2) & $\Gamma\vdash\vp\to\psi$ & (Assumption)\\[1mm]
(3) & $\Gamma\vdash\vp\to\vp\si\psi$ & \eqref{vp-to-psi-si-vp-to-chi-implies-vp-to-psi-si-chi}: (1), (2)\\[2mm]
\end{tabular}

and\\[2mm]
\begin{tabular}{lll}
(1) & $\Gamma\vdash\vp\to\psi$ & (Assumption)\\[1mm]
(2) & $\Gamma\vdash\vp\to\vp$ & \eqref{vp-to-vp}\\[1mm]
(3) & $\Gamma\vdash\vp\to \psi\si\vp$ & \eqref{vp-to-psi-si-vp-to-chi-implies-vp-to-psi-si-chi}: (2), (1)\\[2mm]
\end{tabular}

\eqref{vp-to-psi-and-vp-to-psi-to-chi-implies-vp-to-chi}:\\[1mm]
\begin{tabular}{lll}
(1) & $\Gamma\vdash\vp\to\psi$ & (Assumption)\\[1mm]
(2) & $\Gamma\vdash\vp\to\vp\si\psi$ & \eqref{vp-vp-si-psi-implies-two}: (1)\\[1mm]
(3) & $\Gamma\vdash \vp\to(\psi\to\chi)$ & (Assumption)\\[1mm]
(4) & $\Gamma\vdash\vp\si\psi\to\chi$ & \prule{importation}: (3) \\[1mm]
(5) & $\Gamma\vdash\vp\to\chi$ & \prule{syllogism}: (2), (4)\\[2mm]
\end{tabular}

\eqref{vp-tto-psi-to-chi-and-chi-togamma-imp-vp-tto-psi-to-gamma}:\\[1mm]
\begin{tabular}{lll}
(1) & $\Gamma \vdash \vp\to(\psi\to\chi)$  & (Assumption) \\[1mm]
(2) & $\Gamma \vdash \chi\to\gamma$  & (Assumption) \\[1mm]
(3) & $\Gamma \vdash \vp\si\psi\to\chi$  & \prule{importation}: (1) \\[1mm]
(4) & $\Gamma \vdash \vp\si\psi\to\gamma$  & \prule{syllogism}: (3), (2) \\[1mm]
(5) & $\Gamma \vdash \vp\to (\psi\to\gamma)$  & \prule{exportation}: (4)
\end{tabular}

\,
}

\begin{lemma}
\begin{align}
\vdash & \, (\vp\to(\psi\to\chi))\to(\psi\to(\vp\to\chi)), \label{vp-to-psi-to-chi-to-psi-to-vp-to-chi}\\
\vdash & \, (\vp\si(\vp\to\psi)\to \chi)\to (\vp\si \psi \to\chi), 
\label{vp-si-vp-to-psi-chi-implies}\\
\vdash & \,(\vp\si\psi\to\chi)\to(\vp\to(\psi\to\chi)), \label{vp-si-psi-to-chi-to-vp-to-psi-to-chi}\\
\vdash & \, \big((\vp\to(\psi\to\chi))\si(\vp\to\psi)\big)\si\vp\to\chi, \label{vp-to-psi-to-chi-si-vp-to-psi-si-vp-to-chi}\\
\vdash & \, (\vp\to(\psi\to\chi))\to ((\vp\to\psi)\to(\vp\to\chi)), \label{vp-to-psi-to-chi-to-vp-to-psi-to-vp-to-chi}\\
\vdash & \,(\vp\to\psi)\to((\psi\to\chi)\to(\vp\to\chi)), \label{vp-psi-chi-tranzitivity-ra}\\
\vdash & \,(\vp\to\psi)\to ((\chi\to\vp)\to (\chi\to\psi)), \label{vp-to-psi-tto-chi-to-vp-chi-to-psi}\\
\vdash & \, (\vp\to\psi)\to(\chi\sau\vp\to\chi\sau\psi). \label{expansion-thm}
\end{align}
\end{lemma}
\solution{\,\\[1mm]
\eqref{vp-to-psi-to-chi-to-psi-to-vp-to-chi}: Let $\gamma:=((\vp\to(\psi\to\chi))\si\psi)\si\vp$.\\[1mm]
\begin{tabular}{lll}
(1) & $\vdash \gamma \to (\vp\to(\psi\to\chi))$ & \eqref{weak-si-3-l}\\[1mm]
(2) & $\vdash \gamma \to \vp $ & \eqref{weak-si-2} \\[1mm]
(3) & $\vdash \gamma \to (\psi\to\chi)$ & \eqref{vp-to-psi-and-vp-to-psi-to-chi-implies-vp-to-chi}: (2), (1)\\[1mm]
(4) & $\vdash \gamma \to \psi$ & \eqref{weak-si-3-r}\\[1mm]
(5) & $\vdash \gamma \to \chi$ & \eqref{vp-to-psi-and-vp-to-psi-to-chi-implies-vp-to-chi}: (4), (3)\\[1mm]
(6) & $\vdash ((\vp\to(\psi\to\chi))\si\psi)\to (\vp\to\chi)$ & \prule{exportation}: (5)\\[1mm]
(7) & $\vdash (\vp\to(\psi\to\chi))\to (\psi\to (\vp\to\chi))$ & \prule{exportation}: (6)\\[2mm]
\end{tabular}

\eqref{vp-si-vp-to-psi-chi-implies}:\\[1mm]
\begin{tabular}{lll}
(1) & $\vdash \vp\to\vp$ & \eqref{vp-to-vp}\\[1mm]
(2) & $\vdash \psi \to (\vp\to\psi)$ & \eqref{vp-to-psi-to-vp}\\[1mm]
(3) & $\vdash \vp\si \psi \to \vp\si(\vp\to\psi)$ & \eqref{psi-to-vp-chi-to-gamma-vp-si-chi-to-psi-si-gamma}: (1), (2)\\[1mm]
(4) & $\vdash (\vp\si(\vp\to\psi)\to \chi)\to (\vp\si \psi \to\chi)$ & 
\eqref{vp-to-psi-imp-psi-to-chi-tto-vp-to-chi}: (3)\\[2mm]
\end{tabular}

\eqref{vp-si-psi-to-chi-to-vp-to-psi-to-chi}: Let $\gamma:=((\vp\si\psi\to\chi)\si\vp)\si\psi$. \\[1mm]
\begin{tabular}{lll}
(1) & $\vdash\gamma\to(\vp\si\psi\to\chi)$ & \eqref{weak-si-3-l}\\[1mm]
(2) & $\vdash\gamma\to \vp $ & \eqref{weak-si-3-r}\\[1mm]
(3) & $\vdash\gamma\to \psi $ & \eqref{weak-si-2}\\[1mm]
(4) & $\vdash\gamma\to \vp\si\psi $ &  \eqref{vp-to-psi-si-vp-to-chi-implies-vp-to-psi-si-chi}: (2), (3)\\[1mm]
(5) & $\vdash\gamma\to \chi$ & \eqref{vp-to-psi-and-vp-to-psi-to-chi-implies-vp-to-chi}: (4), (1)\\[1mm]
(6) & $\vdash ((\vp\si\psi\to\chi)\si\vp)\to (\psi\to\chi)$ & \prule{exportation}: (5)\\[1mm]
(7) & $\vdash (\vp\si\psi\to\chi)\to (\vp\to (\psi\to\chi))$ &  \prule{exportation}: (6)\\[2mm]
\end{tabular}

\eqref{vp-to-psi-to-chi-si-vp-to-psi-si-vp-to-chi}: Let $\pi:=\big((\vp\to(\psi\to\chi))\si(\vp\to\psi)\big)\si\vp$.\\[1mm]
\begin{tabular}{lll}
(1) & $\vdash \pi\to\vp$ & \eqref{weak-si-2} \\[1mm]
(2) & $\vdash \pi\to(\vp\to\psi)$ & \eqref{weak-si-3-r}\\[1mm]
(3) & $\vdash \pi\to(\vp\to(\psi\to\chi))$ &  \eqref{weak-si-3-l}\\[1mm]
(4) & $\vdash \pi\to\psi$ & \eqref{vp-to-psi-and-vp-to-psi-to-chi-implies-vp-to-chi}: (1), (2)\\[1mm]
(5) & $\vdash \pi\to(\psi\to\chi)$ & \eqref{vp-to-psi-and-vp-to-psi-to-chi-implies-vp-to-chi}: (1), (3)\\[1mm]
(6) & $\vdash \pi\to\chi$ & \eqref{vp-to-psi-and-vp-to-psi-to-chi-implies-vp-to-chi}: (4), (5)\\[2mm]
\end{tabular}

\eqref{vp-to-psi-to-chi-to-vp-to-psi-to-vp-to-chi}: Let $\pi:=\big((\vp\to(\psi\to\chi))\si(\vp\to\psi)\big)\si\vp$.\\[1mm]
\begin{tabular}{lll}
(1) & $\vdash \pi\to\chi$ & \eqref{vp-to-psi-to-chi-si-vp-to-psi-si-vp-to-chi}\\[1mm]
(2) & $\vdash ((\vp\to(\psi\to\chi))\si(\vp\to\psi)\big) \to (\varphi\to\chi)$ & \prule{exportation}: (1)\\[1mm]
(3) & $\vdash (\vp\to(\psi\to\chi))\to((\vp\to\psi)\to(\vp\to\chi))$ & \prule{exportation}: (2)\\[2mm]
\end{tabular}

\eqref{vp-psi-chi-tranzitivity-ra}:  Let $\gamma:=((\vp\to\psi)\si(\psi\to\chi))\si \vp$.\\[1mm]
\begin{tabular}{lll}
(1) & $\vdash\gamma\to(\vp\to\psi)$ & \eqref{weak-si-3-l}\\[1mm]
(2) & $\vdash\gamma\to\vp$ & \eqref{weak-si-2}\\[1mm]
(3) & $\vdash\gamma\to\psi$ & \eqref{vp-to-psi-and-vp-to-psi-to-chi-implies-vp-to-chi}: (2), (1)\\[1mm]
(4) & $\vdash\gamma\to(\psi\to\chi)$ & \eqref{weak-si-3-r}\\[1mm]
(5) & $\vdash\gamma\to\chi$ & \eqref{vp-to-psi-and-vp-to-psi-to-chi-implies-vp-to-chi}: (3), (4)\\[1mm]
(6) & $\vdash ((\vp\to\psi)\si(\psi\to\chi))\to(\vp\to\chi)$ & \prule{exportation}: (5)\\[1mm]
(7) & $\vdash (\vp\to\psi)\to((\psi\to\chi)\to(\vp\to\chi))$ & \prule{exportation}: (6)\\[2mm]
\end{tabular}

\eqref{vp-to-psi-tto-chi-to-vp-chi-to-psi}:\\[1mm]
\begin{tabular}{lll}
(1) & $\vdash(\chi\to\vp)\to((\vp\to\psi)\to(\chi\to\psi))$ & \eqref{vp-psi-chi-tranzitivity-ra}\\[1mm]
(2) & $\vdash(\vp\to\psi)\to((\chi\to\vp)\to(\chi\to\psi))$ & \eqref{vp-si-vp-to-psi-chi-implies}: (1)\\[2mm]
\end{tabular}

\eqref{expansion-thm}: \\[1mm]
\begin{tabular}{lll}
(1) & $\vdash\vp\to((\vp\to\psi)\to\psi)$ & \eqref{MP-axiom-3}\\[1mm]
(2) & $\vdash\psi\to\chi\sau\psi$ & \eqref{weak-sau-2}\\[1mm]
(3) & $\vdash\vp\to((\vp\to\psi)\to\chi\sau\psi)$ & \eqref{vp-tto-psi-to-chi-and-chi-togamma-imp-vp-tto-psi-to-gamma}: (1), (2)\\[1mm]
(4) & $\vdash\chi\to((\vp\to\psi)\to\chi)$ & \eqref{vp-to-psi-to-vp}\\[1mm]
(5) & $\vdash\chi\to\chi\sau\psi$ & \prule{weakening}\\[1mm]
(6) & $\vdash\chi\to((\vp\to\psi)\to\chi\sau\psi)$ & \prule{syllogism}: (4), (5)\\[1mm]
(7) & $\vdash\chi\sau\vp\to((\vp\to\psi)\to\chi\sau\psi)$ & \eqref{vp-to-chi-and-psi-to-chi-implies-vp-sau-psi-to-chi}: (6), (3)\\[1mm]
(8) & $\vdash(\vp\to\psi)\to(\chi\sau\vp\to\chi\sau\psi)$ & \eqref{vp-to-psi-to-chi--psi-to-vp-to-chi}: (7)
\end{tabular}

\,
}

\begin{lemma}
\begin{align}
\vdash & \, (\vp\si\psi)\si\chi\to\vp\si(\psi\si\chi), \label{si-associativity-1}\\
\vdash & \, \vp\si(\psi\si\chi)\to(\vp\si\psi)\si\chi, \label{si-associativity-2}\\
\vdash & \, (\vp\sau\psi)\sau \chi\to\vp\sau(\psi\sau\chi), \label{sau-associativity-1}\\
\vdash & \, \vp\sau(\psi\sau\chi)\to(\vp\sau\psi)\sau\chi, \label{sau-associativity-2}\\
\vdash & \, (\vp\si\psi)\sau(\vp\si\chi)\to\vp\si(\psi\sau\chi), \label{distrib-si-sau-1}\\
\vdash & \, \vp\si(\psi\sau\chi)\to(\vp\si\psi)\sau(\vp\si\chi), \label{distrib-si-sau-2} \\
\vdash & \, \vp\sau(\psi\si\chi) \to (\vp\sau\psi)\si(\vp\sau\chi), \label{distrib-sau-si-1}\\
\vdash & \, (\vp\sau\psi)\si(\vp\sau\chi) \to \vp\sau(\psi\si\chi). \label{distrib-sau-si-2}
\end{align}
\end{lemma}
\solution{\,\\
\eqref{si-associativity-1}:\\[1mm]
\begin{tabular}{lll}
(1) & $\vdash(\vp\si\psi)\si\chi\to\vp $ & \eqref{weak-si-3-l}\\[1mm]
(2) & $\vdash(\vp\si\psi)\si\chi\to\psi $& \eqref{weak-si-3-r}\\[1mm]
(3) & $\vdash(\vp\si\psi)\si\chi\to\chi $& \eqref{weak-si-2}\\[1mm]
(4) & $\vdash(\vp\si\psi)\si\chi\to \psi\si \chi $ & \eqref{vp-to-psi-si-vp-to-chi-implies-vp-to-psi-si-chi}: (2), (3)\\[1mm]
(5) & $\vdash(\vp\si\psi)\si\chi\to \vp\si(\psi\si \chi)$ & \eqref{vp-to-psi-si-vp-to-chi-implies-vp-to-psi-si-chi}: (1), (4)\\[2mm]
\end{tabular}

\eqref{si-associativity-2}:\\[1mm]
\begin{tabular}{lll}
(1) & $\vdash\vp\si(\psi\si\chi)\to\vp $ & \prule{weakening}\\[1mm]
(2) & $\vdash\vp\si(\psi\si\chi)\to \psi$ & \eqref{weak-si-3-l1}\\[1mm]
(3) & $\vdash\vp\si(\psi\si\chi)\to\chi$ & \eqref{weak-si-3-r1}\\[1mm]
(4) & $\vdash\vp\si(\psi\si\chi)\to\vp\si\psi$ & \eqref{vp-to-psi-si-vp-to-chi-implies-vp-to-psi-si-chi}: (1), (2)\\[1mm]
(5) & $\vdash(\vp\si\psi)\si\chi\to (\vp\si\psi)\si \chi$ & \eqref{vp-to-psi-si-vp-to-chi-implies-vp-to-psi-si-chi}: (3), (4)\\[2mm]
\end{tabular}

\eqref{sau-associativity-1}:\\[1mm]
\begin{tabular}{lll}
(1) & $\vdash \vp \to \vp\sau(\psi\sau\chi)$ & \prule{weakening}\\[1mm]
(2) & $\vdash \psi\to \vp\sau(\psi\sau\chi)$ & \eqref{weak-sau-3-l1} \\[1mm]
(3) & $\vdash \chi\to \vp\sau(\psi\sau\chi)$ & \eqref{weak-sau-3-r1} \\[1mm]
(4) & $\vdash \vp\sau \psi \to \vp\sau(\psi\sau\chi)$ & 
\eqref{vp-to-chi-and-psi-to-chi-implies-vp-sau-psi-to-chi}: (1), (2)\\[1mm]
(5) & $\vdash (\vp\sau\psi)\sau \chi \to \vp\sau(\psi\sau\chi)$ & 
\eqref{vp-to-chi-and-psi-to-chi-implies-vp-sau-psi-to-chi}: (4), (3)\\[2mm]
\end{tabular}

\eqref{sau-associativity-2}:\\[1mm]
\begin{tabular}{lll}
(1) & $\vdash \chi \to (\vp\sau\psi)\sau\chi$ & \eqref{weak-sau-2}\\[1mm]
(2) & $\vdash \vp\to (\vp\sau\psi)\sau\chi$ & \eqref{weak-sau-3-l2} \\[1mm]
(3) & $\vdash \psi\to (\vp\sau\psi)\sau\chi$ & \eqref{weak-sau-3-r2} \\[1mm]
(4) & $\vdash \psi\sau \chi \to (\vp\sau\psi)\sau\chi$ & 
\eqref{vp-to-chi-and-psi-to-chi-implies-vp-sau-psi-to-chi}: (3), (1)\\[1mm]
(5) & $\vdash \vp\sau(\psi\sau\chi)\to (\vp\sau\psi)\sau\chi$ & 
\eqref{vp-to-chi-and-psi-to-chi-implies-vp-sau-psi-to-chi}: (2), (4)\\[2mm]
\end{tabular}

\eqref{distrib-si-sau-1}:\\[1mm]
\begin{tabular}{lll}
(1) & $\vdash\vp\si\psi\to\vp$ & \prule{weakening}\\[1mm]
(2) & $\vdash\vp\si\psi\to\psi$ & \eqref{weak-si-2}\\[1mm]
(3) & $\vdash\psi\to\psi\sau\chi$ & \prule{weakening}\\[1mm]
(4) & $\vdash\vp\si\psi\to\psi\sau\chi$ & \prule{syllogism}: (2), (3)\\[1mm]
(5) & $\vdash\vp\si\psi\to\vp\si(\psi\sau\chi)$ & \eqref{vp-to-psi-si-vp-to-chi-implies-vp-to-psi-si-chi}: (1), (4)\\[1mm]
(6) & $\vdash\vp\si\chi\to\vp$ & \prule{weakening}\\[1mm]
(7) & $\vdash\vp\si\chi\to\chi$ & \eqref{weak-si-2}\\[1mm]
(8) & $\vdash\chi\to\psi\sau\chi$ & \eqref{weak-sau-2}\\[1mm]
(9) & $\vdash\vp\si\chi\to\psi\sau\chi$ & \prule{syllogism}: (7), (8)\\[1mm]
(10) & $\vdash\vp\si\chi\to\vp\si(\psi\sau\chi)$ & \eqref{vp-to-psi-si-vp-to-chi-implies-vp-to-psi-si-chi}: (6), (9)\\[1mm]
(11) & $\vdash(\vp\si\psi)\sau(\vp\si\chi)\to\vp\si(\psi\sau\chi)$ & 
\eqref{vp-to-chi-and-psi-to-chi-implies-vp-sau-psi-to-chi}: (5), (10)\\[2mm]
\end{tabular}

\eqref{distrib-si-sau-2}:\\[1mm]
\begin{tabular}{lll}
(1) & $\vdash \vp\si\psi \to (\vp\si\psi)\sau(\vp\si\chi)$ & \prule{weakening}\\[1mm]
(2) & $\vdash \psi\si\vp \to (\vp\si\psi)\sau(\vp\si\chi)$ & \eqref{vp-si-psi-implies-chi-perm}: (1)\\[1mm]
(3) & $\vdash \psi \to (\vp \to (\vp\si\psi)\sau(\vp\si\chi))$ & \prule{exportation}: (2)\\[1mm]
(4) & $\vdash \vp\si\chi \to (\vp\si\psi)\sau(\vp\si\chi)$ &  \eqref{weak-sau-2}\\[1mm]
(5) & $\vdash \chi\si\vp \to (\vp\si\psi)\sau(\vp\si\chi)$ & \eqref{vp-si-psi-implies-chi-perm}: (4)\\[1mm]
(6) & $\vdash \chi \to (\vp \to (\vp\si\psi)\sau(\vp\si\chi))$ & \prule{exportation}: (5)\\[1mm]
(7) & $\vdash \psi\sau \chi \to (\vp \to (\vp\si\psi)\sau(\vp\si\chi))$ & \eqref{vp-to-chi-and-psi-to-chi-implies-vp-sau-psi-to-chi}: (3), (6)\\[1mm]
(8) & $\vdash (\psi\sau \chi) \si \vp \to (\vp\si\psi)\sau(\vp\si\chi)$ & \prule{importation}: (7)\\[1mm]
(9) & $\vdash \vp \si (\psi\sau \chi) \to (\vp\si\psi)\sau(\vp\si\chi)$ & 
\eqref{vp-si-psi-implies-chi-perm}: (8)\\[2mm]
\end{tabular}

\eqref{distrib-sau-si-1}:\\[1mm]
\begin{tabular}{lll}
(1) & $\vdash\vp\to\vp\sau\psi$ & \prule{weakening}\\[1mm]
(2) & $\vdash\vp\to\vp\sau\chi$ & \prule{weakening}\\[1mm]
(3) & $\vdash\vp\to(\vp\sau\psi)\si(\vp\sau\chi)$ & \eqref{vp-to-psi-si-vp-to-chi-implies-vp-to-psi-si-chi}: (1), (2)\\[1mm]
(4) & $\vdash\psi\si\chi\to\psi$ & \prule{weakening}\\[1mm]
(5) & $\vdash\psi\to\vp\sau\psi$ & \eqref{weak-sau-2}\\[1mm]
(6) & $\vdash\psi\si\chi\to\vp\sau\psi$ & \prule{syllogism}: (4), (5)\\[1mm]
(7) & $\vdash\psi\si\chi\to\chi$ & \eqref{weak-si-2}\\[1mm]
(8) & $\vdash\chi\to\vp\sau\chi$ & \eqref{weak-sau-2}\\[1mm]
(9) & $\vdash\psi\si\chi\to\vp\sau\chi$ & \prule{syllogism}: (7), (8)\\[1mm]
(10) & $\vdash\psi\si\chi\to(\vp\sau\psi)\si(\vp\sau\chi)$ & \eqref{vp-to-psi-si-vp-to-chi-implies-vp-to-psi-si-chi}: (6), (9)\\[1mm]
(11) & $\vdash\vp\sau(\psi\si\chi)\to(\vp\sau\psi)\si(\vp\sau\chi)$ & 
\eqref{vp-to-chi-and-psi-to-chi-implies-vp-sau-psi-to-chi}: (3), (10)\\[2mm]
\end{tabular}

\eqref{distrib-sau-si-2}: Let us denote $\delta:=(\vp\sau\psi)\si(\vp\sau\chi)$. \\[1mm]
\begin{tabular}{lll}
(1) & $\vdash \delta \to \vp\sau\psi$ & \prule{weakening}\\[1mm]
(2) & $\vdash \delta\to \vp\sau\chi$ & \eqref{weak-si-2} \\[1mm]
(3) & $\vdash \vp \to \vp\sau(\psi\si\chi)$ & \prule{weakening}\\[1mm]
(4) & $\vdash \vp\si\psi \to \vp$ & \prule{weakening}\\[1mm]
(5) & $\vdash \vp\si\psi \to \vp\sau(\psi\si\chi)$ & \prule{syllogism}: (4), (3)\\[1mm]
(6) & $\vdash \vp \to (\psi \to \vp\sau(\psi\si\chi))$ & \prule{exportation}: (5)\\[1mm]
(7) & $\vdash \psi\si\chi \to \vp\sau(\psi\si\chi)$ & \eqref{weak-sau-2}\\[1mm]
(8) & $\vdash \chi \si \psi \to \vp\sau(\psi\si\chi)$ & \eqref{vp-si-psi-implies-chi-perm}: (7)\\[1mm]
(9) & $\vdash \chi \to (\psi \to \vp\sau(\psi\si\chi))$ & \prule{exportation}: (8)\\[1mm]
(10) & $\vdash \vp\sau \chi \to (\psi \to \vp\sau(\psi\si\chi))$ & 
\eqref{vp-to-chi-and-psi-to-chi-implies-vp-sau-psi-to-chi}: (6), (9) \\[1mm]
(11) & $\vdash \delta \to (\psi \to \vp\sau(\psi\si\chi))$ & \prule{syllogism}: (2), (10)\\[1mm]
(12) & $\vdash \psi\to (\delta  \to \vp\sau(\psi\si\chi))$ & \eqref{vp-to-psi-to-chi--psi-to-vp-to-chi}: (11)\\[1mm]
(13) &  $\vdash \vp \to \vp\sau(\psi\si\chi)$ & \prule{weakening}\\[1mm]
(14) &  $\vdash \vp\si \delta \to \vp$ & \prule{weakening}\\[1mm]
(15) &  $\vdash \vp\si \delta \to \vp\sau(\psi\si\chi)$ & \prule{syllogism}: (14), (13)\\[1mm]
(16) &  $\vdash \vp \to (\delta \to \vp\sau(\psi\si\chi))$ & \prule{exportation}: (15)\\[1mm]
(17) &  $\vdash \vp\sau\psi \to (\delta \to \vp\sau(\psi\si\chi))$ & 
\eqref{vp-to-chi-and-psi-to-chi-implies-vp-sau-psi-to-chi}: (16), (12) \\[1mm]
(18) &  $\vdash \delta \to (\delta \to \vp\sau(\psi\si\chi))$ & \prule{syllogism}: (1), (17)\\[1mm]
(19) &  $\vdash \delta \si \delta \to \vp\sau(\psi\si\chi)$ &\prule{importation}: (18)\\[1mm]
(20) &  $\vdash \delta \to \delta \si \delta$ & \prule{contraction}\\[1mm]
(21) &  $\vdash \delta \to\vp\sau(\psi\si\chi)$ & \prule{syllogism}: (20), (19) 
\end{tabular}

\,
}

\begin{lemma}
\bea
\Gamma\vdash (\vp\to\psi)\to(\vp\to\chi)  &  \text{implies} & \Gamma\vdash\vp\to(\psi\to\chi), 
\label{Gamma-vp-to-psi-vp-to-chi-tto-vp-to-psi-to-chi}\\
\Gamma\vdash\vp\to(\psi\to\chi)\,\, \text{and}  \,\,\Gamma\vdash\vp\to(\chi\to\gamma) &  \text{imply} &  \Gamma\vdash\vp\to(\psi\to\gamma), 
\label{vp-to-psi-to-chi-vp-to-chi-to-gamma-implies-vp-to-psi-to-gamma}\\
\Gamma\vdash\vp\to\psi\,\, \text{and}  \,\,\Gamma\vdash\chi\to\gamma &  \text{imply} &
\Gamma\vdash(\vp\to\chi)\to(\psi\to\gamma), 
\label{psi-to-vp-chi-to-gamma-vp-to-chi-to-psi-to-gamma}\\
\Gamma\vdash\vp\to(\psi\to\chi)\,\, \text{and}  \,\,\Gamma\vdash\vp\to(\gamma\to\chi) & 
\text{implies} & \Gamma\vdash\vp\to(\psi\sau\gamma\to\chi),\label{vp-to-psi-to-chi-si-vp-to-gamma-to-chi-to-vp-to-psi-sau-gamma-to-chi}\\
\Gamma\vdash\vp\to(\psi\si\chi\to\gamma) & \text{iff} & 
\Gamma\vdash\vp\to(\psi\to(\chi\to\gamma)),
\label{vp-to-psi-si-chi-to-gamma-iff-vp-to-psi-to-chi-to-gamma}\\
\Gamma\vdash \vp\to(\psi\to\chi) & \text{implies} & \Gamma\vdash\vp\to(\gamma\sau\psi\to\gamma\sau\chi).\label{expansion-extra-premise}
\eea
\end{lemma}
\solution{\,\\[1mm]
\eqref{Gamma-vp-to-psi-vp-to-chi-tto-vp-to-psi-to-chi}:\\[1mm]
\begin{tabular}{lll}
(1) & $\Gamma\vdash (\vp\to\psi)\to(\vp\to\chi)$ & (Assumption)\\[1mm]
(2) & $\Gamma\vdash (\vp\to\psi)\si\vp\to\chi$ & \prule{importation}\\[1mm]
(3) & $\Gamma\vdash \vp\si(\vp\to\psi)\to\chi$ & \eqref{vp-si-psi-implies-chi-perm}: (2)\\[1mm]
(4) & $\Gamma\vdash (\vp\si(\vp\to\psi)\to \chi)\to (\vp\si \psi \to\chi)$ & \eqref{vp-si-vp-to-psi-chi-implies}\\[1mm]
(5) & $\Gamma\vdash \vp\si \psi \to\chi$ & \prule{modus ponens}: (3), (4)\\[1mm]
(6) & $\Gamma\vdash \vp \to (\psi \to\chi)$ & \prule{exportation}: (5)\\[2mm]
\end{tabular}\\

\eqref{vp-to-psi-to-chi-vp-to-chi-to-gamma-implies-vp-to-psi-to-gamma}:\\[1mm]
\begin{tabular}{lll}
(1) & $\Gamma\vdash\vp\to(\psi\to\chi)$ & (Assumption)\\[1mm]
(2) & $\Gamma\vdash(\vp\to(\psi\to\chi))\to ((\vp\to\psi)\to(\vp\to\chi))$ & \eqref{vp-to-psi-to-chi-to-vp-to-psi-to-vp-to-chi}\\[1mm]
(3) & $\Gamma\vdash(\vp\to\psi)\to(\vp\to\chi)$ & \prule{modus ponens}: (1), (2)\\[1mm]
(4) & $\Gamma\vdash\vp\to(\chi\to\gamma)$ & (Assumption)\\[1mm]
(5) & $\Gamma\vdash(\vp\to(\chi\to\gamma))\to ((\vp\to\chi)\to(\vp\to\gamma))$ & \eqref{vp-to-psi-to-chi-to-vp-to-psi-to-vp-to-chi}\\[1mm]
(6) & $\Gamma\vdash(\vp\to\chi)\to(\vp\to\gamma)$ & \prule{modus ponens}: (4), (5)\\[1mm]
(7) & $\Gamma\vdash(\vp\to\psi)\to(\vp\to\gamma)$ & \prule{syllogism}: (3), (6)\\[1mm]
(8) & $\Gamma\vdash\vp\to(\psi\to\gamma)$ & \eqref{Gamma-vp-to-psi-vp-to-chi-tto-vp-to-psi-to-chi}: (7)\\[2mm]
\end{tabular}

\eqref{psi-to-vp-chi-to-gamma-vp-to-chi-to-psi-to-gamma}:\\[1mm]
\begin{tabular}{lll}
(1) & $\Gamma\vdash\psi\to\vp$ & (Assumption)\\[1mm]
(2) & $\Gamma\vdash((\vp\to\chi)\si\psi)\to(\psi\to\vp)$ & \eqref{vp-implies-psi-to-vp}: (1)\\[1mm]
(3) & $\Gamma\vdash((\vp\to\chi)\si\psi)\to(\vp\to\chi)$ & \prule{weakening}\\[1mm]
(4) & $\Gamma\vdash((\vp\to\chi)\si\psi)\to(\psi\to\chi)$ & \eqref{vp-to-psi-to-chi-vp-to-chi-to-gamma-implies-vp-to-psi-to-gamma}: (2), (3)\\[1mm]
(5) & $\Gamma\vdash((\vp\to\chi)\si\psi)\to\psi$ & \eqref{weak-si-2}\\[1mm]
(6) & $\Gamma\vdash((\vp\to\chi)\si\psi)\to\chi$ & \eqref{vp-to-psi-and-vp-to-psi-to-chi-implies-vp-to-chi}: (5), (4)\\[1mm]
(7) & $\Gamma\vdash\chi\to\gamma$ & (Assumption)\\[1mm]
(8) & $\Gamma\vdash((\vp\to\chi)\si\psi)\to(\chi\to\gamma)$ & \eqref{vp-implies-psi-to-vp}: (7)\\[1mm]
(9) & $\Gamma\vdash((\vp\to\chi)\si\psi)\to\gamma$ & \eqref{vp-to-psi-and-vp-to-psi-to-chi-implies-vp-to-chi}: (6), (8)\\[1mm]
(10) & $\Gamma\vdash(\vp\to\chi)\to(\psi\to\gamma)$ & \prule{exportation}: (9)\\[2mm]
\end{tabular}

\eqref{vp-to-psi-to-chi-si-vp-to-gamma-to-chi-to-vp-to-psi-sau-gamma-to-chi}:\\
\begin{tabular}{lll}
(1) & $\Gamma\vdash\vp\to(\psi\to\chi)$ & (Assumption)\\[1mm]
(2) & $\Gamma\vdash\psi\to(\vp\to\chi)$ & \eqref{vp-to-psi-to-chi--psi-to-vp-to-chi}: (1)\\[1mm]
(3) & $\Gamma\vdash\vp\to(\gamma\to\chi)$ & (Assumption)\\[1mm]
(4) & $\Gamma\vdash\gamma\to(\vp\to\chi)$ & \eqref{vp-to-psi-to-chi--psi-to-vp-to-chi}: (3)\\[1mm]
(5) & $\Gamma\vdash\psi\sau\gamma\to(\vp\to\chi)$ & \eqref{vp-to-chi-and-psi-to-chi-implies-vp-sau-psi-to-chi}: (2), (4)\\[1mm]
(6) & $\Gamma\vdash\vp\to(\psi\sau\gamma\to\chi)$ & \eqref{vp-to-psi-to-chi--psi-to-vp-to-chi}: (5)\\[2mm]
\end{tabular}

\eqref{vp-to-psi-si-chi-to-gamma-iff-vp-to-psi-to-chi-to-gamma}:\\[1mm]
$(\Ra)$ \\[1mm]
\begin{tabular}{lll}
(1) & $\Gamma\vdash\vp\to(\psi\si\chi\to\gamma)$ & (Assumption)\\[1mm]
(2) & $\Gamma\vdash\vp\si(\psi\si\chi)\to\gamma$ & \prule{importation}: (1)\\[1mm]
(3) & $\Gamma\vdash(\vp\si\psi)\si\chi\to\vp\si(\psi\si\chi)$ & \eqref{si-associativity-1}\\[1mm]
(4) & $\Gamma\vdash(\vp\si\psi)\si\chi\to\gamma$ & \prule{syllogism}: (3), (2)\\[1mm]
(5) & $\Gamma\vdash(\vp\si\psi)\to(\chi\to\gamma)$ & \prule{exportation}: (4)\\[1mm]
(6) & $\Gamma\vdash\vp\to(\psi\to(\chi\to\gamma))$ & \prule{exportation}: (5)\\[2mm]
\end{tabular}

$(\La)$ \\[1mm]
\begin{tabular}{lll}
(1) & $\Gamma\vdash\vp\to(\psi\to(\chi\to\gamma))$ & (Assumption)\\[1mm]
(2) & $\Gamma\vdash \vp\si\psi\to(\chi\to\gamma)$ & \prule{importation}: (1)\\[1mm]
(3) & $\Gamma\vdash(\vp\si\psi)\si\chi\to\gamma$ & \prule{importation}: (2)\\[1mm]
(4) & $\Gamma\vdash\vp\si(\psi\si\chi)\to(\vp\si\psi)\si\chi$ & \eqref{si-associativity-2}\\[1mm]
(5) & $\Gamma\vdash\vp\si(\psi\si\chi)\to\gamma$ & \prule{syllogism}: (4), (3)\\[1mm]
(6) & $\Gamma\vdash\vp\to(\psi\si\chi\to\gamma)$ & \prule{exportation}: (5)\\[2mm]
\end{tabular}

\eqref{expansion-extra-premise}:\\[1mm]
\begin{tabular}{lll}
(1) & $\Gamma\vdash\vp\to(\psi\to\chi)$ & (Assumption)\\[1mm]
(2) & $\Gamma\vdash(\psi\to\chi)\to(\gamma\sau\psi\to\gamma\sau\chi)$ & \eqref{expansion-thm}\\[1mm]
(3) & $\Gamma\vdash\vp\to(\gamma\sau\psi\to\gamma\sau\chi)$ & \prule{syllogism}: (1), (2)
\end{tabular}

\,
}

\begin{lemma}
\begin{align}
\vdash & \, \vp\si\neg\vp\to\psi,\label{ex-falso-and}\\
\vdash & \, \vp\to(\neg\vp\to\psi),\label{ex-falso-impl}\\
\vdash & \, \neg\vp \to(\vp\to\psi),\label{ex-falso-impl-1}\\
\vdash & \,\neg\neg\vp\to\vp, \label{negnegvp-vp}\\
\vdash & \,(\neg\vp\to\psi)\to\vp\sau\psi, \label{axiom-or-2}\\
\vdash & \, (\vp\to\psi)\to\neg\vp\sau\psi, \label{disj-impl-1}\\
\vdash & \, (\vp\to\neg \psi) \to (\psi\to\neg\vp), \label{vptonegpsi-to-psitonegvp}\\
\vdash & \,\vp\sau\psi\to(\neg\vp\to\psi), \label{axiom-or-1}\\
\vdash & \, (\vp\to\psi)\to(\neg\psi\to\neg\vp). \label{contraposition}	
\end{align}
\end{lemma}
\solution{\,\\[1mm]
\eqref{ex-falso-and}:\\[1mm]
\begin{tabular}{lll}
(1) & $\vdash\vp\si(\vp\to\bot)\to\bot$ & \eqref{MP-axiom-2}\\[1mm]
(2) & $\vdash\bot\to\psi$ & \prule{exfalso}\\[1mm]
(3) & $\vdash\vp\si(\vp\to\bot)\to\psi$ & \prule{syllogism}: (1), (2)\\[1mm]
(4) & $\vdash\neg\vp\to(\vp\to\bot)$ & \prule{axiom-not-1}\\[1mm]
(5) & $\vdash\vp\si\neg\vp\to\vp\si(\vp\to\bot)$ & \eqref{Gamma-vp-to-psi-imp-Gamma-vp-si-chi-to-psi-si-chi}: (5)\\[1mm]
(6) & $\vdash\vp\si\neg\vp\to\psi$ & \prule{syllogism}: (5), (3)\\[2mm]
\end{tabular}

\eqref{ex-falso-impl}: Apply \prule{exportation} to \eqref{ex-falso-and}.\\[1mm]
\eqref{ex-falso-impl-1}: Apply \eqref{vp-to-psi-to-chi--psi-to-vp-to-chi} to \eqref{ex-falso-impl}.\\[1mm]
\eqref{negnegvp-vp}:\\[1mm]
\begin{tabular}{lll}
(1) & $\vdash\vp\sau\neg\vp$ & \prule{lem}\\[1mm]
(2) & $\vdash\vp\to(\neg\neg\vp\to\vp)$ & \eqref{vp-to-psi-to-vp}\\[1mm]
(3) & $\vdash\neg\vp\si\neg\neg\vp\to\vp$ & \eqref{ex-falso-and}\\[1mm]
(4) & $\vdash\neg\vp\to(\neg\neg\vp\to\vp)$ & \prule{exportation}: (3)\\[1mm]
(5) & $\vdash\vp\sau\neg\vp\to(\neg\neg\vp\to\vp)$ & \eqref{vp-to-chi-and-psi-to-chi-implies-vp-sau-psi-to-chi}: (2), (4)\\[1mm]
(6) & $\vdash\neg\neg\vp\to\vp$ & \prule{modus ponens}: (1), (5)\\[2mm]
\end{tabular}

\eqref{axiom-or-2}:\\[1mm]
\begin{tabular}{lll}
(1) & $\vdash\vp\to\vp\sau\psi$ & \prule{weakening}\\[1mm]
(2) & $\vdash(\neg\vp\to\psi)\to(\vp\to\vp\sau\psi)$ & \eqref{vp-implies-psi-to-vp}: (1)\\[1mm]
(3) & $\vdash\vp\to((\neg\vp\to\psi)\to\vp\sau\psi)$ & \eqref{vp-to-psi-to-chi--psi-to-vp-to-chi}: (2)\\[1mm]
(4) & $\vdash\neg\vp\si(\neg\vp\to\psi)\to\psi$ & \eqref{MP-axiom-2}\\[1mm]
(5) & $\vdash\psi\to\vp\sau\psi$ & \eqref{weak-sau-2}\\[1mm]
(6) & $\vdash\neg\vp\si(\neg\vp\to\psi)\to\vp\sau\psi$ & \prule{syllogism}: (4), (5)\\[1mm]
(7) & $\vdash\neg\vp\to((\neg\vp\to\psi)\to\vp\sau\psi)$ & \prule{exportation}: (6)\\[1mm]
(8) & $\vdash\vp\sau\neg\vp\to((\neg\vp\to\psi)\to\vp\sau\psi)$ & \eqref{vp-to-chi-and-psi-to-chi-implies-vp-sau-psi-to-chi}: (3), (7)\\[1mm]
(9) & $\vdash\vp\sau\neg\vp$ & \prule{lem}\\[1mm]
(10) & $\vdash(\neg\vp\to\psi)\to\vp\sau\psi$ & \prule{modus ponens}: (9), (8)\\[2mm]
\end{tabular}

\eqref{disj-impl-1}:\\[1mm]
\begin{tabular}{lll}
(1) & $\vdash(\neg\neg\vp\to\psi)\to(\neg\vp\sau\psi)$ & \eqref{axiom-or-2}\\[1mm]
(2) & $\vdash\neg\neg\vp\to\vp$ & \eqref{negnegvp-vp}\\[1mm]
(3) & $\vdash(\vp\to\psi)\to(\neg\neg\vp\to\psi)$ & \eqref{vp-to-psi-imp-psi-to-chi-tto-vp-to-chi}: (2)\\[1mm]
(4) & $\vdash(\vp\to\psi)\to(\neg\vp\sau\psi)$ & \prule{syllogism}: (3), (1)\\[2mm]
\end{tabular}

\eqref{vptonegpsi-to-psitonegvp}:\\[1mm]
\begin{tabular}{lll}
(1) & $\vdash\neg\psi\to(\psi\to\bot)$ & \prule{axiom-not-1}\\[1mm]
(2) & $\vdash(\vp\to\neg\psi)\to(\vp\to(\psi\to\bot))$ & \eqref{Gamma-vp-to-psi-imp-Gamma-chi-to-vp-chi-to-psi}: (1)\\[1mm]
(3) & $\vdash(\vp\to(\psi\to\bot))\to(\psi\to(\vp\to\bot))$ & \eqref{vp-to-psi-to-chi-to-psi-to-vp-to-chi}\\[1mm]
(4) & $\vdash(\vp\to\neg\psi)\to(\psi\to(\vp\to\bot))$ & \prule{syllogism}: (2), (3)\\[1mm]
(5) & $\vdash(\vp\to\bot)\to\neg\vp$ & \prule{axiom-not-2}\\[1mm]
(6) & $\vdash(\psi\to(\vp\to\bot))\to(\psi\to\neg\vp)$ & \eqref{Gamma-vp-to-psi-imp-Gamma-chi-to-vp-chi-to-psi}: (5)\\[1mm]
(7) & $\vdash(\vp\to\neg\psi)\to(\psi\to\neg\vp)$ & \prule{syllogism}: (4), (6)\\[2mm] 
\end{tabular}

\eqref{axiom-or-1}:\\[1mm]
\begin{tabular}{lll}
(1) & $\vdash\vp\to(\neg\vp\to\psi)$ & \eqref{ex-falso-impl}\\[1mm]
(2) & $\vdash\psi\to(\neg\vp\to\psi)$ & \eqref{vp-to-psi-to-vp}\\[1mm]
(3) & $\vdash\vp\sau\psi\to(\neg\vp\to\psi)$ & \eqref{vp-to-chi-and-psi-to-chi-implies-vp-sau-psi-to-chi}: (1), (2)\\[2mm]
\end{tabular}

\eqref{contraposition}:\\[1mm]
\begin{tabular}{lll}
(1) & $\vdash((\vp\to\psi)\si\neg\psi)\si\vp\to(\vp\to\psi)$ & \eqref{weak-si-3-l}\\[1mm]
(2) & $\vdash((\vp\to\psi)\si\neg\psi)\si\vp\to\neg\psi$ & \eqref{weak-si-3-r}\\[1mm]
(3) & $\vdash((\vp\to\psi)\si\neg\psi)\si\vp\to\vp$ & \eqref{weak-si-2}\\[1mm]
(4) & $\vdash((\vp\to\psi)\si\neg\psi)\si\vp\to\psi$ & \eqref{vp-to-psi-and-vp-to-psi-to-chi-implies-vp-to-chi}: (3), (1)\\[1mm]
(5) & $\vdash((\vp\to\psi)\si\neg\psi)\si\vp\to\psi\si\neg\psi$ & \eqref{vp-to-psi-si-vp-to-chi-implies-vp-to-psi-si-chi}: (4), (2)\\[1mm]
(6) & $\vdash\psi\si\neg\psi\to\bot$ & \eqref{ex-falso-and}\\[1mm]
(7) & $\vdash((\vp\to\psi)\si\neg\psi)\si\vp\to\bot$ & \prule{syllogism}: (5), (6)\\[1mm]
(8) & $\vdash((\vp\to\psi)\si\neg\psi)\to(\vp\to\bot)$ & \prule{exportation}: (7)\\[1mm]
(9) & $\vdash(\vp\to\bot)\to\neg\vp$ & \prule{axiom-not-2}\\[1mm]
(10) & $\vdash((\vp\to\psi)\si\neg\psi)\to\neg\vp$ & \prule{syllogism}: (8), (9)\\[1mm]
(11) & $\vdash(\vp\to\psi)\to(\neg\psi\to\neg\vp)$ & \prule{exportation}: (10)
\end{tabular}\

\,
}

\begin{lemma}
\bea
\Gamma\vdash\vp\to\psi & \text{implies} & \Gamma\vdash\neg\psi\to\neg\vp, \label{rec-ax3-rule}\\
\Gamma \vdash \vp\to\neg\psi  & \text{implies} & \Gamma \vdash \psi\to\neg\vp.
\label{Gamma-vptonegpsi-implies-Gamma-psitonegvp}
\eea
\end{lemma}
\solution{\,\\[1mm]
\eqref{rec-ax3-rule}:\\[1mm]
\begin{tabular}{lll}
(1) & $\Gamma\vdash \, \vp \to \psi$ & (Assumption)\\[1mm] 
(2) & $\Gamma\vdash (\vp \to \psi)\to (\neg\psi \to \neg\vp)$ & \eqref{contraposition}\\[1mm]
(3) & $\Gamma\vdash \neg\psi \to \neg \vp$ & \prule{modus ponens}: (1), (2)\\[2mm]
\end{tabular}

\eqref{Gamma-vptonegpsi-implies-Gamma-psitonegvp}:\\[1mm]
\begin{tabular}{lll}
(1) & $\Gamma\vdash\vp\to\neg\psi$ & (Assumption)\\[1mm]
(2) & $\Gamma\vdash(\vp\to\neg \psi) \to (\psi\to\neg\vp)$ & \eqref{vptonegpsi-to-psitonegvp}\\[1mm]
(3) & $\Gamma\vdash\psi\to\neg\vp$ &  \prule{modus ponens}: (1), (2)
\end{tabular}

\,
}

\begin{lemma}
\begin{align}
\vdash & \, \vp\to\neg\neg\vp, \label{dni}\\
\vdash & \,\vp\si\psi\to\neg(\neg\vp\sau\neg\psi), \label{axiom-and-1}\\
\vdash & \,\neg(\neg\vp\sau\neg\psi)\to\vp\si\psi, \label{axiom-and-2}\\
\vdash & \, \neg\vp\sau\psi\to(\vp\to\psi), \label{disj-impl-2}\\
\vdash & \,\neg(\vp\si\psi)\to\neg\vp\sau\neg\psi, \label{de-morgan-and-2}\\
\vdash & \, \neg\vp\sau\neg\psi\to\neg(\vp\si\psi),\label{de-morgan-or-1}\\
\vdash & \, \neg(\vp\sau\psi)\to\neg\vp\si\neg\psi, \label{de-morgan-or-2}\\
\vdash & \, \neg\vp\si\neg\psi\to\neg(\vp\sau\psi), \label{de-morgan-and-1}\\
\vdash &  \, \neg\vp \sau \neg\psi \to (\psi \to \neg\vp). & \label{useful-mem-neg-1}.
\end{align}
\end{lemma}
\solution{\,\\[1mm]
\eqref{dni}:\\[1mm]
\begin{tabular}{lll}
(1) & $\vdash\vp\to((\vp\to\bot)\to\bot)$ & \eqref{MP-axiom-3}\\[1mm]
(2) & $\vdash((\vp\to\bot)\to\bot)\to\neg(\vp\to\bot)$ & \prule{axiom-not-2}\\[1mm]
(3) & $\vdash\vp\to\neg(\vp\to\bot)$ & \prule{syllogism}: (1), (2)\\[1mm]
(4) & $\vdash\neg\vp\to(\vp\to\bot)$ & \prule{axiom-not-1}\\[1mm]
(5) & $\vdash\neg(\vp\to\bot)\to\neg\neg\vp$ & \eqref{rec-ax3-rule}: (4)\\[1mm]
(6) & $\vdash\vp\to\neg\neg\vp$ & \prule{syllogism}: (3), (5)\\[2mm]
\end{tabular}\\

\eqref{axiom-and-1}:\\[1mm]
\begin{tabular}{lll}
(1) & $\vdash\vp\si\psi\to\vp$ & \prule{weakening}\\[1mm]
(2) & $\vdash\neg\vp\to\neg(\vp\si\psi)$ & \eqref{rec-ax3-rule}: (1)\\[1mm]
(3) & $\vdash\vp\si\psi\to\psi$ & \eqref{weak-si-2}\\[1mm]
(4) & $\vdash\neg\psi\to\neg(\vp\si\psi)$ & \eqref{rec-ax3-rule}: (3)\\[1mm]
(5) & $\vdash\neg\vp\sau\neg\psi\to\neg(\vp\si\psi)$ & \eqref{vp-to-chi-and-psi-to-chi-implies-vp-sau-psi-to-chi}: (2), (4)\\[1mm]
(6) & $\vdash\neg\neg(\vp\si\psi)\to\neg(\neg\vp\sau\neg\psi)$ & \eqref{rec-ax3-rule}: (5)\\[1mm]
(7) & $\vdash\vp\si\psi\to\neg\neg(\vp\si\psi)$ & \eqref{dni}\\[1mm]
(8) & $\vdash\vp\si\psi\to\neg(\neg\vp\sau\neg\psi)$ & \prule{syllogism}: (7), (6)\\[2mm]
\end{tabular}

\eqref{axiom-and-2}:\\[1mm]
\begin{tabular}{lll}
(1) & $\vdash(\neg\neg\vp\to\neg\psi)\to(\neg\vp\sau\neg\psi)$ & \eqref{axiom-or-2}\\[1mm]
(2) & $\vdash\neg\neg\vp\to\vp$ & \eqref{negnegvp-vp}\\[1mm]
(3) & $\vdash(\vp\to\neg\psi)\to(\neg\neg\vp\to\neg\psi)$ & \eqref{vp-to-psi-imp-psi-to-chi-tto-vp-to-chi}: (2)\\[1mm]
(4) & $\vdash (\vp\si\psi\to \bot)\to(\vp\to (\psi \to \bot))$ & \eqref{vp-si-psi-to-chi-to-vp-to-psi-to-chi}\\[1mm]
(5) & $\vdash \neg(\vp\si\psi)\to(\vp\si\psi\to \bot)$ & \prule{axiom-not-1}\\[1mm]
(6) & $\vdash \neg(\vp\si\psi)\to (\vp\to (\psi \to \bot))$ & \prule{syllogism}: (5), (4)\\[1mm]
(7) & $\vdash (\psi \to \bot)\to \neg\psi$ & \prule{axiom-not-2}\\[1mm]
(8) & $\vdash (\vp\to (\psi \to \bot))\to (\vp\to\neg\psi)$ & \eqref{Gamma-vp-to-psi-imp-Gamma-chi-to-vp-chi-to-psi}: (7)\\[1mm]
(9) & $\vdash\neg(\vp\si\psi)\to(\vp\to\neg\psi)$ & \prule{syllogism}: (6), (8)\\[1mm]
(10) & $\vdash\neg(\vp\si\psi)\to(\neg\neg\vp\to\neg\psi)$ & \prule{syllogism}: (9), (3)\\[1mm]
(11) & $\vdash\neg(\vp\si\psi)\to(\neg\vp\sau\neg\psi)$ & \prule{syllogism}: (10), (1)\\[1mm]
(12) & $\vdash\neg(\neg\vp\sau\neg\psi)\to\neg\neg(\vp\si\psi)$ & \eqref{rec-ax3-rule}: (11)\\[1mm]
(13) & $\vdash\neg\neg(\vp\si\psi)\to\vp\si\psi$ & \eqref{negnegvp-vp}\\[1mm]
(14) & $\vdash\neg(\neg\vp\sau\neg\psi)\to\vp\si\psi$ & \prule{syllogism}: (12), (13)\\[2mm]
\end{tabular}

\eqref{disj-impl-2}:\\[1mm]
\begin{tabular}{lll}
(1) & $\vdash \neg \vp \to (\vp \to \psi)$ & \eqref{ex-falso-impl-1}\\[1mm]
(2) & $\vdash \psi \to (\vp \to \psi)$ & \eqref{vp-to-psi-to-vp}\\[1mm]
(3) & $\vdash \neg\vp\sau\psi\to(\vp\to\psi)$ & \eqref{vp-to-chi-and-psi-to-chi-implies-vp-sau-psi-to-chi}: (1), (2)\\[2mm]
\end{tabular}\\

\eqref{de-morgan-and-2}:\\[1mm]
\begin{tabular}{lll}
(1) & $\vdash\neg(\neg\vp\sau\neg\psi)\to\vp\si\psi$ & \eqref{axiom-and-2}\\[1mm]
(2) & $\vdash\neg(\vp\si\psi)\to\neg\neg(\neg\vp\sau\neg\psi)$ & \eqref{rec-ax3-rule}: (1)\\[1mm]
(3) & $\vdash\neg\neg(\neg\vp\sau\neg\psi)\to(\neg\vp\sau\neg\psi)$ & \eqref{negnegvp-vp}\\[1mm]
(4) & $\vdash\neg(\vp\si\psi)\to\neg\vp\sau\neg\psi$ & \prule{syllogism}: (2), (3)\\[2mm]
\end{tabular}

\eqref{de-morgan-or-1}:\\[1mm]
\begin{tabular}{lll}
(1) & $\vdash\vp\si\psi\to\vp$ & \prule{weakening}\\[1mm]
(2) & $\vdash\neg\vp\to\neg(\vp\si\psi)$ & \eqref{vp-to-psi-imp-psi-to-chi-tto-vp-to-chi}: (1)\\[1mm]
(3) & $\vdash\vp\si\psi\to\psi$ & \eqref{weak-si-2}\\[1mm]
(4) & $\vdash\neg\psi\to\neg(\vp\si\psi)$ & \eqref{vp-to-psi-imp-psi-to-chi-tto-vp-to-chi}: (3)\\[1mm]
(5) & $\vdash\neg\vp\sau\neg\psi\to\neg(\vp\si\psi)$ & \eqref{vp-to-chi-and-psi-to-chi-implies-vp-sau-psi-to-chi}: (2), (4)\\[2mm]
\end{tabular}\\

\eqref{de-morgan-or-2}:\\[1mm]
\begin{tabular}{lll}
(1) & $\vdash\vp\to\vp\sau\psi$ & \prule{weakening}\\[1mm]
(2) & $\vdash\neg(\vp\sau\psi)\to\neg\vp$ & \eqref{rec-ax3-rule}: (1)\\[1mm]
(3) & $\vdash\psi\to\vp\sau\psi$ & \eqref{weak-sau-2}\\[1mm]
(4) & $\vdash\neg(\vp\sau\psi)\to\neg\psi$ & \eqref{rec-ax3-rule}: (3)\\[1mm]
(5) & $\vdash\neg(\vp\sau\psi)\to\neg\vp\si\neg\psi$ & \eqref{vp-to-psi-si-vp-to-chi-implies-vp-to-psi-si-chi}: (2), (4)\\[2mm]
\end{tabular}

\eqref{de-morgan-and-1}:\\[1mm]
\begin{tabular}{lll}
(1) & $\vdash\neg\vp\si\neg\psi\to\neg\vp$ & \prule{weakening}\\[1mm]
(2) & $\vdash\neg\vp\to(\vp\to\bot)$ & \prule{axiom-not-1}\\[1mm]
(3) & $\vdash\neg\vp\si\neg\psi\to(\vp\to\bot)$ & \prule{syllogism}: (1), (2)\\[1mm]
(4) & $\vdash\neg\vp\si\neg\psi\to\neg\psi$ & \eqref{weak-si-2}\\[1mm]
(5) & $\vdash\neg\psi\to(\psi\to\bot)$ & \prule{axiom-not-1}\\[1mm]
(6) & $\vdash\neg\vp\si\neg\psi\to(\psi\to\bot)$ & \prule{syllogism}: (4), (5)\\[1mm]
(7) & $\vdash\neg\vp\si\neg\psi\to((\vp\sau\psi)\to\bot)$ & \eqref{vp-to-psi-to-chi-si-vp-to-gamma-to-chi-to-vp-to-psi-sau-gamma-to-chi}: (3), (6)\\[1mm]
(8) & $\vdash((\vp\sau\psi)\to\bot)\to\neg(\vp\sau\psi)$ & \prule{axiom-not-2}\\[1mm]
(9) & $\vdash\neg\vp\si\neg\psi\to\neg(\vp\sau\psi)$ & \prule{syllogism}: (7), (8)\\[2mm]
\end{tabular}

\eqref{useful-mem-neg-1}:\\[1mm] 
\begin{tabular}{lll}
(1) & $\vdash\neg\vp\sau\neg\psi \to \neg\psi \sau \neg\vp$ & \prule{permutation}\\[1mm]
(2) & $\vdash \neg\psi \sau \neg\vp \to(\psi \to \neg\vp)$ & \eqref{disj-impl-2}\\[1mm]
(3) & $\vdash \neg\vp \sau \neg\psi \to (\psi \to \neg\vp),$ & \prule{syllogism}: (1), (2)
\end{tabular}

\,
}

\begin{lemma}
\bea
\Gamma\vdash\vp \to \psi \,\, \text{and}  \,\, \Gamma\vdash \neg\vp \to \psi & \text{imply} &
\Gamma\vdash \psi, \label{rule-proof-by-cases}\\
\Gamma\vdash\psi\si\vp\to\chi \,\, \text{and}  \,\, \Gamma\vdash\neg\psi\si\vp\to\chi & \text{imply} &
\Gamma\vdash\vp\to\chi, \label{proof-by-cases-extra-premise}\\
\Gamma\vdash\vp\si\psi\to\chi \,\, \text{and}  \,\, \Gamma\vdash\vp\si\neg\psi\to\chi & \text{imply} &
\Gamma\vdash\vp\to\chi. \label{proof-by-cases-extra-premise-1}
\eea
\end{lemma}
\solution{\,\\[1mm]
\eqref{rule-proof-by-cases}:\\[1mm]
\begin{tabular}{lll}
(1) & $\Gamma\vdash\vp\to\psi$ & (Assumption)\\[1mm]
(2) & $\Gamma\vdash\neg\vp\to\psi$ & (Assumption)\\[1mm]
(3) & $\Gamma\vdash\vp\sau\neg\vp\to\psi$ & \eqref{vp-to-chi-and-psi-to-chi-implies-vp-sau-psi-to-chi}: (1), (2)\\[1mm]
(4) & $\Gamma\vdash\vp\sau\neg\vp$ & \prule{lem}\\[1mm]
(5) & $\Gamma\vdash\psi$ & \prule{modus ponens}: (4), (3)\\[2mm]
\end{tabular}

\eqref{proof-by-cases-extra-premise}:\\[1mm]
\begin{tabular}{lll}
(1) & $\Gamma\vdash \psi\si\vp\to\chi $ & (Assumption)\\[1mm]
(2) & $\Gamma\vdash \psi \to (\vp\to\chi)$ & \prule{exportation}: (1)\\[1mm]
(3) & $\Gamma\vdash \neg\psi\si\vp\to\chi $ & (Assumption)\\[1mm]
(4) & $\Gamma\vdash \neg\psi \to (\vp\to\chi)$ & \prule{exportation}: (3)\\[1mm]
(5) & $\Gamma\vdash \vp\to\chi$ &  \eqref{rule-proof-by-cases}: (2), (4)\\[2mm]
\end{tabular}

\eqref{proof-by-cases-extra-premise-1}:\\[1mm]
\begin{tabular}{lll}
(1) & $\Gamma\vdash \vp\si\psi\to\chi $ & (Assumption)\\[1mm]
(2) & $\Gamma\vdash \psi\si\vp\to\chi $ & \eqref{vp-si-psi-implies-chi-perm}: (1)\\[1mm]
(3) & $\Gamma\vdash \vp\si\neg\psi\to\chi $ & (Assumption)\\[1mm]
(4) & $\Gamma\vdash \neg\psi\si\vp\to\chi $ & \eqref{vp-si-psi-implies-chi-perm}: (3)\\[1mm]
(5) & $\Gamma\vdash \vp\to\chi$ &  \eqref{proof-by-cases-extra-premise}: (2), (4)
\end{tabular}

\,
}

\begin{lemma}
\bea
\Gamma \vdash\vp\dnd\psi &   \text{implies} &  (\Gamma\vdash\vp \text{ iff } \Gamma\vdash\psi),
\label{vpdndpsi-imp-vp-iff-psi}\\
\Gamma \vdash \vp\to\psi \,\, \text{and}  \,\, \Gamma \vdash \psi\to \vp  &  \text{iff}   &   \Gamma \vdash \vp \dnd \psi. 
\label{Gamma-vptopsi-psitovp-iff-vpdndpsi}
\eea
\end{lemma}
\solution{\,\\[1mm]
\eqref{vpdndpsi-imp-vp-iff-psi}: Immediately, by the definition of $\dnd$, \eqref{Gamma-vp-psi-vp-si-psi} 
and \prule{modus ponens}.

\eqref{Gamma-vptopsi-psitovp-iff-vpdndpsi}: Apply \eqref{Gamma-vp-psi-vp-si-psi} and the definition of $\dnd$.\\[1mm]
}

\begin{lemma}
\bea
\Gamma \vdash \vp\dnd\psi  &   \text{implies} &  \Gamma\vdash\psi\dnd\vp,
 \label{sim-symm}\\
\Gamma \vdash\vp\dnd\psi\,\, \text{and}  \,\, \Gamma\vdash \psi\dnd\chi & \text{imply} &  \Gamma\vdash\vp\dnd\chi,
 \label{sim-trans}\\
\Gamma\vdash \vp\dnd\vp'\,\, \text{and}  \,\, \Gamma\vdash\psi\dnd\psi' & \text{imply} &  
\Gamma\vdash \vp\sau\psi\dnd\vp'\sau\psi',\label{sim-pairs-or}\\
\Gamma\vdash\vp\dnd\vp'\,\, \text{and}  \,\, \Gamma\vdash \psi\dnd\psi' & \text{imply} &  
\Gamma\vdash \vp\si\psi \dnd \vp'\si\psi',
\label{sim-pairs-and}\\
\Gamma\vdash\vp\dnd\vp'\,\, \text{and}  \,\, \Gamma\vdash\psi\dnd\psi' & \text{imply} &  
\Gamma\vdash \vp\to \psi\dnd\vp'\to \psi',
\label{sim-pairs-to}\\
\Gamma\vdash \vp\dnd\vp' \text{~and~} \Gamma \vdash \psi\dnd\psi' & \text{implies} &
\Gamma \vdash (\vp\dnd\psi)\to(\vp'\dnd\psi') 
\label{Gamma-vpdndvpp-and-Gamma-psidndpsip-implies-Gamma-vpdndpsitovppdndpsip}\\
\Gamma\vdash \vp\dnd\vp'  \,\, \text{and}  \,\, \Gamma \vdash \vp\to\psi &  \text{imply} & 
\Gamma \vdash \vp'\to\psi,
\label{sim-pairs-to-left-1}\\
\Gamma\vdash \psi\dnd\psi' \,\, \text{and}  \,\, \Gamma \vdash \vp\to\psi &  \text{imply} & 
\Gamma \vdash \vp\to\psi',
\label{sim-pairs-to-right-1}\\
\Gamma \vdash \vp\dnd\vp',\,\,\Gamma \vdash \psi\dnd\psi' \,\, \text{and}  \,\,
\Gamma \vdash \vp\to\psi & \text{implies} & \Gamma\vdash\vp'\to\psi'.
\label{sim-pairs-to-left-right-1}
\eea
\end{lemma}
\solution{\,\\[1mm]
\eqref{sim-symm}:\\[1mm]
\begin{tabular}{lll}
(1) & $\Gamma\vdash \vp\dnd\psi$ & (Assumption)\\[1mm]
(2) & $\Gamma\vdash \vp\to\psi$ & \eqref{Gamma-vptopsi-psitovp-iff-vpdndpsi}: (1)\\[1mm]
(3) & $\Gamma\vdash \psi\to\vp$ & \eqref{Gamma-vptopsi-psitovp-iff-vpdndpsi}: (1)\\[1mm]
(4) & $\Gamma\vdash \psi\dnd\vp $ & \eqref{Gamma-vptopsi-psitovp-iff-vpdndpsi}: (3), (2)\\[2mm]
\end{tabular}

\eqref{sim-trans}:\\[1mm]
\begin{tabular}{lll}
(1) & $\Gamma\vdash \vp\dnd\psi$ & (Assumption)\\[1mm]
(2) & $\Gamma\vdash \psi\dnd\chi$ & (Assumption)\\[1mm]
(3) & $\Gamma\vdash\vp\to\psi$ & \eqref{Gamma-vptopsi-psitovp-iff-vpdndpsi}: (1)\\[1mm]
(4) & $\Gamma\vdash\psi\to\chi$ & \eqref{Gamma-vptopsi-psitovp-iff-vpdndpsi}: (2)\\[1mm]
(5) & $\Gamma\vdash\vp\to\chi$ & \prule{syllogism}: (3), (4)\\[1mm]
(6) & $\Gamma\vdash\psi\to\vp$ & \eqref{Gamma-vptopsi-psitovp-iff-vpdndpsi}: (1)\\[1mm]
(7) & $\Gamma\vdash\chi\to\psi$ & \eqref{Gamma-vptopsi-psitovp-iff-vpdndpsi}: (2)\\[1mm]
(8) & $\Gamma\vdash\chi\to\vp$ & \prule{syllogism}: (7), (6)\\[1mm]
(9) & $\Gamma\vdash \vp \dnd \chi$ & \eqref{Gamma-vptopsi-psitovp-iff-vpdndpsi}: (5), (8)\\[2mm]
\end{tabular}

\eqref{sim-pairs-or}:\\[1mm]
\begin{tabular}{lll}
(1) & $\Gamma\vdash \vp\dnd\vp'$ & (Assumption)\\[1mm]
(2) & $\Gamma\vdash \vp\to\vp'$ & \eqref{Gamma-vptopsi-psitovp-iff-vpdndpsi}: (1)\\[1mm]
(3) & $\Gamma\vdash \psi\dnd\psi'$ & (Assumption)\\[1mm]
(4) & $\Gamma\vdash\psi\to\psi'$ & \eqref{Gamma-vptopsi-psitovp-iff-vpdndpsi}: (3)\\[1mm]
(5) & $\Gamma\vdash\vp\sau\psi\to\vp'\sau\psi'$ & \eqref{vp-to-psi-and-chi-to-gamma-implies-sau}: (2), (4)\\[1mm]
(6) & $\Gamma\vdash\vp'\to\vp$ & \eqref{Gamma-vptopsi-psitovp-iff-vpdndpsi}: (1)\\[1mm]
(7) & $\Gamma\vdash\psi'\to\psi$ & \eqref{Gamma-vptopsi-psitovp-iff-vpdndpsi}: (3)\\[1mm]
(8) & $\Gamma\vdash\vp'\sau\psi'\to\vp\sau\psi$ & \eqref{vp-to-psi-and-chi-to-gamma-implies-sau}: (6), (7)\\[1mm]
(9) & $\Gamma\vdash \vp\sau \psi\dnd \vp'\sau \psi'$ & \eqref{Gamma-vptopsi-psitovp-iff-vpdndpsi}: (5), (8)\\[2mm]
\end{tabular}

\eqref{sim-pairs-and}:\\[1mm]
\begin{tabular}{lll}
(1) & $\Gamma\vdash \vp\dnd\vp'$ & (Assumption)\\[1mm]
(2) & $\Gamma\vdash \vp\to\vp'$ & \eqref{Gamma-vptopsi-psitovp-iff-vpdndpsi}: (1)\\[1mm]
(3) & $\Gamma\vdash \psi\dnd\psi'$ & (Assumption)\\[1mm]
(4) & $\Gamma\vdash\psi\to\psi'$ & \eqref{Gamma-vptopsi-psitovp-iff-vpdndpsi}: (3)\\[1mm]
(5) & $\Gamma\vdash\vp\si\psi\to\vp'\si\psi'$ & \eqref{psi-to-vp-chi-to-gamma-vp-si-chi-to-psi-si-gamma}: (2), (4)\\[1mm]
(6) & $\Gamma\vdash\vp'\to\vp$ & \eqref{Gamma-vptopsi-psitovp-iff-vpdndpsi}: (1)\\[1mm]
(7) & $\Gamma\vdash\psi'\to\psi$ & \eqref{Gamma-vptopsi-psitovp-iff-vpdndpsi}: (3)\\[1mm]
(8) & $\Gamma\vdash\vp'\si\psi'\to\vp\si\psi$ & \eqref{psi-to-vp-chi-to-gamma-vp-si-chi-to-psi-si-gamma}: (6), (7)\\[1mm]
(9) & $\Gamma\vdash \vp\si\psi \dnd \vp'\si\psi'$ & \eqref{Gamma-vptopsi-psitovp-iff-vpdndpsi}: (5), (8) \\[2mm]
\end{tabular}

\eqref{sim-pairs-to}:\\[1mm]
\begin{tabular}{lll}
(1) & $\Gamma\vdash \vp\dnd\vp'$ & (Assumption)\\[1mm]
(2) & $\Gamma\vdash \vp\to\vp'$ & \eqref{Gamma-vptopsi-psitovp-iff-vpdndpsi}: (1)\\[1mm]
(3) & $\Gamma\vdash \psi\dnd\psi'$ & (Assumption)\\[1mm]
(4) & $\Gamma\vdash\psi\to\psi'$ & \eqref{Gamma-vptopsi-psitovp-iff-vpdndpsi}: (3)\\[1mm]
(5) & $\Gamma\vdash(\vp\to\psi)\to(\vp'\to\psi')$ & \eqref{psi-to-vp-chi-to-gamma-vp-to-chi-to-psi-to-gamma}: (2), (4)\\[1mm]
(6) & $\Gamma\vdash\vp'\to\vp$ & \eqref{Gamma-vptopsi-psitovp-iff-vpdndpsi}: (1)\\[1mm]
(7) & $\Gamma\vdash\psi'\to\psi$ & \eqref{Gamma-vptopsi-psitovp-iff-vpdndpsi}: (3)\\[1mm]
(8) & $\Gamma\vdash(\vp'\to\psi')\to(\vp\to\psi)$ & \eqref{psi-to-vp-chi-to-gamma-vp-to-chi-to-psi-to-gamma}: (6), (7)\\[1mm]
(9) & $\Gamma\vdash (\vp\to\psi) \dnd (\vp'\to\psi')$ & \eqref{Gamma-vptopsi-psitovp-iff-vpdndpsi}: (5), (8) \\[2mm]
\end{tabular}

\eqref{Gamma-vpdndvpp-and-Gamma-psidndpsip-implies-Gamma-vpdndpsitovppdndpsip}:\\[1mm]
\begin{tabular}{lll}
(1) & $\Gamma\vdash\vp\dnd\vp'$ & (Assumption)\\[1mm]
(2) & $\Gamma\vdash\vp\to\vp'$ & \eqref{Gamma-vptopsi-psitovp-iff-vpdndpsi}: (1)\\[1mm]
(3) & $\Gamma\vdash\psi\dnd\psi'$ & (Assumption)\\[1mm]
(4) & $\Gamma\vdash\psi\to\psi'$ & \eqref{Gamma-vptopsi-psitovp-iff-vpdndpsi}: (3)\\[1mm]
(5) & $\Gamma\vdash(\vp\to\psi)\to(\vp'\to\psi')$ & \eqref{psi-to-vp-chi-to-gamma-vp-to-chi-to-psi-to-gamma}: (2), (4)\\[1mm]
(6) & $\Gamma\vdash(\psi\to\vp)\to(\psi'\to\vp')$ & \eqref{psi-to-vp-chi-to-gamma-vp-to-chi-to-psi-to-gamma}: (4), (2)\\[1mm]
(7) & $\Gamma\vdash(\vp\dnd\psi)\to(\vp'\dnd\psi')$ & \eqref{psi-to-vp-chi-to-gamma-vp-si-chi-to-psi-si-gamma}: (5), (6)
 and the definition of $\dnd$\\[2mm]
\end{tabular}

\eqref{sim-pairs-to-left-1}:\\[1mm]
\begin{tabular}{lll}
(1) & $\Gamma\vdash\vp\dnd\vp'$ & (Assumption)\\[1mm]
(2) & $\Gamma\vdash\vp\to\psi$ & (Assumption)\\[1mm]
(3) & $\Gamma\vdash\vp'\to\vp$ & \eqref{Gamma-vptopsi-psitovp-iff-vpdndpsi}: (1)\\[1mm]
(4) & $\Gamma\vdash\vp'\to\psi$ & \prule{syllogism}: (2), (1)\\[2mm]
\end{tabular}

\eqref{sim-pairs-to-right-1}:\\[1mm]
\begin{tabular}{lll}
(1) & $\Gamma\vdash\psi\dnd\psi'$ & (Assumption)\\[1mm]
(2) & $\Gamma\vdash\vp\to\psi$ & (Assumption)\\[1mm]
(3) & $\Gamma\vdash\psi\to\psi'$ & \eqref{Gamma-vptopsi-psitovp-iff-vpdndpsi}: (1)\\[1mm]
(4) & $\Gamma\vdash\vp\to\psi'$ &  \prule{syllogism}: (2), (3)\\[2mm]
\end{tabular}

\eqref{sim-pairs-to-left-right-1}:\\[1mm]
\begin{tabular}{lll}
(1) & $\Gamma\vdash\vp\dnd\vp'$ & (Assumption)\\[1mm]
(2) & $\Gamma\vdash\vp'\to\vp$ & \eqref{Gamma-vptopsi-psitovp-iff-vpdndpsi}: (1)\\[1mm]
(3) & $\Gamma\vdash\psi\dnd\psi'$ & (Assumption)\\[1mm]
(4) & $\Gamma\vdash\psi\to\psi'$ & \eqref{Gamma-vptopsi-psitovp-iff-vpdndpsi}: (3) \\[1mm]
(5) & $\Gamma\vdash\vp\to\psi$ & (Assumption)\\[1mm]
(6) & $\Gamma\vdash\vp'\to\psi$ & \prule{syllogism}: (2), (5)\\[1mm]
(7) & $\Gamma\vdash\vp'\to\psi'$ & \prule{syllogism}: (6), (4)
\end{tabular}

\,
}

\begin{lemma}
\begin{align}
\vdash & \, \vp \dnd \vp, \label{sim-refl}\\
\vdash & \, \vp\si\psi \dnd \psi\si\vp, \label{si-comm}\\
\vdash & \, \vp\sau\psi \dnd \psi\sau\vp, \label{sau-comm}\\
\vdash & \,\neg\neg\vp \dnd \vp \label{negnegvp-dnd-vp}\\
\vdash & \,\vp \sau \psi \dnd (\neg\vp\to\psi), \label{vpsaupsi-negvptopsi}\\
\vdash & \, (\vp \to \psi) \dnd (\neg\vp \sau \psi), \label{def-imp-or}\\
\vdash & \,\vp \si \psi \dnd \neg(\neg\vp\sau\neg\psi), \label{vpsipsi-demorgan}\\
\vdash & \, \neg\vp\sau\neg\psi \dnd \neg(\vp\si\psi), \label{de-morgan-or-dnd} \\
\vdash & \, \neg\vp\si\neg\psi \dnd \neg(\vp\sau\psi), \label{de-morgan-and-dnd}\\
\vdash & \, \vp\si (\psi\si\chi)\dnd(\vp\si\psi)\si\chi, \label{si-assoc}\\
\vdash & \, \vp\sau (\psi\sau\chi)\dnd(\vp\sau\psi)\sau\chi, \label{sau-assoc}\\
\vdash & \, \vp\si(\psi\sau\chi)\dnd(\vp\si\psi)\sau(\vp\si\chi), \label{distrib-si-sau}\\
\vdash & \, \vp\sau(\psi\si\chi) \dnd (\vp\sau\psi)\si(\vp\sau\chi). \label{distrib-sau-si}
\end{align}
\end{lemma}
\solution{We use \eqref{Gamma-vptopsi-psitovp-iff-vpdndpsi}.\

\eqref{sim-refl}: Apply \eqref{vp-to-vp}.

\eqref{si-comm}: Apply \prule{permutation}.

\eqref{sau-comm}:  Apply \prule{permutation}.

\eqref{negnegvp-dnd-vp}: Apply \eqref{negnegvp-vp}, \eqref{dni}.

\eqref{vpsaupsi-negvptopsi}: Apply \eqref{axiom-or-2}, \eqref{axiom-or-1}.

\eqref{def-imp-or}: Apply \eqref{disj-impl-1}, \eqref{disj-impl-2}.

\eqref{vpsipsi-demorgan}: Apply \eqref{axiom-and-1}, \eqref{axiom-and-2}.

\eqref{de-morgan-or-dnd}:  Apply \eqref{de-morgan-and-2}, \eqref{de-morgan-or-1}.

\eqref{de-morgan-and-dnd}:  Apply \eqref{de-morgan-or-2}, \eqref{de-morgan-and-1}.

\eqref{si-assoc}: Apply \eqref{si-associativity-1}, \eqref{si-associativity-2}.

\eqref{sau-assoc}: Apply \eqref{sau-associativity-1}, \eqref{sau-associativity-2}.

\eqref{distrib-si-sau}: Apply \eqref{distrib-si-sau-1}, \eqref{distrib-si-sau-2}.

\eqref{distrib-sau-si}: Apply \eqref{distrib-sau-si-1}, \eqref{distrib-sau-si-2}.
}

\begin{lemma}
\begin{align}
\vdash & \, \vp\sau\vp \dnd \vp, \label{sau-idemp} \\
\vdash & \, \vp\si\vp \dnd \vp. \label{si-idemp}
\end{align}
\end{lemma}
\solution{\,\\[1mm]
\eqref{sau-idemp}:\\[1mm]
\begin{tabular}{lll}
(1) & $\vdash\vp\sau\vp\to\vp$ & \prule{contraction}\\[1mm]
(2) & $\vdash\vp\to\vp\sau\vp$ & \prule{weakening}\\[1mm]
(3) & $\vdash\vp\sau\vp\dnd\vp$ & \eqref{Gamma-vptopsi-psitovp-iff-vpdndpsi}: (1), (2)\\[2mm]
\end{tabular}

\eqref{si-idemp}:\\[1mm]
\begin{tabular}{lll}
(1) & $\vdash\vp\si\vp\to\vp$ & \prule{weakening}\\[1mm]
(2) & $\vdash\vp\to\vp\si\vp$ & \prule{contraction}\\[1mm]
(3) & $\vdash\vp\si\vp\dnd\vp$ & \eqref{Gamma-vptopsi-psitovp-iff-vpdndpsi}: (1), (2)
\end{tabular}

\,
}

\blem
\begin{align}
\vdash &  \,  (\vp\dnd \chi) \si (\psi \dnd \chi) \to (\vp\dnd \psi). \label{dnd-tranz}
\end{align}
\elem
\solution{\,\\[1mm]
\begin{tabular}{lll}
(1) & $\vdash(\vp\dnd \chi) \si (\psi \dnd \chi)\to(\vp\dnd \chi)$ & \prule{weakening}\\[1mm]
(2) & $\vdash(\vp\dnd \chi)\to(\vp\to\chi)$ & \prule{weakening} and the definition of $\dnd$\\[1mm]
(3) & $\vdash(\vp\dnd \chi) \si (\psi \dnd \chi)\to(\vp\to\chi)$ & \prule{syllogism}: (1), (2)\\[1mm]
(4) & $\vdash(\vp\dnd \chi) \si (\psi \dnd \chi)\to(\psi \dnd \chi)$ & \eqref{weak-si-2}\\[1mm]
(5) & $\vdash(\psi \dnd \chi)\to(\chi\to\psi)$ & \eqref{weak-si-2} and the definition of $\dnd$\\[1mm]
(6) & $\vdash(\vp\dnd \chi) \si (\psi \dnd \chi)\to(\chi\to\psi)$ & \prule{syllogism}: (4), (5)\\[1mm]
(7) & $\vdash(\vp\dnd \chi) \si (\psi \dnd \chi)\to(\vp\to\chi)\si(\chi\to\psi)$ & \eqref{vp-to-psi-si-vp-to-chi-implies-vp-to-psi-si-chi}: (3), (6)\\[1mm]
(8) & $\vdash(\vp\to\chi)\to((\chi\to\psi)\to(\vp\to\psi))$ & \eqref{vp-psi-chi-tranzitivity-ra}\\[1mm]
(9) & $\vdash(\vp\to\chi)\si(\chi\to\psi)\to(\vp\to\psi)$ & \prule{importation}: (8)\\[1mm]
(10) & $\vdash(\vp\dnd \chi) \si (\psi \dnd \chi)\to(\vp\to\psi)$ & \prule{syllogism}: (7), (9)\\[1mm]
(11) & $\vdash(\vp\dnd \chi) \si (\psi \dnd \chi)\to(\psi\to\vp)$ & Similarly\\[1mm]
(12) & $\vdash(\vp\dnd \chi) \si (\psi \dnd \chi)\to(\vp\dnd\psi)$ & 
\eqref{vp-to-psi-si-vp-to-chi-implies-vp-to-psi-si-chi}: (10), (11) and the definition of $\dnd$
\end{tabular}

\,
}

\blem
\bea
\Gamma \vdash \vp\dnd \chi \,\, \text{and}  \,\, \Gamma\vdash \psi \dnd \chi & \text{imply} &  
\Gamma\vdash \vp\dnd \psi. \label{sim-trans-2}
\eea
\elem
\solution{\,\\[1mm]
\begin{tabular}{lll}
(1) & $\Gamma \vdash \vp\dnd \chi$ & (Assumption)\\[1mm]
(2) & $\Gamma \vdash \psi \dnd \chi$ & (Assumption)\\[1mm]
(3) & $\Gamma \vdash (\vp\dnd \chi) \si (\psi \dnd \chi)$ & \eqref{vp-to-psi-si-vp-to-chi-implies-vp-to-psi-si-chi}: (1), (2)\\[1mm]
(4) & $\Gamma \vdash (\vp\dnd \chi) \si (\psi \dnd \chi) \to (\vp\dnd \psi)$ & \eqref{dnd-tranz}\\[1mm]
(5) & $\Gamma \vdash \vp\dnd \psi$ & \prule{modus ponens}: (3), (4)
\end{tabular}

\,
}

\begin{lemma}
\bea
\Gamma\vdash \vp\to(\psi\dnd\chi)\,\, \text{and}  \,\,\Gamma\vdash\vp\to(\chi\dnd\gamma) & 
\text{imply} & \Gamma\vdash\vp\to(\psi\dnd\gamma),
\label{vp-to-psi-dnd-chi-vp-dnd-chi-to-gamma-implies-vp-to-psi-dnd-gamma}\\
\Gamma\vdash \vp\to(\psi\dnd\chi)\,\, \text{and}  \,\,\Gamma\vdash \chi\dnd\gamma & 
\text{imply} & \Gamma\vdash\vp\to(\psi\dnd\gamma).
\label{vp-to-psi-dnd-chi-and-chi-dnd-gamma-implies-vp-to-psi-dnd-gamma} 
\eea
\end{lemma}
\solution{\,\\[1mm]
\eqref{vp-to-psi-dnd-chi-vp-dnd-chi-to-gamma-implies-vp-to-psi-dnd-gamma}:\\[1mm]
\begin{tabular}{lll}
(1) & $\Gamma\vdash \vp\to(\psi\dnd\chi)$ & (Assumption)\\[1mm]
(2) & $\Gamma\vdash (\psi\dnd\chi) \to (\psi\to\chi) $ & \prule{weakening}  and the definition of $\dnd$\\[1mm]
(3) & $\Gamma\vdash (\psi\dnd\chi) \to (\chi\to\psi) $ & \eqref{weak-si-2}  and the definition of $\dnd$\\[1mm]
(4) & $\Gamma\vdash \vp\to (\psi\to\chi) $ & \prule{syllogism}: (1), (2)\\[1mm]
(5) & $\Gamma\vdash \vp\to (\chi\to\psi) $ & \prule{syllogism}: (1), (3)\\[1mm]
(6) & $\Gamma\vdash \vp\to(\chi\dnd\gamma)$ & (Assumption)\\[1mm]
(7) & $\Gamma\vdash (\chi\dnd\gamma) \to (\chi\to\gamma) $ & \prule{weakening} and the definition of $\dnd$\\[1mm]
(8) & $\Gamma\vdash (\chi\dnd\gamma) \to (\gamma\to\chi) $ & \eqref{weak-si-2} and the definition of $\dnd$\\[1mm]
(9) & $\Gamma\vdash \vp\to (\chi\to\gamma) $ & \prule{syllogism}: (6), (7)\\[1mm]
(10) & $\Gamma\vdash \vp\to (\gamma\to\chi) $ & \prule{syllogism}: (6), (8)\\[1mm]
(11) & $\Gamma\vdash \vp\to (\psi\to\gamma) $ & 
\eqref{vp-to-psi-to-chi-vp-to-chi-to-gamma-implies-vp-to-psi-to-gamma}: (4), (9)\\[1mm]
(12) & $\Gamma\vdash \vp\to (\gamma\to\psi) $ & 
\eqref{vp-to-psi-to-chi-vp-to-chi-to-gamma-implies-vp-to-psi-to-gamma}: (10), (5)\\[1mm]
(13) & $\Gamma\vdash \vp\to (\psi\dnd\gamma) $ &  \eqref{vp-to-psi-si-vp-to-chi-implies-vp-to-psi-si-chi}: (11), (12) and the definition of $\dnd$\\[1mm]
\end{tabular}

\eqref{vp-to-psi-dnd-chi-and-chi-dnd-gamma-implies-vp-to-psi-dnd-gamma}:\\[1mm]
\begin{tabular}{lll}
(1) & $\Gamma\vdash \vp\to(\psi\dnd\chi)$ & (Assumption)\\[1mm]
(2) & $\Gamma\vdash \chi\dnd\gamma $ & (Assumption)\\[1mm]
(3) & $\Gamma\vdash \vp\to(\chi\dnd\gamma)$ & \eqref{vp-implies-psi-to-vp}: (2)\\[1mm]
(4) & $\Gamma\vdash \vp\to(\psi\dnd\gamma)$ & 
\eqref{vp-to-psi-dnd-chi-vp-dnd-chi-to-gamma-implies-vp-to-psi-dnd-gamma}: (1), (3)
\end{tabular}

\,
}

\begin{lemma}\label{lemma-vpdndpsi-neg-or-sau-si}
\begin{align}
\vdash & \,(\vp \dnd \psi) \to (\neg\vp \dnd \neg \psi), \label{vpdndpsi-neg}\\
\vdash & \,(\vp \dnd \psi) \to ((\vp \to \chi) \dnd  (\psi \to \chi)), \label{vpdndpsi-to-r}\\
\vdash & \,(\vp \dnd \psi) \to ((\chi \to \vp) \dnd  (\chi  \to \psi)), \label{vpdndpsi-to-l}\\
\vdash & \,(\vp \dnd \psi) \to (\vp \sau \chi \dnd  \psi \sau \chi), \label{vpdndpsi-sau-r}\\
\vdash & \,(\vp \dnd \psi) \to (\chi \sau \vp \dnd  \chi \sau \psi), \label{vpdndpsi-sau-l}\\
\vdash & \,(\vp \dnd \psi) \to (\vp \si \chi \dnd  \psi\si \chi), \label{vpdndpsi-si-r}\\
\vdash & \,(\vp \dnd \psi) \to (\chi \si \vp \dnd  \chi \si \psi). \label{vpdndpsi-si-l}
\end{align}
\end{lemma}
\solution{\,\\[1mm]
\eqref{vpdndpsi-neg}:\\[1mm]
\begin{tabular}{lll}
(1) & $\vdash (\vp \dnd \psi) \to (\vp \to \psi)$ &  \prule{weakening} 
and the definition of $\dnd$ \\[1mm]
(2) & $\vdash (\vp \to \psi) \to (\neg \psi\to \neg\vp)$  & \eqref{contraposition} \\[1mm]
(3) & $\vdash (\vp \dnd \psi) \to  (\neg \psi\to \neg\vp)$ & \prule{syllogism}: (1), (2) \\[1mm]
(4) & $\vdash (\vp \dnd \psi) \to (\psi \to \vp)$ & \eqref{weak-si-2} and the definition of $\dnd$\\[1mm]
(5) & $\vdash (\psi \to \vp) \to (\neg \vp \to \neg\psi)$  & \eqref{contraposition} \\[1mm]
(6) & $\vdash (\vp \dnd \psi) \to   (\neg \vp \to \neg\psi)$ & \prule{syllogism}: (4), (5) \\[1mm]
(7) & $\vdash (\vp \dnd \psi) \to  (\neg\vp \dnd \neg \psi)$ & 
\eqref{vp-to-psi-si-vp-to-chi-implies-vp-to-psi-si-chi}: (6), (3) and the definition of $\dnd$\\[2mm]
\end{tabular}

\eqref{vpdndpsi-to-r}:\\[1mm]
\begin{tabular}{lll}
(1) & $\vdash (\vp \dnd \psi) \to (\vp \to \psi)$ &  \prule{weakening} and the definition of $\dnd$ \\[1mm]
(2) & $\vdash (\vp \to \psi) \to ((\psi \to \chi) \to  (\vp \to \chi))$  & \eqref{vp-psi-chi-tranzitivity-ra} \\[1mm]
(3) & $\vdash (\vp \dnd \psi) \to  ((\psi \to \chi) \to  (\vp \to \chi))$ & \prule{syllogism}: (1), (2) \\[1mm]
(4) & $\vdash (\vp \dnd \psi) \to (\psi \to \vp)$ & \eqref{weak-si-2} and the definition of $\dnd$\\[1mm]
(5) & $\vdash (\psi \to \vp) \to ((\vp \to \chi) \to  (\psi \to \chi))$  & \eqref{vp-psi-chi-tranzitivity-ra}  \\[1mm]
(6) & $\vdash (\vp \dnd \psi) \to  ((\vp \to \chi) \to  (\psi \to \chi))$ & \prule{syllogism}: (4), (5) \\[1mm]
(7) & $\vdash (\vp \dnd \psi) \to  ((\vp \to \chi) \dnd  (\psi \to \chi)) $ & 
\eqref{vp-to-psi-si-vp-to-chi-implies-vp-to-psi-si-chi}: (6), (3) and the definition of $\dnd$\\[2mm]
\end{tabular}

\eqref{vpdndpsi-to-l}:\\[1mm]
\begin{tabular}{lll}
(1) & $\vdash (\vp \dnd \psi) \to (\vp \to \psi)$ &  \prule{weakening} and the definition of $\dnd$ \\[1mm]
(2) & $\vdash (\vp \to \psi) \to ((\chi \to \vp) \to  (\chi \to \psi))$  & \eqref{vp-to-psi-tto-chi-to-vp-chi-to-psi} \\[1mm]
(3) & $\vdash (\vp \dnd \psi) \to ((\chi \to \vp) \to  (\chi \to \psi))$ & \prule{syllogism}: (1), (2) \\[1mm]
(4) & $\vdash (\vp \dnd \psi) \to (\psi \to \vp)$ & \eqref{weak-si-2} and the definition of $\dnd$\\[1mm]
(5) & $\vdash (\psi \to \vp) \to ((\chi \to \psi) \to  (\chi \to \vp))$  & \eqref{vp-to-psi-tto-chi-to-vp-chi-to-psi}  \\[1mm]
(6) & $\vdash (\vp \dnd \psi) \to   ((\chi \to \psi) \to  (\chi \to \vp))$ & \prule{syllogism}: (4), (5) \\[1mm]
(7) & $\vdash (\vp \dnd \psi) \to  ((\chi \to \vp) \dnd  (\chi  \to \psi))$ & 
\eqref{vp-to-psi-si-vp-to-chi-implies-vp-to-psi-si-chi}: (3), (6) and the definition of $\dnd$\\[2mm]
\end{tabular}

\eqref{vpdndpsi-sau-r}:\\[1mm]
\begin{tabular}{lll}
(1) & $\vdash  (\vp \dnd \psi) \to (\neg\vp \dnd \neg \psi)$ &  \eqref{vpdndpsi-neg}\\[1mm]
(2) & $\vdash (\neg\vp \dnd \neg \psi) \to ((\neg\vp\to \chi) \dnd (\neg\psi\to \chi))$ 
& \eqref{vpdndpsi-to-r}\\[1mm]
(3) & $\vdash (\vp \dnd \psi) \to ((\neg\vp\to \chi) \dnd (\neg\psi\to \chi))$ 
& \prule{syllogism}: (1), (2) \\[1mm]
(4) & $\vdash (\neg\vp\to \chi) \dnd \vp \sau \chi$  & \eqref{vpsaupsi-negvptopsi}\\[1mm]
(5) & $\vdash (\neg\psi\to \chi) \dnd \psi \sau \chi$  & \eqref{vpsaupsi-negvptopsi}\\[1mm]
(6) & $\vdash ((\neg\vp\to \chi) \dnd (\neg\psi\to \chi)) \to (\vp \sau \chi \dnd \psi \sau \chi )$ 
& \eqref{Gamma-vpdndvpp-and-Gamma-psidndpsip-implies-Gamma-vpdndpsitovppdndpsip}: (4), (5)\\[1mm]
(7) & $\vdash (\vp \dnd \psi) \to (\vp \sau \chi \dnd \psi \sau \chi)$ 
& \prule{syllogism}: (3), (6) \\[2mm]
\end{tabular}

\eqref{vpdndpsi-sau-l}:\\[1mm]
\begin{tabular}{lll}
(1) & $\vdash  (\vp \dnd \psi) \to ((\neg\chi\to \vp) \dnd (\neg\chi \to \psi))$ 
& \eqref{vpdndpsi-to-l}\\[1mm]
(2) & $\vdash (\neg\chi\to \vp)\dnd \chi \sau \vp$  & \eqref{vpsaupsi-negvptopsi}\\[1mm]
(3) & $\vdash (\neg\chi \to \psi) \dnd  \chi \sau  \psi$  & \eqref{vpsaupsi-negvptopsi}\\[1mm]
(4) & $\vdash ((\neg\chi\to \vp) \dnd (\neg\chi \to \psi)) \to (\chi \sau \vp \dnd \chi \sau  \psi)$ 
& \eqref{Gamma-vpdndvpp-and-Gamma-psidndpsip-implies-Gamma-vpdndpsitovppdndpsip}: (4), (5)\\[1mm]
(5) & $\vdash (\vp \dnd \psi) \to (\chi \sau \vp \dnd \chi \sau  \psi)$ 
& \prule{syllogism}: (1), (4) \\[2mm]
\end{tabular}

\eqref{vpdndpsi-si-r}:\\[1mm]
\begin{tabular}{lll}
(1) & $\vdash  (\vp \dnd \psi) \to (\neg\vp \dnd \neg \psi)$ &  \eqref{vpdndpsi-neg}\\[1mm]
(2) & $\vdash (\neg\vp \dnd \neg \psi) \to (\neg\vp\sau \neg\chi \dnd \neg\psi\sau \neg\chi)$ 
& \eqref{vpdndpsi-sau-r}\\[1mm]
(3) & $\vdash (\vp \dnd \psi)\to (\neg\vp\sau \neg\chi \dnd \neg\psi\sau \neg\chi)$ 
& \prule{syllogism}: (1), (2) \\[1mm]
(4) & $\vdash (\neg\vp\sau \neg\chi \dnd \neg\psi\sau \neg\chi) \to 
(\neg(\neg\vp\sau \neg\chi) \dnd \neg(\neg\psi\sau \neg\chi))$ & \eqref{vpdndpsi-neg}\\[1mm]
(5) & $\vdash (\vp \dnd \psi)\to (\neg(\neg\vp\sau \neg\chi) \dnd \neg(\neg\psi\sau \neg\chi))$ 
& \prule{syllogism}: (3), (4) \\[1mm]
(6) & $\vdash \neg(\neg\vp\sau \neg\chi) \dnd \vp \si \chi$  & \eqref{vpsipsi-demorgan}\\[1mm]
(7) & $\vdash \neg(\neg\psi\sau \neg\chi) \dnd \psi \si \chi$  & \eqref{vpsipsi-demorgan}\\[1mm]
(8) & $\vdash (\neg(\neg\vp\sau \neg\chi) \dnd \neg(\neg\psi\sau \neg\chi)) \to (\vp \si \chi \dnd 
\psi \si \chi )$ 
& \eqref{Gamma-vpdndvpp-and-Gamma-psidndpsip-implies-Gamma-vpdndpsitovppdndpsip}: (6), (7)\\[1mm]
(9) & $\vdash (\vp \dnd \psi) \to (\vp \si \chi \dnd \psi \si \chi)$ 
& \prule{syllogism}: (5), (8) \\[2mm]
\end{tabular}

\eqref{vpdndpsi-si-l}:\\[1mm]
\begin{tabular}{lll}
(1) & $\vdash  (\vp \dnd \psi) \to (\neg\vp \dnd \neg \psi)$ &  \eqref{vpdndpsi-neg}\\[1mm]
(2) & $\vdash (\neg\vp \dnd \neg \psi) \to (\neg\chi\sau \neg\vp \dnd \neg\chi \sau \neg\psi)$ 
& \eqref{vpdndpsi-sau-l}\\[1mm]
(3) & $\vdash (\vp \dnd \psi)\to (\neg\chi\sau \neg\vp \dnd \neg\chi \sau \neg\psi)$ 
& \prule{syllogism}: (1), (2) \\[1mm]
(4) & $\vdash (\neg\chi\sau \neg\vp \dnd \neg\chi \sau \neg\psi) \to 
(\neg(\neg\chi\sau \neg\vp) \dnd \neg(\neg\chi\sau \neg\psi))$ & \eqref{vpdndpsi-neg}\\[1mm]
(5) & $\vdash (\vp \dnd \psi)\to (\neg(\neg\chi\sau \neg\vp) \dnd \neg(\neg\chi\sau \neg\psi))$ 
& \prule{syllogism}: (3), (4) \\[1mm]
(6) & $\vdash \neg(\neg\chi\sau \neg\vp) \dnd \chi\si \vp $  & \eqref{vpsipsi-demorgan}\\[1mm]
(7) & $\vdash \neg(\neg\chi\sau \neg\psi) \dnd \chi\si\psi$   & \eqref{vpsipsi-demorgan}\\[1mm]
(8) & $\vdash (\neg(\neg\chi\sau \neg\vp) \dnd \neg(\neg\chi\sau \neg\psi))\to (\chi\si \vp \dnd 
\chi\si \psi)$ 
& \eqref{Gamma-vpdndvpp-and-Gamma-psidndpsip-implies-Gamma-vpdndpsitovppdndpsip}: (6), (7)\\[1mm]
(9) & $\vdash (\vp \dnd \psi) \to (\chi\si \vp \dnd \chi\si \psi)$ 
& \prule{syllogism}: (5), (8) 
\end{tabular}

\,
}

\begin{lemma}\label{lemma-Gamma-vpdndpsi-neg-or-sau-si}
\begin{align}
\Gamma \vdash \vp \dnd \psi  & \quad \text{implies} 
\quad \Gamma \vdash \neg\vp \dnd \neg \psi, \label{Gamma-dnd-cong-neg}\\
\Gamma \vdash \vp \dnd \psi  & \quad \text{implies} 
\quad \Gamma \vdash (\vp \to \chi) \dnd  (\psi \to \chi), \label{sim-pairs-to-left}\\
\Gamma \vdash \vp \dnd \psi  & \quad \text{implies} 
\quad \Gamma \vdash  (\chi \to \vp) \dnd  (\chi  \to \psi), \label{sim-pairs-to-right}\\
\Gamma \vdash \vp \dnd \psi  & \quad \text{implies} 
\quad \Gamma \vdash \vp \sau \chi \dnd  \psi \sau \chi, \label{sim-pairs-or-left}\\
\Gamma \vdash \vp \dnd \psi  & \quad \text{implies} 
\quad \Gamma \vdash  \chi \sau \vp \dnd  \chi \sau \psi, \label{sim-pairs-or-right}\\
\Gamma \vdash \vp \dnd \psi  & \quad \text{implies} 
\quad \Gamma \vdash  \vp \si \chi \dnd  \psi\si \chi, \label{sim-pairs-and-left}\\
\Gamma \vdash \vp \dnd \psi  & \quad \text{implies} 
\quad \Gamma \vdash \chi \si \vp \dnd  \chi \si \psi. \label{sim-pairs-and-right}
\end{align}
\end{lemma}
\solution{It is an immediate consequence of Lemma \ref{lemma-vpdndpsi-neg-or-sau-si}.

\,
}

\subsection{First-order connectives}

\begin{lemma}
\begin{align}
\vdash & \, \forall x\vp \to \vp, \label{forall-quant-x}\\
\vdash  & \, \vp \to \exists x\vp, \label{exists-quant-x}\\
\vdash & \, \forall x\vp \to \exists x\vp. \label{forall-vp-to-exists-vp}
\end{align}
\end{lemma}
\solution{\,\\[1mm]
\eqref{forall-quant-x}:  Apply  \prule{$\forall$-Quantifier} with $y:=x$, as $x$ is free 
for $x$ in $\vp$ and $\subf{x}{x}{\vp}=\vp$.\\[1mm]
\eqref{exists-quant-x}:  Apply  \prule{$\exists$-Quantifier} with $y:=x$, $x$ is free 
for $x$ in $\vp$ and $\subf{x}{x}{\vp}=\vp$.\\[1mm]
\eqref{forall-vp-to-exists-vp}: \\[1mm]
\begin{tabular}{lll}
(1) & $\vdash\forall x\vp \to \vp$ & \eqref{forall-quant-x}\\[1mm]
(2) & $\vdash\vp \to \exists x\vp $  & \eqref{exists-quant-x}\\[1mm]
(3) & $\vdash \forall x\vp \to \exists x\vp$  & \prule{syllogism}: (1), (2)
\end{tabular}

\,
}

\begin{lemma}
\begin{align}
\vdash & \, \exists x \vp \to \neg \forall x \neg \vp, \label{existsxvp-to-notforallxnotvp}\\
\vdash & \, \forall  x \neg\vp \to \neg \exists x \vp, \label{forallxnegvp-to-negexistsxvp}\\
\vdash & \, (\exists x \vp \to \psi) \to \forall x (\vp\to \psi), \quad \text{if~}x\notin FV(\psi) \label{existsxvp-to-psi-tto-forallx-vptopsi}
\end{align}
\end{lemma}
\solution{\,\\[1mm]
\eqref{existsxvp-to-notforallxnotvp}: \\[1mm]
\begin{tabular}{lll}
(1) & $\vdash \forall x \neg \vp \to \neg\vp$ & \eqref{forall-quant-x}\\[1mm]
(2) & $\vdash\vp\to \neg \forall x \neg \vp$ & \eqref{Gamma-vptonegpsi-implies-Gamma-psitonegvp}: (1)\\[1mm]
(3) & $\vdash \exists x \vp\to \neg \forall x \neg \vp$  & \prule{$\exists$-quantifier rule}: (2), as 
$x\notin FV(\neg \forall x \neg \vp)$ \\[2mm]
\end{tabular}

\eqref{forallxnegvp-to-negexistsxvp}: \\[1mm]
\begin{tabular}{lll}
(1) & $\vdash \exists x \vp\to \neg \forall x \neg \vp$ & \eqref{existsxvp-to-notforallxnotvp}\\[1mm]
(2) & $\vdash \forall  x \neg\vp \to \neg \exists x \vp$ & \eqref{Gamma-vptonegpsi-implies-Gamma-psitonegvp}: (1)\\[2mm]
\end{tabular}

\eqref{existsxvp-to-psi-tto-forallx-vptopsi}: \\[1mm]
\begin{tabular}{lll}
(1) & $\vdash \vp\to \exists x \vp$ & \eqref{exists-quant-x} \\[1mm]
(2) & $\vdash(\exists x \vp \to \psi) \to(\vp\to \psi)$ & \eqref{vp-to-psi-imp-psi-to-chi-tto-vp-to-chi}: (1) \\[1mm]
(3) & $\vdash(\exists x \vp \to \psi) \to \forall x (\vp\to \psi)$ & \prule{$\forall$-quantifier rule}: (2), as $x\notin FV(\exists x \vp \to \psi)$
\end{tabular}

\,
}

\begin{lemma}
\begin{align}
\Gamma \vdash \vp \to \psi & \quad \text{implies} \quad \Gamma \vdash \forall x \vp \to \forall x \psi, 
\label{vp-ra-psi-forall-vp-ra-forall-psi}\\
\Gamma \vdash \vp \dnd \psi & \quad \text{implies} \quad  \Gamma \vdash \forall x \vp \dnd \forall x \psi,
\label{vp-dnd-psi-forall-vp-dnd-forall-psi}\\
\Gamma \vdash \vp \to \psi & \quad \text{implies} \quad  \Gamma \vdash \exists x \vp \to \exists x \psi,
\label{vp-ra-psi-exists-vp-ra-exists-psi}\\
\Gamma \vdash \vp \dnd \psi & \quad \text{implies} \quad  \Gamma \vdash \exists x \vp \dnd \exists x \psi. 
\label{vp-dnd-psi-exists-vp-dnd-exists-psi}
\end{align}
\end{lemma}
\solution{\,\\[1mm]
\eqref{vp-ra-psi-forall-vp-ra-forall-psi}:\\[1mm]
\begin{tabular}{lll}
(1) & $\Gamma \vdash \forall x \vp \to \vp$  & \eqref{forall-quant-x} \\[1mm]
(2) & $\Gamma \vdash \vp \to \psi$ & (Assumption) \\[1mm]
(3) & $\Gamma \vdash \forall x\vp \to \psi$  & \prule{syllogism}: (1), (2)\\[1mm]
(4) & $\Gamma \vdash \forall x \vp \to \forall x \psi$  & \prule{$\forall$-quantifier rule}: (3), as $x\notin FV(\forall x \vp)$\\[2mm]
\end{tabular}

\eqref{vp-dnd-psi-forall-vp-dnd-forall-psi}:\\[1mm]
\begin{tabular}{lll}
(1) & $\Gamma \vdash \vp \dnd \psi$ & (Assumption) \\[1mm]
(2) & $\Gamma \vdash \vp\to\psi$  & \eqref{Gamma-vp-psi-vp-si-psi}: (1) \\[1mm]
(3) & $\Gamma \vdash \psi\to\vp$  & \eqref{Gamma-vp-psi-vp-si-psi}: (1) \\[1mm]
(4) & $\Gamma \vdash \forall x \vp\to\forall x \psi$  & \eqref{vp-ra-psi-forall-vp-ra-forall-psi}: (2) \\[1mm]
(5) & $\Gamma \vdash \forall x \psi \to\forall x \vp$  & \eqref{vp-ra-psi-forall-vp-ra-forall-psi}: (3) \\[1mm]
(6) & $\Gamma \vdash\forall x \vp\dnd\forall x \psi$  & \eqref{Gamma-vp-psi-vp-si-psi}: (4), (5)\\[2mm]
\end{tabular}

\eqref{vp-ra-psi-exists-vp-ra-exists-psi}:\\[1mm]
\begin{tabular}{lll}
(1) & $\Gamma \vdash \vp \to \psi$ & (Assumption) \\[1mm]
(2) & $\Gamma \vdash \psi\to \exists x\psi$  & \eqref{exists-quant-x} \\[1mm]
(3) & $\Gamma \vdash \vp \to \exists x\psi$  & \prule{syllogism}: (1), (2)\\[1mm]
(4) & $\Gamma \vdash \exists x \vp \to \exists x\psi$  & \prule{$\exists$-quantifier rule}: (3),
as $x\notin FV(\exists x \psi)$\\[2mm]
\end{tabular}

\eqref{vp-dnd-psi-exists-vp-dnd-exists-psi}:\\[1mm]
\begin{tabular}{lll}
(1) & $\Gamma \vdash \vp \dnd \psi$ & (Assumption) \\[1mm]
(2) & $\Gamma \vdash \vp\to\psi$  & \eqref{Gamma-vp-psi-vp-si-psi}: (1) \\[1mm]
(3) & $\Gamma \vdash \psi\to\vp$  & \eqref{Gamma-vp-psi-vp-si-psi}: (1) \\[1mm]
(4) & $\Gamma \vdash \exists x \vp \to \exists x \psi$  & \eqref{vp-ra-psi-exists-vp-ra-exists-psi}: (2) \\[1mm]
(5) & $\Gamma \vdash \exists x \psi \to \exists x \vp$  & \eqref{vp-ra-psi-exists-vp-ra-exists-psi}: (3) \\[1mm]
(6) & $\Gamma \vdash\exists x \vp\dnd \exists x \psi$  & \eqref{Gamma-vp-psi-vp-si-psi}: (4), (5)
\end{tabular}

\,
}

\begin{lemma}
\begin{align}
\vdash & \, \forall  x \vp \to \neg \exists x \neg \vp, \label{forallxvp-to-notexistsxnotvp}\\
\vdash & \,  \neg \exists x \neg \vp \dnd  \forall x \neg\neg  \vp, \label{forallxnegnegvpforall-dnd-negexistsxnegvp}\\
\vdash & \, \neg\exists x\neg\vp\to\forall x\vp. \label{axiom-forall-2-Gc}
\end{align}
\end{lemma}
\solution{\,\\[1mm]
\eqref{forallxvp-to-notexistsxnotvp}:\\[1mm]
\begin{tabular}{lll}
(1) & $\vdash \vp\to \neg\neg\vp$  & \eqref{dni}\\[1mm]
(2) & $\vdash \forall x \vp\to \forall x \neg\neg\vp$  & \eqref{vp-ra-psi-forall-vp-ra-forall-psi}: (1)\\[1mm]
(3) & $\vdash  \forall x \neg\neg\vp \to \neg \exists x \neg \vp$  & \eqref{forallxnegvp-to-negexistsxvp}: (1)\\[1mm]
(4) & $\vdash \forall  x \vp \to \neg \exists x \neg \vp$ & \prule{syllogism}: (2), (3)\\[2mm]
\end{tabular}

\eqref{forallxnegnegvpforall-dnd-negexistsxnegvp}:\\[1mm]
\begin{tabular}{lll}
(1) & $\vdash(\exists x\neg\vp\to\bot)\to\forall x(\neg\vp\to\bot)$ & \eqref{existsxvp-to-psi-tto-forallx-vptopsi}, as $x\notin FV(\bot)$\\[1mm]
(2) & $\vdash\neg\exists x\neg\vp\to(\exists x\neg\vp\to\bot)$ & \prule{axiom-not-1}\\[1mm]
(3) & $\vdash\neg\exists x\neg\vp\to\forall x(\neg\vp\to\bot)$ & \prule{syllogism}: (2), (1)\\[1mm]
(4) & $\vdash(\neg\vp\to\bot)\to\neg\neg\vp$ & \prule{axiom-not-2}\\[1mm]
(5) & $\vdash\forall x(\neg\vp\to\bot)\to\forall x\neg\neg\vp$ & \eqref{vp-ra-psi-forall-vp-ra-forall-psi}: (4)\\[1mm]
(6) & $\vdash\neg\exists x\neg\vp\to\forall x\neg\neg\vp$ & \prule{syllogism}: (3), (5)\\[1mm]
(7) & $\vdash\forall x\neg\neg\vp\to\neg\exists x\neg\vp$ & \eqref{forallxnegvp-to-negexistsxvp}\\[1mm]
(8) & $\vdash \forall x \neg\neg\vp  \dnd \neg \exists x \neg \vp$ & \eqref{Gamma-vp-psi-vp-si-psi}: (6), (7) and the definition of $\dnd$\\[2mm]
\end{tabular}

\eqref{axiom-forall-2-Gc}:\\[1mm]
\begin{tabular}{lll}
(1) & $\vdash\neg \exists x \neg \vp \dnd \forall x \neg\neg  \vp$ & \eqref{forallxnegnegvpforall-dnd-negexistsxnegvp}\\[1mm]
(2) & $\vdash\neg\exists x\neg\vp\to\forall x\neg\neg\vp$ & \eqref{Gamma-vp-psi-vp-si-psi}: (1) and the definition of $\dnd$\\[1mm]
(3) & $\vdash\neg\neg\vp\to\vp$ & \eqref{negnegvp-vp}\\[1mm]
(4) & $\vdash\forall x\neg\neg\vp\to\forall x\vp$ & \eqref{vp-ra-psi-forall-vp-ra-forall-psi}: (5)\\[1mm]
(5) & $\vdash\neg\exists x\neg\vp\to\forall x\vp$ & \prule{syllogism}: (2), (4)
\end{tabular}

\,
}

\begin{lemma}
\begin{align}
\vdash  \, \vp \dnd \forall x \vp  & \quad  \text{if } x\notin FV(\vp), \label{vp-dnd-forall-vp}\\
\vdash  \,  \vp \dnd \exists x \vp & \quad \text{if } x\notin FV(\vp). \label{vp-dnd-exists-vp}
\end{align}
\end{lemma}
\solution{\,\\[1mm]
\eqref{vp-dnd-forall-vp}: \\[1mm]
\begin{tabular}{lll}
(1) & $\vdash \vp \to \vp$  & \eqref{vp-to-vp}\\[1mm]
(2) & $\vdash \vp \to \forall x\vp$  & \prule{$\forall$-quantifier rule}: (1), as $x\notin FV(\vp)$\\[1mm]
(3) & $\vdash \forall x\vp \to \vp$  & \eqref{forall-quant-x} \\[1mm]
(4) & $\vdash \vp \dnd \forall x\vp$ & \eqref{Gamma-vp-psi-vp-si-psi}: (2), (3)\\[2mm]
\end{tabular}

\eqref{vp-dnd-exists-vp}: \\[1mm]
\begin{tabular}{lll}
(1) & $\vdash \vp \to \vp$  & \eqref{vp-to-vp}\\[1mm]
(2) & $\vdash \exists x \vp \to \vp$  & \prule{$\exists$-quantifier rule}: (1), as $x\notin FV(\vp)$\\[1mm]
(3) & $\vdash \vp \to \exists x\vp$  & \eqref{exists-quant-x} \\[1mm]
(4) & $\vdash \vp \dnd \exists x \vp$ & \eqref{Gamma-vp-psi-vp-si-psi}: (2), (3)
\end{tabular}

\,
}

\blem\label{forallxvp-dnd-subfxyvp-Gc}
Let $\vp$ be a pattern and $x$, $y$ be element variables such that 
\bce 
$x$ does not occur bound in $\vp$ and $y$ does not occur in $\vp$.
\ece
Then 
\be 
\item\label{forallxvp-to-subfxyvp-Gc} 
$\vdash \,\forall x \vp \to \forall y\subf{x}{y}{\vp}$;
\item\label{subfxyvp-to-forallxvp-Gc}
$\vdash \,\forall y\subf{x}{y}{\vp} \to \forall x \vp$;
\item\label{forallxvp-dns-subfxyvp-Gc}
$\vdash \,\forall x \vp \dnd \forall y\subf{x}{y}{\vp}$.
\ee
\elem
\solution{The case $x=y$ is obvious, as $\subf{x}{x}{\vp}=\vp$. Assume in the sequel that $x$, $y$ 
are distinct.

\be
\item As, by hypothesis, $y$ does not occur in $\vp$, we have, by Lemma \ref{lemma-useful-y-noccur-vp-imp-y-freefor-x-abstractML}, 
that $x$ is free for $y$ in $\vp$. We get that \\

\begin{tabular}{lll}
(1) & $\vdash \forall x \vp \to \subf{x}{y}{\vp}$ & \prule{$\forall$-Quantifier}\\[1mm]
(2) & $\vdash \forall y\forall x \vp \to \forall y\subf{x}{y}{\vp}$ & \eqref{vp-ra-psi-forall-vp-ra-forall-psi}: (1)\\[1mm]
(3) & $\vdash \forall x \vp \to \forall y\forall x \vp $ & \eqref{vp-dnd-forall-vp} as $y$ does not occur in  $\forall x \vp$, hence 
$y \notin \FVE(\forall x \vp)$\\ 
&&  and \eqref{Gamma-vp-psi-vp-si-psi}\\[1mm]
(4) & $\vdash \forall x \vp \to \forall y\subf{x}{y}{\vp}$ & \prule{syllogism}: (3), (2) \\[2mm]
\end{tabular}

\item Let us denote $\psi:=\subf{x}{y}{\vp}$. By hypothesis, $y$ does not occur in $\vp$. In particular, $y$ does not occur bound 
in $\vp$. We can apply Lemma \ref{subf-lemma-useful-subfxy-1-abstractML} to get that $y$ does not occur bound in $\psi$.

By hypothesis, $x$ does not occur bound in $\vp$. We can apply Lemma \ref{subf-lemma-useful-subfxy-3-abstractML} to get that 
$x$ does not occur in $\psi$.

Apply \eqref{forallxvp-to-subfxyvp} with $x:=y$, $y:=x$, $\vp:=\psi$ to get that 
$$\vdash \forall y \psi \to \forall x \subf{y}{x}{\psi},$$
that is 
$$\vdash \forall y\subf{x}{y}{\vp} \to \forall x \subf{y}{x}{\subf{x}{y}{\vp}}.$$
As, by hypothesis, $y$ does not occur in $\vp$, we can apply Lemma \ref{subf-lemma-useful-subfxy-2-abstractML} to get 
$\subf{y}{x}{\subf{x}{y}{\vp}}=\vp$. Thus, \eqref{subfxyvp-to-forallxvp} follows. \\

\item Apply \eqref{forallxvp-to-subfxyvp}, \eqref{subfxyvp-to-forallxvp}, the definition of $\dnd$ and \eqref{Gamma-vp-psi-vp-si-psi}.
\ee
}

\subsection{Application}\label{thms-rules-application}

\begin{proposition}
\begin{align}
\Gamma\vdash\vp\dnd\psi \,\, \text{and}  \,\, \Gamma \vdash \chi\dnd\gamma &
\quad \text{implies} \quad \Gamma \vdash \appln{\vp}{\chi} \dnd \appln{\psi}{\gamma}, 
\label{rule-iff-compat-in-app-Gc}\\
\Gamma\vdash\vp\dnd\psi  &
\quad \text{implies} \quad \Gamma \vdash \appln{\vp}{\chi} \dnd \appln{\psi}{\chi}, 
\label{rule-iff-compat-in-app-left-Gc}\\
\Gamma \vdash \chi\dnd\gamma  &
\quad \text{implies} \quad \Gamma \vdash \appln{\vp}{\chi}\dnd \appln{\vp}{\gamma} .
\label{rule-iff-compat-in-app-right-Gc}
\end{align}
\end{proposition}
\solution{
\eqref{rule-iff-compat-in-app-Gc}:\\[1mm]
\begin{tabular}{lll}
(1) & $\Gamma\vdash\vp\dnd\psi$ & (Assumption)\\[1mm]
(2) & $\Gamma\vdash\vp\to\psi$ & \eqref{Gamma-vp-psi-vp-si-psi}: (1) and the definition of $\dnd$\\[1mm]
(3) & $\Gamma\vdash\appln{\vp}{\chi}\to\appln{\psi}{\chi}$ & \prule{framing-left}: (2)\\[1mm]
(4) & $\Gamma\vdash\chi\dnd\gamma$ & (Assumption)\\[1mm]
(5) & $\Gamma\vdash\chi\to\gamma$ & \eqref{Gamma-vp-psi-vp-si-psi}: (4) and the definition of $\dnd$\\[1mm]
(6) & $\Gamma\vdash\appln{\psi}{\chi}\to\appln{\psi}{\gamma}$ & \prule{framing-right}: (5)\\[1mm]
(7) & $\Gamma\vdash\appln{\vp}{\chi}\to\appln{\psi}{\gamma}$ & \prule{syllogism}: (3), (6)\\[1mm]
(8) & $\Gamma\vdash\vp\dnd\psi$ & (Assumption)\\[1mm]
(9) & $\Gamma\vdash\psi\to\vp$ & \eqref{Gamma-vp-psi-vp-si-psi}: (9) and the definition of $\dnd$\\[1mm]
(10) & $\Gamma\vdash\appln{\psi}{\gamma}\to\appln{\vp}{\gamma}$ & \prule{framing-left}: (9)\\[1mm]
(11) & $\Gamma\vdash\gamma\to\chi$ & \eqref{Gamma-vp-psi-vp-si-psi}: (4) and the definition of $\dnd$\\[1mm]
(12) & $\Gamma\vdash\appln{\vp}{\gamma}\to\appln{\vp}{\chi}$ & \prule{framing-right}: (11)\\[1mm]
(13) & $\Gamma\vdash\appln{\psi}{\gamma}\to\appln{\vp}{\chi}$ & \prule{syllogism}: (10), (12)\\[1mm]
(14) & $\Gamma\vdash\appln{\vp}{\chi}\dnd\appln{\psi}{\gamma}$ & \eqref{Gamma-vp-psi-vp-si-psi}: (7), (13) and the definition of $\dnd$\\[2mm]
\end{tabular}

\eqref{rule-iff-compat-in-app-left-Gc} We can apply \eqref{rule-iff-compat-in-app-Gc} with $\gamma:=\chi$,  as 
$\vdash \chi \dnd \chi$, by \eqref{sim-refl}.\\[1mm]
\eqref{rule-iff-compat-in-app-right} Apply \eqref{rule-iff-compat-in-app-Gc} with $\psi:=\vp$, as 
$\vdash \vp \dnd \vp$, by \eqref{sim-refl}.

\,
}

\subsection{A congruence}

Let $\Gamma$ be a set of patterns.

\bprop\label{cong-dnd-Gc}
Define the binary relation $\eqreplGdnd$ on the set of patterns as follows: for all patterns $\vp$, $\psi$,
\begin{align*}
\eqreplGdnda{\vp}{\psi} \quad \text{iff} \quad \Gamma \vdash \vp \dnd\psi.
\end{align*}
Then $\eqreplGdnd$ is a congruence that is also an equivalence relation.
\eprop
\solution{We have that: \\[1mm]
$\eqreplGdnd$ is compatible with $\neg$, by \eqref{Gamma-dnd-cong-neg}.\\
$\eqreplGdnd$ is compatible with $\sau$, by \eqref{sim-pairs-or-left} and \eqref{sim-pairs-or-right}.\\
$\eqreplGdnd$ is compatible with $\si$, by \eqref{sim-pairs-and-left} and \eqref{sim-pairs-and-right}.\\
$\eqreplGdnd$ is compatible with $\to$, by \eqref{sim-pairs-to-left} and \eqref{sim-pairs-to-right}.\\
$\eqreplGdnd$ is compatible with $\forall$,  by \eqref{vp-dnd-psi-forall-vp-dnd-forall-psi}.\\
$\eqreplGdnd$ is compatible with $\exists$,  by \eqref{vp-dnd-psi-exists-vp-dnd-exists-psi}. \\
$\eqreplGdnd$ is compatible with $\Appl$,  by \eqref{rule-iff-compat-in-app-left-Gc} and 
\eqref{rule-iff-compat-in-app-right-Gc}.

Thus, $\eqreplGdnd$ is a congruence.\\[1mm]
Furthermore, \\[1mm]
$\eqreplGdnd$ is reflexive, by \eqref{sim-refl}. \\
$\eqreplGdnd$ is symmetric, by \eqref{sim-symm}. \\
$\eqreplGdnd$ is transitive, by \eqref{sim-trans}.

Thus, $\eqreplGdnd$ is also an equivalence relation. 
}
\subsection{Replacement theorems}

Let $\Gamma$ be a set of patterns.

\bthm[Replacement Theorem for  contexts]\label{replacement-thm-contexts-Gc}$\,$\\
For any context  $\fcontext$ and any patterns $\vp,\psi$,
\bce
$\Gamma \vdash \vp \dnd\psi$ implies $\Gamma \vdash \fcontext[\vp] \dnd \fcontext[\psi]$.
\ece
\ethm
\solution{Apply Proposition \ref{cong-dnd-Gc} and 
Theorem \ref{replacement-thm-contexts-abstractML}.
}

\bthm[Replacement Theorem]\label{replacement-thm-Gc}$\,$\\
Let $\vp,\psi, \chi,\theta$ be patterns such that $\vp$ is a subpattern of $\chi$ and $\theta$ is obtained
from $\chi$ by replacing one or more occurrences of $\vp$ with $\psi$. Then 
\bce
$\Gamma \vdash \vp \dnd\psi$ implies $\Gamma \vdash  \chi \dnd \theta$.
\ece
\ethm 
\solution{Apply Proposition \ref{cong-dnd-Gc} and 
Theorem \ref{replacement-thm-more-occurence-abstractML}.
}
\subsection{Bounded substitution theorem}

\bthm\label{thm-vp-dnd-subbxybvp-Gc}
For any pattern $\vp$ and any variables $x$, $y$ such that $y$ does not occur in $\vp$
\begin{align*}
\vdash \vp\dnd \subbf{x}{y}{\vp}. 
\end{align*}
\ethm
\solution{Apply Proposition \ref{cong-dnd-Gc},  
Lemma \ref{forallxvp-dnd-subfxyvp-Gc}.\eqref{forallxvp-dns-subfxyvp-Gc} 
and Theorem \ref{thm-vp-cong-subbxybvp-abstractML}.
}

\subsection{Application more}

\begin{lemma}
For every patterns $\vp$, $\psi$,
\begin{align}
\appln{(\exists x\vp)}{\psi} \to \exists x(\appln{\vp}{\psi}) & 
 \quad \text{if~} x\notin \FVE(\psi),\label{propagation-exists-1-H-Gc}\\[1mm]
 \appln{\psi}{(\exists x\vp)} \to \exists x(\appln{\psi}{\vp}) & 
 \quad \text{if~} x\notin \FVE(\psi). \label{propagation-exists-2-H-Gc}
\end{align}
\end{lemma}
\solution{Assume that $x\notin \FVE(\psi)$ and let $y$ be a new variable, distinct from $x$  
and not occuring in $\psi$. Let us denote $\delta:=\subb{x}{y}{\psi}$. 
By Lemma \ref{subb-lemma-useful-x-noccur-free-vp-noccur-subbxyvp-abstractML}, we have that 
$x$ does not  occur in $\delta$. \\[1mm]
\eqref{propagation-exists-1-H-Gc}: \\[1mm]
\bt{lll}
(1) & $\vdash \appln{(\exists x\vp)}{\delta} \to \exists x(\appln{\vp}{\delta})$ & \prule{Propagation$_\exists$} \\[1mm]
(2) & $\vdash \psi \dnd \delta$ & Theorem \ref{thm-vp-dnd-subbxybvp-Gc}\\[1mm]
(3) & $\vdash \bigg(\appln{(\exists x\vp)}{\psi} \to \exists x(\appln{\vp}{\psi})\bigg)  \dnd
\bigg(\appln{(\exists x\vp)}{\delta} \to \exists x(\appln{\vp}{\delta})\bigg)$ & 
Theorem \ref{replacement-thm-Gc}: (2)\\[1mm]
(4) & $\vdash \bigg(\appln{(\exists x\vp)}{\delta} \to \exists x(\appln{\vp}{\delta})\bigg) \to 
\bigg(\appln{(\exists x\vp)}{\psi} \to \exists x(\appln{\vp}{\psi})\bigg)$ & \eqref{vpdndpsi-imp-vp-iff-psi}: (3) \\[1mm]
(5) & $\vdash\appln{(\exists x\vp)}{\psi} \to \exists x(\appln{\vp}{\psi})$ &  
\prule{modus ponens}: (1), (4)\\[2mm]
\et

\eqref{propagation-exists-2-H-Gc}: \\[1mm]
\bt{lll}
(1) & $\vdash \appln{\delta}{(\exists x\vp)} \to \exists x(\appln{\delta}{\vp})$ & \prule{Propagation$_\exists$} \\[1mm]
(2) & $\vdash \psi \dnd \delta$ & Theorem \ref{thm-vp-dnd-subbxybvp-Gc}\\[1mm]
(3) & $\vdash \bigg(\appln{\psi}{(\exists x\vp)} \to \exists x(\appln{\psi}{\vp})\bigg)  \dnd
\bigg(\appln{\delta}{(\exists x\vp)} \to \exists x(\appln{\delta}{\vp})\bigg)$ & 
Theorem \ref{replacement-thm-Gc}: (2)\\[1mm]
(4) & $\vdash \bigg(\appln{\delta}{(\exists x\vp)} \to \exists x(\appln{\delta}{\vp})\bigg) \to 
\bigg(\appln{\psi}{(\exists x\vp)} \to \exists x(\appln{\psi}{\vp})\bigg)$ & \eqref{vpdndpsi-imp-vp-iff-psi}: (3) \\[1mm]
(5) & $\vdash\appln{\psi}{(\exists x\vp)} \to \exists x(\appln{\psi}{\vp})$ &  
\prule{modus ponens}: (1), (4)
\et

\,
}

\section{Equivalence of the proof systems $\appMLGc$ and $\appMLnow$}

In the sequel, we prove that the proof system $\cG^c$
is equivalent with the proof system $\appMLnow$, given in Appendix \ref{appML-now}.

\subsection{Some theorems and derived rules in $\appMLnow$}

In this section we write simply $\vdash$ instead of $\vdash_\appMLnow$.

\subsubsection{Propositional connectives}

In the following, $\Gamma$ is an arbitrary set of patterns and $\vp$, $\psi$, $\chi$ are arbitrary patterns.

\bprop
\bea
\Gamma\vdash\vp\to\psi  \,\, \text{and}  \,\, \Gamma\vdash\psi\to\chi & \text{imply} &
 \Gamma \vdash \vp\to\chi,
\label{syllogism'}\\
\Gamma\vdash\vp\to\psi & \text{implies} &\Gamma\vdash\chi\sau\vp\to\chi\sau\psi,
\label{expansion'}\\
\Gamma\vdash \vp\to(\psi\to\chi) \,\, \text{and}  \,\, \Gamma\vdash\vp\to\psi & \text{imply} & 
\Gamma\vdash\vp\to\chi, \label{ax2-rule'}\\
\Gamma\vdash \vp\to(\psi\to\chi) & \text{implies} & \Gamma\vdash\psi\to(\vp\to\chi),
\label{premise-comm'}\\
\Gamma\vdash \vp\to\psi & \text{implies} & \Gamma\vdash(\chi\to\vp)\to(\chi\to\psi), \label{extra-premise-alt'}\\
\Gamma\vdash  \vp\to(\psi\to\chi)\,\, \text{and}  \,\,\Gamma\vdash\chi\to\gamma & \text{imply} & 
\Gamma\vdash  \vp\to(\psi\to\gamma). \label{imp-trans-alt'}
\eea
\eprop
\solution{\,\\[1mm]
\eqref{syllogism'}:\\[1mm]
\begin{tabular}{lll}
(1) & $\Gamma\vdash(\vp\to(\psi\to\chi))\to((\vp\to\psi)\to(\vp\to\chi))$ & \prule{axiom-2}\\[1mm]
(2) & $\Gamma\vdash(\psi\to\chi)\to(\vp\to(\psi\to\chi))$ & \prule{axiom-1}\\[1mm]
(3) & $\Gamma\vdash\psi\to\chi$ & (Assumption)\\[1mm]
(4) & $\Gamma\vdash\vp\to(\psi\to\chi)$ & \prule{modus-ponens}: (3), (2)\\[1mm]
(5) & $\Gamma\vdash(\vp\to\psi)\to(\vp\to\chi)$ & \prule{modus-ponens}: (4), (1)\\[1mm]
(6) & $\Gamma\vdash\vp\to\psi$ & (Assumption)\\[1mm]
(7) & $\Gamma\vdash\vp\to\chi$ & \prule{modus-ponens}: (6), (5) \\[2mm]
\end{tabular}

\eqref{expansion'}:\\[1mm]
\begin{tabular}{lll}
(1) & $\Gamma\vdash\vp\to\psi$ & (Assumption)\\[1mm]
(2) & $\Gamma\vdash(\vp\to\psi)\to(\neg\chi\to(\vp\to\psi))$ & \prule{axiom-1}\\[1mm]
(3) & $\Gamma\vdash\neg\chi\to(\vp\to\psi)$ & \prule{modus-ponens}: (1), (2)\\[1mm]
(4) & $\Gamma\vdash(\neg\chi\to(\vp\to\psi))\to((\neg\chi\to\vp)\to(\neg\chi\to\psi))$ & \prule{axiom-2}\\[1mm]
(5) & $\Gamma\vdash(\neg\chi\to\vp)\to(\neg\chi\to\psi)$ & \prule{modus-ponens}: (3), (4)\\[1mm]
(6) & $\Gamma\vdash\chi\sau\vp\to(\neg\chi\to\vp)$ & \prule{axiom-or-1}\\[1mm]
(7) & $\Gamma\vdash\chi\sau\vp\to(\neg\chi\to\psi)$ & \eqref{syllogism'}: (6), (5)\\[1mm]
(8) & $\Gamma\vdash(\neg\chi\to\psi)\to\chi\sau\psi$ & \prule{axiom-or-2}\\[1mm]
(9) & $\Gamma\vdash\chi\sau\vp\to\chi\sau\psi$ & \eqref{syllogism'}: (7), (8)\\[2mm]
\end{tabular}

\eqref{ax2-rule'}:\\[1mm]
\begin{tabular}{lll}
(1) & $\Gamma\vdash(\vp\to(\psi\to\chi))\to((\vp\to\psi)\to(\vp\to\chi))$ & \prule{axiom-2}\\[1mm]
(2) & $\Gamma\vdash\vp\to(\psi\to\chi)$ & (Assumption)\\[1mm]
(3) & $\Gamma\vdash(\vp\to\psi)\to(\vp\to\chi)$ & \prule{modus-ponens}: (2), (1)\\[1mm]
(4) & $\Gamma\vdash\vp\to\psi$ & (Assumption)\\[1mm]
(5) & $\Gamma\vdash\vp\to\chi$ & \prule{modus-ponens}: (4), (3) \\[2mm]
\end{tabular}

\eqref{premise-comm'}:\\[1mm]
\begin{tabular}{lll}
(1) & $\Gamma\vdash(\vp\to(\psi\to\chi))\to((\vp\to\psi)\to(\vp\to\chi))$ & \prule{axiom-2}\\[1mm]
(2) & $\Gamma\vdash\vp\to(\psi\to\chi)$ & (Assumption)\\[1mm]
(3) & $\Gamma\vdash(\vp\to\psi)\to(\vp\to\chi)$ & \prule{modus-ponens}: (2), (1)\\[1mm]
(4) & $\Gamma\vdash(\psi\to((\vp\to\psi)\to(\vp\to\chi)))\to $ & \prule{axiom-2}\\[1mm]
& \quad $\to((\psi\to(\vp\to\psi))\to(\psi\to(\vp\to\chi)))$ \\[1mm]
(5) & $\Gamma\vdash((\vp\to\psi)\to(\vp\to\chi))\to(\psi\to$ & \prule{axiom-1}\\[1mm]
& \quad $\to((\vp\to\psi)\to(\vp\to\chi)))$ \\[1mm]
(6) & $\Gamma\vdash\psi\to((\vp\to\psi)\to(\vp\to\chi))$ & \prule{modus-ponens}: (3), (5)\\[1mm]
(7) & $\Gamma\vdash(\psi\to(\vp\to\psi))\to(\psi\to(\vp\to\chi))$ & \prule{modus-ponens}: (6), (4)\\[1mm]
(8) & $\Gamma\vdash\psi\to(\vp\to\psi)$ & \prule{axiom-1}\\[1mm]
(9) & $\Gamma\vdash\psi\to(\vp\to\chi)$ & \prule{modus-ponens}: (8), (7) \\[2mm]
\end{tabular}

\eqref{extra-premise-alt'}:\\[1mm]
\begin{tabular}{lll}
(1) & $\Gamma\vdash\vp\to\psi$ & (Assumption)\\[1mm]
(2) & $\Gamma\vdash(\vp\to\psi)\to(\chi\to(\vp\to\psi))$ & \prule{axiom-1}\\[1mm]
(3) & $\Gamma\vdash\chi\to(\vp\to\psi)$ & \prule{modus-ponens}: (1), (2)\\[1mm]
(4) & $\Gamma\vdash(\chi\to(\vp\to\psi))\to((\chi\to\vp)\to(\chi\to\psi))$ & \prule{axiom-2}\\[1mm]
(5) & $\Gamma\vdash(\chi\to\vp)\to(\chi\to\psi)$ & \prule{modus-ponens}: (3), (4) \\[2mm]
\end{tabular}

\eqref{imp-trans-alt'}:\\[1mm]
\begin{tabular}{lll}
(1) & $\Gamma\vdash\vp\to(\psi\to\chi)$ & (Assumption)\\[1mm]
(2) & $\Gamma\vdash(\vp\to(\psi\to\chi))\to((\vp\to\psi)\to(\vp\to\chi))$ & \prule{axiom-2}\\[1mm]
(3) & $\Gamma\vdash(\vp\to\psi)\to(\vp\to\chi)$ & \prule{modus-ponens}: (1), (2)\\[1mm]
(4) & $\Gamma\vdash\chi\to\gamma$ & (Assumption)\\[1mm]
(5) & $\Gamma\vdash(\chi\to\gamma)\to(\vp\to(\chi\to\gamma))$ & \prule{axiom-1}\\[1mm]
(6) & $\Gamma\vdash\vp\to(\chi\to\gamma)$ & \prule{modus-ponens}: (4), (5)\\[1mm]
(7) & $\Gamma\vdash(\vp\to(\chi\to\gamma))\to((\vp\to\chi)\to(\vp\to\gamma))$ & \prule{axiom-2}\\[1mm]
(8) & $\Gamma\vdash(\vp\to\chi)\to(\vp\to\gamma)$ & \prule{modus-ponens}: (6), (7)\\[1mm]
(9) & $\Gamma\vdash(\vp\to\psi)\to(\vp\to\gamma)$ & \eqref{syllogism'}: (3), (8)\\[1mm]
(10) & $\Gamma\vdash(\psi\to((\vp\to\psi)\to(\vp\to\gamma)))\to$ & \prule{axiom-2}\\[1mm]
& \quad $\to((\psi\to(\vp\to\psi))\to(\psi\to(\vp\to\gamma)))$ \\[1mm]
(11) & $\Gamma\vdash((\vp\to\psi)\to(\vp\to\gamma))\to$ & \prule{axiom-1}\\[1mm]
& \quad $((\psi\to(\vp\to\psi))\to(\vp\to\gamma))$ \\[1mm]
(12) & $\Gamma\vdash(\psi\to((\vp\to\psi))\to(\psi\to(\vp\to\gamma)))$ & \prule{modus-ponens}: (9), (11)\\[1mm]
(13) & $\Gamma\vdash\psi\to(\vp\to\psi)$ & \prule{axiom-1}\\[1mm]
(14) & $\Gamma\vdash\psi\to(\vp\to\gamma)$ & \prule{modus-ponens}: (13), (12)\\[1mm]
(15) & $\Gamma\vdash\vp\to(\psi\to\gamma)$ & \eqref{premise-comm'}: (14)
\end{tabular}

\,
}

\bprop
\begin{align}
\vdash &  \vp\to\vp, \label{imp-reflexivity'}\\
\vdash & \vp\sau\neg\vp. \label{lem'}
\end{align}
\eprop
\solution{\,\\[1mm]
\eqref{imp-reflexivity'}:\\[1mm]
\begin{tabular}{lll}
(1) & $\vdash(\vp\to((\vp\to\vp)\to\vp))\to((\vp\to(\vp\to\vp))\to(\vp\to\vp))$ & \prule{axiom-2}\\[1mm]
(2) & $\vdash\vp\to((\vp\to\vp)\to\vp)$ & \prule{axiom-1}\\[1mm]
(3) & $\vdash(\vp\to(\vp\to\vp))\to(\vp\to\vp)$ & \prule{modus-ponens}: (2), (1)\\[1mm]
(4) & $\vdash\vp\to(\vp\to\vp)$ & \prule{axiom-1}\\[1mm]
(5) & $\vdash\vp\to\vp$ & \prule{modus-ponens}: (4), (3)\\[2mm]
\end{tabular}

\eqref{lem'}:\\[1mm]
\begin{tabular}{lll}
(1) & $\vdash\neg\vp\to\neg\vp$ & \eqref{imp-reflexivity'}\\[1mm]
(2) & $\vdash(\neg\vp\to\neg\vp)\to\vp\sau\neg\vp$ & \prule{axiom-or-2}\\[1mm]
(3) & $\vdash\vp\sau\neg\vp$ & \prule{modus-ponens}: (1), (2)
\end{tabular}

\,
}

\bprop
\begin{align}
\vdash & \,(\vp\to\psi)\to((\psi\to\chi)\to(\vp\to\chi)). \label{imp-trans'}
\end{align}
\eprop
\solution{\,\\[1mm]
\begin{tabular}{lll}
(1) & $\vdash(\vp\to(\psi\to\chi))\to((\vp\to\psi)\to(\vp\to\chi))$ & \prule{axiom-2}\\[1mm]
(2) & $\vdash(\psi\to\chi)\to(\vp\to(\psi\to\chi))$ & \prule{axiom-1}\\[1mm]
(3) & $\vdash(\psi\to\chi)\to((\vp\to\psi)\to(\vp\to\chi))$ & \eqref{syllogism'}: (2), (1)\\[1mm]
(4) & $\vdash(\vp\to\psi)\to((\psi\to\chi)\to(\vp\to\chi))$ & \eqref{premise-comm'}: (3)
\end{tabular}

\,
}

\bprop
\bea
\Gamma\vdash \vp\to\psi  & \text{implies} & \Gamma\vdash(\psi\to\chi)\to(\vp\to\chi). 
\label{extra-premise'}
\eea
\eprop
\solution{\,\\[1mm]
\begin{tabular}{lll}
(1) & $\Gamma\vdash(\vp\to\psi)\to((\psi\to\chi)\to(\vp\to\chi))$ & \eqref{imp-trans'}\\[1mm]
(2) & $\Gamma\vdash\vp\to\psi$ & (Assumption)\\[1mm]
(3) & $\Gamma\vdash(\psi\to\chi)\to(\vp\to\chi)$ & \prule{modus-ponens}: (2), (1)
\end{tabular}

\,
}

\bprop
\begin{align}
\vdash & \, (\vp\to\psi)\to(\neg\psi\to\neg\vp). \label{rec-ax3'}
\end{align}
\eprop
\solution{\,\\[1mm]
\begin{tabular}{lll}
(1) & $\vdash(\vp\to\psi)\to((\psi\to\bot)\to(\vp\to\bot))$ & \eqref{imp-trans'}\\[1mm]
(2) & $\vdash(\vp\to\bot)\to\neg\vp$ & \prule{axiom-not-2}\\[1mm]
(3) & $\vdash(\vp\to\psi)\to((\psi\to\bot)\to\neg\vp)$ & \eqref{imp-trans-alt'}: (1), (2)\\[1mm]
(4) & $\vdash\neg\psi\to(\psi\to\bot)$ & \prule{axiom-not-1}\\[1mm]
(5) & $\vdash((\psi\to\bot)\to\neg\vp)\to(\neg\psi\to\neg\vp)$ & \eqref{extra-premise'}: (4)\\[1mm]
(6) & $\vdash(\vp\to\psi)\to(\neg\psi\to\neg\vp)$ & \eqref{syllogism'}: (3), (5)
\end{tabular}

\,
}

\bprop
\bea
\Gamma\vdash\vp\to\psi  & \text{implies} &  \Gamma\vdash\neg\psi\to\neg\vp. \label{rec-ax3-rule'}
\eea
\eprop
\solution{\,\\[1mm]
\eqref{rec-ax3-rule'}:\\
\begin{tabular}{lll}
(1) & $\Gamma\vdash \, \vp \to \psi$ & (Assumption)\\[1mm] 
(2) & $\Gamma\vdash (\vp \to \psi)\to (\neg\psi \to \neg\vp)$ & \eqref{rec-ax3'}\\[1mm]
(3) & $\Gamma\vdash \neg\psi \to \neg \vp$ & \prule{modus-ponens}: (2), (1)
\end{tabular}

\,
}

\bprop
\begin{align}
\vdash & \bot\to\vp, \label{exfalso'}\\
\vdash & \, \vp\to((\vp\to\psi)\to\psi), \label{vp-to-vp-to-psi-to-psi'}\\
\vdash & \, \vp\to\neg\neg\vp, \label{dni-pg'}\\
\vdash & \, (\neg\vp\to\neg\psi)\to(\psi\to\vp), \label{ax3'}\\
\vdash & \, \vp\to\vp\sau\psi, \label{weak-disj-1'}\\
\vdash & \, \psi\to\vp\sau\psi, \label{weak-disj-2'}\\
\vdash & \, \vp\si\psi\to\vp, \label{weak-conj-1'}\\
\vdash & \, \vp\si\psi\to\psi, \label{weak-conj-2'} \\
\vdash & \, \vp\sau\psi\to\psi\sau\vp, \label{perm-disj'}\\
\vdash & \, \vp\si\psi\to\psi\si\vp. \label{perm-conj'}
\end{align}
\eprop
\solution{\,\\[1mm]
\eqref{exfalso'}:\\[1mm]
\begin{tabular}{lll}
(1) & $\vdash\neg\neg\vp\to\vp$ & \prule{axiom-3}\\[1mm]
(2) & $\vdash(\bot\to\neg\neg\vp)\to(\bot\to\vp)$ & \eqref{extra-premise-alt'}: (1) \\[1mm]
(3) & $\vdash\bot\to((\vp\to\bot)\to\bot)$ & \prule{axiom-1} \\[1mm]
(4) & $\vdash((\vp\to\bot)\to\bot)\to\neg(\vp\to\bot)$ & \prule{axiom-not-2}\\[1mm]
(5) & $\vdash\bot\to\neg(\vp\to\bot)$ & \eqref{syllogism'}: (3), (4)\\[1mm]
(6) & $\vdash\neg\vp\to(\vp\to\bot)$ & \prule{axiom-not-1}\\[1mm]
(7) & $\vdash\neg(\vp\to\bot)\to\neg\neg\vp$ & \eqref{rec-ax3-rule'}: (6)\\[1mm]
(8) & $\vdash\bot\to\neg\neg\vp$ & \eqref{syllogism'}: (5), (7)\\[1mm]
(4) & $\vdash\bot\to\vp$ & \prule{modus-ponens}: (2), (8) \\[2mm]
\end{tabular}

\eqref{vp-to-vp-to-psi-to-psi'}:\\[1mm]
\begin{tabular}{lll}
(1) & $\vdash(\vp\to\psi)\to(\vp\to\psi)$ & \eqref{imp-reflexivity'}\\[1mm]
(2) & $\vdash\vp\to((\vp\to\psi)\to\psi)$ & \eqref{premise-comm'}: (1)\\[2mm]
\end{tabular}

\eqref{dni-pg'}:\\[1mm]
\begin{tabular}{lll}
(1) & $\vdash\vp\to((\vp\to\bot)\to\bot)$ & \eqref{vp-to-vp-to-psi-to-psi'}\\[1mm]
(2) & $\vdash((\vp\to\bot)\to\bot)\to\neg(\vp\to\bot)$ & \prule{axiom-not-2}\\[1mm]
(3) & $\vdash\vp\to\neg(\vp\to\bot)$ & \eqref{syllogism'}: (1), (2)\\[1mm]
(4) & $\vdash\neg\vp\to(\vp\to\bot)$ & \prule{axiom-not-1}\\[1mm]
(5) & $\vdash\neg(\vp\to\bot)\to\neg\neg\vp$ & \eqref{rec-ax3-rule'}: (4)\\[1mm]
(6) & $\vdash\vp\to\neg\neg\vp$ & \eqref{syllogism'}: (3), (5)\\[2mm]
\end{tabular}

\eqref{ax3'}:\\[1mm]
\begin{tabular}{lll}
(1) & $\vdash(\neg\vp\to\neg\psi)\to(\neg\neg\psi\to\neg\neg\vp)$ & \eqref{rec-ax3'}\\[1mm]
(2) & $\vdash\psi\to\neg\neg\psi$ & \eqref{dni-pg'}\\[1mm]
(3) & $\vdash(\neg\neg\psi\to\neg\neg\vp)\to(\psi\to\neg\neg\vp)$ & \eqref{extra-premise'}: (2)\\[1mm]
(4) & $\vdash(\neg\vp\to\neg\psi)\to(\psi\to\neg\neg\vp)$ & \eqref{syllogism'}: (1), (3)\\[1mm]
(5) & $\vdash\neg\neg\vp\to\vp$ & \prule{axiom-3}\\[1mm]
(6) & $\vdash(\neg\vp\to\neg\psi)\to(\psi\to\vp)$ & \eqref{imp-trans-alt'}: (4), (5)\\[2mm]
\end{tabular}

\eqref{weak-disj-1'}:\\[1mm]
\begin{tabular}{lll}
(1) & $\vdash\vp\to\neg\neg\vp$ & \eqref{dni-pg'}\\[1mm]
(2) & $\vdash\neg\neg\vp\to(\neg\vp\to\bot)$ & \prule{axiom-not-1}\\[1mm]
(3) & $\vdash\vp\to(\neg\vp\to\bot)$ & \eqref{syllogism'}: (1), (2)\\[1mm]
(4) & $\vdash\bot\to\psi$ & \eqref{exfalso'}\\[1mm]
(5) & $\vdash\vp\to(\neg\vp\to\psi)$ & \eqref{imp-trans-alt'}: (3), (4)\\[1mm]
(6) & $\vdash(\neg\vp\to\psi)\to\vp\sau\psi$ & \prule{axiom-or-2}\\[1mm]
(7) & $\vdash\vp\to\vp\sau\psi$ & \eqref{syllogism'}: (5), (6)\\[2mm]
\end{tabular}

\eqref{weak-disj-2'}:\\[1mm]
\begin{tabular}{lll}
(1) & $\vdash\psi\to(\neg\vp\to\psi)$ & \prule{axiom-1}\\[1mm]
(2) & $\vdash(\neg\vp\to\psi)\to\vp\sau\psi$ & \prule{axiom-or-2}\\[1mm]
(3) & $\vdash\psi\to\vp\sau\psi$ & \eqref{syllogism'}: (2), (3)\\[2mm]
\end{tabular}

\eqref{weak-conj-1'}:\\[1mm]
\begin{tabular}{lll}
(1) & $\vdash\neg\vp\to\neg\vp\sau\neg\psi$ & \eqref{weak-disj-1'}\\[1mm]
(2) & $\vdash(\neg\vp\to\neg\vp\sau\neg\psi)\to(\neg(\neg\vp\sau\neg\psi)\to\neg\neg\vp)$ & \eqref{rec-ax3'}\\[1mm]
(3) & $\vdash\neg(\neg\vp\sau\neg\psi)\to\neg\neg\vp$ & \prule{modus-ponens}: (1), (2)\\[1mm]
(4) & $\vdash\neg\neg\vp\to\vp$ & \prule{axiom-3}\\[1mm]
(5) & $\vdash\neg(\neg\vp\sau\neg\psi)\to\vp$ & \eqref{syllogism'}: (3), (4)\\[1mm]
(6) & $\vdash\vp\si\psi\to\neg(\neg\vp\sau\neg\psi)$ & \prule{axiom-and-1}\\[1mm]
(7) & $\vdash\vp\si\psi\to\vp$ & \eqref{syllogism'}: (6), (5)\\[2mm]
\end{tabular}

\eqref{weak-conj-2'}:\\[1mm]
\begin{tabular}{lll}
(1) & $\vdash\neg\psi\to\neg\vp\sau\neg\psi$ & \eqref{weak-disj-2'}\\[1mm]
(2) & $\vdash(\neg\psi\to\neg\vp\sau\neg\psi)\to(\neg(\neg\vp\sau\neg\psi)\to\neg\neg\psi)$ & \eqref{rec-ax3'}\\[1mm]
(3) & $\vdash\neg(\neg\vp\sau\neg\psi)\to\neg\neg\psi$ & \prule{modus-ponens}: (1), (2)\\[1mm]
(4) & $\vdash\neg\neg\psi\to\psi$ & \prule{axiom-3}\\[1mm]
(5) & $\vdash\neg(\neg\vp\sau\neg\psi)\to\psi$ & \eqref{syllogism'}: (3), (4)\\[1mm]
(6) & $\vdash\vp\si\psi\to\neg(\neg\vp\sau\neg\psi)$ & \prule{axiom-and-1}\\[1mm]
(7) & $\vdash\vp\si\psi\to\psi$ & \eqref{syllogism'}: (6), (5)\\[2mm]
\end{tabular}

\eqref{perm-disj'}:\\[1mm]
\begin{tabular}{lll}
(1) & $\vdash(\neg\vp\to\psi)\to(\neg\psi\to\neg\neg\vp)$ & \eqref{rec-ax3'}\\[1mm]
(2) & $\vdash\neg\neg\vp\to\vp$ & \prule{axiom-3}\\[1mm]
(3) & $\vdash(\neg\vp\to\psi)\to(\neg\psi\to\vp)$ & \eqref{imp-trans-alt'}: (1), (2)\\[1mm]
(4) & $\vdash\vp\sau\psi\to(\neg\vp\to\psi)$ & \prule{or-axiom-1}\\[1mm]
(5) & $\vdash\vp\sau\psi\to(\neg\psi\to\vp)$ & \eqref{syllogism'}: (4), (3)\\[1mm]
(6) & $\vdash(\neg\psi\to\vp)\to\psi\sau\vp$ & \prule{or-axiom-2}\\[1mm]
(7) & $\vdash\vp\sau\psi\to\psi\sau\vp$ & \eqref{syllogism'}: (5), (6)\\[2mm]
\end{tabular}

\eqref{perm-conj'}:\\[1mm]
\begin{tabular}{lll}
(1) & $\vdash(\neg\psi\sau\neg\vp\to\neg\vp\sau\neg\psi)\to$ & \eqref{rec-ax3'}\\[1mm]
& \quad $(\neg(\neg\vp\sau\neg\psi)\to\neg(\neg\psi\sau\neg\vp))$ \\[1mm]
(2) & $\vdash\neg\psi\sau\neg\vp\to\neg\vp\sau\neg\psi$ & \eqref{perm-disj'}\\[1mm]
(3) & $\vdash\neg(\neg\vp\sau\neg\psi)\to\neg(\neg\psi\sau\neg\vp)$ & \prule{modus-ponens}: (2), (1)\\[1mm]
(4) & $\vdash\vp\si\psi\to\neg(\neg\vp\sau\neg\psi)$ & \prule{axiom-and-1}\\[1mm]
(5) & $\vdash\vp\si\psi\to\neg(\neg\psi\sau\neg\vp)$ & \eqref{syllogism'}: (4), (3)\\[1mm]
(6) & $\vdash\neg(\neg\psi\sau\neg\vp)\to\psi\si\vp$ & \prule{axiom-and-2}\\[1mm]
(7) & $\vdash\vp\si\psi\to\psi\si\vp$ & \eqref{syllogism'}: (5), (6)
\end{tabular}

\,
}

\bprop
\bea
\Gamma\vdash \, \neg\neg\vp \to \psi   & \text{implies} & \Gamma \vdash \vp \to \psi,
\label{rule-negnegvp-premise-vp-conc'}\\
\Gamma\vdash \, \neg\psi\to\neg(\vp\to\vp) & \text{implies} & \Gamma\vdash\psi,
\label{contra'}\\
\Gamma\vdash \, \vp\to(\psi\to\chi) & \text{implies} & \Gamma\vdash\vp\si\psi\to\chi. \label{importation'}
\eea
\eprop
\solution{\,\\[1mm] 
\eqref{rule-negnegvp-premise-vp-conc'}:\\[1mm]
\begin{tabular}{lll}
(1) & $\Gamma\vdash \, \neg\neg\vp \to \psi$ & (Assumption)\\[1mm] 
(2) & $\Gamma\vdash \, \vp \to \neg\neg\vp$ & \eqref{dni-pg'}\\[1mm]
(3) & $\Gamma\vdash \vp \to \psi$ & \eqref{syllogism'}: (1), (2)\\[2mm]
\end{tabular}

\eqref{contra'}:\\[1mm]
\begin{tabular}{lll}
(1) & $\Gamma\vdash\neg\psi\to\neg(\vp\to\vp)$ & (Assumption)\\[1mm]
(2) & $\Gamma\vdash(\neg\psi\to\neg(\vp\to\vp))\to((\vp\to\vp)\to\psi)$ & \eqref{ax3'}\\[1mm]
(3) & $\Gamma\vdash(\vp\to\vp)\to\psi$ & \prule{modus-ponens}: (1), (2)\\[1mm]
(4) & $\Gamma\vdash\vp\to\vp$ & \eqref{imp-reflexivity'}\\[1mm]
(5) & $\Gamma\vdash\psi$ & \prule{modus-ponens}: (3), (4)\\[2mm]
\end{tabular}

\eqref{importation'}:\\[1mm]
\begin{tabular}{lll}
(1) & $\Gamma\vdash\vp\to(\psi\to\chi)$ & (Assumption)\\[1mm]
(2) & $\Gamma\vdash\vp\si\psi\to\vp$ & \eqref{weak-conj-1'}\\[1mm]
(3) & $\Gamma\vdash\vp\si\psi\to(\psi\to\chi)$ & \eqref{syllogism'}: (2), (1)\\[1mm]
(4) & $\Gamma\vdash\vp\si\psi\to\psi$ & \eqref{weak-conj-2'}\\[1mm]
(5) & $\Gamma\vdash\vp\si\psi\to\chi$ & \eqref{ax2-rule'}: (3), (4)
\end{tabular}

\,
}

\bprop
\begin{align}
\vdash \, \vp\to(\neg\vp\to\psi). \label{ex-falso-impl'}
\end{align}
\eprop
\solution{\,\\[1mm] 
\begin{tabular}{lll}
(1) & $\vdash\neg\vp\to(\vp\to\bot)$ & \prule{axiom-not-1}\\[1mm]
(2) & $\vdash((\vp\to\bot)\to\psi)\to(\neg\vp\to\psi)$ & \eqref{extra-premise'}: (1)\\[1mm]
(3) & $\vdash(\vp\to((\vp\to\bot)\to\psi))\to(\vp\to(\neg\vp\to\psi))$ & \eqref{extra-premise-alt'}: (2)\\[1mm]
(4) & $\vdash\bot\to\psi$ & \eqref{exfalso'}\\[1mm]
(5) & $\vdash((\vp\to\bot)\to\bot)\to((\vp\to\bot)\to\psi)$ & \eqref{extra-premise-alt'}: (4)\\[1mm]
(6) & $\vdash(\vp\to((\vp\to\bot)\to\bot))\to(\vp\to((\vp\to\bot)\to\psi))$ & \eqref{extra-premise-alt'}: (5)\\[1mm]
(7) & $\vdash\vp\to((\vp\to\bot)\to\bot)$ & \eqref{vp-to-vp-to-psi-to-psi'}\\[1mm]
(8) & $\vdash\vp\to((\vp\to\bot)\to\psi)$ & \prule{modus-ponens}: (7), (6)\\[1mm]
(9) & $\vdash\vp\to(\neg\vp\to\psi)$ & \prule{modus-ponens}: (8), (3)
\end{tabular}

\,
}

\bprop
\begin{align}
\vdash & \vp\sau\vp\to\vp, \label{contr-disj'}\\
\vdash & \, \vp\to\vp\si\vp. \label{contr-conj'}
\end{align}
\eprop
\solution{\,\\[1mm] 
\eqref{contr-disj'}:\\[1mm] 
\begin{tabular}{lll}
(1)& $\vdash(\neg\vp\to\vp)\si\neg\vp\to\neg\vp$ & \eqref{weak-conj-2'}\\[1mm]
(2)& $\vdash(\neg\vp\to\vp)\si\neg\vp\to(\neg\vp\to\vp)$ & \eqref{weak-conj-1'}\\[1mm]
(3)& $\vdash(\neg\vp\to\vp)\si\neg\vp\to\vp$ & \eqref{ax2-rule'}: (1), (2)\\[1mm]
(4)& $\vdash\vp\to(\neg\vp\to\neg(\vp\to\vp))$ & \eqref{ex-falso-impl'}\\[1mm]
(5)& $\vdash(\vp\to(\neg\vp\to\neg(\vp\to\vp)))\to$ & \prule{axiom-1}\\[1mm]
&\quad $\to((\neg\vp\to\vp)\si\neg\vp\to(\vp\to(\neg\vp\to\neg(\vp\to\vp))))$ & \prule{axiom-1}\\[1mm]
(6)& $\vdash(\neg\vp\to\vp)\si\neg\vp\to(\vp\to(\neg\vp\to\neg(\vp\to\vp)))$ & \prule{modus-ponens}: (4), (5)\\[1mm]
(7)& $\vdash(\neg\vp\to\vp)\si\neg\vp\to(\neg\vp\to\neg(\vp\to\vp))$ & \eqref{ax2-rule'}: (6), (3)\\[1mm]
(8)& $\vdash(\neg\vp\to\vp)\si\neg\vp\to\neg(\vp\to\vp)$ & \eqref{ax2-rule'}: (7), (1)\\[1mm]
(9)& $\vdash\neg(\neg(\neg\vp\to\vp)\sau\neg\neg\vp)\to(\neg\vp\to\vp)\si\neg\vp$ & \prule{axiom-and-2}\\[1mm]
(10)& $\vdash\neg(\neg(\neg\vp\to\vp)\sau\neg\neg\vp)\to\neg(\vp\to\vp)$ & \eqref{syllogism'}: (9), (8)\\[1mm]
(11)& $\vdash\neg(\neg\vp\to\vp)\sau\neg\neg\vp$ & \eqref{contra'}: (10)\\[1mm]
(12)& $\vdash\neg(\neg\vp\to\vp)\sau\neg\neg\vp\to(\neg\neg(\neg\vp\to\vp)\to\neg\neg\vp)$ & \prule{axiom-or-1}\\[1mm]
(13)& $\vdash\neg\neg(\neg\vp\to\vp)\to\neg\neg\vp$ & \prule{modus-ponens}: (11), (12)\\[1mm]
(14)& $\vdash(\neg\vp\to\vp)\to\neg\neg(\neg\vp\to\vp)$ & \eqref{dni-pg'}\\[1mm]
(15)& $\vdash(\neg\vp\to\vp)\to\neg\neg\vp$ & \eqref{syllogism'}: (14), (13)\\[1mm]
(16)& $\vdash\neg\neg\vp\to\vp$ & \prule{axiom-3}\\[1mm]
(17)& $\vdash(\neg\vp\to\vp)\to\vp$ & \eqref{syllogism'}: (15), (16)\\[1mm]
(18)& $\vdash\vp\sau\vp\to(\neg\vp\to\vp)$ & \prule{axiom-or-1}\\[1mm]
(19)& $\vdash\vp\sau\vp\to\vp$ & \eqref{syllogism'}: (18), (17)\\[2mm]
\end{tabular}

\eqref{contr-conj'}:\\[1mm] 
\begin{tabular}{lll}
(1) & $\Gamma\vdash\neg\vp\sau\neg\vp\to\neg\vp$ & \eqref{contr-disj'}\\[1mm]
(2) & $\Gamma\vdash(\neg\vp\sau\neg\vp\to\neg\vp)\to(\neg\neg\vp\to\neg(\neg\vp\sau\neg\vp))$ & \eqref{rec-ax3'}\\[1mm]
(3) & $\Gamma\vdash\neg\neg\vp\to\neg(\neg\vp\sau\neg\vp)$ & \prule{modus-ponens}: (1), (2)\\[1mm]
(4) & $\Gamma\vdash\vp\to\neg\neg\vp$ & \eqref{dni-pg'}\\[1mm]
(5) & $\Gamma\vdash\vp\to\neg(\neg\vp\sau\neg\vp)$ & \eqref{syllogism'}: (4), (3)\\[1mm]
(6) & $\Gamma\vdash\neg(\neg\vp\sau\neg\vp)\to\vp\si\vp$ & \prule{axiom-and-2}\\[1mm]
(7) & $\Gamma\vdash\vp\to\vp\si\vp$ & \eqref{syllogism'}: (5), (6)
\end{tabular}

\,
}

\blem
\begin{align}
\vdash & \, \vp\to(\psi\to\vp\si\psi). \label{and-intro'}
\end{align}
\elem
\solution{\,\\[1mm]
\begin{tabular}{lll}
(1) & $\vdash \neg\vp \sau \neg\psi \to (\neg\neg\vp\to \neg\psi)$ & \prule{axiom-or-1} \\[1mm]
(2) & $\vdash \vp\to \neg\neg\vp$ & \eqref{dni-pg'} \\[1mm]
(3) & $\vdash (\neg\neg\vp\to \neg\psi) \to (\vp\to \neg\psi)$ & \eqref{extra-premise'}: (2) \\[1mm]
(4) & $\vdash \neg\vp \sau \neg\psi \to (\vp\to \neg\psi)$ &  \eqref{syllogism'}: (1), (3)\\[1mm]
(5) & $\vdash \vp\to (\neg\vp \sau \neg\psi \to  \neg\psi)$ &  \eqref{premise-comm'}: (4)\\[1mm]
(6) & $\vdash (\neg\vp \sau \neg\psi \to  \neg\psi)\to (\neg\neg\psi\to \neg(\neg\vp \sau \neg\psi))$
 & \eqref{rec-ax3'}\\[1mm]
(7) & $\vdash \vp\to (\neg\neg\psi\to \neg(\neg\vp \sau \neg\psi))$ & \eqref{syllogism'}: (5), (6)\\[1mm]
(8) & $\vdash \psi \to \neg\neg\psi$ & \eqref{dni-pg'} \\[1mm]
(9) & $\vdash (\neg\neg\psi \to \neg(\neg\vp \sau \neg\psi)) \to (\psi \to \neg(\neg\vp \sau \neg\psi))$ 
& \eqref{extra-premise'}: (8) \\[1mm]
(10) & $\vdash \vp\to (\psi\to \neg(\neg\vp \sau \neg\psi))$ & \eqref{syllogism'}: (7), (9)\\[1mm]
(11) & $\vdash\neg(\neg\vp \sau \neg\psi) \to \vp\si\psi$ & \prule{and-axiom-2}\\[1mm]
(12) & $\vdash \vp\to(\psi\to\vp\si\psi)$ & \eqref{imp-trans-alt'}: (10), (11)
\end{tabular}

\,
}

\bprop
\bea
\Gamma\vdash\vp\si\psi\to\chi & \text{implies} &\Gamma\vdash\vp\to(\psi\to\chi).
\label{exportation'}
\eea
\eprop
\solution{\,\\[1mm]
\begin{tabular}{lll}
(1) & $\Gamma\vdash\vp\si\psi\to\chi$ & (Assumption)\\[1mm]
(2) & $\Gamma\vdash\vp\to(\psi\to\vp\si\psi)$ & \eqref{and-intro'}\\[1mm]
(3) & $\Gamma\vdash\vp\to(\psi\to\chi)$ & \eqref{imp-trans-alt'}: (1), (2)
\end{tabular}

\,
}

\subsubsection{First-order connectives}

\bprop
For every pattern $\vp$ and element variables $x,y$ such that $x$ is free for $y$ in $\vp$,
\begin{align}
\vdash\forall x\vp\to\subf{x}{y}{\vp}. \label{forall-x-subf'}
\end{align}
\eprop
\solution{\,\\[1mm]
\begin{tabular}{lll}
(1) & $\vdash\subf{x}{y}{\neg\vp}\to\exists x\neg\vp$ & \prule{$\exists$-Quantifier}, as x is free for y in $\neg\vp$\\[1mm]
(2) & $\vdash\neg\exists x\neg\vp\to\neg\subf{x}{y}{\neg\vp}$ & \eqref{rec-ax3-rule'}: (1)\\[1mm]
(3) & $\vdash\forall x\vp\to\neg\exists x\neg\vp$ & \prule{axiom-forall-1}\\[1mm]
(4) & $\vdash\forall x\vp\to\neg\subf{x}{y}{\neg\vp}$ & \eqref{syllogism'}: (3), (2)\\[1mm]
(5) & $\vdash\neg\subf{x}{y}{\neg\vp}\to\neg\neg\subf{x}{y}{\vp}$ & by the 
definition by recursion of $\subfxy$\\[1mm]
(6) & $\vdash\forall x\vp\to\neg\neg\subf{x}{y}{\vp}$ & \eqref{syllogism'}: (4), (5)\\[1mm]
(7) & $\vdash\neg\neg\subf{x}{y}{\vp}\to\subf{x}{y}{\vp}$ & \prule{axiom-3}\\[1mm]
(8) & $\vdash\forall x\vp\to\subf{x}{y}{\vp}$ & \eqref{syllogism'}: (6), (7)
\end{tabular}

\,
}

\bprop
For every pattern $\vp$ and element variable $x\notin \FVE(\vp)$,
\bea
\Gamma\vdash\vp\to \psi & \text{implies} & \Gamma \vdash \vp \to  \forall x \psi.
\label{forall-x-rule'}
\eea
\eprop
\solution{\,\\[1mm]
\begin{tabular}{lll}
(1) & $\Gamma\vdash\vp\to \psi$ & (Assumption)\\[1mm]
(2) & $\Gamma\vdash \neg\psi \to \neg\vp$ & \eqref{rec-ax3-rule'}: (1)\\[1mm]
(3) & $\Gamma\vdash \exists x\neg\psi \to \neg\vp$ & \prule{$\exists$-quantifier rule}: (2), 
as $x\notin \FVE(\neg\vp)$ \\[1mm]
(4) & $\Gamma\vdash \neg\neg\vp \to \neg\exists x\neg\psi $ & \eqref{rec-ax3-rule'}: (3)\\[1mm]
(5) & $\Gamma\vdash \vp \to \neg\exists x\neg\psi $ & \eqref{rule-negnegvp-premise-vp-conc'}: (4)\\[1mm]
(6) & $\Gamma\vdash \neg\exists x\neg\psi \to \forall x\psi $ & \prule{axiom-forall-2}\\[1mm]
(7) & $\Gamma\vdash \vp \to \forall x\psi$ & \eqref{syllogism'}: (4), (5)
\end{tabular}

\,
}

\subsection{$\appMLGc \equivps \appMLnow$}

We use Proposition \ref{absps1-weakerps-absps2} to prove that  $\appMLnow \weakerps \appMLGc$ and 
$\appMLGc \weakerps \appMLnow$. 

\bthm\label{P-weaker-Gc}
$\appMLnow \weakerps \appMLGc$.
\ethm
\solution{
We prove first that $\vdash_{\appMLGc} \varphi$ for every axiom $\varphi$  of $\appMLnow$:\\[1mm]
\prule{axiom-1}: By \eqref{vp-to-psi-to-vp}.

\prule{axiom-2}: By \eqref{vp-to-psi-to-chi-to-vp-to-psi-to-vp-to-chi}.

\prule{axiom-3}: By \eqref{negnegvp-vp}.

\prule{axiom-not-1}: common.

\prule{axiom-not-2}: common.

\prule{axiom-or-1}: By \eqref{axiom-or-1}.

\prule{axiom-or-2}: By \eqref{axiom-or-2}.

\prule{axiom-and-1}: By \eqref{axiom-and-1}.

\prule{axiom-and-2}: By \eqref{axiom-and-2}.

\prule{$\exists$-Quantifier}: common.

\prule{axiom-forall-1}: By \eqref{forallxvp-to-notexistsxnotvp}.

\prule{axiom-forall-2}: By \eqref{axiom-forall-2-Gc}.

\prule{Propagation$_\bot$}, \prule{Propagation$_\vee$}: common. 

\prule{Propagation$^*_\exists$}: By \eqref{propagation-exists-1-H-Gc}, \eqref{propagation-exists-2-H-Gc}\\

Let $\Gamma$ be a set of patterns. We prove now that $\amlthmgen{\Gamma}{\appMLGc}$ is closed to 
the deduction rules of $\appMLnow$:\\[1mm]
\prule{modus ponens}: common.

\prule{$\exists$-quantifier rule}: common.

\prule{framing-left}, \prule{framing-right}: common.
}

\bthm\label{Gc-weaker-P}
$\appMLGc \weakerps \appMLnow$.
\ethm
\solution{
We prove first that $\vdash_\appMLnow \varphi$ for every axiom $\varphi$ of $\appMLGc$:\\[1mm]
\prule{contraction}: \eqref{contr-disj'}, \eqref{contr-conj'}.

\prule{weakening}: By \eqref{weak-disj-1'}\, \eqref{weak-conj-1'}.

\prule{permutation}: By \eqref{perm-disj'}\, \eqref{perm-conj'}.

\prule{ex falso quodlibet}: By \eqref{exfalso'}.

\prule{lem}: By  \eqref{lem'}.

\prule{axiom-not-1}: common.

\prule{axiom-not-2}: common.

\prule{$\exists$-Quantifier}: common.

\prule{$\forall$-Quantifier}: By \eqref{forall-x-subf'}.

\prule{Propagation$_\bot$}, \prule{Propagation$_\vee$}: common. 

\prule{Propagation$_\exists$}: is a particular case of \prule{Propagation$^*_\exists$}, as  
$x$ does not occur in $\psi$ implies $x\notin \FVE(\psi)$. \\

Let $\Gamma$ be a set of patterns. We prove now that $\amlthmgen{\Gamma}{\appMLnow}$ is closed to 
the deduction rules of $\appMLGc$:\\[1mm]
\prule{modus ponens}: common.

\prule{syllogism}: By \eqref{syllogism'}.

\prule{exportation}: By \eqref{exportation'}.

\prule{importation}: By \eqref{importation'}.

\prule{expansion}: By \eqref{expansion'}.

\prule{$\exists$-quantifier rule}: common.

\prule{$\forall$-quantifier rule}: By \eqref{forall-x-rule'}.

\prule{framing-left}, \prule{framing-right}: common.
}

\bthm\label{Gc-equivps-P}
$\appMLGc \equivps \appMLnow$.
\ethm
\solution{Apply Theorems \ref{Gc-weaker-P}, \ref{P-weaker-Gc} and  
Proposition \ref{absps1-equivps-absps2}.
}

\chapter{Applicative matching logic with application and definedness  $\appMLdef$}\label{AML-Tdef}


The logic $\appMLdef$ is defined as follows:
\be
\item $\lc_{\appMLdef}$: 
$\defsymb \in \Sigma$ ($\defsymb$ is called the \textit{definedness symbol}) \\
$\PropConstants=\{\bot\}$, $\PropUnary=\{\neg\}$, $\PropBinary=\{\to, \si, \sau, \Appl\}$, 
$\FolQ=\{\forall, \exists\}$, 	$\Equal=\SolQ=\svar=\emptyset$.
\item Its proof system, $\appMLnewGc$, is given in Appendix \ref{AML-proof-system-new-Gc}.
\ee

We write simply $\vdash$ instead of $\vdash_{\appMLnewGc}$. As in Chapter \ref{AML-Gc}, we use the infix notation for patterns: we write $\vp\to\psi$, $\vp\sau\psi$, $\vp\si\psi$, 
$\appln{\vp}{\psi}$.

We introduce the derived  connective $\dnd$, $\defmapn$, $\totmapn$, $\eqdef$ by the following abbreviations:
\begin{align*}
\vp \dnd \psi & := (\vp \to \psi) \si (\psi \to \vp),\\
\defmap{\vp} & := \appln{\defsymb}{\vp},\\
\totmap{\vp} & := \neg\ceil{\neg\vp},\\
\vp \eqdef \psi & := \floor{\vp\dnd\psi}.
\end{align*}
For every pattern $\vp$, $\defmap{\vp}$ is called the \textit{definedness of $\vp$}  and 
$\totmap{\vp}$ is called the \textit{totality of $\vp$}.\\

We use the same conditions on the precedence of connectives/quantifiers as in Chapter \ref{AML-Gc}. We assume, moreover, that 
$\eqdef$  has higher precedence than $\ra$, $\si$, $\sau$, $\dnd$.

\newpage

\section{Theorems and derived rules in $\appMLnewGc$ - I}

In the following, $\Gamma$ is an arbitrary set of patterns and $\vp$, $\psi$, $\chi$ are arbitrary patterns.

\subsection{Propositional connectives}

\blem
\begin{align}
\vdash &  \,  \psi \to (\psi\si\vp)\sau(\psi\si\neg\vp). \label{psi-topsisivp-sau-psisinegvp}
\end{align}
\elem
\solution{\,\\[1mm]
\begin{tabular}{lll}
(1) & $\vdash \psi\to \psi$ & \eqref{sim-refl} and \eqref{Gamma-vptopsi-psitovp-iff-vpdndpsi} \\[1mm]
(2) & $\vdash \vp\sau\neg\vp$ & \prule{lem}\\[1mm]
(3) & $\vdash \psi\to\vp\sau\neg\vp$ & \eqref{vp-implies-psi-to-vp}: (2)\\[1mm]
(4) & $\vdash \psi\to \psi\si(\vp\sau\neg\vp)$ & \eqref{vp-to-psi-si-vp-to-chi-implies-vp-to-psi-si-chi}: (1), (3)\\[1mm]
(5) & $\vdash \psi\si(\vp\sau\neg\vp)\to(\psi\si\vp)\sau(\psi\si\neg\vp)$ & 
\eqref{distrib-si-sau} and \eqref{Gamma-vptopsi-psitovp-iff-vpdndpsi}\\[1mm]
(6) & $\vdash \psi\to(\psi\si\vp)\sau(\psi\si\neg\vp)$ & \prule{syllogism}: (4), (5)
\end{tabular}

\,
}

\blem
\begin{align}
\vdash & \, \neg (\psi \si \neg\vp) \dnd \neg \psi \sau\vp, \label{neg-psisinegvp-dnd-negpsisauvp} \\
\vdash & \, \neg(\neg\vp \sau \psi) \dnd \vp  \si\neg\psi, \label{neg-negvpsaupsi-dnd-vpsinegpsi} \\
\vdash & \, \neg(\vp\to \psi) \dnd \vp \si \neg\psi, \label{neg-vptopsi-dnd-vpsinegpsi} \\
\vdash & \, \neg(\vp\to \psi) \dnd \neg\psi \si \vp, \label{neg-vptopsi-dnd-negpsisivp} \\
\vdash & \, \neg \vp \sau \neg \psi \dnd (\psi \to \neg \vp). \label{negvp-sau-negpsi--dnd--vp-to-negpsi} 
\end{align}
\elem
\solution{\,\\[1mm]
\eqref{neg-psisinegvp-dnd-negpsisauvp}:\\[1mm]
\begin{tabular}{lll}
(1) & $\vdash\neg\psi\sau\neg\neg\vp\dnd\neg(\psi\si\neg\vp)$ & \eqref{de-morgan-or-dnd}\\[1mm]
(2) & $\vdash\neg(\psi\si\neg\vp)\dnd(\neg\psi\sau\neg\neg\vp)$ & \eqref{sim-symm}: (1)\\[1mm]
(3) & $\vdash\neg\neg\vp\dnd\vp$ & \eqref{negnegvp-dnd-vp}\\[1mm]
(4) & $\vdash\neg\psi\sau\neg\neg\vp\dnd\neg\psi\sau\vp$ & \eqref{sim-pairs-or-right}: (3)\\[1mm]
(5) & $\vdash\neg(\psi\si\neg\vp)\dnd\neg\psi\sau\vp$ & \eqref{sim-trans}: (2), (4)\\[2mm]
\end{tabular}

\eqref{neg-negvpsaupsi-dnd-vpsinegpsi}:\\[1mm]
\begin{tabular}{lll}
(1) & $\vdash \neg(\neg\vp \sau \psi) \dnd \neg\neg\vp \si \neg\psi$ & \eqref{sim-symm}: \eqref{de-morgan-and-dnd}\\[1mm]
(2) & $\vdash \neg\neg\vp \dnd \vp$ & \eqref{negnegvp-dnd-vp}\\[1mm]
(3) & $\vdash \neg\neg\vp \si \neg\psi \dnd  \vp  \si \neg\psi$ & \eqref{sim-pairs-and-left}: (2)\\[1mm]
(4) & $\vdash \neg(\neg\vp \sau \psi) \dnd \vp  \si \neg\psi$ & \eqref{sim-trans}: (1), (3)\\[2mm]
\end{tabular}

\eqref{neg-vptopsi-dnd-vpsinegpsi}:\\[1mm]
\begin{tabular}{lll}
(1) & $\vdash (\vp\to \psi) \dnd \neg\vp \sau \psi$ & \eqref{def-imp-or} \\[1mm]
(2) & $\vdash \neg(\vp\to \psi) \dnd \neg(\neg\vp \sau \psi)$ & \eqref{Gamma-dnd-cong-neg}: (1)\\[1mm]
(3) & $\vdash \neg(\neg\vp \sau \psi) \dnd \vp  \si\neg\psi$ & \eqref{neg-negvpsaupsi-dnd-vpsinegpsi}\\[1mm]
(4) & $\vdash \neg(\vp\to \psi) \dnd \vp  \si\neg\psi$ & \eqref{sim-trans}: (2), (3)\\[2mm]
\end{tabular}

\eqref{neg-vptopsi-dnd-negpsisivp}:\\[1mm]
\begin{tabular}{lll}
(1) & $\vdash \neg(\vp\to \psi) \dnd \vp \si \neg\psi$ & \eqref{neg-vptopsi-dnd-vpsinegpsi} \\[1mm]
(2) & $\vdash \vp \si \neg\psi \dnd \neg\psi \si \vp$ & \eqref{si-comm} \\[1mm]
(3) & $\vdash \neg(\vp\to \psi) \dnd \neg\psi \si \vp$ & \eqref{sim-trans}: (1), (2)\\[2mm]
\end{tabular}

\eqref{negvp-sau-negpsi--dnd--vp-to-negpsi}:\\[1mm]
\begin{tabular}{lll}
(1) & $\vdash (\psi \to \neg \vp) \dnd \neg \psi  \sau \neg \vp$ & \eqref{def-imp-or} \\[1mm]
(2) & $\vdash \neg \psi  \sau \neg \vp \dnd \neg \vp \sau \neg \psi$  & \eqref{sau-comm}\\[1mm]
(3) & $\vdash  (\psi \to \neg \vp) \dnd \neg \vp \sau \neg \psi$   &\eqref{sim-trans}: (1), (2)\\[1mm]
(4) & $\vdash \neg \vp \sau \neg \psi \dnd (\psi \to \neg \vp)$ & \eqref{sim-symm}: (3)\\[2mm]
\end{tabular}
}

\blem
\begin{align}
\vdash &  \, \neg(\vp\dnd \psi)\dnd(\vp\si\neg \psi)\sau(\neg \vp\si \psi), \label{useful-mem-neg-2}\\
\vdash &  \, (\vp\dnd \psi) \to (\vp \sau \psi \dnd \vp \si \psi), \label{vpdndpsi-to-vpsaupsi-dnd-vpsipsi}\\
\vdash & \, \vp \sau \psi \to \neg(\vp\dnd \psi)\sau(\vp\si \psi). \label{useful-mem-intro}
\end{align}
\elem
\solution{\,\\[1mm]
\eqref{useful-mem-neg-2}:\\[1mm] 
\begin{tabular}{lll}
(1) & $\vdash\neg(\vp\dnd\psi) \dnd \neg(\vp\to \psi)\sau \neg(\psi\to \vp)$ &  
the definition of $\dnd$ and \eqref{sim-symm}: \eqref{de-morgan-or-dnd}\\[1mm]
(2) & $\vdash \neg(\vp\to \psi) \dnd \vp \si \neg\psi$ & \eqref{neg-vptopsi-dnd-vpsinegpsi}\\[1mm]
(3) & $\vdash \neg(\psi\to \vp) \dnd \neg \vp\si \psi$ & \eqref{neg-vptopsi-dnd-negpsisivp} \\[1mm]
(4) & $\vdash \neg(\vp\to \psi)\sau \neg(\psi\to \vp) \dnd (\vp\si\neg \psi)\sau(\neg \vp\si \psi)$ & \eqref{sim-pairs-or}: (2), (3)\\[1mm]
(5) & $\vdash \neg(\vp\dnd \psi) \dnd (\vp\si\neg \psi)\sau(\neg \vp\si \psi)$ & \eqref{sim-trans}: (1),(4)\\[2mm] 	
\end{tabular}

\eqref{vpdndpsi-to-vpsaupsi-dnd-vpsipsi}: \\[1mm] 
\begin{tabular}{lll}
(1) & $\vdash (\vp\dnd \psi) \to (\vp \sau \psi \dnd \psi \sau \psi) $ & \eqref{vpdndpsi-sau-r}\\[1mm] 
(2) & $\vdash (\vp\dnd \psi) \to (\vp \sau \psi \dnd \psi)  $ & \eqref{vp-to-psi-dnd-chi-and-chi-dnd-gamma-implies-vp-to-psi-dnd-gamma}: (1), \eqref{sau-idemp}\\[1mm] 	
(3) & $\vdash (\vp\dnd \psi) \to (\vp \si \psi \dnd \psi \si \psi)$ & \eqref{vpdndpsi-si-r}\\[1mm] 
(4) & $\vdash (\vp\dnd \psi) \to (\vp \sau \psi \dnd \psi)$ & \eqref{vp-to-psi-dnd-chi-and-chi-dnd-gamma-implies-vp-to-psi-dnd-gamma}: (1), \eqref{si-idemp}\\[1mm] 	
(5) & $\vdash (\vp\dnd \psi) \to (\vp \sau \psi \dnd \psi) \si (\vp \si \psi \dnd \psi)$ & \eqref{vp-to-psi-si-vp-to-chi-implies-vp-to-psi-si-chi}: (2),(4)\\[1mm] 
(6) & $\vdash  (\vp \sau \psi \dnd \vp) \si (\vp \si \psi \dnd \vp) \to  (\vp \sau \psi \dnd \vp \si \psi)$ & \eqref{dnd-tranz}\\[1mm] 
(7) & $\vdash (\vp\dnd \psi) \to (\vp \sau \psi \dnd \vp \si \psi)$ & \prule{syllogism}: (5),(6)\\[2mm] 
\end{tabular}

\eqref{useful-mem-intro}:\\[1mm]	
\begin{tabular}{lll}
(1) & $\vdash (\vp\dnd \psi) \to (\vp \sau \psi \dnd \vp \si \psi)$ & \eqref{vpdndpsi-to-vpsaupsi-dnd-vpsipsi}\\[1mm]
(2) & $\vdash (\vp \sau \psi \dnd \vp \si \psi) \to (\vp \sau \psi \to \vp \si \psi)$ & \prule{weakening} and the 
definition of $\dnd$\\[1mm]
(3) & $\vdash (\vp\dnd \psi) \to (\vp \sau \psi \to \vp \si \psi)$ & \prule{syllogism}: (1), (2)\\[1mm]
(3) & $\vdash \vp \sau \psi \to (\vp\dnd \psi \to \vp \si \psi)$ & \eqref{vp-to-psi-to-chi--psi-to-vp-to-chi}: (1)\\[1mm]
(4) & $\vdash  (\vp\dnd \psi \to \vp \si \psi) \dnd \neg(\vp\dnd \psi)\sau(\vp\si \psi)$ & 
\eqref{def-imp-or} \\[1mm]
(5) & $\vdash \vp \sau \psi \to \neg(\vp\dnd \psi)\sau(\vp\si \psi)$ & 
\eqref{sim-pairs-to-right-1}: (4), (3)
\end{tabular}

\, 
}

\subsection{First-order}

\begin{lemma}
\begin{align}
\Gamma \vdash \vp \to \psi & \quad \text{implies} 
\quad \Gamma \vdash \forall x \vp \to \forall x \psi, \label{vp-ra-psi-forall-vp-ra-forall-psi-M}\\
\Gamma \vdash \vp \dnd \psi & \quad \text{implies} 
\quad \Gamma \vdash \forall x \vp \dnd \forall x \psi, \label{vp-ra-psi-forall-vp-dnd-forall-psi-M}\\
\Gamma \vdash \vp \to (\psi \to \chi) & \quad \text{implies} 
\quad \Gamma \vdash \forall x \vp \to (\forall x \psi \to \forall x \chi). \label{vp-ra-psi-ra-chi-forall-M}
\end{align}
\end{lemma}
\solution{\,\\
\eqref{vp-ra-psi-forall-vp-ra-forall-psi-M}:\\[1mm]
\begin{tabular}{lll}
(1) & $\Gamma \vdash \vp \to \psi$ & (Assumption) \\[1mm]
(2) & $\Gamma \vdash \forall x(\vp \to \psi)$  & \prule{gen}: (1)\\[1mm]
(3) & $\Gamma \vdash \forall x(\vp \to \psi) \to  (\forall x\vp \to \forall x\psi)$  & \prule{\Monku}\\[1mm]
(4) & $\Gamma \vdash \forall x\vp \to \forall x\psi$ & \prule{modus ponens}: (2), (3) \\[2mm]
\end{tabular}

\eqref{vp-ra-psi-forall-vp-dnd-forall-psi-M}:\\[1mm]
\begin{tabular}{lll}
(1) & $\Gamma \vdash \vp \dnd \psi$ & (Assumption) \\[1mm]
(2) & $\Gamma \vdash \vp\to\psi$  & \eqref{Gamma-vp-psi-vp-si-psi}: (1) 
and the definition of $\dnd$\\[1mm]
(3) & $\Gamma \vdash \psi\to\vp$  & \eqref{Gamma-vp-psi-vp-si-psi}: (1) and the definition of $\dnd$\\[1mm]
(4) & $\Gamma \vdash \forall x \vp\to\forall x \psi$  & \eqref{vp-ra-psi-forall-vp-ra-forall-psi-M}: (2) \\[1mm]
(5) & $\Gamma \vdash \forall x \psi \to\forall x \vp$  & \eqref{vp-ra-psi-forall-vp-ra-forall-psi-M}: (3) \\[1mm]
(6) & $\Gamma \vdash \forall x \psi\dnd\forall x \vp$  & \eqref{Gamma-vp-psi-vp-si-psi}: (4), (5)
 and the definition of $\dnd$ \\[2mm]
\end{tabular}

\eqref{vp-ra-psi-ra-chi-forall-M}:\\[1mm]
\begin{tabular}{lll}
(1) & $\Gamma \vdash \vp \to (\psi \to \chi)$ & (Assumption) \\[1mm]
(2) & $\Gamma \vdash \forall x\vp \to \forall x(\psi \to \chi)$ & \eqref{vp-ra-psi-forall-vp-ra-forall-psi-M}: (1) \\[1mm]
(3) & $\Gamma \vdash \forall x(\psi \to \chi) \to (\forall x\psi \to \forall x\chi)$  & \prule{\Monku}\\[1mm]
(4) & $\Gamma \vdash \forall x \vp \to (\forall x\psi \to \forall x\chi)$  & \prule{syllogism}: (2), (3)
\end{tabular}

\,
}

\begin{lemma}
\begin{align}
\Gamma \vdash \vp \to \psi & \quad \text{implies} 
\quad \Gamma \vdash \exists x \vp \to \exists x \psi, \label{vp-ra-psi-exists-vp-ra-exists-psi-M}\\
\Gamma \vdash \vp \dnd \psi & \quad \text{implies} 
\quad \Gamma \vdash \exists x \vp \dnd \exists x \psi. \label{vp-ra-psi-exists-vp-dnd-exists-psi-M}
\end{align}
\end{lemma}
\solution{ \,\\

\eqref{vp-ra-psi-exists-vp-ra-exists-psi-M}:\\[1mm]
\begin{tabular}{lll}
(1) & $\Gamma \vdash \vp \to \psi$ & (Assumption) \\[1mm]
(2) & $\Gamma \vdash \neg \psi\to \neg\vp$  & \eqref{rec-ax3-rule}: (1) \\[1mm]
(3) & $\Gamma \vdash \forall x \neg \psi\to \forall x \neg\vp$  & 
\eqref{vp-ra-psi-forall-vp-ra-forall-psi-M}: (2)\\[1mm]
(4) & $\Gamma \vdash \neg\forall x \neg\vp \to \neg\forall x \neg\psi$  & \eqref{rec-ax3-rule}: (3) \\[1mm]
(5) & $\Gamma \vdash \exists x \vp \to \neg\forall x \neg\vp$  & \prule{axiom-exists-1} \\[1mm]
(6) & $\Gamma \vdash \exists x \vp \to \neg\forall x \neg\psi$  & \prule{syllogism}: (5), (4) \\[1mm]
(7) & $\Gamma \vdash \neg\forall x \neg\psi \to \exists x \psi$  & \prule{axiom-exists-2} \\[1mm]
(8) & $\Gamma \vdash \exists x \vp \to \exists x \psi$  & \prule{syllogism}: (6), (7) \\[2mm]
\end{tabular}

\eqref{vp-ra-psi-exists-vp-dnd-exists-psi-M}:\\[1mm]
\begin{tabular}{lll}
(1) & $\Gamma \vdash \vp \dnd \psi$ & (Assumption) \\[1mm]
(2) & $\Gamma \vdash \vp\to\psi$  & \eqref{Gamma-vp-psi-vp-si-psi}: (1) and the definition of $\dnd$  \\[1mm]
(3) & $\Gamma \vdash \psi\to\vp$  & \eqref{Gamma-vp-psi-vp-si-psi}: (1) and the definition of $\dnd$ \\[1mm]
(4) & $\Gamma \vdash \exists x \vp\to\exists x \psi$  & \eqref{vp-ra-psi-exists-vp-ra-exists-psi-M}: (2) \\[1mm]
(5) & $\Gamma \vdash \exists x \psi \to\exists x \vp$  & \eqref{vp-ra-psi-exists-vp-ra-exists-psi-M}: (3) \\[1mm]
(6) & $\Gamma \vdash\exists x \psi\dnd\exists x \vp$  & \eqref{Gamma-vp-psi-vp-si-psi}: (4), (5)
 and the definition of $\dnd$ 
\end{tabular}

\,
}

\begin{lemma}
\begin{align}
\vdash \exists x \vp \dnd  \neg \forall x\neg \vp, \label{existsxvp-dnd-negforallxngvp} \\
\vdash \forall x \vp \to \neg \exists x\neg \vp, \label{forallxvp-to-negexistsxngvp} \\
\vdash \neg \exists x\neg \vp  \to  \forall x \vp, \label{negexistsxngvp-to-forallxvp} \\
\vdash \forall x \vp \dnd \neg \exists x\neg \vp. \label{forallxvp-dnd-negexistsxngvp}
\end{align}
\end{lemma}
\solution{\,\\
\eqref{forallxvp-dnd-negexistsxngvp}: Apply \prule{axiom-exists-1}, \prule{axiom-exists-2},  
\eqref{Gamma-vp-psi-vp-si-psi} and the definition of $\dnd$.\\

\eqref{forallxvp-to-negexistsxngvp}: \\[1mm]
\begin{tabular}{lll}
(1) & $ \vdash \exists x \neg \vp \to \neg \forall x\neg \neg \vp $  & \prule{axiom-exists-1} \\[1mm]
(2) & $ \vdash \neg\neg \forall x\neg \neg \vp  \to \neg\exists x \neg \vp$  & 
\eqref{rec-ax3-rule}: (1) \\[1mm]
(3) & $ \vdash \forall x\neg \neg \vp \to \neg\neg \forall x\neg \neg \vp$ & \eqref{Gamma-vptopsi-psitovp-iff-vpdndpsi} 
and \eqref{negnegvp-dnd-vp} \\[1mm]
(4) & $ \vdash \forall x\neg \neg \vp  \to \neg\exists x \neg \vp $  & \prule{syllogism}: (3), (2) \\[1mm]
(5) & $ \vdash \vp \to\neg \neg \vp $ & \eqref{Gamma-vptopsi-psitovp-iff-vpdndpsi} and \eqref{negnegvp-dnd-vp} \\[1mm]
(6) & $ \vdash \forall x\vp \to  \forall x\neg \neg \vp $ & \eqref{vp-ra-psi-forall-vp-ra-forall-psi-M}: (5) \\[1mm]
(7) & $ \vdash  \forall x \vp \to  \neg\exists x \neg \vp$  & \prule{syllogism}: (6), (4) \\[2mm]
\end{tabular}

\eqref{negexistsxngvp-to-forallxvp}: \\[1mm]
\begin{tabular}{lll}
(1) & $ \vdash \neg \forall x\neg \neg \vp \to \exists x \neg \vp  $  & \prule{axiom-exists-2} \\[1mm]
(2) & $ \vdash \neg\exists x \neg \vp \to  \neg\neg \forall x\neg \neg \vp $  &
\eqref{rec-ax3-rule}: (1)\\[1mm]
(3) & $ \vdash \neg\neg \forall x\neg \neg \vp \to  \forall x\neg \neg \vp$ & \eqref{Gamma-vptopsi-psitovp-iff-vpdndpsi} and 
\eqref{negnegvp-dnd-vp} \\[1mm]
(4) & $ \vdash \neg\exists x \neg \vp \to  \forall x\neg \neg \vp $  & \prule{syllogism}: (2), (3) \\[1mm]
(5) & $ \vdash \neg \neg \vp \to \vp$ & \eqref{Gamma-vptopsi-psitovp-iff-vpdndpsi} and 
\eqref{negnegvp-dnd-vp} \\[1mm]
(6) & $ \vdash \forall x\neg \neg \vp \to \forall x\vp$ & \eqref{vp-ra-psi-forall-vp-ra-forall-psi-M}: (5) \\[1mm]
(7) & $ \vdash \neg\exists x \neg \vp \to   \forall x \vp $  & \prule{syllogism}: (4), (6) \\[2mm]
\end{tabular}

\eqref{forallxvp-dnd-negexistsxngvp}: Apply \eqref{forallxvp-to-negexistsxngvp}, 
\eqref{negexistsxngvp-to-forallxvp},  
\eqref{Gamma-vp-psi-vp-si-psi} and the definition of $\dnd$.
}

\begin{lemma}
\begin{align}
\vdash  \neg \exists x \vp\dnd  \forall x\neg \vp, \label{negexistsxvp-dnd-forallxnegvp} \\
\vdash \neg \forall x \vp \dnd   \exists x\neg \vp. \label{negforallxvp-dnd-existsxnegvp} 
\end{align}
\end{lemma}
\solution{\,\\[1mm]
\eqref{negexistsxvp-dnd-forallxnegvp}: \\[1mm]
\begin{tabular}{lll}
(1) & $\vdash \exists x \vp \dnd  \neg \forall x\neg \vp $ & \eqref{existsxvp-dnd-negforallxngvp} \\[1mm]
(2) & $\vdash \neg \neg\forall x\neg \vp \dnd \neg\exists x \vp$ & \eqref{Gamma-dnd-cong-neg}: (1) \\[1mm]
(3) & $\vdash \neg \neg \forall x\neg \vp \dnd \forall x\neg \vp$ & \eqref{negnegvp-dnd-vp}\\[1mm]
(4) & $\vdash \forall x\neg \vp \dnd \neg \neg \forall x\neg \vp$ & \eqref{sim-symm}: (3)\\[1mm]
(5) & $\vdash   \forall x\neg \vp \dnd \neg \exists x\vp$ & \prule{syllogism}: (4), (2)\\[1mm]
(6) & $\vdash \neg \exists x \vp\dnd  \forall x\neg \vp$ & \eqref{sim-symm}: (5)\\[2mm]
\end{tabular}

\eqref{negforallxvp-dnd-existsxnegvp} : \\[1mm]
\begin{tabular}{lll}
(1) & $\vdash \forall x \vp \dnd  \neg \exists x\neg \vp $ & \eqref{forallxvp-dnd-negexistsxngvp} \\[1mm]
(2) & $\vdash \neg \neg \exists x\neg \vp \dnd \neg \forall x \vp$ & \eqref{Gamma-dnd-cong-neg}: (1) \\[1mm]
(3) & $\vdash \neg \neg \exists x\neg \vp \dnd \exists x\neg \vp$ & \eqref{negnegvp-dnd-vp}\\[1mm]
(4) & $\vdash \exists x\neg \vp \dnd \neg \neg \exists x\neg \vp$ & \eqref{sim-symm}: (3)\\[1mm]
(5) & $\vdash \exists x\neg \vp \dnd \neg \forall x\vp$& \prule{syllogism}: (4), (2)\\[2mm]
(6) & $\vdash \neg \forall x \vp \dnd  \exists x\neg \vp$ & \eqref{sim-symm}: (5)\\[2mm]
\end{tabular}

\, 
}

\begin{lemma}
\begin{align}
\vdash & \, \forall x(\vp\si \psi)\to \forall x\vp \si \forall x \psi,  \label{forallx-vp-si-psi-to-forallxvp-si-forallxpsi-M}\\
\vdash & \, \forall x\vp \si \forall x \psi \to  \forall x(\vp\si \psi), \label{forallxvp-si-forallxpsi-to-forallx-vp-si-psi-M}\\
\vdash & \, \forall x\vp \si \forall x \psi \dnd  \forall x(\vp\si \psi). \label{forallxvp-si-forallxpsi-dnd-forallx-vp-si-psi-M}
\end{align}
\end{lemma}
\solution{$\,$\\
\eqref{forallx-vp-si-psi-to-forallxvp-si-forallxpsi-M}: \\[1mm]
\begin{tabular}{lll}
(1) & $\vdash \vp\si \psi \to \vp$ & \prule{weakening} \\[1mm]
(2) & $\vdash\forall x (\vp\si \psi) \to  \forall x \vp$ & \eqref{vp-ra-psi-forall-vp-ra-forall-psi-M}: (1) \\[1mm]
(3) & $\vdash \vp\si \psi \to \psi$ & \eqref{weak-si-2} \\[1mm]
(4) & $\vdash\forall x (\vp\si \psi) \to \forall x \psi$ & \eqref{vp-ra-psi-forall-vp-ra-forall-psi-M}: (3) \\[1mm]
(5) & $\vdash \forall x (\vp\si \psi) \to \forall x \psi \si \forall x\vp$ & 
\eqref{vp-to-psi-si-vp-to-chi-implies-vp-to-psi-si-chi}: (3), (4)\\[2mm]
\end{tabular}

\eqref{forallxvp-si-forallxpsi-to-forallx-vp-si-psi-M}: \\[1mm]
\begin{tabular}{lll}
(1) & $\vdash \vp \to (\psi \to \vp \si \psi)$ & \eqref{vp-to-psi-to-vpsipsi}\\[1mm]
(2) & $\vdash \forall x \vp  \to \forall x (\psi \to \vp \si \psi)$ & \eqref{vp-ra-psi-forall-vp-ra-forall-psi-M}: (1)\\[1mm]
(3) & $\vdash \forall x (\psi \to \vp \si \psi) \to (\forall x \psi  \to \forall x (\vp \si \psi))$ & \prule{\Monku}\\[1mm]
(4) & $\vdash \forall x \vp  \to (\forall x \psi  \to \forall x (\vp \si \psi))$ & \prule{syllogism}: (2), (3)\\[1mm]
(5) & $\vdash \forall x \vp \si \forall x \psi  \to \forall x(\vp\si \psi)$ & \prule{importation}: (5) \\[1mm]
\end{tabular}

\eqref{forallxvp-si-forallxpsi-dnd-forallx-vp-si-psi-M} follows from \eqref{forallx-vp-si-psi-to-forallxvp-si-forallxpsi-M}, 
\eqref{forallxvp-si-forallxpsi-to-forallx-vp-si-psi-M},
the definition of $\dnd$ and \eqref{Gamma-vp-psi-vp-si-psi}.\\
}

\begin{lemma}
\begin{align}
\vdash \exists x \vp \to \vp  \quad  \text{if $x$ does not occur in $\vp$}. \label{exists xvp-to-vpp}
\end{align}
\end{lemma}
\solution{\,\\[1mm]
\begin{tabular}{lll}
(1) & $\vdash \neg\vp \to \forall x \neg\vp$ & \prule{\Monku}, as $x$ does not occur in $\neg\vp$ \\[1mm]
(2) & $\vdash \neg\forall x \neg\vp \to \neg\neg\vp$ & \eqref{rec-ax3-rule}: (1)  \\[1mm]
(3) & $\vdash \exists x \vp  \to \neg\forall x \neg\vp $ &  \eqref{existsxvp-dnd-negforallxngvp}\\[1mm]
(4) & $\vdash \exists x \vp \to \neg\neg\vp$ & \prule{syllogism}: (3), (2) \\[1mm]
(5) & $\vdash  \neg\neg\vp \to \vp $ & \eqref{Gamma-vptopsi-psitovp-iff-vpdndpsi} and \eqref{negnegvp-dnd-vp}\\[1mm]
(6) & $\vdash \exists x \vp \to \vp$ & \prule{syllogism}: (4), (5)
\end{tabular}

\,
}

\begin{lemma}
Let $z$ be a variable such that $z$ does not occur in $\chi$.
\begin{align}
\Gamma \vdash  \chi \to (\vp \to \psi) & \quad \text{implies} 
\quad \Gamma \vdash   \chi  \to (\forall z \vp \to \forall z \psi), \label{chi-vppsi-chi-forallxvppsi-M}\\
\Gamma \vdash   \chi  \to (\vp \dnd \psi) & \quad \text{implies} 
\quad \Gamma \vdash   \chi  \to (\forall z \vp \dnd \forall z \psi), \label{chi-vppsi-chi-forallxvppsi-dnd-M}\\
\Gamma \vdash   \chi  \to (\vp \to \psi) & \quad \text{implies} 
\quad \Gamma \vdash   \chi  \to (\exists z \vp \to \exists z \psi), \label{chi-vppsi-chi-existsxvppsi-M}\\
\Gamma \vdash   \chi  \to (\vp \dnd \psi) & \quad \text{implies} 
\quad \Gamma \vdash   \chi  \to (\exists z \vp \dnd \exists z \psi). \label{chi-vppsi-chi-existsxvppsi-dnd-M}
\end{align}
\end{lemma}
\solution{\,\\[1mm]
\eqref{chi-vppsi-chi-forallxvppsi-M}:\\[1mm]
\begin{tabular}{lll}
(1) & $\Gamma \vdash  \chi  \to (\vp \to \psi)$  & (Assumption)\\[1mm]
(2) & $\Gamma \vdash \forall z \chi  \to \forall z(\vp \to \psi)$  
& \eqref{vp-ra-psi-forall-vp-ra-forall-psi-M}: (1) \\[1mm]
(3) & $\Gamma \vdash \forall z(\vp \to \psi) \to  (\forall z\vp \to \forall z\psi)$  & \prule{\Monku}\\[1mm]
(4) & $\Gamma \vdash \forall z \chi  \to  (\forall z\vp \to \forall z\psi)$  & 
\prule{syllogism}: (2), (3)\\[1mm]
(5) & $\Gamma \vdash  \chi  \to \forall z\chi$  
& \prule{{\Monkd}}, as $z$ does not occur in $\chi$  \\[1mm]
(6) & $\Gamma \vdash  \chi  \to (\forall z \vp \to \forall z \psi)$  & 
\prule{syllogism}: (5), (4) \\[2mm]
\end{tabular}

\eqref{chi-vppsi-chi-forallxvppsi-dnd-M}:\\[1mm]
\begin{tabular}{lll}
(1) & $\Gamma \vdash \chi \to (\vp \dnd \psi)$  & (Assumption)\\[1mm]
(2) & $\Gamma \vdash (\vp \dnd \psi) \to (\vp \to \psi)$  & \prule{weakening} and the definition of $\dnd$\\[1mm]
(3) & $\Gamma \vdash \chi  \to (\vp \to \psi)$  & \prule{syllogism}: (1), (2)\\[1mm]
(4) & $\Gamma \vdash \chi  \to (\forall z\vp \to \forall z\psi)$  & \eqref{chi-vppsi-chi-forallxvppsi-M}: (3)\\[1mm]
(5) & $\Gamma \vdash (\vp \dnd \psi) \to (\psi \to \vp)$  & \eqref{weak-si-2} and the definition of $\dnd$\\[1mm]
(6) & $\Gamma \vdash \chi  \to (\psi \to \vp)$  & \prule{syllogism}: (4), (5)\\[1mm]
(7) & $\Gamma \vdash \chi \to (\forall z \psi \to \forall z \vp)$  & \eqref{chi-vppsi-chi-forallxvppsi-M}: (6)\\[1mm]
(8) & $\Gamma \vdash \chi \to (\forall z \vp \dnd \forall z \psi)$  & 
\eqref{vp-to-psi-si-vp-to-chi-implies-vp-to-psi-si-chi}: (4), (7)\\[2mm]
\end{tabular}

\eqref{chi-vppsi-chi-existsxvppsi-M}:\\[1mm]
\begin{tabular}{lll}
(1) & $\Gamma \vdash \chi   \to (\vp \to \psi)$  
& (Assumption)\\[1mm]
(2) & $\Gamma \vdash (\vp \to \psi) \to (\neg\psi\to\neg\vp)$ & 
\eqref{contraposition} \\[1mm]
(3) & $\Gamma \vdash \chi   \to (\neg\psi \to \neg\vp)$  
& \prule{syllogism}: (1), (2)\\[1mm]
(4) & $\Gamma \vdash \chi \to (\forall z \neg\psi \to \forall z \neg\vp)$  
& \eqref{chi-vppsi-chi-forallxvppsi-M}: (3)\\[1mm]
(5) & $\Gamma \vdash (\forall z \neg\psi \to \forall z \neg\vp) \to (\neg\forall z \neg\vp \to \neg\forall z\neg\psi)$ & 
\eqref{contraposition} \\[1mm]
(6) & $\Gamma \vdash \chi   \to (\neg\forall z \neg\vp \to \neg\forall z \neg\psi)$ 
& \prule{syllogism}: (4), (5)\\[1mm]
(7) & $\Gamma \vdash \neg\forall z\neg\vp\to\exists z\vp$ & \prule{axiom-exists-2}\\[1mm]
(8) & $\Gamma \vdash \neg\forall z\neg\psi\to\exists z\psi$ & \prule{axiom-exists-2}\\[1mm]
(9) & $\Gamma \vdash (\neg\forall z \neg\vp \to \neg\forall z \neg\psi) \to 
(\exists z \vp \to \exists z \psi)$ & \eqref{psi-to-vp-chi-to-gamma-vp-to-chi-to-psi-to-gamma}: (7), (8)\\[1mm]
(10) & $\Gamma \vdash \chi  \to (\exists z \vp \to \exists z \psi)$ & \prule{syllogism}: (6), (9)\\[2mm]
\end{tabular}

\eqref{chi-vppsi-chi-existsxvppsi-dnd-M}: \\[1mm]
\begin{tabular}{lll}
(1) & $\Gamma \vdash \chi \to (\vp \dnd \psi)$  & (Assumption)\\[1mm]
(2) & $\Gamma \vdash (\vp \dnd \psi) \to (\vp \to \psi)$  & \prule{weakening} and the definition of $\dnd$\\[1mm]
(3) & $\Gamma \vdash \chi  \to (\vp \to \psi)$  & \prule{syllogism}: (1), (2)\\[1mm]
(4) & $\Gamma \vdash \chi  \to (\exists z\vp \to \exists z\psi)$  & \eqref{chi-vppsi-chi-existsxvppsi-M}: (3)\\[1mm]
(5) & $\Gamma \vdash (\vp \dnd \psi) \to (\psi \to \vp)$  & \eqref{weak-si-2}\\[1mm]
(6) & $\Gamma \vdash \chi  \to (\psi \to \vp)$  & \prule{syllogism}: (1), (2)\\[1mm]
(7) & $\Gamma \vdash \chi  \to (\exists z \psi \to \exists z \vp)$  & \eqref{chi-vppsi-chi-existsxvppsi-M}: (6)\\[1mm]
(8) & $\Gamma \vdash \chi  \to (\exists z \vp \dnd \exists z \psi)$  & 
\eqref{vp-to-psi-si-vp-to-chi-implies-vp-to-psi-si-chi}: (4), (7)
\end{tabular}

\,
}

\begin{lemma}
Let $z$ be a variable such that $z$ does not occur free in $\chi$.
\begin{align}
\Gamma \vdash  \chi \to (\vp \to \psi) & \quad \text{implies} 
\quad \Gamma \vdash   \chi  \to (\forall z \vp \to \forall z \psi), \label{chi-vppsi-chi-forallxvppsi-M-NFV}\\
\Gamma \vdash   \chi  \to (\vp \dnd \psi) & \quad \text{implies} 
\quad \Gamma \vdash   \chi  \to (\forall z \vp \dnd \forall z \psi), \label{chi-vppsi-chi-forallxvppsi-dnd-M-NFV}\\
\Gamma \vdash   \chi  \to (\vp \to \psi) & \quad \text{implies} 
\quad \Gamma \vdash   \chi  \to (\exists z \vp \to \exists z \psi), \label{chi-vppsi-chi-existsxvppsi-M-NFV}\\
\Gamma \vdash   \chi  \to (\vp \dnd \psi) & \quad \text{implies} 
\quad \Gamma \vdash   \chi  \to (\exists z \vp \dnd \exists z \psi). \label{chi-vppsi-chi-existsxvppsi-dnd-M-NFV}
\end{align}
\end{lemma}
\solution{Assume that $z\notin \FVE(\chi)$ and let $y$ be a new variable, distinct from $z$  
and not occuring in $\chi$. Let us denote $\delta:=\subb{z}{y}{\chi}$. 
By Lemma \ref{subb-lemma-useful-x-noccur-free-vp-noccur-subbxyvp-abstractML}, we have that 
$z$ does not  occur in $\delta$. \\[1mm]
\eqref{chi-vppsi-chi-forallxvppsi-M-NFV}: \\[1mm]
\bt{lll}
(1) & $\Gamma \vdash \chi \to (\vp \to \psi)$ & (Assumption) \\[1mm]
(2) & $\Gamma \vdash \chi \dnd \delta$ & Theorem \ref{thm-vp-dnd-subbxybvp}\\[1mm]
(3) & $\Gamma \vdash (\chi \to (\vp \to \psi)) \dnd (\delta \to (\vp \to \psi))$ & 
Theorem \ref{replacement-thm-NEW-dnd}: (2) \\[1mm]
(4) & $\Gamma \vdash  \delta \to (\vp \to \psi)$ & \eqref{vpdndpsi-imp-vp-iff-psi}: (1), (3) \\[1mm]
(5) & $\Gamma \vdash  \delta \to (\forall z \vp \to \forall z \psi)$ & 
\eqref{chi-vppsi-chi-forallxvppsi-M}: (4)\\[1mm]
(6) & $\Gamma \vdash (\chi \to (\forall z \vp \to \forall z \psi)) \dnd (\delta \to (\forall z \vp \to \forall z \psi))$ & 
Theorem \ref{replacement-thm-NEW-dnd}: (2) \\[1mm]
(7) & $\Gamma \vdash  \chi \to (\forall z \vp \to \forall z \psi)$ 
& \eqref{vpdndpsi-imp-vp-iff-psi}: (5), (6) \\[2mm]
\et

\eqref{chi-vppsi-chi-forallxvppsi-dnd-M-NFV} - \eqref{chi-vppsi-chi-existsxvppsi-dnd-M-NFV} are proved similary, using 
\eqref{chi-vppsi-chi-forallxvppsi-dnd-M} - \eqref{chi-vppsi-chi-existsxvppsi-dnd-M}.
}
\subsection{Definedness, totality and equality}\label{section-def}

We give detailed proofs of some  basic theorems and derived rules for  definedness, totality, equality.

\begin{lemma}
\begin{align}
\vdash  \, \bot \dnd \ceil{\bot}. \label{bot-dnd-ceilbot}
\end{align}
\end{lemma}
\solution{Apply \prule{ex falso quodlibet} and \prule{\axceilbot}.
}

\begin{lemma}
\begin{align}
\Gamma\vdash\vp\to\psi & \quad \text{implies} \quad \Gamma\vdash\ceil{\vp}\to\ceil{\psi},
\label{lemma-imp-compat-in-ceil}\\
\Gamma\vdash\vp\dnd\psi & \quad \text{implies} \quad \Gamma\vdash\ceil{\vp}\dnd\ceil{\psi}, 
\label{rule-iff-compat-in-ceil}\\
\Gamma\vdash\vp\to\psi & \quad \text{implies} \quad \Gamma\vdash\floor{\vp}\to\floor{\psi}, 
\label{lemma-imp-compat-in-floor}\\
\Gamma\vdash\vp \dnd \psi & \quad \text{implies} \quad \Gamma\vdash\floor{\vp} \dnd \floor{\psi}. 
\label{lemma-iff-compat-in-floor}
\end{align}
\end{lemma}
\solution{\, \\[1mm]
\eqref{lemma-imp-compat-in-ceil}: By \prule{framing-right} and the definition of $\defmapn$.

\eqref{rule-iff-compat-in-ceil}: By \eqref{lemma-imp-compat-in-ceil} and the definition of $\dnd$.

\eqref{lemma-imp-compat-in-floor}:\\[1mm]
\begin{tabular}{lll}
(1) & $\Gamma\vdash\vp\to\psi$ & (Assumption)\\[1mm]
(2) & $\Gamma\vdash\neg\psi\to\neg\vp$ & \eqref{rec-ax3-rule}: (1) \\[1mm]
(3) & $\Gamma\vdash\ceil{\neg\psi}\to\ceil{\neg\vp}$ & \eqref{lemma-imp-compat-in-ceil}: (2)\\[1mm]
(4) & $\Gamma\vdash\neg\ceil{\neg\vp}\to\neg\ceil{\neg\psi}$ & \eqref{rec-ax3-rule}: (3) \\[1mm]
(5) & $\Gamma\vdash\floor{\vp}\to\floor{\psi}$ & (4) and the definition of $\totmapn$ \\[2mm]
\end{tabular}

\eqref{lemma-iff-compat-in-floor}: By \eqref{lemma-imp-compat-in-floor} and the definition of $\dnd$.
}

\begin{lemma}
\begin{align}
\vdash  \,  \neg\floor{\vp} \dnd  \ceil{\neg\vp}, \label{negfloorvp-dnd-ceilnegvp} \\
\vdash  \, \neg\ceil{\vp} \dnd   \floor{\neg\vp}, \label{negceilvp-dnd-negfloornegfloor} \\
\vdash  \, \ceil{\vp}\dnd \neg \floor{\neg\vp}.  \label{lemma-ceil-negfloornegfloor}
\end{align}
\end{lemma}
\solution{
By definition, $\floor{\vp}=\neg\ceil{\neg\vp}$, hence $\neg\floor{\vp}=\neg\neg\ceil{\neg\vp}$.
Apply \eqref{negnegvp-dnd-vp} to get \eqref{negfloorvp-dnd-ceilnegvp}.

As, by  definition, $\floor{\neg\vp}=\neg\ceil{\neg\neg\vp}$, we get that \\[1mm]
\begin{tabular}{lll}
(1) & $\Gamma\vdash \vp \dnd \neg\neg\vp$ & \eqref{negnegvp-dnd-vp} \\[1mm]
(2) & $\Gamma\vdash \ceil{\vp} \dnd \ceil{\neg\neg\vp} $ & \eqref{rule-iff-compat-in-ceil}: (1) \\[1mm]
(3) & $\Gamma\vdash \neg \ceil{\vp} \dnd \floor{\neg\vp}$ & \eqref{rec-ax3-rule}: (2) \\[2mm]
\end{tabular}

Hence, \eqref{negceilvp-dnd-negfloornegfloor} holds. Furthermore, \\[1mm]
\begin{tabular}{lll}
(1) & $\Gamma\vdash \neg \ceil{\vp} \dnd \floor{\neg\vp}$ & \eqref{negceilvp-dnd-negfloornegfloor} \\[1mm]
(2) & $\Gamma\vdash \neg\neg\ceil{\vp} \dnd \neg\floor{\neg\vp}$ & \eqref{rec-ax3-rule}: (1) \\[1mm]
(3) & $\Gamma\vdash \ceil{\vp} \dnd \neg\neg\ceil{\vp}$ & \eqref{negnegvp-dnd-vp} \\[1mm]
(4) & $\Gamma\vdash \ceil{\vp}\dnd \neg\floor{\neg\vp}$ & \eqref{sim-trans}: (3), (2)\\[2mm]
\end{tabular}

Hence, \eqref{lemma-ceil-negfloornegfloor} also holds. 
}

\begin{lemma}
\begin{align}
\vdash  \, \floor{\vp}\to\vp. \label{lemma-floor-elim-alt}
\end{align}
\end{lemma}
\solution{\,\\[1mm]
\begin{tabular}{lll}
(1) & $\vdash\neg\vp\to\ceil{\neg\vp}$ & \prule{\axvptoceilvp}\\[1mm]
(2) & $\vdash\neg\ceil{\neg\vp}\to\neg\neg\vp$ & \eqref{rec-ax3-rule}: (1)\\[1mm]
(3) & $\vdash\neg\neg\vp\to\vp$ & \eqref{Gamma-vptopsi-psitovp-iff-vpdndpsi} and \eqref{negnegvp-dnd-vp}\\[1mm]
(4) & $\vdash\neg\ceil{\neg\vp}\to\vp$ & \prule{syllogism}: (2), (3)\\[1mm]
(5) & $\vdash\floor{\vp}\to\vp$ & (4) and the definition of $\totmapn$
\end{tabular}

\,
}

\begin{lemma}
\begin{align}
\Gamma\vdash\vp & \quad \text{iff} \quad \Gamma\vdash\floor{\vp}. \label{lemma-floor-intro-iff}
\end{align}
\end{lemma}
\solution{\,\\[1mm]
$(\Ra)$: \\[1mm]
\begin{tabular}{lll}
(1) & $\Gamma\vdash\vp$ & (Assumption)\\[1mm]
(2) & $\Gamma\vdash\vp\to\neg\neg\vp$ & \eqref{Gamma-vptopsi-psitovp-iff-vpdndpsi} and \eqref{negnegvp-dnd-vp}\\[1mm]
(3) & $\Gamma\vdash\neg\neg\vp$ & \prule{modus-ponens}: (1), (2)\\[1mm]
(4) & $\Gamma\vdash\neg\neg\vp \to (\neg\vp\to \bot)$ & \prule{axiom-not-1}\\[1mm]
(5) & $\Gamma\vdash \neg\vp\to \bot$ & \prule{modus-ponens}: (3), (4)\\[1mm]
(6) & $\Gamma\vdash\ceil{\neg\vp}\to\ceil{\bot}$ & \eqref{lemma-imp-compat-in-ceil}: (5) \\[1mm]
(7) & $\Gamma\vdash\ceil{\bot}\to\bot$ & \prule{Propagation$_\bot$} and the definition of 
$\defmapn$\\[1mm]
(8) & $\Gamma\vdash\ceil{\neg\vp}\to\bot$ & \prule{syllogism}: (6), (7)\\[1mm]
(9) & $\Gamma\vdash (\ceil{\neg\vp}\to\bot) \to \neg\ceil{\neg\vp}$ & \prule{axiom-not-2}\\[1mm]
(10) & $\Gamma\vdash  \neg\ceil{\neg\vp}$ & \prule{modus-ponens}: (8), (9)\\[1mm]
(9) & $\Gamma\vdash\floor{\vp}$ & the definitions of $\totmapn$\\[2mm]
\end{tabular}

$(\La)$: \\[1mm]
\bt{lll}
(1) & $\Gamma\vdash \floor{\vp}$ & (Assumption)\\[1mm]
(2) & $\Gamma\vdash \floor{\vp}\to\vp$ & \eqref{lemma-floor-elim-alt}\\[1mm]
(1) & $\Gamma\vdash \vp$ & \prule{modus-ponens}: (1), (2)
\et

\,
}

\begin{lemma}\label{rule-vp-floor-vp}
Let $n\geq 1$ and $\vp_1, \ldots, \vp_n, \psi$ be patterns. The following are equivalent:
\be
\item \big($\Gamma\vdash\vp_1$, $\Gamma\vdash\vp_2$, \ldots, $\Gamma\vdash\vp_n$\big) imply $\Gamma\vdash\psi$;
\item \big($\Gamma\vdash\floor{\vp_1}$, $\Gamma\vdash\floor{\vp_2}$, \ldots, $\Gamma\vdash\floor{\vp_n}$\big) imply $\Gamma\vdash\floor{\psi}$.
\ee
\end{lemma}
\solution{By \eqref{lemma-floor-intro-iff}, we have that (for all $i=1,\ldots,n$, $\Gamma\vdash\vp_i$ iff 
$\Gamma\vdash \floor{\vp_i}$) and  ($\Gamma\vdash\psi$ iff $\Gamma\vdash \floor{\psi}$).
}

\begin{lemma}
\begin{align}
\vdash & \, \ceil{\vp\sau\psi} \to \ceil{\vp}\sau\ceil{\psi},  \label{lemma-ceil-compat-in-or-1}\\
\vdash & \, \ceil{\vp}\sau\ceil{\psi} \to \ceil{\vp\sau\psi}, \label{lemma-ceil-compat-in-or-2}\\
\vdash & \, \ceil{\vp\sau\psi}\dnd\ceil{\vp}\sau\ceil{\psi}, \label{lemma-ceil-compat-in-or}\\
\vdash & \, \floor{\vp\si\psi}\to\floor{\vp}\si\floor{\psi}, \label{lemma-floor-compat-in-and-1} \\
\vdash & \, \floor{\vp}\si\floor{\psi}  \to \floor{\vp\si\psi}, \label{lemma-floor-compat-in-and-2} \\
\vdash & \, \floor{\vp\si\psi}\dnd\floor{\vp}\si\floor{\psi}. \label{lemma-floor-compat-iff-and}
\end{align}
\end{lemma}
\solution{\,\\[1mm]
\eqref{lemma-ceil-compat-in-or-1}: By \prule{Propagation$_\vee$} and 
the definition of $\defmapn$. 

\eqref{lemma-ceil-compat-in-or-2}:\\[1mm]
\begin{tabular}{lll}
(1) & $\vdash\vp\to\vp\sau\psi$ & \prule{weakening}\\[1mm]
(2) & $\vdash\ceil{\vp}\to\ceil{\vp\sau\psi}$ & \prule{framing-right}: (1) and the definition of 
$\defmapn$\\[1mm]
(3) & $\vdash\psi\to\vp\sau\psi$ & \eqref{weak-sau-2}\\[1mm]
(4) & $\vdash\ceil{\psi}\to\ceil{\vp\sau\psi}$ & \prule{framing-right}: (3) and the definition of 
$\defmapn$\\[1mm]
(5) & $\vdash\ceil{\vp}\sau\ceil{\psi}\to\ceil{\vp\sau\psi}$ & 
\eqref{vp-to-chi-and-psi-to-chi-implies-vp-sau-psi-to-chi}: (2), (4) \\[2mm]
\end{tabular}

\eqref{lemma-ceil-compat-in-or}: Apply \eqref{lemma-ceil-compat-in-or-1}, \eqref{lemma-ceil-compat-in-or-2}, 
\eqref{Gamma-vp-psi-vp-si-psi} and the definition of $\dnd$. 

\eqref{lemma-floor-compat-in-and-1}:\\[1mm]
\begin{tabular}{lll}
(1) & $\vdash\ceil{\neg\vp}\sau\ceil{\neg\psi}\to\ceil{\neg\vp\sau\neg\psi}$ & 
 \eqref{lemma-ceil-compat-in-or-2}\\[1mm]
(2) & $\vdash\neg\vp\sau\neg\psi\to\neg(\vp\si\psi)$ & \eqref{Gamma-vptopsi-psitovp-iff-vpdndpsi} and \eqref{de-morgan-or-dnd}\\[1mm]
(3) & $\vdash\ceil{\neg\vp\sau\neg\psi}\to\ceil{\neg(\vp\si\psi)}$ & \prule{framing-right}: (2) 
and the definition of $\defmapn$\\[1mm]
(4) & $\vdash\ceil{\neg\vp}\sau\ceil{\neg\psi}\to\ceil{\neg(\vp\si\psi)}$ & \prule{syllogism}: (2), (4)\\[1mm]
(5) & $\vdash\neg\ceil{\neg(\vp\si\psi)}\to\neg(\ceil{\neg\vp}\sau\ceil{\neg\psi})$ & 
\eqref{rec-ax3-rule}: (5)\\[1mm]
(6) & $\vdash\neg(\ceil{\neg\vp}\sau\ceil{\neg\psi})\to\neg{\ceil{\neg\vp}}\si\neg{\ceil{\neg\psi}}$ & 
\eqref{Gamma-vptopsi-psitovp-iff-vpdndpsi} and\eqref{de-morgan-or-dnd}\\[1mm]
(7) & $\vdash\floor{\vp\si\psi}\to\floor{\vp}\si\floor{\psi}$ & \prule{syllogism}: (6), (7) and the 
definition of $\totmapn$ \\[2mm]
\end{tabular}

\eqref{lemma-floor-compat-in-and-2}:\\[1mm]
\begin{tabular}{lll}
(1) & $\vdash\neg\ceil{\neg\vp}\si\neg\ceil{\neg\psi}\to\neg(\ceil{\neg\vp}\sau\ceil{\neg\psi})$ & 
\eqref{Gamma-vptopsi-psitovp-iff-vpdndpsi} and \eqref{de-morgan-and-dnd} \\[1mm]
(2) & $\vdash\neg(\vp\si\psi)\to\neg\vp\sau\neg\psi$ & \eqref{Gamma-vptopsi-psitovp-iff-vpdndpsi} and 
\eqref{de-morgan-and-dnd} \\[1mm]
(3) & $\vdash\ceil{\neg(\vp\si\psi)}\to\ceil{\neg\vp\sau\neg\psi}$ & \prule{framing-right}: (2)\\[1mm]
(4) & $\vdash\ceil{\neg\vp\sau\neg\psi}\to\ceil{\neg\vp}\sau\ceil{\neg\psi}$ & \eqref{lemma-ceil-compat-in-or-1} \\[1mm]
(5) & $\vdash\ceil{\neg(\vp\si\psi)}\to\ceil{\neg\vp}\sau\ceil{\neg\psi}$ & \prule{syllogism}: (3), (4)\\[1mm]
(6) & $\vdash\neg(\ceil{\neg\vp}\sau\ceil{\neg\psi})\to\neg\ceil{\neg(\vp\si\psi)}$ & 
\eqref{rec-ax3-rule}: (5)\\[1mm]
(7) & $\vdash\neg\ceil{\neg\vp}\si\neg\ceil{\neg\psi}\to\neg\ceil{\neg(\vp\si\psi)}$ & \prule{syllogism}: (1), (6)\\[1mm]
(8) & $\vdash\floor{\vp}\si\floor{\psi}\to\floor{\vp\si\psi}$ & definition of $\totmapn$ \\[2mm]
\end{tabular}

\eqref{lemma-floor-compat-iff-and}: Apply \eqref{lemma-floor-compat-in-and-1}, \eqref{lemma-floor-compat-in-and-2}, 
\eqref{Gamma-vp-psi-vp-si-psi} and the definition of $\dnd$. 
}

\begin{lemma}
\begin{align}
\vdash & \, \ceil{x\sau \vp}.  \label{ceil-xsauvp}
\end{align}
\end{lemma}
\solution{\,\\[1mm]
\begin{tabular}{lll}
(1) & $\vdash\ceil{x}$ & \prule{def}\\[1mm]
(2) & $\vdash\ceil{x}\to\ceil{x}\sau\ceil{\vp}$ & \prule{weakening}\\[1mm]
(3) & $\vdash\ceil{x}\sau\ceil{\vp}$ & \prule{modus-ponens}: (1), (2)\\[1mm]
(4) & $\vdash\ceil{x}\sau\ceil{\vp}\to\ceil{x\sau \vp}$ & \eqref{lemma-ceil-compat-in-or-2}\\[1mm]
(5) & $\vdash\ceil{x\sau \vp}$ & \prule{modus-ponens}: (3), (4)
\end{tabular}

\,
}

\subsubsection{Equality}\label{section-eq}

\begin{lemma}
\begin{align}
\vdash \vp \eqdef \psi \to (\vp \dnd \psi). \label{equal-imp-dnd}
\end{align}
\end{lemma}
\solution{
It follows immediately by \eqref{lemma-floor-elim-alt} and the definition of $\eqdef$.
}

\begin{lemma}
\bea
\Gamma  \vdash \vp \dnd  \psi & \text{iff} &  \Gamma \vdash \vp \eqdef \psi. \label{Gamma-equal-iff-dnd}
\eea
\end{lemma}
\solution{It follows immediately from \eqref{lemma-floor-intro-iff} and the definition of $\eqdef$.
}

\begin{lemma}
\begin{align}
\vdash \vp \eqdef \vp .\label{eq-refl}
\end{align}
\end{lemma}
\solution{We have that \\[1mm]
\begin{tabular}{lll}
(1) & $\vdash \vp\dnd\vp$ & \eqref{sim-refl}\\[1mm]
(2) & $\vdash \floor{\vp\dnd\vp}$ & \eqref{lemma-floor-intro-iff}: (1)\\[1mm]
(3) & $\vdash \vp \eqdef \vp$ & (2) and the definition of $\eqdef$
\end{tabular}

\,
}

\begin{lemma}
\bea
\Gamma \vdash \vp \eqdef \psi & \text{implies} & \Gamma \vdash \psi \eqdef \vp, \label{eq-symm}\\
\Gamma \vdash \vp \eqdef \psi \text{ and } \Gamma \vdash \psi \eqdef \chi & \text{imply}  & 
\Gamma \vdash \vp \eqdef \chi. \label{eq-trans}
\eea
\end{lemma}
\solution{
Apply Lemma \ref{rule-vp-floor-vp} and \eqref{sim-symm}, \eqref{sim-trans}.
}

\begin{lemma}
\bea
\Gamma\vdash\vp \eqdef \psi & \text{implies} & \Gamma\vdash (\neg\vp) \eqdef (\neg\psi),
\label{lemma-eq-cong-not}\\[1mm]
\Gamma\vdash\vp \eqdef \psi & \text{implies} & \Gamma\vdash (\vp\sau\chi) \eqdef (\psi \sau \chi), 
\label{lemma-eq-cong-or-1}\\
\Gamma\vdash\psi \eqdef \psi  & \text{implies} & \Gamma\vdash (\chi\sau\vp) \eqdef (\chi\sau\psi), 
\label{lemma-eq-cong-or-2}\\
\Gamma\vdash \vp \eqdef \psi & \text{implies} & \Gamma\vdash (\vp \si \chi)  \eqdef (\psi \si \chi), 
\label{lemma-eq-cong-and-1}\\[1mm]
\Gamma\vdash \psi \eqdef \psi & \text{implies} & \Gamma\vdash (\chi\si\vp) \eqdef (\chi\si\psi), 
\label{lemma-eq-cong-and-2} \\[1mm]
\Gamma\vdash \vp \eqdef \psi  & \text{implies} & \Gamma\vdash (\vp\to\chi) \eqdef (\psi \to \chi), 
\label{lemma-eq-cong-imp-1}\\
\Gamma\vdash \vp \eqdef \psi & \text{implies} & \Gamma\vdash (\chi\to\vp) \eqdef (\chi\to\psi), 
\label{lemma-eq-cong-imp-2} \\[1mm]
\Gamma\vdash\vp \eqdef \psi & \text{implies} & \Gamma\vdash (\forall x\vp) \eqdef (\forall x\psi), 
\label{lemma-eq-cong-forall} \\[1mm]
\Gamma\vdash\vp \eqdef \psi & \text{implies} & \Gamma\vdash (\exists x\vp) \eqdef (\exists x\psi), 
\label{lemma-eq-cong-exists} \\[1mm]
\Gamma\vdash\vp \eqdef \psi & \text{implies} & \Gamma\vdash (\appln{\vp}{\chi}) \eqdef (\appln{\psi}{\chi}),
\label{lemma-eq-cong-app-1}\\
\Gamma\vdash\vp \eqdef \psi & \text{implies} & \Gamma\vdash (\appln{\chi}{\vp}) \eqdef (\appln{\chi}{\psi}).
\label{lemma-eq-cong-app-2}
\eea
\end{lemma}
\solution{Apply Lemma \ref{rule-vp-floor-vp}, the definition of $\eqdef$ and 
\eqref{Gamma-dnd-cong-neg}-\eqref{sim-pairs-to-right} to get 
\eqref{lemma-eq-cong-not}-\eqref{lemma-eq-cong-imp-2}, \eqref{vp-ra-psi-forall-vp-dnd-forall-psi-M} to get  \eqref{lemma-eq-cong-forall},
\eqref{vp-ra-psi-exists-vp-dnd-exists-psi-M}  to get \eqref{lemma-eq-cong-exists}, 
\eqref{rule-iff-compat-in-app-left} to get \eqref{lemma-eq-cong-app-1}, and 
\eqref{rule-iff-compat-in-app-right} to get \eqref{lemma-eq-cong-app-2}.
}

\begin{lemma}
Let $x$, $y$, be variables. 
\begin{align}
\Gamma \vdash x \eqdef y \to (\vp \to \psi) & \quad \text{implies} 
\quad \Gamma \vdash x \eqdef y \to (\appln{\vp}{\chi} \to \appln{\psi}{\chi}),
\label{xeqytovppsi-imp-xeqy-vpchitopsichi}\\
\Gamma \vdash x \eqdef y \to (\vp \dnd \psi) & \quad \text{implies} 
\quad \Gamma \vdash x \eqdef y \to (\appln{\vp}{\chi} \dnd \appln{\psi}{\chi}),
\label{xeqytovpdndpsi-imp-xeqy-vpchidndpsichi}\\
\Gamma \vdash x \eqdef y \to (\vp \to \psi) & \quad \text{implies} 
\quad \Gamma \vdash x \eqdef y \to (\appln{\chi}{\vp} \to \appln{\chi}{\psi}),
\label{xeqytovppsi-imp-xeqy-chivptochipsi}\\
\Gamma \vdash x \eqdef y \to (\vp \dnd \psi) & \quad \text{implies} 
\quad \Gamma \vdash x \eqdef y \to (\appln{\chi}{\vp} \dnd \appln{\chi}{\psi}).
\label{xeqytovpdndpsi-imp-xeqy-chivpndchipsi}
\end{align}
\end{lemma}
\solution{Apply \eqref{floordelta-imp-vpchitopsichi}-\eqref{floordelta-imp-chivpndchipsi} and the fact 
that $x \eqdef y=\floor{x \dnd y}$.
}

\subsection{Application}

\begin{proposition}
\begin{align}
\Gamma\vdash\vp\dnd\psi \,\, \text{and}  \,\, \Gamma \vdash \chi\dnd\gamma &
\quad \text{implies} \quad \Gamma \vdash \appln{\vp}{\chi} \dnd \appln{\psi}{\gamma}, 
\label{rule-iff-compat-in-app}\\
\Gamma\vdash\vp\dnd\psi  &
\quad \text{implies} \quad \Gamma \vdash \appln{\vp}{\chi} \dnd \appln{\psi}{\chi}, 
\label{rule-iff-compat-in-app-left}\\
\Gamma \vdash \chi\dnd\gamma  &
\quad \text{implies} \quad \Gamma \vdash \appln{\vp}{\chi}\dnd \appln{\vp}{\gamma} .
\label{rule-iff-compat-in-app-right}
\end{align}
\end{proposition}
\solution{\,\\[1mm]
\eqref{rule-iff-compat-in-app}:\\[1mm]
\begin{tabular}{lll}
(1) & $\Gamma\vdash\vp\dnd\psi$ & (Assumption)\\[1mm]
(2) & $\Gamma\vdash\vp\to\psi$ & \eqref{Gamma-vp-psi-vp-si-psi}: (1) and the definition of $\dnd$\\[1mm]
(3) & $\Gamma\vdash\appln{\vp}{\chi}\to\appln{\psi}{\chi}$ & \prule{framing-left}: (2)\\[1mm]
(4) & $\Gamma\vdash\chi\dnd\gamma$ & (Assumption)\\[1mm]
(5) & $\Gamma\vdash(\chi\dnd\gamma)\to(\chi\to\gamma)$ & \prule{weakening} and the definition of $\dnd$\\[1mm]
(6) & $\Gamma\vdash\chi\to\gamma$ & \prule{modus-ponens}: (4), (5)\\[1mm]
(7) & $\Gamma\vdash\appln{\psi}{\chi}\to\appln{\psi}{\gamma}$ & \prule{framing-right}: (6)\\[1mm]
(8) & $\Gamma\vdash\appln{\vp}{\chi}\to\appln{\psi}{\gamma}$ & \prule{syllogism}: (3), (7)\\[1mm]
(9) & $\Gamma\vdash\vp\dnd\psi$ & (Assumption)\\[1mm]
(10) & $\Gamma\vdash\psi\to\vp$ & \eqref{Gamma-vp-psi-vp-si-psi}: (9) and the definition of $\dnd$\\[1mm]
(11) & $\Gamma\vdash\appln{\psi}{\gamma}\to\appln{\vp}{\gamma}$ & \prule{framing-left}: (10)\\[1mm]
(12) & $\Gamma\vdash\chi\dnd\gamma$ & (Assumption)\\[1mm]
(13) & $\Gamma\vdash(\chi\dnd\gamma)\to(\gamma\to\chi)$ & \prule{weakening} and the definition of $\dnd$\\[1mm]
(14) & $\Gamma\vdash\gamma\to\chi$ & \prule{modus-ponens}: (12), (13)\\[1mm]
(15) & $\Gamma\vdash\appln{\vp}{\gamma}\to\appln{\vp}{\chi}$ & \prule{framing-right}: (14)\\[1mm]
(16) & $\Gamma\vdash\appln{\psi}{\gamma}\to\appln{\vp}{\chi}$ & \prule{syllogism}: (11), (15)\\[1mm]
(17) & $\Gamma\vdash\appln{\vp}{\chi}\dnd\appln{\psi}{\gamma}$ & \eqref{Gamma-vp-psi-vp-si-psi}: (8), (16) and the definition of $\dnd$\\[2mm]
\end{tabular}

\eqref{rule-iff-compat-in-app-left} We can apply \eqref{rule-iff-compat-in-app} with $\gamma:=\chi$,  as 
$\vdash \chi \dnd \chi$, by \eqref{sim-refl}.\\

\eqref{rule-iff-compat-in-app-right} Apply \eqref{rule-iff-compat-in-app} with $\psi:=\vp$, as 
$\vdash \vp \dnd \vp$, by \eqref{sim-refl}.
}

\begin{lemma}
\begin{align}
& \vdash  \, \appln{\bot}{\vp} \to \bot, \label{propagation-bot-l} \\
& \vdash  \, \appln{\vp}{\bot} \to \bot. \label{propagation-bot-r}
\end{align}
\end{lemma}
\solution{\,\\[1mm]
\eqref{propagation-bot-l}:\\[1mm]
\begin{tabular}{lll}
(1) & $\vdash \ceil{\bot} \dnd  \bot $ & \eqref{bot-dnd-ceilbot}\\[1mm]
(2) & $\vdash \appln{\ceil{\bot}}{\vp} \dnd \appln{\bot}{\vp} $ & \eqref{rule-iff-compat-in-app-left}: (1) \\[1mm]
(3) & $\vdash \appln{\ceil{\bot}}{\vp}  \to \ceil{\bot}$  &  \prule{\propagationdef} \\[1mm]
(4) & $\vdash  \appln{\bot}{\vp}  \to \bot$ & \eqref{sim-pairs-to-left-right-1} : (2), (1), (3)\\[2mm]
\end{tabular}

\eqref{propagation-bot-r}:\\[1mm]
\begin{tabular}{lll}
(1) & $\vdash \ceil{\bot} \dnd  \bot $ & \eqref{bot-dnd-ceilbot}\\[1mm]
(2) & $\vdash \appln{\vp}{\ceil{\bot}} \dnd \appln{\vp}{\bot}$ & \eqref{rule-iff-compat-in-app-right}: (1) \\[1mm]
(3) & $\vdash \appln{\vp}{\ceil{\bot}}  \to \ceil{\bot}$  &  \prule{\propagationdef} \\[1mm]
(4) & $\vdash  \appln{\vp}{\bot}  \to \bot$ & \eqref{sim-pairs-to-left-right-1} : (2), (1), (3)
\end{tabular}

\,
}

\begin{lemma}
\begin{align}
\Gamma \vdash \floor{\delta} \to (\vp \to \psi) & \quad \text{implies} 
\quad \Gamma \vdash \floor{\delta} \to (\appln{\vp}{\chi} \to \appln{\psi}{\chi}),
\label{floordelta-imp-vpchitopsichi}\\
\Gamma \vdash \floor{\delta} \to (\vp \dnd \psi) & \quad \text{implies} 
\quad \Gamma \vdash \floor{\delta} \to (\appln{\vp}{\chi} \dnd \appln{\psi}{\chi}),
\label{floordelta-imp-vpchidndpsichi}\\
\Gamma \vdash \floor{\delta} \to (\vp \to \psi) & \quad \text{implies} 
\quad \Gamma \vdash \floor{\delta} \to (\appln{\chi}{\vp} \to \appln{\chi}{\psi}),
\label{floordelta-imp-chivptochipsi}\\
\Gamma \vdash \floor{\delta} \to (\vp \dnd \psi) & \quad \text{implies} 
\quad \Gamma \vdash \floor{\delta} \to (\appln{\chi}{\vp} \dnd \appln{\chi}{\psi}).
\label{floordelta-imp-chivpndchipsi}
\end{align}
\end{lemma}
\solution{\,\\[1mm]
\eqref{floordelta-imp-vpchitopsichi}: \\[1mm]
\begin{tabular}{lll}
(1) & $\Gamma\vdash \floor{\delta} \to (\vp \to \psi)$ & (Induction hypothesis)\\[1mm]
(2) & $\Gamma\vdash \vp  \to (\floor{\delta}\to \psi)$ & \eqref{vp-to-psi-to-chi--psi-to-vp-to-chi}: (1)\\[1mm]
(3) & $\Gamma\vdash (\floor{\delta}\to \psi) \to \neg\floor{\delta} \sau\psi$ & 
\eqref{Gamma-vp-psi-vp-si-psi}: \eqref{def-imp-or} and the definition of $\dnd$ \\[1mm]
(4) & $\Gamma\vdash \vp  \to \neg\floor{\delta} \sau\psi$ & \prule{syllogism}: (2), (3)\\[1mm]
(5) & $\Gamma\vdash \neg\floor{\delta} \to  \ceil{\neg\delta}$ & \eqref{Gamma-vp-psi-vp-si-psi}: \eqref{negfloorvp-dnd-ceilnegvp}
and the definition of $\dnd$ \\[1mm]
(6) & $\Gamma\vdash \neg\floor{\delta} \sau \psi \to \ceil{\neg\delta} \sau \psi$ & \prule{expansion}: (5)\\[1mm]
(7) & $\Gamma\vdash \vp  \to \ceil{\neg\delta} \sau \psi$ & \prule{syllogism}: (4), (6)\\[1mm]
(8) & $\Gamma\vdash \appln{\vp}{\chi}   \to \appln{(\ceil{\neg\delta}\sau\psi)}{\chi}$ & 
\prule{framing-right}: (7)\\[1mm]
(9) & $\Gamma\vdash \appln{(\ceil{\neg\delta}\sau\psi)}{\chi} \to (\appln{\ceil{\neg\delta}}{\chi} \sau 
\appln{\psi}{\chi})$ & \prule{Propagation$_\vee$}\\[1mm]
(10) & $\Gamma\vdash \appln{\vp}{\chi}  \to  \appln{\ceil{\neg\delta}}{\chi} \sau
\appln{\psi}{\chi}$ & \prule{syllogism}: (8), (9)\\[1mm]
(11) & $\Gamma\vdash \appln{\ceil{\neg\delta}}{\chi} \to \ceil{\neg\delta}$ &  \prule{\propagationdef}\\[1mm]
(12) & $\Gamma\vdash \appln{\ceil{\neg\delta}}{\chi}\sau  \appln{\psi}{\chi} \to \ceil{\neg\delta}\sau  \appln{\psi}{\chi}$ 
& \prule{expansion}: (11)\\[1mm]
(13) & $\Gamma\vdash \appln{\vp}{\chi}  \to \ceil{\neg\delta}\sau  \appln{\psi}{\chi}$ &  \prule{syllogism}: (10), (12) \\[1mm]
(14) & $\Gamma\vdash \ceil{\neg\delta}\sau  \appln{\psi}{\chi} \to (\floor{\delta} \to  \appln{\psi}{\chi})$ & 
\eqref{Gamma-vp-psi-vp-si-psi}: \eqref{vpsaupsi-negvptopsi} and the definitions of $\dnd$,  $\totmapn$\\[1mm]
(15) & $\Gamma\vdash \appln{\vp}{\chi}  \to (\floor{\delta} \to  \appln{\psi}{\chi})$ &  
\prule{syllogism}: (13), (14)\\[1mm]
(16) & $\Gamma\vdash  \floor{\delta} \to ( \appln{\vp}{\chi}\to  \appln{\psi}{\chi})$ &  
\eqref{vp-to-psi-to-chi--psi-to-vp-to-chi}: (11) \\[2mm]
\end{tabular}

\eqref{floordelta-imp-vpchidndpsichi}: \\[1mm]
\begin{tabular}{lll}
(1) & $\Gamma \vdash \floor{\delta}  \to (\vp \dnd \psi)$  & (Assumption)\\[1mm]
(2) & $\Gamma \vdash \floor{\delta}  \to (\vp \to \psi)$   & \eqref{vp-to-psi-si-vp-to-chi-implies-vp-to-psi-si-chi}: (1) and the definition of $\dnd$\\[1mm]
(3) & $\Gamma \vdash \floor{\delta}  \to (\appln{\chi}{\vp}\to \appln{\chi}{\psi})$  
& \eqref{floordelta-imp-vpchitopsichi}: (2)\\[1mm]
(4) & $\Gamma \vdash \floor{\delta}  \to (\psi \to \vp)$  & \eqref{vp-to-psi-si-vp-to-chi-implies-vp-to-psi-si-chi}: (1)  and the definition of $\dnd$\\[1mm]
(5) & $\Gamma \vdash \floor{\delta}  \to (\appln{\chi}{\psi} \to \appln{\chi}{\vp}) $  & 
\eqref{floordelta-imp-vpchitopsichi}: (4)\\[1mm]
(6) & $\Gamma \vdash \floor{\delta}  \to (\appln{\chi}{\vp} \dnd \appln{\chi}{\psi})$  
& \eqref{vp-to-psi-si-vp-to-chi-implies-vp-to-psi-si-chi}: (3), (5) and the definition of $\dnd$\\[2mm]
\end{tabular}

\eqref{floordelta-imp-chivptochipsi}: \\[1mm]
\begin{tabular}{lll}
(1) & $\Gamma\vdash \floor{\delta} \to (\vp \to \psi)$ & (Induction hypothesis)\\[1mm]
(2) & $\Gamma\vdash \vp  \to (\floor{\delta}\to \psi)$ & \eqref{vp-to-psi-to-chi--psi-to-vp-to-chi}: (1)\\[1mm]
(3) & $\Gamma\vdash (\floor{\delta}\to \psi) \to \neg\floor{\delta} \sau\psi$ & 
\eqref{Gamma-vp-psi-vp-si-psi}: \eqref{def-imp-or} and the definition of $\dnd$ \\[1mm]
(4) & $\Gamma\vdash \vp  \to \neg\floor{\delta} \sau\psi$ & \prule{syllogism}: (2), (3)\\[1mm]
(5) & $\Gamma\vdash \neg\floor{\delta} \to  \ceil{\neg\delta}$ & \eqref{Gamma-vp-psi-vp-si-psi}: \eqref{negfloorvp-dnd-ceilnegvp}
and the definition of $\dnd$ \\[1mm]
(6) & $\Gamma\vdash \neg\floor{\delta} \sau \psi \to \ceil{\neg\delta} \sau \psi$ & \prule{expansion}: (5)\\[1mm]
(7) & $\Gamma\vdash \vp  \to \ceil{\neg\delta} \sau \psi$ & \prule{syllogism}: (4), (6)\\[1mm]
(8) & $\Gamma\vdash \appln{\chi} {\vp}  \to \appln{\chi}{(\ceil{\neg\delta}\sau\psi)}$ & 
\prule{framing-left}: (7)\\[1mm]
(9) & $\Gamma\vdash \appln{\chi} {(\ceil{\neg\delta}\sau\psi)} \to (\appln{\chi} {\ceil{\neg\delta}} \sau 
\appln{\chi}{\psi})$ & \prule{Propagation$_\vee$}\\[1mm]
(10) & $\Gamma\vdash \appln{\chi}{\vp} \to  \appln{\chi}{\ceil{\neg\delta}}\sau
\appln{\chi}{\psi}$ & \prule{syllogism}: (8), (9)\\[1mm]
(11) & $\Gamma\vdash \appln{\chi}{\ceil{\neg\delta}}\to \ceil{\neg\delta}$ &  \prule{\propagationdef}\\[1mm]
(12) & $\Gamma\vdash \appln{\chi}{\ceil{\neg\delta}} \sau \appln{\chi}{\psi} \to \ceil{\neg\delta}\sau  \appln{\chi}{\psi}$ 
& \prule{expansion}: (11)\\[1mm]
(13) & $\Gamma\vdash \appln{\chi}{\vp}  \to \ceil{\neg\delta}\sau  \appln{\chi}{\psi}$ &  \prule{syllogism}: (10), (12) \\[1mm]
(14) & $\Gamma\vdash \ceil{\neg\delta}\sau  \appln{\chi}{\psi} \to (\floor{\delta} \to  \appln{\chi}{\psi})$ & 
\eqref{Gamma-vp-psi-vp-si-psi}: \eqref{vpsaupsi-negvptopsi} and the definitions of $\dnd$, $\totmapn$ \\[1mm]
(15) & $\Gamma\vdash \appln{\chi}{\vp} \to (\floor{\delta} \to  \appln{\chi}{\psi})$ &  
\prule{syllogism}: (13), (14)\\[1mm]
(16) & $\Gamma\vdash  \floor{\delta} \to (\appln{\chi}{\vp}\to  \appln{\chi}{\psi})$ &  
\eqref{vp-to-psi-to-chi--psi-to-vp-to-chi}: (11) \\[2mm]
\end{tabular}

\eqref{floordelta-imp-chivpndchipsi}: \\[1mm]
\begin{tabular}{lll}
(1) & $\Gamma \vdash \floor{\delta}  \to (\vp \dnd \psi)$  & (Assumption)\\[1mm]
(2) & $\Gamma \vdash \floor{\delta}  \to (\vp \to \psi)$   & \eqref{vp-to-psi-si-vp-to-chi-implies-vp-to-psi-si-chi}: (1) and the definition of $\dnd$\\[1mm]
(3) & $\Gamma \vdash \floor{\delta}  \to (\appln{\chi}{\vp} \to \appln{\chi}{\psi}) $  
& \eqref{floordelta-imp-chivptochipsi}: (2)\\[1mm]
(4) & $\Gamma \vdash \floor{\delta}  \to (\psi \to \vp)$  & \eqref{vp-to-psi-si-vp-to-chi-implies-vp-to-psi-si-chi}: (1)  and the definition of $\dnd$\\[1mm]
(5) & $\Gamma \vdash \floor{\delta}  \to (\appln{\chi}{\psi} \to \appln{\chi}{\vp}) $  & 
\eqref{floordelta-imp-chivptochipsi}: (4)\\[1mm]
(6) & $\Gamma \vdash \floor{\delta}  \to (\appln{\chi}{\psi} \dnd \appln{\chi}{\vp})$  
& \eqref{vp-to-psi-si-vp-to-chi-implies-vp-to-psi-si-chi}: (3), (5) and the definition of $\dnd$
\end{tabular}

\,
}

\subsection{Two congruences}

Let $\Gamma$ be a set of patterns.

\bprop\label{NEW-cong-dnd}
Define the binary relation $\eqreplGdnd$ on the set of patterns as follows: for all patterns $\vp$, $\psi$,
\begin{align*}
\eqreplGdnda{\vp}{\psi} \quad \text{iff} \quad \Gamma \vdash \vp \dnd\psi.
\end{align*}
Then $\eqreplGdnd$ is a congruence that is also an equivalence relation.
\eprop
\solution{We have that: \\[1mm]
$\eqreplGdnd$ is compatible with $\neg$, by \eqref{Gamma-dnd-cong-neg}.\\
$\eqreplGdnd$ is compatible with $\sau$, by \eqref{sim-pairs-or-left} and \eqref{sim-pairs-or-right}.\\
$\eqreplGdnd$ is compatible with $\si$, by \eqref{sim-pairs-and-left} and \eqref{sim-pairs-and-right}.\\
$\eqreplGdnd$ is compatible with $\to$, by \eqref{sim-pairs-to-left} and \eqref{sim-pairs-to-right}.\\
$\eqreplGdnd$ is compatible with $\forall$,  by \eqref{vp-ra-psi-forall-vp-dnd-forall-psi-M}.\\
$\eqreplGdnd$ is compatible with $\exists$,  by \eqref{vp-ra-psi-exists-vp-dnd-exists-psi-M}.\\
$\eqreplGdnd$ is compatible with $\Appl$,  by \eqref{rule-iff-compat-in-app-left} and \eqref{rule-iff-compat-in-app-right}.

Thus, $\eqreplGdnd$ is a congruence.\\[1mm]
Furthermore, \\[1mm]
$\eqreplGdnd$ is reflexive, by \eqref{sim-refl}.\\
$\eqreplGdnd$ is symmetric, by \eqref{sim-symm}.\\
$\eqreplGdnd$ is transitive, by \eqref{sim-trans}.

Thus, $\eqreplGdnd$ is also an equivalence relation. 
}

Let $\Gamma$ be a set of patterns.

\bprop\label{NEW-cong-eqdef}
Define the binary relation $\eqreplGdnd$ on the set of patterns as follows: for all patterns $\vp$, $\psi$,
\begin{align*}
{\vp}\eqreplGdnd{\psi} \quad \text{iff} \quad \Gamma \vdash \vp \eqdef \psi.
\end{align*}
Then $\eqreplGdnd$ is a congruence that is also an equivalence relation.
\eprop
\solution{We have that: \\[1mm]
$\eqreplGdnd$ is compatible with $\neg$, by \eqref{lemma-eq-cong-not}.\\
$\eqreplGdnd$ is compatible with $\sau$, by \eqref{lemma-eq-cong-or-1} and \eqref{lemma-eq-cong-or-2}.\\
$\eqreplGdnd$ is compatible with $\si$, by \eqref{lemma-eq-cong-and-1} and \eqref{lemma-eq-cong-and-2}.\\
$\eqreplGdnd$ is compatible with $\to$, by \eqref{lemma-eq-cong-imp-1} and \eqref{lemma-eq-cong-imp-2}.\\
$\eqreplGdnd$ is compatible with $\forall$,  by \eqref{lemma-eq-cong-forall}.\\
$\eqreplGdnd$ is compatible with $\exists$,  by \eqref{lemma-eq-cong-exists}.\\
$\eqreplGdnd$ is compatible with $\Appl$,  by \eqref{lemma-eq-cong-app-1} and \eqref{lemma-eq-cong-app-2}.

Thus, $\eqreplGdnd$ is a congruence.\\[1mm]
Furthermore, \\[1mm]
$\eqreplGdnd$ is reflexive, by \eqref{eq-refl}.\\
$\eqreplGdnd$ is symmetric, by \eqref{eq-symm}.\\
$\eqreplGdnd$ is transitive, by \eqref{eq-trans}.

Thus, $\eqreplGdnd$ is also an equivalence relation.
}
\subsection{Replacement theorems}\label{section-replacement-thms}

Let $\Gamma$ be a set of patterns.

\bthm[Replacement Theorem for  contexts for $\dnd$]\label{replacement-thm-contexts-NEW-dnd}\,\\
For any context  $\fcontext$ and any patterns $\vp,\psi$,
\bce
$\Gamma \vdash \vp \dnd\psi$ implies $\Gamma \vdash \fcontext[\vp] \dnd \fcontext[\psi]$.
\ece
\ethm
\solution{Apply Proposition \ref{NEW-cong-dnd} and Theorem \ref{replacement-thm-contexts-abstractML}.
}

\bthm[Replacement Theorem for $\dnd$]\label{replacement-thm-NEW-dnd}\,\\
Let $\vp,\psi, \chi,\theta$ be patterns such that $\vp$ is a subpattern of $\chi$ and $\theta$ is obtained
from $\chi$ by replacing one or more occurrences of $\vp$ with $\psi$. Then 
\bce
$\Gamma \vdash \vp \dnd\psi$ implies $\Gamma \vdash  \chi \dnd \theta$.
\ece
\ethm 
\solution{Apply Proposition \ref{NEW-cong-dnd} and Theorem \ref{replacement-thm-more-occurence-abstractML}.
}

\bthm[Replacement Theorem for  contexts for $\eqdef$]\label{replacement-thm-contexts-MG-eqdef}\,\\
For any context  $\fcontext$ and any patterns $\vp,\psi$,
\bce
$\Gamma \vdash \vp \eqdef \psi$ implies $\Gamma \vdash \fcontext[\vp] \eqdef  \fcontext[\psi]$.
\ece
\ethm
\solution{Apply Proposition \ref{NEW-cong-eqdef} and Theorem \ref{replacement-thm-contexts-abstractML}.
}

\bthm[Replacement Theorem for $\eqdef$]\label{replacement-thm-NEW-eqdef} \,\\
Let $\vp,\psi, \chi,\theta$ be patterns such that $\vp$ is a subpattern of $\chi$ and $\theta$ is obtained
from $\chi$ by replacing one or more occurrences of $\vp$ with $\psi$. Then 
\bce
$\Gamma \vdash \vp \eqdef \psi$ implies $\Gamma \vdash  \chi \eqdef  \theta$.
\ece
\ethm 
\solution{Apply Proposition \ref{NEW-cong-eqdef} and Theorem \ref{replacement-thm-more-occurence-abstractML}.
}

\section{Theorems and derived rules in $\appMLnewGc$ - II}

\subsection{Free substitution}\label{section-free-subst}

\begin{proposition}\label{subf-one-occurence-x-y}
Let $\vp$, $\psi$ be patterns and $x$, $y$ be element variables such that 
\be
\item $x$ is free for $y$ in $\vp$;
\item $\psi$ is obtained from $\vp$ by replacing a free occurence of $x$ in $\vp$ with $y$.
\ee
Then 
$$
\vdash x \eqdef y \to (\vp \dnd \psi).
$$
\end{proposition}
\solution{ Assume first that $x = y$. Then  $x$ is free for $x$ in $\vp$ (by Lemma \ref{subf-lemma-useful-x-freefor-x-abstractML})
and $\psi = \vp$. We get that \\[1mm]
\begin{tabular}{lll}
(1) & $\vdash \vp \dnd \vp$ & \eqref{sim-refl} \\[1mm]
(2) & $\vdash (\vp \dnd \vp) \to (x \eqdef x \to (\vp \dnd  \vp))$ &  \eqref{vp-to-psi-to-vp}\\[1mm]
(3) & $\vdash  x \eqdef x \to (\vp \dnd  \vp)$ &  \prule{modus ponens}: (1), (2)\\[2mm]
\end{tabular}

Assume in the sequel that $x$, $y$ are distinct. 

If $x$ does not occur free in $\vp$, then $\psi = \vp$. We get that  \\[1mm]
\begin{tabular}{lll}
(1) & $\vdash \vp \dnd \vp$ & \eqref{sim-refl} \\[1mm]
(2) & $\vdash (\vp \dnd \vp) \to (x \eqdef y \to (\vp \dnd  \vp))$ &   
\eqref{vp-to-psi-to-vp}\\[1mm]
(3) & $\vdash  x \eqdef y \to (\vp \dnd  \vp)$ &  \prule{modus ponens}: (1), (2)\\[2mm]
\end{tabular}

Assume in the following that $x$ occurs free in $\vp$. The proof is by induction on $\vp$.

\be
\item $\vp$ is an atomic pattern. As $x$ occurs free in $\vp$, we must have $\vp = x$. Obviously,
$x$ is free for $y$ in $x$. Thus, 
$\psi=y$.
Then, by \eqref{equal-imp-dnd}, $ \vdash x \eqdef y \to (x \dnd y)$, that is 
$$
\vdash x \eqdef y \to (\vp \dnd \psi).
$$
\item $\vp=\neg \chi$.  As $x$ occurs free in $\vp$, we have that  $x$ occurs free in $\chi$. 
As $x$ is free for $y$ in $\vp$, we have that $x$ is free for $y$ in $\chi$.  Furthermore, 
$\psi=\neg \delta$, where $\delta$ is obtained from $\chi$ by replacing a free occurence of 
$x$ in $\chi$ with $y$. We get that \\[1mm]
\begin{tabular}{lll}
(1) & $\vdash x \eqdef y \to (\chi \dnd \delta)$ & induction hypothesis on $\chi$ \\[1mm]
(2) & $\vdash (\chi \dnd \delta) \to  (\vp \dnd \psi)$ & \eqref{vpdndpsi-neg} \\[1mm]
(3) & $\vdash x \eqdef y \to (\vp \dnd  \psi)$ &   \prule{syllogism}: (1), (2)
\end{tabular}

\item $\vp= \chi \pbin \sigma$, where $\pbin\in \{\to, \si,\sau\}$. 
As $x$ occurs free in $\vp$, we have that $x$ occurs free in $\chi$ or $x$ occurs free in $\sigma$.
As $x$ is free for $y$ in $\vp$, we have that $x$ is free for $y$ in $\chi$ and $x$ is free for $y$ in $\sigma$. 
We have two cases:
\be
\item $\psi$ is obtained from $\vp$ by replacing a free occurence of $x$ in $\chi$ with $y$. Then 
$\psi=\delta \pbin \sigma$, where $\delta$ is obtained from $\chi$ by replacing a free occurence of 
$x$ in $\chi$ with $y$. We get that \\[1mm]
\begin{tabular}{lll}
(1) & $\vdash x \eqdef y \to (\chi \dnd \delta)$ & induction hypothesis on $\chi$ \\[1mm]
(2) & $\vdash (\chi \dnd \delta) \to  (\vp \dnd \psi)$ &   \eqref{vpdndpsi-to-r} if $\pbin=\to$; 
\eqref{vpdndpsi-sau-r}  if $\pbin=\sau$;  \eqref{vpdndpsi-si-r}  if $\pbin=\si$ \\[1mm]
(3) & $\vdash x \eqdef y \to (\vp \dnd \psi)$ &   \prule{syllogism}: (1), (2)
\end{tabular}
\item $\psi$ is obtained from $\vp$ by replacing a free occurence of $x$ in $\sigma$ with $y$. Then 
$\psi=\chi \pbin \delta$, where $\delta$ is obtained from $\sigma$ by replacing a free occurence 
of $x$ in $\sigma$ with $y$. We get that \\[1mm]
\begin{tabular}{lll}
(1) & $\vdash x\eqdef y \to (\sigma \dnd \delta)$ & induction hypothesis on $\sigma$ \\[1mm]
(2) & $\vdash (\sigma \dnd \delta) \to  (\vp \dnd \psi)$ & \eqref{vpdndpsi-to-l} if $\pbin=\to$; 
\eqref{vpdndpsi-sau-l} if $\pbin=\sau$;  \eqref{vpdndpsi-si-l} if $\pbin=\si$ \\[1mm]
(3) & $\vdash x \eqdef y \to (\vp \dnd \psi)$ &   \prule{syllogism}: (1), (2)
\end{tabular}
\ee
\item $\vp= \appln{\chi}{\sigma}$. As $x$ occurs free in $\vp$, we have that $x$ occurs free in 
$\chi$ or $x$ occurs free in $\sigma$. As $x$ is free for $y$ in $\vp$, we have that $x$ is 
free for $y$ in $\chi$ and $x$ is free for $y$ in $\sigma$. 
We have two cases:
\be
\item $\psi$ is obtained from $\vp$ by replacing a free occurence of $x$ in $\chi$ with $y$. Then 
$\psi=\appln{\delta}{\sigma}$, where $\delta$ is obtained from $\chi$ by replacing a free occurence 
of $x$ in $\chi$ with $y$. We get that \\[1mm]
\begin{tabular}{lll}
(1) & $\vdash x \eqdef y \to (\chi \dnd \delta)$ & induction hypothesis on $\chi$ \\[1mm]
(2) & $\vdash x \eqdef y \to (\vp \dnd \psi)$ &   \eqref{xeqytovpdndpsi-imp-xeqy-vpchidndpsichi}: (1) 
\end{tabular}
\item $\psi$ is obtained from $\vp$ by replacing a free occurence of $x$ in $\sigma$ with $y$. Then 
$\psi=\appln{\chi}{\delta}$, where $\delta$ is obtained from $\sigma$ by replacing a free occurence 
of $x$ in $\sigma$ with $y$. We get that \\[1mm]
\begin{tabular}{lll}
(1) & $\vdash x\eqdef y \to (\sigma \dnd \delta)$ & induction hypothesis on $\sigma$ \\[1mm]
(2) & $\vdash x \eqdef y \to (\vp \dnd \psi)$ &   \eqref{xeqytovpdndpsi-imp-xeqy-chivpndchipsi}: (1)
\end{tabular}
\ee
\item $\vp = Q z \chi$, where $Q\in \{\forall, \exists\}$.  As $x\in \FVE(\vp)=\FVE(\chi)\setminus \{z\}$, 
we have that $x \neq z$ and  $x$ occurs free in $\chi$. As $x$ is free for 
$y$ in $\vp$, we must have that $y \neq z$, hence $x$ is free for $y$ in $\chi$. 
Then $\psi=Q z \delta$, where $\delta$ is obtained 
from $\chi$ by replacing a free occurence of $x$ in $\chi$ with $y$. We get that \\[1mm]
\begin{tabular}{lll}
(1) & $\vdash x \eqdef y \to (\chi \dnd \delta)$ & induction hypothesis on $\chi$ \\[1mm]
(2) & $\vdash x \eqdef y \to (\vp \dnd \psi)$ &   \eqref{chi-vppsi-chi-forallxvppsi-dnd-M} or 
\eqref{chi-vppsi-chi-existsxvppsi-dnd-M}: (1) with $\chi:=x \eqdef y$, as $z\ne x,y$
\end{tabular}
\ee
}

\begin{proposition}[Substitution of equal variables]\label{subf-more-occurences-x-y}\,\\
Let $\vp$, $\psi$ be patterns and $x$, $y$ be element variables such that 
\be
\item $x$ is free for $y$ in $\vp$;
\item $\psi$ is obtained from $\vp$ by replacing one or more free occurences of $x$ in $\vp$ with $y$.
\ee
Then
$$
\vdash x \eqdef y \to (\vp \dnd \psi).
$$
\end{proposition}
\solution{Let $n$ be the number of free occurences of  $x$ in $\vp$ that are replaced with $y$. The proof is by induction on $n$.

$n=1$: Apply Proposition \ref{subf-one-occurence-x-y}.

$n\Ra n+1$: Let $\psi'$ be the pattern obtained by replacing $n$ occurences of $x$ in $\vp$ by $y$. Thus, we can apply the induction hypothesis
to get that 
\beq\label{psip-n-occ-vp}
\vdash x \eqdef y  \to (\vp\dnd\psi').
\eeq
Remark that $\psi$ is obtained from $\psi'$ by replacing one occurence of $x$ in $\psi'$ with $y$. 
Furthermore, as $x$ is free for $y$ in $\vp$, obviously, $x$ is free for $y$ in $\psi'$. Thus, we can apply Proposition~\ref{subf-one-occurence-x-y}
to get that 
\beq\label{psip-1-occ-psi}
\vdash x \eqdef y  \to (\psi'\dnd\psi).
\eeq
Appply now \eqref{vp-to-psi-dnd-chi-vp-dnd-chi-to-gamma-implies-vp-to-psi-dnd-gamma}, \eqref{psip-n-occ-vp}, \eqref{psip-1-occ-psi} to get the conclusion.
}

\begin{proposition}\label{subf-x-y}
Let $\vp$ be a pattern and $x$, $y$ be element variables such that $x$ is free for $y$ in $\vp$.

Then 
$$
\vdash x \eqdef y \to \left(\vp \dnd \subf{x}{y}{\vp}\right).
$$
\end{proposition}
\solution{Apply Proposition \ref{subf-more-occurences-x-y}.
}

\begin{proposition}\label{forallxvp-subfxyvp-1}
Let $\vp$ be a pattern and $x$, $y$ be distinct element variables such that 
\be
\item $x$ does not occur bound in $\vp$;
\item $x$ is free for $y$ in $\vp$.
\ee
Then 
$$
\vdash \forall x \vp \to \subf{x}{y}{\vp}.
$$
\end{proposition}
\solution{Remark first that, by Lemma \ref{subf-lemma-useful-subfxy-3-abstractML}, we have 
that $x$ does not occur in $\subf{x}{y}{\vp}$. We get that \\[1mm]
\begin{tabular}{lll}
(1) & $\vdash x \eqdef y \to \left(\vp \dnd \subf{x}{y}{\vp}\right)$ & Proposition \ref{subf-x-y}\\[1mm]
(2) & $\vdash x \eqdef y \to \left(\vp \to \subf{x}{y}{\vp}\right)$ & \eqref{vp-to-psi-si-vp-to-chi-implies-vp-to-psi-si-chi}: (1) \\[1mm]
(3) & $\vdash  \vp \to \left(x \eqdef y \to \subf{x}{y}{\vp}\right)$ &  \eqref{vp-to-psi-to-chi--psi-to-vp-to-chi}: (2) \\[1mm]
(4) & $\vdash (x \eqdef y \to \subf{x}{y}{\vp})\to(\neg\subf{x}{y}{\vp}\to\neg(x \eqdef y))$ & 
\eqref{contraposition}\\[1mm]
(5) & $\vdash\vp\to(\neg\subf{x}{y}{\vp}\to\neg(x \eqdef y))$ & \prule{syllogism}: (3), (4)\\[1mm]
(6) & $\vdash \forall x\vp  \to (\forall x\neg \subf{x}{y}{\vp} \to \forall x\neg(x \eqdef y))$ & \eqref{vp-ra-psi-ra-chi-forall-M}: (5)\\[1mm]
(7) & $\vdash (\forall x\neg \subf{x}{y}{\vp} \to \forall x\neg(x \eqdef y)) \to $
& \eqref{contraposition}\\[1mm]
& \quad $\to (\neg\forall x\neg(x \eqdef y) \to \neg\forall x\neg \subf{x}{y}{\vp})$ \\[1mm]
(8) & $\vdash \forall x\vp \to (\neg\forall x\neg(x \eqdef y) \to \neg\forall x\neg \subf{x}{y}{\vp})$ & \prule{syllogism}: (6), (7)\\[1mm]
(9) & $\vdash \neg\forall x\neg(x \eqdef y	) \to (\forall x\vp  \to \neg\forall x\neg \subf{x}{y}{\vp})$ & 
\eqref{vp-to-psi-to-chi--psi-to-vp-to-chi}: (8) \\[1mm]
(10) & $\vdash \exists x (x \eqdef y) \to \neg\forall x\neg(x \eqdef y)$ & \prule{axiom-exists-1}\\[1mm]
(11) & $\vdash \exists x (x \eqdef y) \to (\forall x\vp  \to \neg\forall x\neg \subf{x}{y}{\vp})$ & \prule{syllogism}: (10), (9)\\[1mm]
(12) & $\vdash \exists x(x \eqdef y)$ & \prule{\Monkt}\\[1mm]
(13) & $\vdash \forall x\vp  \to \neg\forall x\neg \subf{x}{y}{\vp}$ & \prule{modus ponens}: (12), (11)\\[1mm]
(14) & $\vdash \neg \subf{x}{y}{\vp} \to \forall x\neg \subf{x}{y}{\vp}$ & \prule{\Monkd}, \\
&&  as $x$ does not occur in $\subf{x}{y}{\vp}$\\[1mm]
(15) & $\vdash \neg\forall x\neg \subf{x}{y}{\vp}  \to \neg\neg \subf{x}{y}{\vp}$ &  
\eqref{rec-ax3-rule}: (14) \\[1mm]
(16) & $\vdash \neg\neg \subf{x}{y}{\vp} \to  \subf{x}{y}{\vp}$ & 
\eqref{Gamma-vptopsi-psitovp-iff-vpdndpsi} and \eqref{negnegvp-dnd-vp}\\[1mm]
(17) & $\vdash \neg\forall x\neg \subf{x}{y}{\vp}  \to  \subf{x}{y}{\vp}$ & \prule{syllogism}: (15), (16)\\[1mm]
(18) & $\vdash \forall x \vp \to \subf{x}{y}{\vp}$ & \prule{syllogism}: (13), (17) 
\end{tabular}

\,
}

\blem\label{forallxvp-dnd-subfxyvp}
Let $\vp$ be a pattern and $x$, $y$ be element variables such that 
\bce 
$x$ does not occur bound in $\vp$ and $y$ does not occur in $\vp$.
\ece
Then 
\be 
\item\label{forallxvp-to-subfxyvp} 
$\vdash \,\forall x \vp \to \forall y\subf{x}{y}{\vp}$;
\item\label{subfxyvp-to-forallxvp}
$\vdash \,\forall y\subf{x}{y}{\vp} \to \forall x \vp$;
\item\label{forallxvp-dns-subfxyvp}
$\vdash \,\forall x \vp \dnd \forall y\subf{x}{y}{\vp}$.
\ee
\elem
\solution{The case $x=y$ is obvious, as $\subf{x}{x}{\vp}=\vp$. Assume in the sequel that $x$, $y$ 
are distinct.

\be
\item As, by hypothesis, $y$ does not occur in $\vp$, we have, by Lemma \ref{lemma-useful-y-noccur-vp-imp-y-freefor-x-abstractML}, 
that $x$ is free for $y$ in $\vp$. We get that \\[1mm]
\begin{tabular}{lll}
(1) & $\vdash \forall x \vp \to \subf{x}{y}{\vp}$ & Proposition \ref{forallxvp-subfxyvp-1}\\[1mm]
(2) & $\vdash \forall y\forall x \vp \to \forall y\subf{x}{y}{\vp}$ & 
\eqref{vp-ra-psi-forall-vp-ra-forall-psi-M}: (1)\\[1mm]
(3) & $\vdash \forall x \vp \to \forall y\forall x \vp $ & 
\prule{\Monkd}, as $y$ does not occur in  $\forall x \vp$\\[1mm]
(4) & $\vdash \forall x \vp \to \forall y\subf{x}{y}{\vp}$ & \prule{syllogism}: (3), (2) \\[2mm]
\end{tabular}

\item Let us denote $\psi:=\subf{x}{y}{\vp}$. By hypothesis, $y$ does not occur in $\vp$. In particular, $y$ does not occur bound 
in $\vp$. We can apply Lemma \ref{subf-lemma-useful-subfxy-1-abstractML} to get that $y$ does not occur bound in $\psi$.

By hypothesis, $x$ does not occur bound in $\vp$. We can apply Lemma \ref{subf-lemma-useful-subfxy-3-abstractML} to get that 
$x$ does not occur in $\psi$.

Apply \eqref{forallxvp-to-subfxyvp} with $x:=y$, $y:=x$, $\vp:=\psi$ to get that 
$$\vdash \forall y \psi \to \forall x \subf{y}{x}{\psi},$$
that is 
$$\vdash \forall y\subf{x}{y}{\vp} \to \forall x \subf{y}{x}{\subf{x}{y}{\vp}}.$$
As, by hypothesis, $y$ does not occur in $\vp$, we can apply Lemma \ref{subf-lemma-useful-subfxy-2-abstractML} to get 
$\subf{y}{x}{\subf{x}{y}{\vp}}=\vp$. Thus, \eqref{subfxyvp-to-forallxvp} follows. \\

\item Apply \eqref{forallxvp-to-subfxyvp}, \eqref{subfxyvp-to-forallxvp},
the definition of $\dnd$ and \eqref{Gamma-vp-psi-vp-si-psi}.
\ee
}

\subsection{Bounded substitution theorem}\label{section-bounded-subst-thm}

\bthm\label{thm-vp-dnd-subbxybvp}
For any pattern $\vp$ and any variables $x$, $y$ such that $y$ does not occur in $\vp$
\begin{align*}
\vdash \vp\dnd \subbf{x}{y}{\vp}. 
\end{align*}
\ethm
\solution{Apply Proposition \ref{NEW-cong-dnd},  
Lemma \ref{forallxvp-dnd-subfxyvp}.\eqref{forallxvp-dns-subfxyvp} 
and Theorem \ref{thm-vp-cong-subbxybvp-abstractML}.
}

\subsection{Universal and existential specification}\label{section-univ-exist-spec}

\begin{proposition}\label{forallxvp-subfxyvp}
Let $\vp$ be a pattern and $x$, $y$ be distinct element variables such that $x$ is free for $y$ in $\vp$.

Then 
$$
\vdash \forall x \vp \to \subf{x}{y}{\vp}.
$$
\end{proposition}
\solution{Let $z$ be a new variable, different from $x$, $y$  and not occuring in $\vp$. We get, by Theorem \ref{thm-vp-dnd-subbxybvp}, that \\
\beq 
\vdash \vp\dnd \subb{x}{z}{\vp}. \label{forallxvp-subfxyvp-u1}
\eeq
Apply \eqref{forallxvp-subfxyvp-u1} and \eqref{vp-ra-psi-forall-vp-dnd-forall-psi-M} to get that
\beq 
\vdash \forall x\vp\dnd \forall x\subb{x}{z}{\vp} \label{forallxvp-subfxyvp-u2}
\eeq

Let us denote $\psi:=\subb{x}{z}{\vp}$. As $x\ne z$, we can apply Lemma \ref{subf-lemma-useful-x-noccur-bound-subbxy-abstractML}
to get that 
\bce 
(*) \quad $x$ does not occur bound in $\psi$. 
\ece

Furthermore, as $x$ is free for $y$ in $\vp$, we can apply 
Lemma \ref{subf-lemma-useful-xfreeyvp-xfreeysubbxzvp} to get that 
\bce 
(**) \quad $x$ is free for $y$ in $\psi$.
\ece
Using (*), (**) and the fact that $x\ne y$, an application of Proposition \ref{forallxvp-subfxyvp-1} gives us
$$
\vdash \forall x\psi \to \subf{x}{y}{\psi},
$$
that is 
\beq 
\vdash \forall x\subb{x}{z}{\vp} \to \subf{x}{y}{\subb{x}{z}{\vp}} \label{forallxvp-subfxyvp-u3}
\eeq

Let us denote $\chi:=\subf{x}{y}{\subb{x}{z}{\vp}}$. As $x\ne y,z$, we have, by Lemma \ref{x-noccur-subfxysubbxzvp-abstractML}, that 
$x$ does not occur in $\chi$. Thus,  we can apply Theorem \ref{thm-vp-dnd-subbxybvp} to get that 
$$
\vdash \chi \dnd \subb{z}{x}{\chi},
$$
that is, 
$$
\vdash \subf{x}{y}{\subb{x}{z}{\vp}} \dnd \subb{z}{x}{\subf{x}{y}{\subb{x}{z}{\vp}}}.
$$

As $x$, $y$, $z$ are distinct element variables and $z$ does not occur in $\vp$, we have, by Lemma \ref{subb-xyz-2-abstractML}, that
$\subb{z}{x}{\subf{x}{y}{\subb{x}{z}{\vp}}}=\subf{x}{y}{\vp}$. It follows that 

\beq 
\vdash \subf{x}{y}{\subb{x}{z}{\vp}} \dnd \subf{x}{y}{\vp}.	 \label{forallxvp-subfxyvp-u4}
\eeq

Apply \eqref{forallxvp-subfxyvp-u2}, \eqref{forallxvp-subfxyvp-u3} and \eqref{forallxvp-subfxyvp-u4} to conclude that 
$$
\vdash \forall x \vp \to \subf{x}{y}{\vp}.
$$
}

\begin{corollary}\label{forallxvp-subfxyvp-ynot}
Let $\vp$ be a pattern and $x$, $y$ be distinct element variables such that $y$ does not  occur in $\vp$.

Then 
$$
\vdash \forall x \vp \to \subf{x}{y}{\vp}.
$$
\end{corollary}
\solution{As $y$ does not  occur in $\vp$, we have, by Lemma \ref{subf-lemma-useful-subfxy-3-abstractML}, 
that $x$ is free for $y$ in $\vp$. Apply Proposition \ref{forallxvp-subfxyvp}.
}

\bthm[Universal Specification]\label{forallxvp-subfxyvp-xfreey}\,\\
Let $\vp$ be a pattern and $x$, $y$ be element variables such that $x$ is free for $y$ in $\vp$.

Then 
$$
\vdash \forall x \vp \to \subf{x}{y}{\vp}.
$$
\ethm
\solution{Let $z$ be a new variable, distinct from $x$, $y$  and not occuring in $\vp$. 
We get that \\[1mm]
\begin{tabular}{lll}
(1) & $\vdash \forall x\vp \to \subf{x}{z}{\vp}$ & Corollary \ref{forallxvp-subfxyvp-ynot}, 
as $z \neq x$ and $z$ does not occur in $\vp$ \\[1mm]
(2) & $\vdash \forall z\forall x\vp \to \forall z\subf{x}{z}{\vp}$ & \eqref{vp-ra-psi-forall-vp-ra-forall-psi-M}: (1) \\[1mm]
(3) & $\vdash \forall x\vp  \to \forall z\forall x\vp$ & \prule{\Monkd}, as $z$ does not 
occur in $\forall x\vp$ \\[1mm]
(4) & $\vdash \forall x\vp  \to \forall z\subf{x}{z}{\vp}$ & \prule{syllogism}: (3), (2)\\[1mm]
\end{tabular}

Let $\psi:=\subf{x}{z}{\vp}$. As $z$ does not occur in $\vp$ and, by hypothesis, $x$ is free for $y$ in $\vp$, we get from 
Lemma \ref{subf-us-xyz-1-abstractML} that 
\bce 
(*) \quad $z$ is free for $y$ in $\psi$.
\ece

By (*) and the fact that $z\neq y$, we can apply Proposition \ref{forallxvp-subfxyvp} to get that 
$$ \vdash  \forall z\psi \to \subf{z}{y}{\psi}.$$
It follows that \\[1mm]
\begin{tabular}{lll}
(5) & $\vdash  \forall z\subf{x}{z}{\vp} \to \subf{z}{y}{\subf{x}{z}{\vp}}$ & by the definition of $\psi$\\
(6) & $\vdash  \forall z\subf{x}{z}{\vp}  \to \subf{x}{y}{\vp}$ & by 
Lemma \ref{subfzy-subfxzvp=subfxyvp-abstractML}, as $z$ does not occur in $\vp$\\
(7) & $\vdash \forall x\vp \to \subf{x}{y}{\vp}$ & \prule{syllogism}: (4), (6).
\end{tabular}

\,
}

\bthm[Existential Specification]\label{subfxyvp-existsxvp-xfreey}\,\\
Let $\vp$ be a pattern and $x$, $y$ be element variables such that $x$ is free for $y$ in $\vp$.

Then 
$$
\vdash \subf{x}{y}{\vp} \to \exists x \vp.
$$
\ethm
\solution{We have that \\[1mm]
\begin{tabular}{lll}
(1) & $\vdash \forall x \neg\vp \to \subf{x}{y}{\neg\vp}$ &  Theorem \ref{forallxvp-subfxyvp-xfreey}, as $x$ is free for $y$ in $\neg\vp$ \\[1mm]
(2) & $\vdash \forall x \neg\vp \to \neg\subf{x}{y}{\vp}$ & as $\subf{x}{y}{\neg\vp}=\neg\subf{x}{y}{\vp}$ \\[1mm]
(3) & $\vdash \neg\neg\subf{x}{y}{\vp} \to \neg\forall x \neg\vp $ &  \eqref{rec-ax3-rule}: (2)\\[1mm]
(4) & $\vdash \subf{x}{y}{\vp} \to \neg\neg\subf{x}{y}{\vp}$ &  
\eqref{Gamma-vptopsi-psitovp-iff-vpdndpsi} and \eqref{negnegvp-dnd-vp}\\[1mm]
(5) & $\vdash \subf{x}{y}{\vp} \to \neg\forall x \neg\vp$ &  \prule{syllogism}: (4), (3)\\[1mm]
(6) & $\vdash  \neg\forall x \neg\vp \to \exists x \vp$ &  \prule{axiom-exists-2}\\[1mm]
(7) & $\vdash \subf{x}{y}{\vp} \to \exists x \vp$ &  \prule{syllogism}: (5), (6)
\end{tabular}

\,
}

\subsection{More first-order}

\begin{lemma}
\begin{align}
\vdash & \, \forall x\vp \to \vp, \label{forall-quant-x-M}\\
\vdash  & \, \vp \to \exists x\vp, \label{exists-quant-x-M}\\
\vdash & \, \forall x\vp \to \exists x\vp. \label{forall-vp-to-exists-vp-M}
\end{align}
\end{lemma}
\solution{\,\\
\eqref{forall-quant-x-M}:  Apply  Theorem \ref{forallxvp-subfxyvp-xfreey} with $y:=x$, as $x$ is free 
for $x$ in $\vp$ and $\subf{x}{x}{\vp}=\vp$.

\eqref{exists-quant-x-M}:  Apply  Theorem \ref{subfxyvp-existsxvp-xfreey} with $y:=x$,  as $x$ is free 
for $x$ in $\vp$ and $\subf{x}{x}{\vp}=\vp$.

\eqref{forall-vp-to-exists-vp-M}: \\[1mm]
\begin{tabular}{lll}
(1) & $\vdash\forall x\vp \to \vp$ & \eqref{forall-quant-x-M}\\[1mm]
(2) & $\vdash\vp \to \exists x\vp $  & \eqref{exists-quant-x-M}\\[1mm]
(3) & $\vdash \forall x\vp \to \exists x\vp$  & \prule{syllogism}: (1), (2) 
\end{tabular}

\,
}

\begin{lemma}
\begin{align}
\vdash & \, \vp \dnd \forall x \vp  \quad \text{if~} x\notin  \FVE(\vp), \label{vp-dnd-forall-vp-M}\\
\vdash & \,  \vp \dnd \exists x \vp   \quad \text{if~} x\notin  \FVE(\vp). \label{vp-dnd-exists-vp-M}
\end{align}
\end{lemma}
\solution{Assume that $x\notin \FVE(\vp)$ and let $y$ be a new variable, distinct from $x$  and not occuring in $\vp$. 
We get that \\[1mm]
\begin{tabular}{lll}
(1) & $\vdash \forall x\vp \to \vp $ & \eqref{forall-quant-x-M}\\[1mm]
(2) & $\vdash \vp\dnd \subb{x}{y}{\vp}$ & Theorem \ref{thm-vp-dnd-subbxybvp}\\[1mm]
(3) & $\vdash \subb{x}{y}{\vp} \to \vp $ & \eqref{Gamma-vp-psi-vp-si-psi}: (1) and the definition of $\dnd$\\[1mm]
(4) & $\vdash \forall x \subb{x}{y}{\vp} \to \forall x \vp $  & \eqref{vp-ra-psi-forall-vp-ra-forall-psi-M}: (3) \\[1mm]
(5) & $\vdash \subb{x}{y}{\vp}  \to \forall x \subb{x}{y}{\vp}$  & \prule{\Monkd} and  Lemma \ref{subb-lemma-useful-x-noccur-free-vp-noccur-subbxyvp-abstractML} \\[1mm]
(6) & $\vdash \subbf{x}{y}{\vp}  \to \forall x \vp $  & \prule{syllogism}: (5), (4)\\[1mm]
(7) & $\vdash \vp \to \subbf{x}{y}{\vp} $ & \eqref{Gamma-vp-psi-vp-si-psi}: (1) and the 
definition of $\dnd$\\[1mm]
(8) & $\vdash \vp \to \forall x \vp $  & \prule{syllogism}: (7), (6)\\[1mm]
(9) & $\vdash \vp \dnd \forall x \vp $  &  \eqref{Gamma-vp-psi-vp-si-psi}: (1), (8) and 
the definition of $\dnd$\\[2mm]
\end{tabular}

\eqref{vp-dnd-exists-vp-M}:\\[1mm]
\begin{tabular}{lll}
(1) & $\vdash \forall x \neg \vp \dnd  \neg \vp $ & \eqref{forall-quant-x-M}\\[1mm]
(2) & $\vdash \neg \forall x \neg \vp  \dnd \neg\neg \vp$ & \eqref{Gamma-dnd-cong-neg}: (1)\\[1mm]
(3) & $\vdash \neg\neg \vp \dnd \vp$ & \eqref{negnegvp-dnd-vp}\\[1mm]
(4) & $\vdash \neg \forall x \neg \vp  \dnd \vp$ & \eqref{sim-trans}: (2), (3)\\[1mm]
(5) & $\vdash  \exists x\vp \dnd \neg \forall x \neg \vp$ & \eqref{existsxvp-dnd-negforallxngvp}\\[1mm]
(6) & $\vdash  \exists x\vp \dnd \vp$ & \eqref{sim-trans}: (5), (4) 
\et

\,
}

\begin{lemma}
\begin{align}
\vdash &  \,  \forall x(\vp \to \psi) \to (\vp \to \forall x\psi) \quad \text{if~} x\notin \FVE(\vp). 
\label{forallxvptopsi-to-forallxvp}
\end{align}
\end{lemma}
\solution{We get that \\[1mm]
\begin{tabular}{lll}
(1) & $\vdash \forall x(\vp \to \psi)\to (\forall x\vp \to \forall x\psi)$ & \prule{\Monku} \\[1mm]
(2) & $\vdash \vp \to\forall x\vp $ & \eqref{Gamma-vptopsi-psitovp-iff-vpdndpsi}: \eqref{vp-dnd-forall-vp-M}\\[1mm]
(3) & $\vdash (\forall x\vp \to \forall x\psi) \to (\vp \to \forall x\psi)$ & \eqref{vp-to-psi-imp-psi-to-chi-tto-vp-to-chi}: (2)\\[1mm]
(4) & $\vdash \forall x(\vp \to \psi) \to (\vp \to \forall x\psi)$ & \prule{syllogism}: (1), (3)
\end{tabular}

\,
}

\begin{lemma}
\begin{align}
\Gamma \vdash \vp \to \psi & \quad \text{implies} 
\quad \Gamma \vdash  \vp \to \forall x\psi \quad \text{if~}x\notin \FVE(\vp).
\label{forall-quant-rule}
\end{align}
\end{lemma}
\solution{We get that \\[1mm]
\begin{tabular}{lll}
(1) & $\Gamma \vdash \vp \to \psi $ & (Assumption) \\[1mm]
(2) & $\Gamma \vdash \forall x(\vp \to \psi)$ & \prule{gen}: (1) \\[1mm]
(3) & $\Gamma \vdash \forall x(\vp \to \psi) \to (\vp \to \forall x\psi)$ & \eqref{forallxvptopsi-to-forallxvp}\\[1mm]
(4) & $\Gamma \vdash \vp \to \forall x\psi$ & \prule{modus ponens}: (3), (2)
\end{tabular}

\,
}

\begin{lemma}
\begin{align}
\Gamma\vdash\vp\to \psi & \quad \text{implies} 
\quad  \Gamma \vdash \exists x\vp \to   \psi
\quad \text{if~}x\notin \FVE(\psi). \label{exists-quant-rule}
\end{align}
\end{lemma}
\solution{We get that \\[1mm]
\begin{tabular}{lll}
(1) & $\Gamma\vdash\vp\to \psi$ & (Assumption)\\[1mm]
(2) & $\Gamma\vdash \neg\psi \to \neg\vp$ & \eqref{rec-ax3-rule}: (1)\\[1mm]
(3) & $\Gamma\vdash \neg\psi \to \forall x\neg\vp$ &  \eqref{forall-quant-rule}: (2), 
as $x\notin \FVE(\neg\psi)$ \\[1mm]
(4) & $\Gamma\vdash \neg\forall x\neg\vp \to \neg\neg\psi$ & \eqref{rec-ax3-rule}:  (3)\\[1mm]
(5) & $\Gamma\vdash \neg\neg\psi \to \psi$ & \eqref{Gamma-vptopsi-psitovp-iff-vpdndpsi} and \eqref{negnegvp-dnd-vp} \\[1mm]
(6) & $\Gamma\vdash \neg\forall x\neg\vp \to \psi $ & \prule{syllogism}: (4), (5)\\[1mm]
(7) & $\Gamma\vdash \exists x\vp \to \neg\forall x\neg\vp $ & \prule{axiom-exists-1} \\[1mm]
(8) & $\Gamma\vdash \exists x\vp \to \psi$ &  \prule{syllogism}: (7), (6)
\end{tabular}

\,
}

\begin{lemma}
\begin{align}
\vdash & \, \exists x(\vp\si \psi)\to \vp \si \exists x \psi \quad \text{if~}x\notin FV(\vp), \label{existsx-vp-si-psi-to-vp-si-existsxpsi}\\
\vdash & \, \vp \si \exists x \psi \to  \exists x(\vp\si \psi) \quad \text{if~}x\notin FV(\vp), 
\label{vp-si-existsxpsi-to-existsx-vp-si-psi}\\
\vdash & \, \vp \si \exists x \psi \dnd  \exists x(\vp\si \psi) \quad \text{if~}x\notin FV(\vp).
\label{vp-si-existsxpsi-dnd-existsx-vp-si-psi}
\end{align}
\end{lemma}
\solution{\,\\
\eqref{existsx-vp-si-psi-to-vp-si-existsxpsi}: \\[1mm]
\begin{tabular}{lll}
(1) & $\vdash \vp\to \vp$ & \eqref{Gamma-vptopsi-psitovp-iff-vpdndpsi} and \eqref{sim-refl} \\[1mm]
(2) & $\vdash \psi \to \exists x \psi$ & \eqref{exists-quant-x-M} \\[1mm]
(3) & $\vdash \vp\si \psi \to \vp\si \exists x \psi$ & \eqref{psi-to-vp-chi-to-gamma-vp-si-chi-to-psi-si-gamma}: (1), (2)\\[1mm]
(4) & $\vdash \exists x(\vp\si \psi)\to \vp \si \exists x \psi$ & \eqref{exists-quant-rule}: (3), 
as $x\notin FV(\vp\si \exists x\psi)$ \\[2mm]
\end{tabular}

\eqref{vp-si-existsxpsi-to-existsx-vp-si-psi}: \\[1mm]
\begin{tabular}{lll}
(1) & $\vdash \vp\si\psi \to \exists x (\vp\si\psi)$ & \eqref{exists-quant-x-M} \\[1mm]
(2) & $\vdash \psi \si \vp\to \exists x (\vp\si\psi)$ & \eqref{vp-si-psi-implies-chi-perm}: (1) \\[1mm]
(3) & $\vdash \psi \to (\vp \to \exists x (\vp\si\psi))$ & \prule{exportation} \\[1mm]
(4) & $\vdash \exists x\psi \to (\vp \to \exists x (\vp\si\psi))$ & \eqref{exists-quant-rule}: (3) \quad as 
$x\notin FV(\vp\to \exists x(\vp \si \psi))$ \\[1mm]
(5) & $\vdash \exists x \psi \si \vp \to \exists x (\vp\si\psi)$ & \prule{importation}\\[1mm]
(6) & $\vdash \vp \si  \exists x\psi \to \exists x (\vp\si\psi)$ & \eqref{vp-si-psi-implies-chi-perm}: (5) \\[2mm]
\end{tabular}

\eqref{vp-si-existsxpsi-dnd-existsx-vp-si-psi} follows from \eqref{existsx-vp-si-psi-to-vp-si-existsxpsi}, 
\eqref{vp-si-existsxpsi-to-existsx-vp-si-psi}, \eqref{Gamma-vp-psi-vp-si-psi} and the definition of $\dnd$.
}

\begin{lemma}
\begin{align}
\vdash & \, \forall x (\vp\to \psi)\to (\exists x \vp \to \psi) \quad \text{if~}x\notin FV(\psi). \label{forallx-vptopsi-tto-existsxvp-to-psi}
\end{align}
\end{lemma}
\solution{\,\\[1mm]
\eqref{forallx-vptopsi-tto-existsxvp-to-psi}: \\[1mm]
\begin{tabular}{lll}
(1) & $\vdash \psi \to (\forall x(\vp\to\psi)\to \psi)$ & \eqref{vp-to-psi-to-vp} \\[1mm]
(2) & $\vdash \exists x \psi \to (\forall x(\vp\to\psi)\to \psi)$ &  \eqref{exists-quant-rule}: (1), as 
$x\notin FV(\forall x(\vp\to\psi)\to \psi)$  \\[1mm]
(3) & $\forall x (\vp\to \psi)\to (\exists x \vp \to \psi)$ & \eqref{vp-to-psi-to-chi--psi-to-vp-to-chi}: (2)
\end{tabular}

\,
}

\begin{lemma}
Let $n\geq 0$ and $x_1, \ldots, x_n$ be variables. Then 
\begin{align}
\vdash \forall x_1 \ldots  \forall x_n \vp \to \vp. \label{forallxnvp-vp}
\end{align}
\end{lemma}
\solution{\,\\[1mm]
The proof is by induction on $n$. The case $n=0$ is obvious. For $n=1$ apply \eqref{forall-quant-x-M}.\\
$n\Ra n+1$: \\[1mm]
\begin{tabular}{lll}
(1) & $\vdash\forall x_2\ldots\forall x_n\vp\ra \vp$ & (Induction hypothesis) \\
(2) & $\vdash \forall x_1\forall x_2\ldots\forall x_n\vp\to \forall x_2\ldots\forall x_n\vp$ & 
\eqref{forall-quant-x-M}\\
(3) & $\vdash \forall x_1\ldots\forall x_n\vp\ra \vp$ & \prule{syllogism}: (2), (1)
\end{tabular}

\,
}

\subsection{Definedness, totality, equality}\label{section-def-tot-eq}

\begin{lemma}
Let $x$ be a variable. Then 
\begin{align}
\vdash & \, \exists x\ceil{\vp} \to \ceil{\exists x\vp}, \label{existsxceilvp-to-ceilexistsxvp}\\
\vdash & \, \ceil{\exists x\vp} \to\exists x\ceil{\vp}, \label{ceilexistsxvp-to-existsxceilvp}\\
\vdash & \, \exists x\ceil{\vp} \dnd \ceil{\exists x\vp}, \label{existsxceilvp-dnd-ceilexistsxvp}\\
\vdash & \, \forall x\floor{\vp} \dnd \floor{\forall x \vp}. \label{forallxfloor-dnd-floorforallx}
\end{align}
\end{lemma}
\solution{\,\\[1mm]
\eqref{existsxceilvp-to-ceilexistsxvp}:\\[1mm]
\begin{tabular}{lll}
(1) & $\vp\to\exists x\vp$ & \eqref{exists-quant-x-M}\\[1mm]
(2) & $\ceil{\vp}\to\ceil{\exists x\vp}$ & \eqref{lemma-imp-compat-in-ceil}: (1)\\[1mm]
(3) & $\exists x\ceil{\vp}\to\ceil{\exists x\vp}$ & \eqref{exists-quant-rule}: (2), as $x\notin FV(\ceil{\exists x \vp})$\\[2mm]
\end{tabular}

\eqref{ceilexistsxvp-to-existsxceilvp}: Remark that, by definition, 
\bce 
$\ceil{\exists x\vp}=\appln{\defsymb}{\exists x\vp}$ and $\exists x\ceil{\vp}=\exists x(\appln{\defsymb}{\vp})$.
\ece
As $x$ does not occur in $\defsymb$, we can apply \prule{propagation$_\exists$} with $\psi:=\defsymb$ to get 
\eqref{ceilexistsxvp-to-existsxceilvp}.

\eqref{existsxceilvp-dnd-ceilexistsxvp} Apply \eqref{existsxceilvp-to-ceilexistsxvp}, 
\eqref{ceilexistsxvp-to-existsxceilvp} and the definition of $\dnd$.

Remark that, by definition, 
\bce 
$\floor{\forall x \vp}=\neg  \ceil{\neg\forall x \vp}$ and $\forall x\floor{\vp}=\forall x \neg\ceil{\neg \vp}$.
\ece

\eqref{forallxfloor-dnd-floorforallx}:\\[1mm]
\begin{tabular}{lll}
(1) & $\vdash \neg \exists x \ceil{\neg \vp} \dnd \forall x\floor{\vp}$ & \eqref{negexistsxvp-dnd-forallxnegvp}\\[1mm]
(2) & $\vdash \exists x \ceil{\neg \vp} \dnd  \ceil{\exists x\neg  \vp}$ & \eqref{existsxceilvp-dnd-ceilexistsxvp}\\[1mm]
(3) & $\vdash \neg \ceil{\exists x\neg  \vp} \dnd \neg\exists x \ceil{\neg \vp}$ & \eqref{Gamma-dnd-cong-neg}: (2)\\[1mm]
(4) & $\vdash \neg \ceil{\exists x\neg  \vp} \dnd \forall x\floor{\vp}$ & \prule{syllogism}: (3), (1)\\[1mm]
(5) & $\vdash \neg\forall x\vp \dnd \exists x\neg  \vp$ & \eqref{negforallxvp-dnd-existsxnegvp}\\[1mm]
(6) & $\vdash \ceil{\neg\forall x\vp} \dnd \ceil{\exists x\neg  \vp}$ & \eqref{rule-iff-compat-in-ceil}: (5)\\[1mm]
(7) & $\vdash\neg \ceil{\exists x\neg  \vp}\dnd \floor{\forall x \vp}$ &\eqref{Gamma-dnd-cong-neg}: (6)\\[1mm]
(8) & $\vdash \forall x\floor{\vp} \dnd \neg \ceil{\exists x\neg  \vp}$ & \eqref{sim-symm}: (4)\\[1mm]
(9) & $\vdash \forall x\floor{\vp} \dnd \floor{\forall x \vp}$ & \prule{syllogism}: (8), (7)
\end{tabular}

\,
}

\begin{lemma}
Let $n\geq 0$ and $x_1, \ldots, x_n$ be variables. Then 
\begin{align}
\vdash & \, \forall x_1 \ldots \forall x_n \floor{\vp} \dnd \floor{\forall x_1 \ldots \forall x_n \vp}.
\label{forallxnfloor-dnd-floorforallxn}
\end{align}
\end{lemma}
\solution{\,\\[1mm]
The proof is by induction on $n$. The case $n=0$ is obvious. For $n=1$ apply 
\eqref{forallxfloor-dnd-floorforallx}.

$n\Ra n+1$:\\[1mm]
\bt{lll}
(1) & $\vdash \forall x_2 \ldots \forall x_{n+1} \floor{\vp} \dnd 
\floor{\forall x_2 \ldots \forall x_{n+1} \vp}$ 
& (Induction hypothesis)\\[1mm]
(2) & $\vdash \forall x_1 \forall x_2 \ldots \forall x_{n+1}\floor{\vp} \dnd \forall x_1 \floor{\forall x_2 \ldots \forall x_{n+1} \vp}$ 
& \eqref{vp-ra-psi-forall-vp-dnd-forall-psi-M}: (1)\\[1mm]
(3) & $\vdash \forall x_1 \floor{\forall x_2 \ldots \forall x_{n+1} \vp}
\dnd \floor{\forall x_1 \forall x_2 \ldots \forall x_{n+1}\vp}$ & \eqref{forallxfloor-dnd-floorforallx} for
$\vp:=\forall x_2 \ldots \forall x_{n+1} \vp$ \\[1mm]
(4) & $\vdash \forall x_1 \ldots \forall x_{n+1} \floor{\vp} \dnd \floor{\forall x_1 \ldots \forall x_{n+1}\vp}$ 
& \prule{syllogism}: (2), (3)
\et
 
\, 
}

\begin{lemma}
Let $n\geq 0$ and $x_1, \ldots, x_n$ be variables. Then 
\begin{align}
\vdash & \, \forall x_1 \ldots \forall x_n (\vp \eqdef \psi) \dnd \floor{\forall x_1 \ldots \forall x_n (\vp \dnd \psi)}.
\label{forallxnvpeqdefpsi-dnd-floorforallxnvpdndpsi}
\end{align}
\end{lemma}
\solution{Apply \eqref{forallxnfloor-dnd-floorforallxn} and the fact 
that $\vp \eqdef \psi=\floor{\vp \dnd \psi}$.
}

\subsection{Application}\label{thms-rules-application-more}

\begin{lemma}
For every patterns $\vp$, $\psi$,
\begin{align}
\appln{(\exists x\vp)}{\psi} \to \exists x(\appln{\vp}{\psi}) & 
 \quad \text{if~} x\notin \FVE(\psi),\label{propagation-exists-1-H}\\[1mm]
 \appln{\psi}{(\exists x\vp)} \to \exists x(\appln{\psi}{\vp}) & 
 \quad \text{if~} x\notin \FVE(\psi). \label{propagation-exists-2-H}
\end{align}
\end{lemma}
\solution{Assume that $x\notin \FVE(\psi)$ and let $y$ be a new variable, distinct from $x$  
and not occuring in $\psi$. Let us denote $\delta:=\subb{x}{y}{\psi}$. 
By Lemma \ref{subb-lemma-useful-x-noccur-free-vp-noccur-subbxyvp-abstractML}, we have that 
$x$ does not  occur in $\delta$. \\[2mm]
\eqref{propagation-exists-1-H}: \\[1mm]
\bt{lll}
(1) & $\vdash \appln{(\exists x\vp)}{\delta} \to \exists x(\appln{\vp}{\delta})$ & \prule{Propagation$_\exists$}\\[1mm]
(2) & $\vdash \psi \dnd \delta$ & Theorem \ref{thm-vp-dnd-subbxybvp}\\[1mm]
(3) & $\vdash \bigg(\appln{(\exists x\vp)}{\psi} \to \exists x(\appln{\vp}{\psi})\bigg)  \dnd
\bigg(\appln{(\exists x\vp)}{\delta} \to \exists x(\appln{\vp}{\delta})\bigg)$ & 
Theorem \ref{replacement-thm-NEW-dnd}: (2)\\[1mm]
(4) & $\vdash \bigg(\appln{(\exists x\vp)}{\delta} \to \exists x(\appln{\vp}{\delta})\bigg) \to 
\bigg(\appln{(\exists x\vp)}{\psi} \to \exists x(\appln{\vp}{\psi})\bigg)$ & \eqref{vpdndpsi-imp-vp-iff-psi}: (3) \\[1mm]
(5) & $\vdash\appln{(\exists x\vp)}{\psi} \to \exists x(\appln{\vp}{\psi})$ &  
\prule{modus ponens}: (1), (4)\\[2mm]
\et

\eqref{propagation-exists-2-H}: \\[1mm]
\bt{lll}
(1) & $\vdash \appln{\delta}{(\exists x\vp)} \to \exists x(\appln{\delta}{\vp})$ & \prule{Propagation$_\exists$}\\[1mm]
(2) & $\vdash \psi \dnd \delta$ & Theorem \ref{thm-vp-dnd-subbxybvp}\\[1mm]
(3) & $\vdash \bigg(\appln{\psi}{(\exists x\vp)} \to \exists x(\appln{\psi}{\vp})\bigg)  \dnd
\bigg(\appln{\delta}{(\exists x\vp)} \to \exists x(\appln{\delta}{\vp})\bigg)$ & 
Theorem \ref{replacement-thm-NEW-dnd}: (2)\\[1mm]
(4) & $\vdash \bigg(\appln{\delta}{(\exists x\vp)} \to \exists x(\appln{\delta}{\vp})\bigg) \to 
\bigg(\appln{\psi}{(\exists x\vp)} \to \exists x(\appln{\psi}{\vp})\bigg)$ & \eqref{vpdndpsi-imp-vp-iff-psi}: (3) \\[1mm]
(6) & $\vdash\appln{\psi}{(\exists x\vp)} \to \exists x(\appln{\psi}{\vp})$ &  
\prule{modus ponens}: (1), (4)
\et

\,
}

\subsection{Substitutivity of equal patterns}\label{section-subst-equal-patterns}

\bthm\label{thm-subst-equal-one}\,\\
Let $\vp,\psi, \chi,\theta$ be patterns such that $\vp$ is a subpattern of $\chi$ and $\theta$ is obtained
from $\chi$ by replacing one occurrence of $\vp$ with $\psi$. 
Let $x_1, \ldots x_n$ be variables with the following property:
\[ (*) \quad (\FVE(\vp)\cup \FVE(\psi))\cap \BVE(\chi)\se \{x_1, \ldots, x_n\}.\]
Then 
\bce
$\vdash \forall x_1 \ldots  \forall x_n (\vp \eqdef \psi) \to (\chi \dnd \theta)$.
\ece
\ethm
\solution{If $\vp=\chi$, then $\theta=\psi$. We get that \\[1mm]
\begin{tabular}{lll}
(1) & $\vdash \forall x_1 \ldots  \forall x_n (\vp \eqdef \psi) \to \vp \eqdef \psi$ & 
\eqref{forallxnvp-vp} \\[1mm]
(2) & $\vdash \vp \eqdef \psi \to  (\vp \dnd \psi)$ & \eqref{equal-imp-dnd} \\[1mm]
(3) & $\vdash \forall x_1 \ldots  \forall x_n (\vp \eqdef \psi) \to (\vp \dnd \psi)$ &  \prule{syllogism}: (1), (2)\\[2mm]
\end{tabular}

Assume in the sequel that $\vp \ne \chi$. The proof is by induction on $\chi$.
\be
\item $\chi$ is an atomic pattern. Then one must have $\vp=\chi$.
\item $\chi=\neg \sigma$. As $\vp \ne \chi$, we must have that $\vp$ is a subpattern of $\sigma$.
Thus, $\theta=\neg \delta$, where \\$\delta$ is obtained from $\sigma$ by replacing an occurence of 
$\vp$ in $\sigma$ with $\psi$. Furthermore, 
\[(\FVE(\vp)\cup \FVE(\psi))\cap \BVE(\sigma)=(\FVE(\vp)\cup \FVE(\psi))\cap \BVE(\chi)\se \{x_1, \ldots, x_n\}.\]
Hence, (*) holds for $\sigma$ instead of $\chi$. 
We get that

\begin{tabular}{lll}
(1) & $\vdash \forall x_1 \ldots  \forall x_n (\vp \eqdef \psi) \to (\sigma \dnd \delta)$ & 
induction hypothesis on $\sigma$ \\[1mm]
(2) & $\vdash (\sigma \dnd \delta) \to  (\chi \dnd \theta)$ & \eqref{vpdndpsi-neg} \\[1mm]
(3) & $\vdash \forall x_1 \ldots  \forall x_n (\vp \eqdef \psi) \to (\chi \dnd \theta)$ &  \prule{syllogism}: (1), (2)
\end{tabular}

\item $\chi= \sigma_1 \pbin \sigma_2$, where $\pbin\in \{\to, \si,\sau\}$. 
As  $\vp \ne \chi$, we must have that $\vp$ is a subpattern of $\sigma_1$ or that $\vp$ is a subpattern of  $\sigma_2$.
Furthermore, for all $i=1,2$, 
\[(\FVE(\vp)\cup \FVE(\psi))\cap \BVE(\sigma_i) \se (\FVE(\vp)\cup \FVE(\psi))\cap \BVE(\chi)\se \{x_1, \ldots, x_n\}.\]
Hence,  (*) holds for $\sigma_1$  or $\sigma_2$ instead of $\chi$. 

We have two cases:
\be
\item $\theta=\delta \pbin \sigma_2$, where $\delta$ is obtained from $\sigma_1$ by replacing 
an occurence of $\vp$ in $\sigma_1$ with $\psi$. We get that \\
\begin{tabular}{lll}
(1) & $\vdash \forall x_1 \ldots  \forall x_n (\vp \eqdef \psi) \to (\sigma_1 \dnd \delta)$ & induction hypothesis on $\sigma_1$ \\[1mm]
(2) & $\vdash (\sigma_1 \dnd \delta) \to  (\chi \dnd \theta)$ &   \eqref{vpdndpsi-to-r} if $\pbin=\to$; 
\eqref{vpdndpsi-sau-r}  if $\pbin=\sau$;  \eqref{vpdndpsi-si-r}  if $\pbin=\si$ \\[1mm]
(3) & $\vdash \forall x_1 \ldots  \forall x_n (\vp \eqdef \psi) \to (\chi \dnd \theta)$ &   \prule{syllogism}: (1), (2)
\end{tabular}
\item $\theta=\sigma_1\pbin \delta $, where $\delta$ is obtained from $\sigma_2$ by replacing 
an occurence of $\vp$ in $\sigma_2$ with $\psi$. We get that \\
\begin{tabular}{lll}
(1) & $\vdash \forall x_1 \ldots  \forall x_n (\vp \eqdef \psi) \to (\sigma_2 \dnd \delta)$ & 
induction hypothesis on $\sigma_2$ \\[1mm]
(2) & $\vdash (\sigma_2 \dnd \delta) \to  (\chi \dnd \theta)$ &   \eqref{vpdndpsi-to-l} if $\pbin=\to$; 
\eqref{vpdndpsi-sau-l} if $\pbin=\sau$;  \eqref{vpdndpsi-si-l} if $\pbin=\si$ \\[1mm]
(3) & $\vdash \forall x_1 \ldots  \forall x_n (\vp \eqdef \psi) \to (\chi \dnd \theta)$ &   
\prule{syllogism}: (1), (2)
\end{tabular}
\ee
\item $\chi= \appln{\sigma_1}{\sigma_2}$. 
As  $\vp \ne \chi$, we must have that $\vp$ is a subpattern of $\sigma_1$ or that $\vp$ is a subpattern of  $\sigma_2$.
Furthermore, for all $i=1,2$, 
\[(\FVE(\vp)\cup \FVE(\psi))\cap \BVE(\sigma_i) \se (FV(\vp)\cup FV(\psi))\cap \BVE(\chi)\se \{x_1, \ldots, x_n\}.\]
Hence,  (*) holds for $\sigma_1$  or $\sigma_2$ instead of $\chi$. 

We have two cases:
\be
\item $\theta= \appln{\delta}{\sigma_2}$, where $\delta$ is obtained from $\sigma_1$ by replacing 
an occurence of $\vp$ in $\sigma_1$ with $\psi$. We get that \\
\begin{tabular}{lll}
(1) & $\vdash \forall x_1 \ldots  \forall x_n (\vp = \psi) \to (\sigma_1 \dnd \delta)$ & 
induction hypothesis on $\sigma_1$ \\[1mm]
(2) & $\vdash \floor{\forall x_1 \ldots \forall x_n (\vp \dnd \psi)}\to (\sigma_1 \dnd \delta)$ 
& \eqref{sim-pairs-to-left-1}: \eqref{forallxnvpeqdefpsi-dnd-floorforallxnvpdndpsi} and (1)\\[1mm]
(3) & $\vdash \floor{\forall x_1 \ldots \forall x_n (\vp \dnd \psi)}\to (\chi \dnd \theta)$ & \eqref{floordelta-imp-vpchidndpsichi}: (2)\\[1mm]
(4) & $\vdash \forall x_1 \ldots  \forall x_n (\vp = \psi) \to (\chi \dnd \theta)$ 
& \eqref{sim-pairs-to-left-1}: \eqref{forallxnvpeqdefpsi-dnd-floorforallxnvpdndpsi} and (3)
\end{tabular}
\item $\theta=\appln{\sigma_1}{\delta}$, where $\delta$ is obtained from $\sigma_2$ by replacing 
an occurence of $\vp$ in $\sigma_2$ with $\psi$. We get that \\
\begin{tabular}{lll}
(1) & $\vdash \forall x_1 \ldots  \forall x_n (\vp = \psi) \to (\sigma_2 \dnd \delta)$ & 
induction hypothesis on $\sigma_2$ \\[1mm]
(2) & $\vdash \floor{\forall x_1 \ldots \forall x_n (\vp \dnd \psi)}\to (\sigma_2 \dnd \delta)$ 
& \eqref{sim-pairs-to-left-1}: \eqref{forallxnvpeqdefpsi-dnd-floorforallxnvpdndpsi} and (1)\\[1mm]
(3) & $\vdash \floor{\forall x_1 \ldots \forall x_n (\vp \dnd \psi)}\to (\chi \dnd \theta)$ & \eqref{floordelta-imp-chivpndchipsi}: (2)\\[1mm]
(4) & $\vdash \forall x_1 \ldots  \forall x_n (\vp = \psi) \to (\chi \dnd \theta)$ 
& \eqref{sim-pairs-to-left-1}: \eqref{forallxnvpeqdefpsi-dnd-floorforallxnvpdndpsi} and (3)
\end{tabular}
\ee
\item $\chi = Qz\sigma$, where $Q\in \{\forall, \exists\}$. As $\vp \ne \chi$, we must have that
$\vp$ is a subpattern of $\sigma$. \\
Thus, $\theta=Qz\delta$, where $\delta$ is obtained from $\sigma$ by replacing an occurence of 
$\vp$ in $\sigma$ with $\psi$.

\claim 
$z\notin \FVE(\forall x_1 \ldots  \forall x_n(\vp \dnd\psi))$.\\
\pclaim 
We have two cases:
\be
\item $z\notin \FVE(\vp)\cup \FVE(\psi)$. Then the conclusion is obvious.
\item $z\in\FVE(\vp)\cup \FVE(\psi)$. As $z\in \BVE(\chi)$, we have  that 
$z\in (\FVE(\vp)\cup \FVE(\psi))\cap BV(\chi)$. \\
Hence, by (*), $z\in \{x_1, \ldots, x_n\}$. Thus, 
$z\notin \FVE(\forall x_1 \ldots  \forall x_n(\vp \dnd\psi))$. \hfill $\blacksquare$ \\
\ee

We get that \\
\bt{lll}
(1) & $\vdash \forall x_1 \ldots  \forall x_n (\vp \dnd\psi) \to (\sigma \dnd \delta)$ & 
induction hypothesis on $\sigma$, as $BV(\sigma)\se BV(\chi)$, \\ 
&&  hence (*) 
holds for $\sigma$ instead of $\chi$ \\[1mm]
(2) & $\vdash \forall x_1 \ldots  \forall x_n (\vp \dnd\psi) \to (\chi \dnd \theta)$ & (1) and 
\eqref{chi-vppsi-chi-forallxvppsi-dnd-M-NFV} if $Q=\forall$; 
\eqref{chi-vppsi-chi-existsxvppsi-dnd-M-NFV} if $Q=\exists$
\et
\ee
}

\bthm[Substitutivity of equal patterns]\label{thm-subst-equal-more}\,\\
Let $\vp,\psi, \chi,\theta$ be patterns such that $\vp$ is a subpattern of $\chi$ and $\theta$ is obtained
from $\chi$ by replacing one or more occurrences of $\vp$ with $\psi$. 
Let $x_1, \ldots x_n$ be variables with the following property:
\[ (*) \quad (\FVE(\vp)\cup \FVE(\psi))\cap \BVE(\chi)\se \{x_1, \ldots, x_n\}.\]
Then 
\bce
$\vdash \forall x_1 \ldots  \forall x_n (\vp \eqdef \psi) \to (\chi \dnd \theta)$.
\ece
\ethm
\solution{\,\\
 The proof is by induction on the number $n$ of occurrences of $\vp$ in $\chi$ thar are replaced with $\psi$.

$n=1$: Apply Theorem \ref{thm-subst-equal-one}.

$n=n+1$: Let $\theta'$ be the pattern obtained by replacing $n$ occurrences of $\vp$ in $\chi$ with $\psi$. Then 
$\theta$ is obtained from $\theta'$ by replacing one occurence of $\vp$ in $\theta'$ with $\psi$.
We get that:\\[1mm]
\begin{tabular}{lll}
(1) & $\vdash\forall x_1 \ldots  \forall x_n (\vp \eqdef \psi) \to (\chi \dnd \theta')$ & (Induction hypothesis) \\[1mm]
(2) & $\vdash\forall x_1 \ldots  \forall x_n (\vp \eqdef \psi) \to (\theta' \dnd \theta)$ & Theorem \ref{thm-subst-equal-one} \\[1mm]
(3) & $\vdash\forall x_1 \ldots  \forall x_n (\vp \eqdef \psi) \to (\chi \dnd \theta)$ & 
\eqref{vp-to-psi-dnd-chi-vp-dnd-chi-to-gamma-implies-vp-to-psi-dnd-gamma}: (1), (2)
\end{tabular}

\,
}

The following is an immediate corollary of Theorem \ref{thm-subst-equal-more}.

\bprop
Let $\vp,\psi, \chi,\theta$ be patterns such that $\vp$ is a  
subpattern of $\chi$ and $\theta$ is obtained from $\chi$ by replacing one or more occurrences of $\vp$ with $\psi$. 
\be
\item If $\chi$ is quantifier-free, then 
\bce
$\vdash  \vp \eqdef \psi \to (\chi \dnd \theta)$.
\ece
\item Let $\{x_1, \ldots, x_n\}=\FVE(\vp)\cup \FVE(\psi)$.
 Then 
\bce
$\vdash  \forall x_1 \ldots  \forall x_n(\vp \dnd\psi) \to (\chi \dnd \theta)$.
\ece
\item Let $\{x_1, \ldots, x_n\}=\BVE(\chi)$.
 Then 
\bce
$\vdash  \forall x_1 \ldots  \forall x_n(\vp \dnd\psi) \to (\chi \dnd \theta)$.
\ece
\ee
\eprop

\subsection{A deduction theorem}\label{section-deduction-thm}

We prove a deduction theorem similar with \cite[Theorem 4.2]{Che23}, the deduction theorem 
for matching $\mu$-logic.

Let $\Gamma\cup\{\vp\}$ be a set of patterns. For every pattern $\psi$, let us denote

\bce 
\bt{llll}
(*) & $\Gamma\cup\{\vp\} \vdashgen \psi$ & iff & there exists a $\Gamma\cup\{\vp\}$-proof of $\psi$ 
that contains no applications of \\
&&& \prule{gen} involving a variable which occurs free in $\vp$.
\et
\ece

\bthm[Deduction Theorem]\label{deduction-theorem}\,\\
Assume that $\Gamma\cup\{\vp\} \vdashgen \psi$. Then $\Gamma\vdash \floor{\vp}\ra\psi$.
\ethm
\solution{Let $\theta_1, \ldots, \theta_n=\psi$ be a $\Gamma\cup\{\vp\}$-proof 
of $\psi$ as in (*).  We prove by induction on $i$ that for all $i=1,\ldots,n$,  
\[\Gamma\vdash \floor{\vp}\ra\theta_i.\]  
As a consequence, $\Gamma\vdash \floor{\vp}\ra\theta_n=\psi$. \\

$i=1$: We have the following cases:
\be
\item  $\theta_1$ is an axiom or  $\theta_1\in \Gamma$.  We get that \\[1mm]
\begin{tabular}{lll}
(1) & $\Gamma\vdash\theta_1$ &   \\[1mm]
(2) & $\Gamma\vdash \theta_1\ra(\vp\ra\theta_1)$  & \eqref{vp-to-psi-to-vp}\\[1mm]
(3) & $\Gamma\vdash\vp\ra\theta_1$ & \prule{modus ponens}:   (1), (2)\\[1mm]
(4) & $\Gamma\vdash \floor{\vp}\ra \vp$ & \eqref{lemma-floor-elim-alt}\\[1mm]
(5) & $\Gamma\vdash \floor{\vp}\ra \theta_1$ &  \prule{syllogism}: (4), (5)\\[1mm]
\end{tabular}

\item  $\theta_1=\vp$. Apply \eqref{lemma-floor-elim-alt}.
\ee

Assume that the induction hypothesis is true for all $j=1, \ldots, i$. We have the following cases for $\theta_{i+1}$:

\be
\item $\theta_{i+1}$ is an axiom or  $\theta_{i+1}\in \Gamma\cup\{\vp\}$. Then, as in the case $i=1$, 
$\Gamma\vdash \floor{\vp}\ra\vp_{i+1}$.

\item $\theta_{i+1}$ is obtained by applying a propositional deduction rule.
\be 
\item \prule{modus ponens}: There exist $j,k< i+1$ such that $\theta_k=\theta_j \to \theta_{i+1}$. 

We get that \\[1mm]
\begin{tabular}{lll}
(1) & $\Gamma\vdash \floor{\vp} \ra \theta_j$ & (Induction hypothesis)\\[1mm]
(2) & $\Gamma\vdash \floor{\vp}\ra \theta_k$ & (Induction hypothesis)\\[1mm]
(3) & $\Gamma\vdash \floor{\vp} \ra \theta_{i+1}$ &  \eqref{vp-to-psi-and-vp-to-psi-to-chi-implies-vp-to-chi}: (1), (2)\\[2mm]
\end{tabular}

\item \prule{syllogism}: There exist $j, k < i + 1$  and patterns $\alpha$, $\beta$, $\chi$ such that 
$\theta_j=\alpha\to\beta$, $\theta_k=\beta\to\chi$ and $\theta_{i + 1}=\alpha\to\chi$.\\
We get that \\[1mm]
\begin{tabular}{lll}
(1) & $\Gamma\vdash\floor{\vp} \ra \theta_j$ & (Induction hypothesis)\\[1mm]
(2) & $\Gamma\vdash\floor{\vp} \ra \theta_k$ & (Induction hypothesis)\\[1mm]
(3) & $\Gamma\vdash\floor{\vp} \ra \theta_{i + 1}$ & \eqref{vp-to-psi-to-chi-vp-to-chi-to-gamma-implies-vp-to-psi-to-gamma}: (1), (2)\\[2mm]
\end{tabular}

\item \prule{exportation}: There exists $j < i + 1$  and patterns $\alpha$, $\beta$, $\chi$ such that
 $\theta_j=\alpha \si \beta \to \chi$ and $\theta_{i + 1}=\alpha\to(\beta \to \chi)$.\\
We get that \\[1mm]
\begin{tabular}{lll}
(1) & $\Gamma\vdash\floor{\vp} \ra \theta_j$ & (Induction hypothesis)\\[1mm]
(2) & $\Gamma\vdash\floor{\vp} \ra \theta_{i + 1}$ & \eqref{vp-to-psi-si-chi-to-gamma-iff-vp-to-psi-to-chi-to-gamma}: (1)\\[2mm]
\end{tabular}

\item \prule{importation}: There exists $j < i + 1$  and patterns $\alpha$, $\beta$, $\chi$ such that
 $\theta_j=\alpha\to(\beta \to \chi)$ and $\theta_{i + 1}=\alpha \si \beta \to \chi$.\\
We get that \\[1mm]
\begin{tabular}{lll}
(1) & $\Gamma\vdash\floor{\vp} \ra \theta_j$ & (Induction hypothesis)\\[1mm]
(2) & $\Gamma\vdash\floor{\vp} \ra \theta_{i + 1}$ & \eqref{vp-to-psi-si-chi-to-gamma-iff-vp-to-psi-to-chi-to-gamma}: (1)\\[2mm]
\end{tabular}

\item \prule{expansion}: There exists $j < i + 1$  and patterns $\alpha$, $\beta$, $\chi$ such that
$\theta_j=\alpha \to \beta$ and $\theta_{i + 1}=\chi \sau \alpha \to \chi \sau \beta$.\\
We get that \\[1mm]
\begin{tabular}{lll}
(1) & $\Gamma\vdash\floor{\vp} \ra \theta_j$ & (Induction hypothesis)\\[1mm]
(3) & $\Gamma\vdash\floor{\vp} \ra \theta_{i + 1}$ & \eqref{expansion-extra-premise}: (1)\\[2mm]
\end{tabular}
\ee

\item $\theta_{i+1}$ is obtained by applying \prule{gen} that does not involve a variable which occurs free in $\vp$.
Thus, there exists $j<i+1$  such that $\theta_{i+1}=\forall x \theta_j$ and $x\notin FV(\vp)$.

We get that \\[1mm]
\begin{tabular}{lll}
(1) & $\Gamma\vdash \floor{\vp} \ra \theta_j$ & (Induction hypothesis)\\[1mm]
(2) & $\Gamma\vdash \forall x(\floor{\vp} \ra \theta_j)$ & \prule{gen}: (1) \\[1mm]
(3) & $\Gamma\vdash \forall x(\floor{\vp} \ra \theta_j) \to  (\floor{\vp} \to \theta_{i+1})$ & 
\eqref{forallxvptopsi-to-forallxvp}, as $FV(\floor{\vp})=FV(\vp)$, so $x\notin FV(\floor{\vp})$ \\[1mm]
(4) & $\Gamma\vdash \floor{\vp} \to \theta_{i+1}$ & \prule{modus ponens}: (2), (3) \\[1mm]
\end{tabular}

\item $\theta_{i+1}$ is obtained by applying one of the application deduction rules:

\be
\item \prule{framing-left}: There exists $j < i + 1$  and patterns $\alpha$, $\beta$, $\chi$  such that 
$\theta_j=\alpha \to \beta$ and 
 $\theta_{i+1}=\appln{\alpha}{\chi} \to \appln{\beta}{\chi}$ for some pattern $\chi$. 

We get that \\[1mm]
\begin{tabular}{lll}
(1) & $\Gamma\vdash \floor{\vp} \ra \theta_j$ & (Induction hypothesis)\\[1mm]
(2) & $\Gamma\vdash \floor{\vp} \to \theta_{i+1}$ &  \eqref{floordelta-imp-vpchitopsichi}: (1)
\end{tabular}

\item \prule{framing-right}: There exists $j < i + 1$  and patterns $\alpha$, $\beta$, $\chi$  such that 
$\theta_j=\alpha \to \beta$ and 
 $\theta_{i+1}=\appln{\chi}{\alpha} \to \appln{\chi}{\beta}$ for some pattern $\chi$. 

We get that \\[1mm]
\begin{tabular}{lll}
(1) & $\Gamma\vdash \floor{\vp} \ra \theta_j$ & (Induction hypothesis)\\[1mm]
(2) & $\Gamma\vdash \floor{\vp} \to \theta_{i+1}$ &  \eqref{floordelta-imp-vpchidndpsichi}: (1)
\end{tabular}

\ee

\ee
}

\bprop
Assume that $\vp$ is a sentence. Then 
\bce
$\Gamma\cup\{\vp\} \vdash \psi$ iff $\Gamma\vdash \floor{\vp}\ra\psi$.
\ece
\eprop
\solution{
($\Ra$) As $\vp$ is a sentence, we have that $\Gamma\cup\{\vp\} \vdash \psi$ iff $\Gamma\cup\{\vp\} \vdashgen \psi$. Apply now
Theorem \ref{deduction-theorem}.

($\La$) We get that \\[1mm]
\begin{tabular}{lll}
(1) & $\Gamma\vdash \floor{\vp} \ra\psi$ &  (Assumption)\\[1mm]
(2) &  $\Gamma\cup\{\vp\}\vdash \vp\ra\psi$ & \\[1mm]
(3) &  $\Gamma\cup\{\vp\}\vdash \vp$ & \\[1mm]
(4) & $\Gamma\cup\{\vp\}\vdash \psi$ & \prule{modus ponens}: (2), (3)\\[2mm]
\end{tabular}

}

\subsection{Application contexts}\label{section-app-contexts}

We refer to \cite[Section 2.9]{Leu25} for the definition and basic properties of contexts that shall be studied
in this section, referred to as \textit{application contexts}.
We prove for $\appMLdef$ properties of application contexts  that were obtained in 
\cite[pp. 46-48, 53]{Che23} for matching $\mu$-logic. We remark the fact that some of these 
properties (see \cite[Proposition 3.3, Lemma 3.14]{Che23}) are proved in \cite{Che23} by using 
the so-called \prule{singleton variable}, an axiom in the proof system for matching $\mu$-logic 
from \cite{Che23}. Our proofs are simpler, as they do not need \prule{singleton variable}. \\

In the following, $C$ is an application context.

\blem
\begin{align}
\Gamma\vdash \vp\to\psi \text{ implies } & \Gamma\vdash C[\vp] \to C[\psi], \label{framing-contexts-implies}\\[1mm]
\Gamma\vdash \vp\dnd\psi \text{ implies } & \Gamma\vdash C[\vp] \dnd C[\psi]. \label{framing-contexts-dnd}
\end{align}
\elem
\solution{ \, \\[1mm]
\eqref{framing-contexts-implies}:
The proof is by induction on the application context $C$.
\be
\item $C=\Box$. The conclusions are obvious, as $\Box[\delta]=\delta$ for every pattern $\delta$. 
\item $C = Appl_\Box C_1\chi$ where $C_1$ is an application context. Then $C[\delta]=\appln{C_1[\delta]}\chi$ 
for every pattern $\delta$. We get that \\[1mm]
\begin{tabular}{lll}
(1) & $\Gamma\vdash \vp\to\psi$ & (Assumption)\\[1mm]
(2) & $\Gamma\vdash C_1[\vp]\to C_1[\psi]$ & (Induction hypothesis): (1)\\[1mm]
(3) & $\Gamma\vdash C[\vp] \to C[\psi]$ & \prule{framing-left}: (2)
\end{tabular}
\item If C = $Appl_\Box\chi C_1$ where $C_1$ is an application context. Then $C[\delta]=\appln{\chi}{C_1[\delta]}$ 
for every pattern $\delta$. We get that \\[1mm]
\begin{tabular}{lll}
(1) & $\Gamma\vdash\vp\to\psi$ & (Assumption)\\[1mm]
(2) & $\Gamma\vdash C_1[\vp]\to C_1[\psi]$ & (Induction hypothesis): (1)\\[1mm]
(3) & $\Gamma\vdash C[\vp] \to C[\psi]$ & \prule{framing-right}: (2)\\[2mm]
\end{tabular}
\ee

\eqref{framing-contexts-dnd}: Apply \eqref{framing-contexts-implies} and \eqref{Gamma-vptopsi-psitovp-iff-vpdndpsi}.
}

\blem
\begin{align}
\vdash \,  C[\ceil{\vp} ] \to \ceil{\vp}. \label{context-Cceilvp-ceilvp}
\end{align}
\elem
\solution{The proof is by induction on the application context $C$.
\be
\item $C=\Box$. Obviously, as $\Box[\ceil{\vp} ]=\ceil{\vp} $. 
\item $C = Appl_\Box C_1\psi$ where $C_1$ is an application context. 
Then $C[\ceil{\vp}]=\appln{C_1[\ceil{\vp}]}\psi$. We get that \\[1mm]
\begin{tabular}{lll}
(1) & $\Gamma\vdash C_1[\ceil{\vp}]\to \ceil{\vp} $ & (Induction hypothesis) \\[1mm]
(2) & $\Gamma\vdash C[\ceil{\vp}] \to \appln{\ceil{\vp}}\psi$ & \prule{framing-left}: (2) \\[1mm]
(3) & $\Gamma\vdash \appln{\ceil{\vp}}\psi \to \ceil{\vp}$ & \prule{\propagationdef}\\[1mm]
(4) & $\Gamma\vdash C[\ceil{\vp}] \to \ceil{\vp}$ & \prule{syllogism}: (2), (3)\\[1mm]
\end{tabular}
\item If C = $Appl_\Box\chi C_1$ where $C_1$ is an application context. 
Then $C[\ceil{\vp}]=\appln{\chi}{C_1[\ceil{\vp}]}$. We get that \\[1mm]
\begin{tabular}{lll}
(1) & $\Gamma\vdash C_1[\ceil{\vp}]\to \ceil{\vp} $ & (Induction hypothesis) \\[1mm]
(2) & $\Gamma\vdash C[\ceil{\vp}] \to \appln{\psi}{\ceil{\vp}}$ & \prule{framing-right}: (2) \\[1mm]
(3) & $\Gamma\vdash \appln{\psi}{\ceil{\vp}} \to \ceil{\vp}$ & \prule{\propagationdef} \\[1mm]
(4) & $\Gamma\vdash C[\ceil{\vp}] \to \ceil{\vp}$ & \prule{syllogism}: (2), (3) \\[1mm]
\end{tabular}
\ee
}

\blem
\begin{align}
\vdash & \, C[\vp] \to \ceil{\vp}.	 \label{context-Cvp-ceilvp}
\end{align}
\elem
\solution{\, \\[1mm]
\begin{tabular}{lll}
(1) & $\vdash \vp\to \ceil{\vp} $ & \prule{\axvptoceilvp} \\[1mm]
(2) & $\vdash C[\vp] \to C[\ceil{\vp}]$ & \eqref{framing-contexts-implies}: (1) \\[1mm]
(3) & $\vdash C[\ceil{\vp} ] \to \ceil{\vp}$ & \eqref{context-Cceilvp-ceilvp}\\[1mm]
(4) & $\vdash C[\vp] \to \ceil{\vp}$ & \prule{syllogism}: (2), (3)\\[2mm]
\end{tabular}

}

\blem
\begin{align}
& \vdash C[\bot]\dnd \bot, \label{context-bot} \\[1mm]
& \vdash C[\vp\sau\psi] \dnd C[\vp] \sau C[\psi], \label{context-sau}\\[1mm]
& \vdash C[\exists x\vp]\dnd \exists x C[\vp] \quad \text{if x } \notin FV(C[\exists x\vp]). 
\label{context-exists}
\end{align}
\elem
\solution{
We prove \eqref{context-bot}-\eqref{context-exists} by induction on the application context C.
\be
\item $C = \Box$. The conclusions are obvious, as $\Box[\delta]=\delta$ for every pattern $\delta$. 
\item $C = Appl_\Box C_1\chi$ where $C_1$ is an application context. Then $C[\delta]=\appln{C_1[\delta]}\chi$ 
for every pattern $\delta$. 

\eqref{context-bot}:\\[1mm]
\begin{tabular}{lll}
(1) & $\vdash C_1[\bot] \dnd \bot$ & (Induction hypothesis)\\[1mm]
(2) & $\vdash C_1[\bot] \to \bot$ & \eqref{Gamma-vp-psi-vp-si-psi}: (1) and the definition of $\dnd$\\[1mm]
(3) & $\vdash C[\bot] \to \appln{\bot}{\chi}$ & \prule{framing-left}: (2) \\[1mm]
(4) & $\vdash \appln{\bot}{\chi}\to\bot$ & \eqref{propagation-bot-l} \\[1mm]
(5) & $\vdash C[\bot] \to \bot$ & \prule{syllogism}: (3), (4)\\[1mm]
(6) & $\vdash \bot \to C[\bot]$ & \prule{exfalso}\\[1mm]
(7) & $\vdash C[\bot] \dnd \bot$ & \eqref{Gamma-vp-psi-vp-si-psi}: (5), (6) and 
the definition of $\dnd$\\[2mm]
\end{tabular}

\eqref{context-sau}:\\[1mm]
\begin{tabular}{lll}
(1) & $\vdash C_1[\vp\sau\psi]\dnd C_1[\vp]\sau C_1[\psi]$ & (Induction hypothesis)\\[1mm]
(2) & $\vdash C_1[\vp\sau\psi]\to C_1[\vp]\sau C_1[\psi]$ & \eqref{Gamma-vp-psi-vp-si-psi}: (1) and the definition of $\dnd$\\[1mm]
(3) & $\vdash C[\vp\sau\psi] \to \appln{(C_1[\vp]\sau C_1[\psi])}{\chi}$ & \prule{framing-left}: (2) \\[1mm]
(4) & $\vdash \appln{(C_1[\vp]\sau C_1[\psi])}{\chi} \to C[\vp]\sau C[\psi]$ & \prule{propagation$_\vee$} \\[1mm]
(5) & $\vdash C[\vp\sau\psi]\to C[\vp]\sau C[\psi]$ & \prule{syllogism}: (3), (4)\\[1mm]
(6) & $\vdash \vp\to\vp\sau\psi$ & \prule{weakening}\\[1mm]
(7) & $\vdash C[\vp] \to C[\vp\sau\psi]$ & \eqref{framing-contexts-implies}: (6)\\[1mm]
(8) & $\vdash \psi\to\vp\sau\psi$ & \eqref{weak-sau-2}\\[1mm]
(9) & $\vdash C[\psi] \to C[\vp\sau\psi]$ & \eqref{framing-contexts-implies}: (8)\\[1mm]
(10)& $\vdash C[\vp]\sau C[\psi] \to C[\vp\sau\psi]$ & \eqref{vp-to-chi-and-psi-to-chi-implies-vp-sau-psi-to-chi}: (6), (8)\\[1mm]
(11)& $\vdash C[\vp\sau\psi] \dnd C[\vp]\sau C[\psi]$ & \eqref{Gamma-vp-psi-vp-si-psi}: (5), (10)\\[2mm]
\end{tabular}

\eqref{context-exists}: Remark that $C[\exists x\vp]=\appln{C_1[\exists x\vp]}\chi$. Hence, 
$FV(C[\exists x\vp])=FV(C_1[\exists x\vp])\cup FV(\chi)$. As $x\notin FV(C[\exists x\vp])$, we get that 
$x\notin FV(C_1[\exists x\vp])$ and $x\notin FV(\chi)$. We get that  \\[1mm]
\begin{tabular}{lll}
(1) & $\vdash C_1[\exists x\vp] \dnd \exists xC_1[\vp]$ & (Induction hypothesis)\\[1mm]
(2) & $\vdash C_1[\exists x\vp] \to \exists xC_1[\vp]$ & \eqref{Gamma-vp-psi-vp-si-psi}: (1) and the definition of $\dnd$\\[1mm]
(3) & $\vdash C[\exists x\vp] \to \appln{(\exists xC_1[\vp])}{\chi}$ & \prule{framing-left}: (2)\\[1mm]
(4) & $\vdash \appln{(\exists xC_1[\vp])}{\chi} \to \exists x C[\vp]$ & \eqref{propagation-exists-1-H}, as $x\notin FV(\chi)$ \\[1mm]
(5) & $\vdash C[\exists x\vp] \to \exists x C[\vp]$ & \prule{syllogism}: (3), (4)\\[1mm]
(6) & $\vdash \vp\to\exists x\vp$ & \eqref{exists-quant-x-M}\\[1mm]
(7) & $\vdash C[\vp] \to C[\exists x\vp]$ & \eqref{framing-contexts-implies}: (6)\\[1mm]
(8) & $\vdash \exists x C[\vp] \to C[\exists x\vp]$ & \eqref{exists-quant-rule}: (7), as $x\notin FV(C[\exists x\vp])$\\[1mm]
(9)& $\vdash C[\exists x\vp] \dnd \exists xC[\vp]$ & \eqref{Gamma-vp-psi-vp-si-psi}: (5), (8)\\[2mm]
\end{tabular}

\item $C = Appl_\Box \chi C_1$, where $C_1$ is an application context. Then $C[\delta]=\appln{\chi}{C_1[\delta]}$ 
for every pattern $\delta$. 

\eqref{context-bot}:\\[1mm]
\begin{tabular}{lll}
(1) & $\vdash C_1[\bot] \dnd \bot$ & (Induction hypothesis)\\[1mm]
(2) & $\vdash C_1[\bot] \to \bot$ & \eqref{Gamma-vp-psi-vp-si-psi}: (1) and the definition of $\dnd$\\[1mm]
(3) & $\vdash C[\bot] \to \appln{\chi}{\bot}$ & \prule{framing-right}: (2) \\[1mm]
(4) & $\vdash \appln{\chi}{\bot}\to\bot$ & \eqref{propagation-bot-r}\\[1mm]
(5) & $\vdash C[\bot] \to \bot$ & \prule{syllogism}: (3), (4)\\[1mm]
(6) & $\vdash \bot \to C[\bot]$ & \prule{exfalso}\\[1mm]
(7) & $\vdash C[\bot] \dnd \bot$ & \eqref{Gamma-vp-psi-vp-si-psi}: (5), (6) and the definition 
of $\dnd$\\[2mm]
\end{tabular}

\eqref{context-sau}:\\[1mm]
\begin{tabular}{lll}
(1) & $\vdash C_1[\vp\sau\psi]\dnd C_1[\vp]\sau C_1[\psi]$ & (Induction hypothesis)\\[1mm]
(2) & $\vdash C_1[\vp\sau\psi]\to C_1[\vp]\sau C_1[\psi]$ & \eqref{Gamma-vp-psi-vp-si-psi}: (1) and the definition of $\dnd$\\[1mm]
(3) & $\vdash C[\vp\sau\psi] \to \appln{\chi}{(C_1[\vp]\sau C_1[\psi])}$ & \prule{framing-right}: (2) \\[1mm]
(4) & $\vdash \appln{\chi}{(C_1[\vp]\sau C_1[\psi])} \to C[\vp]\sau C[\psi]$ & \prule{propagation$_\vee$} \\[1mm]
(5) & $\vdash C[\vp\sau\psi]\to C[\vp]\sau C[\psi]$ & \prule{syllogism}: (3), (4)\\[1mm]
(6) & $\vdash \vp\to\vp\sau\psi$ & \prule{weakening}\\[1mm]
(7) & $\vdash C[\vp] \to C[\vp\sau\psi]$ & \eqref{framing-contexts-implies}: (6)\\[1mm]
(8) & $\vdash \psi\to\vp\sau\psi$ & \eqref{weak-sau-2}\\[1mm]
(9) & $\vdash C[\psi] \to C[\vp\sau\psi]$ & \eqref{framing-contexts-implies}: (8)\\[1mm]
(10)& $\vdash C[\vp]\sau C[\psi] \to C[\vp\sau\psi]$ & \eqref{vp-to-chi-and-psi-to-chi-implies-vp-sau-psi-to-chi}: (6), (8)\\[1mm]
(11)& $\vdash C[\vp\sau\psi] \dnd C[\vp]\sau C[\psi]$ & \eqref{Gamma-vp-psi-vp-si-psi}: (5), (10)\\[2mm]
\end{tabular}

\eqref{context-exists}: Remark that $C[\exists x\vp]=\appln{\chi}{C_1[\exists x\vp]}$. Hence, 
$FV(C[\exists x\vp])=FV(\chi) \cup FV(C_1[\exists x\vp])$. As $x\notin FV(C[\exists x\vp])$, we get that 
$x\notin FV(\chi)$ and $x\notin FV(C_1[\exists x\vp])$. We get that  \\[1mm]
\begin{tabular}{lll}
(1) & $\vdash C_1[\exists x\vp] \dnd \exists xC_1[\vp]$ & (Induction hypothesis)\\[1mm]
(2) & $\vdash C_1[\exists x\vp] \to \exists xC_1[\vp]$ & \eqref{Gamma-vp-psi-vp-si-psi}: (1) and the definition of $\dnd$\\[1mm]
(3) & $\vdash C[\exists x\vp] \to \appln{\chi}{(\exists xC_1[\vp])}$ & \prule{framing-right}: (2)\\[1mm]
(4) & $\vdash \appln{\chi}{(\exists xC_1[\vp])} \to \exists x C[\vp]$ & \eqref{propagation-exists-2-H}, as $x\notin FV(\chi)$ \\[1mm]
(5) & $\vdash C[\exists x\vp] \to \exists x C[\vp]$ & \prule{syllogism}: (3), (4)\\[1mm]
(6) & $\vdash \vp\to\exists x\vp$ & \eqref{exists-quant-x-M}\\[1mm]
(7) & $\vdash C[\vp] \to C[\exists x\vp]$ & \eqref{framing-contexts-implies}: (6)\\[1mm]
(8) & $\vdash \exists x C[\vp] \to C[\exists x\vp]$ & \eqref{exists-quant-rule}: (7), as $x\notin FV(C[\exists x\vp])$\\[1mm]
(9)& $\vdash C[\exists x\vp] \dnd \exists xC[\vp]$ & \eqref{Gamma-vp-psi-vp-si-psi}: (5), (8)
\end{tabular}
\ee
}

\blem
\bea
\Gamma \vdash C[\vp\sau\psi] & \text{ iff } & \Gamma \vdash C[\vp] \sau C[\psi], \label{context-sau-rule} \\
\Gamma \vdash C[\exists x\vp] & \text{ iff } & \Gamma \vdash \exists x C[\vp].  \quad \text{if x } \notin FV(C[\exists x\vp]) 
 \label{context-exists-rule}
\eea
\elem
\solution{Apply \eqref{context-sau}, \eqref{context-exists} 
and \eqref{vpdndpsi-imp-vp-iff-psi}.
}

\blem
\bea 
\Gamma\vdash \vp & \text{ implies } & \Gamma\vdash \neg C[\neg\vp]. \label{context-cp-imp-negCnegvp}
\eea
\elem
\solution{\, \\[1mm]
\begin{tabular}{lll}
(1) & $\Gamma\vdash \vp$ & (Assumption)\\[1mm]
(2) & $\Gamma\vdash \vp\to \neg\neg\vp$ & \eqref{Gamma-vptopsi-psitovp-iff-vpdndpsi} and \eqref{negnegvp-dnd-vp}\\[1mm]
(3) & $\Gamma\vdash \neg\neg\vp$ & \prule{modus-ponens}: (1), (2)\\[1mm]
(4) & $\Gamma\vdash \neg\neg\vp \to (\neg\vp\to\bot)$ & \prule{axiom-not-1}\\[1mm]
(5) & $\Gamma\vdash \neg\vp\to\bot$ & \prule{modus-ponens}: (3), (4)\\[1mm]
(6) & $\Gamma\vdash C[\neg\vp] \to C[\bot]$ & \eqref{framing-contexts-implies}: (5)\\[1mm]
(7) & $\Gamma\vdash C[\bot] \dnd \bot$ & \eqref{context-bot}\\[1mm]
(8) & $\Gamma\vdash C[\bot]\to\bot$ & \eqref{Gamma-vptopsi-psitovp-iff-vpdndpsi}: (6)\\[1mm]
(9) & $\Gamma\vdash C[\neg\vp] \to \bot$ & \prule{syllogism}: (6), (8)\\[1mm]
(10) & $\Gamma\vdash(C[\neg\vp] \to \bot) \to \neg C[\neg\vp] $ & \prule{axiom-not-2}\\[1mm]
(11) & $\Gamma\vdash \neg C[\neg\vp] $ & \prule{modus-ponens}: (9), (10)
\end{tabular}

\,
}

\subsection{Membership}\label{section-membership}

The derived  connective $\in$ (called \textit{membership}) is defined as follows: for every 
variable $x$ and pattern $\vp$, 
\begin{align*}
x \in \vp & := \ceil{x\si\vp}.
\end{align*}
We assume that $\in$ has higher precedence than $\ra$, $\si$, $\sau$, $\dnd$.

We give in the sequel proofs of theorems and derived rules for membership, by following the proofs from 
\cite[pp. 51-53]{Che23} for matching $\mu$-logic. Most of the proofs  from \cite{Che23} make use of the
\prule{Singleton Variable} axiom. We show that we can replace it with a simpler axiom, \prule{\singletons}.

\begin{lemma}
\begin{align}
\vdash & \, (x\in \vp) \dnd \ceil{\vp \si x}. \label{xinvp-char-equiv}
\end{align}
\end{lemma}
\solution{\,\\[1mm]
\begin{tabular}{lll}
(1) & $\vdash x\si \vp \dnd \vp \si x$ & \eqref{si-comm}\\[1mm]
(2) & $\vdash (x\in \vp)\dnd \ceil{\vp \si x}$ & \eqref{rule-iff-compat-in-ceil}: (1) and the definition of $\in$
\end{tabular}

\,
}

\begin{lemma}
\bea
\Gamma \vdash \vp \to \psi &  \text{implies} & \Gamma\vdash x\in \vp \to x\in \psi, \label{lemma-in-compat-to-vppsi}\\
\Gamma \vdash \vp \dnd \psi &  \text{implies} & \Gamma\vdash x\in \vp \dnd x\in \psi, \label{lemma-in-compat-dnd-vppsi}\\
\Gamma \vdash x \to y &  \text{implies} & \Gamma\vdash x\in \vp \to y\in \vp, \label{lemma-in-compat-to-xy}\\
\Gamma \vdash x \dnd y &  \text{implies} & \Gamma\vdash x\in \vp \dnd y\in \vp, \label{lemma-in-compat-dnd-xy}\\
\Gamma \vdash x \to y \,\, \text{and} \,\,  \vdash \vp \to \psi &  \text{imply} & \Gamma\vdash x\in \vp \to y\in \psi,
 \label{lemma-in-compat-to-xy-vppsi}\\
\Gamma \vdash x \dnd y \,\, \text{and} \,\,  \vdash \vp \dnd \psi & \text{imply} & \Gamma\vdash x\in \vp \dnd y\in \psi.
 \label{lemma-in-compat-dnd-xy-vppsi}
\eea
\end{lemma}
\solution{\,\\[1mm]
\eqref{lemma-in-compat-to-vppsi}:\\[1mm]
\begin{tabular}{lll}
(1) & $\Gamma\vdash \vp \to \psi$ & (Assumption)\\[1mm]
(2) & $\Gamma\vdash x \si \vp \to x \si \psi$ & \eqref{sim-pairs-and-right}: (1)\\[1mm]
(3) & $\Gamma\vdash x\in \vp \to x\in \psi$ & \eqref{lemma-imp-compat-in-ceil}: (2)\\[2mm]
\end{tabular}

\eqref{lemma-in-compat-dnd-vppsi}: By \eqref{lemma-in-compat-to-vppsi} and the definition of $\dnd$.\\[2mm]
\eqref{lemma-in-compat-to-xy}:\\[1mm]
\begin{tabular}{lll}
(1) & $\Gamma\vdash  x \to y$ & (Assumption)\\[1mm]
(2) & $\Gamma\vdash x \si \vp \to y \si \vp$ & \eqref{sim-pairs-and-left}: (1)\\[1mm]
(3) & $\Gamma\vdash x\in \vp \to  y \in \vp$ & \eqref{lemma-imp-compat-in-ceil}: (2)\\[2mm]
\end{tabular}

\eqref{lemma-in-compat-dnd-xy}: By \eqref{lemma-in-compat-to-xy} and the definition of $\dnd$.\\[2mm]
\eqref{lemma-in-compat-to-xy-vppsi}:\\[1mm]
\begin{tabular}{lll}
(1) & $\Gamma\vdash \vp \to \psi$ & (Assumption)\\[1mm]
(2) & $\Gamma\vdash x\in \vp \to x\in \psi$ & \eqref{lemma-in-compat-to-vppsi}: (2)\\[1mm]
(3) & $\Gamma\vdash  x \to y$ & (Assumption)\\[1mm]
(4) & $\Gamma\vdash x\in \psi \to y\in \psi$ & \eqref{lemma-in-compat-to-xy}: (2)\\[1mm]
(5) & $\Gamma\vdash x\in \vp \to y\in \psi$ & \prule{syllogism}: (2), (4)\\[2mm]
\end{tabular}

\eqref{lemma-in-compat-dnd-xy-vppsi}:\\[1mm]
\begin{tabular}{lll}
(1) & $\Gamma\vdash \vp \dnd \psi$ & (Assumption)\\[1mm]
(2) & $\Gamma\vdash x\in \vp \dnd x\in \psi$ & \eqref{lemma-in-compat-dnd-vppsi}: (2)\\[1mm]
(3) & $\Gamma\vdash  x \dnd y$ & (Assumption)\\[1mm]
(4) & $\Gamma\vdash x\in \psi \dnd y\in \psi$ & \eqref{lemma-in-compat-dnd-xy}: (2)\\[1mm]
(5) & $\Gamma\vdash x\in \vp \dnd y\in \psi$ & \eqref{sim-trans}: (2), (4)
\end{tabular}

\, 
}

\begin{lemma}
\bea
\Gamma \vdash \vp &  \text{implies} & \Gamma\vdash\forall x(x\in\vp). \label{mem-intro}
\eea
\end{lemma}
\solution{\,\\[1mm]
\begin{tabular}{lll}
(1) & $\Gamma\vdash\vp$ & (Assumption)\\[1mm]
(2) & $\Gamma\vdash\vp\to(x\to\vp)$ & \eqref{vp-to-psi-to-vp}\\[1mm]
(3) & $\Gamma\vdash x\to\vp$ & \prule{modus-ponens}: (1), (2)\\[1mm]
(4) & $\Gamma\vdash x\to x$ & \eqref{sim-refl} and \eqref{Gamma-vptopsi-psitovp-iff-vpdndpsi}\\[1mm]
(5) & $\Gamma\vdash x\to x\si\vp$ & \eqref{vp-to-psi-si-vp-to-chi-implies-vp-to-psi-si-chi}: (4), (3)\\[1mm]
(6) & $\Gamma\vdash\ceil{x} \to x\in\vp$ & \eqref{lemma-imp-compat-in-ceil}: (5)  and the definition of $\in$ \\[1mm]
(7) & $\Gamma\vdash\ceil{x}$ & \prule{def}\\[1mm]
(8) & $\Gamma\vdash x\in\vp$ & \prule{modus-ponens}: (7), (6)\\[1mm]
(9) & $\Gamma\vdash\forall x(x\in\vp)$ & \prule{gen}: (8)
\end{tabular}

\,
}

\begin{lemma}
\begin{align}
\vdash & \, x\in y \dnd y\in x, \label{xiny-dnd-yinx}\\
\vdash & \, x\eqdef y \to x\in y. \label{xeqdefy-to-xiny}
\end{align}
\end{lemma}
\solution{\,\\[1mm]
\eqref{xiny-dnd-yinx}:\\[1mm]
\begin{tabular}{lll}
(1) & $\vdash x\si y \dnd y \si x$ & \eqref{si-comm}\\[1mm]
(2) & $\vdash x\in y \dnd y\in x$ & \eqref{rule-iff-compat-in-ceil}: (1) and the definition of $\in$\\[2mm]
\end{tabular}

\eqref{xeqdefy-to-xiny}:\\[1mm]
\begin{tabular}{lll}
(1) & $\vdash x\sau y \to (\neg(x\dnd y)\sau(x\si y))$ & \eqref{useful-mem-intro} \\[1mm]
(2) & $\vdash\ceil{x\sau y}\to\ceil{\neg(x\dnd y)\sau(x\si y)}$ & \eqref{lemma-imp-compat-in-ceil}: (1)\\[1mm]
(3) & $\vdash\ceil{x\sau y}$ & \eqref{ceil-xsauvp}\\[1mm]
(4) & $\vdash\ceil{\neg(x\dnd y)\sau(x\si y)}$ & \prule{modus-ponens}: (1), (3)\\[1mm]
(5) & $\vdash\ceil{\neg(x\dnd y)\sau(x\si y)}\to$& \eqref{lemma-ceil-compat-in-or-1} \\[1mm]
& $\quad \to \ceil{\neg(x\dnd y)}\sau\ceil{x\si y}$ \\[1mm]
(6) & $\vdash\ceil{\neg(x\dnd y)}\sau\ceil{x\si y}$ & \prule{modus-ponens}: (4), (5)\\[1mm]
(7) & $\vdash(\ceil{\neg(x\dnd y)}\sau\ceil{x\si y})\to$ &
 \eqref{vpsaupsi-negvptopsi} and \eqref{Gamma-vptopsi-psitovp-iff-vpdndpsi}\\[1mm]
& $\quad \to  (\neg\ceil{\neg(x\dnd y)}\to\ceil{x\si y})$ \\[1mm]
(8) & $\vdash \neg\ceil{\neg(x\dnd y)}\to\ceil{x\si y}$ & \prule{modus-ponens}: (6), (17) \\[1mm]
(9) & $\vdash x\eqdef y \to x\in y$ & (8) and the definitions of $\eqdef$ and $\in$
\end{tabular}

\,
}

\begin{lemma}
\begin{align}
\vdash & \,  x\eqdef y  \dnd \neg(x \in \neg y) \si  \neg(y\in \neg x). \label{xeqdefy-dnd-negxinnegy-si-negyinnegx}
\end{align}
\end{lemma}
\solution{\,\\[1mm]
\begin{tabular}{lll}
(1) & $\vdash \neg(x\dnd y)\dnd (x\si\neg y)\sau(\neg x\si y)$ & \eqref{useful-mem-neg-2}\\[1mm]
(2) & $\vdash\ceil{\neg(x\dnd y)}\dnd \ceil{(x\si\neg y)\sau(\neg x\si y)}$ & \eqref{rule-iff-compat-in-ceil}: (1) \\[1mm]
(3) & $\vdash \ceil{(x\si\neg y)\sau(\neg x\si y)} \dnd (x \in \neg y) \sau \ceil{\neg x\si y}$ & 
\eqref{lemma-ceil-compat-in-or} and the 
definition of $\in$\\[1mm]
(4) & $\vdash (x \in \neg y) \sau \ceil{\neg x\si y} \dnd (x \in \neg y) \sau (y\in \neg x)$ & 
\eqref{sim-pairs-or-right}: \eqref{xinvp-char-equiv} \\[1mm]
(5) & $\vdash \ceil{(x\si\neg y)\sau(\neg x\si y)} \dnd(x \in \neg y) \sau (y\in \neg x)$ & \eqref{sim-trans}: (3), (4) \\[1mm]
(6) & $\vdash\ceil{\neg(x\dnd y)} \dnd (x \in \neg y) \sau (y\in \neg x)$ & \eqref{sim-trans}: (2), (5) \\[1mm]
(7) & $\vdash \neg((x \in \neg y) \sau (y\in \neg x)) \dnd x\eqdef y$ & \eqref{Gamma-dnd-cong-neg}: (6)
and the definition of $\eqdef$\\[1mm]
(8) & $\vdash \neg(x \in \neg y) \si  \neg(y\in \neg x) \dnd \neg((x \in \neg y) \sau (y\in \neg x))$ & \eqref{de-morgan-and-dnd}\\[1mm]
(9) & $\vdash \neg(x \in \neg y) \si  \neg(y\in \neg x) \dnd x\eqdef y$ & \eqref{sim-trans}: (8), (7)\\[1mm]
(10) & $\vdash   x\eqdef y  \dnd \neg(x \in \neg y) \si  \neg(y\in \neg x)$ & \eqref{sim-symm}: (10)
\end{tabular}

\,
}

\begin{lemma}
\begin{align}
\vdash & \, x\in \vp \sau x \in \neg\vp,\label{xinvp-sau-xinnegvp}\\
\vdash & \,\neg(x\in\vp) \to x\in\neg\vp, \label{neg-xinvp-to-xinnegvp}\\
\vdash & \,\neg(x\in\neg\vp) \to x\in\vp, \label{neg-xinnegvp-to-xinvp}\\
\vdash & \, x\in(\vp\sau\psi) \dnd x\in\vp \sau x\in\psi, \label{xinvpsaupsi-dnd-xinvpsauxinpsi}\\
\vdash & \, x\in(\vp\sau\psi) \eqdef x\in\vp \sau x\in\psi. \label{xinvpsaupsi-eqdef-xinvpsauxinpsi}
\end{align}
\end{lemma}
\solution{\,\\[1mm]
\eqref{xinvp-sau-xinnegvp}:\\[1mm]
\begin{tabular}{lll}
(1) & $\vdash x\to(x\si\vp)\sau(x\si\neg\vp)$ &  \eqref{psi-topsisivp-sau-psisinegvp} \\[1mm]
(2) & $\vdash \ceil{x}\to\ceil{(x\si\vp)\sau(x\si\neg\vp)}$ & \eqref{lemma-imp-compat-in-ceil}: (1)\\[1mm]
(3) & $\vdash \ceil{x}$ & \prule{def}\\[1mm]
(4) & $\vdash \ceil{(x\si\vp)\sau(x\si\neg\vp)}$ & \prule{modus-ponens}: (3), (2)\\[1mm]
(5) & $\vdash \ceil{(x\si\vp)\sau(x\si\neg\vp)}\to \ceil{x\si\vp}\sau\ceil{x\si\neg\vp}$ & \eqref{lemma-ceil-compat-in-or-1}\\[1mm]
(6) & $\vdash \ceil{x\si\vp}\sau\ceil{x\si\neg\vp}$ & \prule{modus-ponens}: (4), (5) \\
(7) & $\vdash x\in \vp \sau x \in \neg\vp$ & (6) and the definition of $\in$\\[2mm]
\end{tabular}

\eqref{neg-xinvp-to-xinnegvp}:\\[1mm]
\begin{tabular}{lll}
(1) & $\vdash  x\in \vp \sau x \in \neg\vp$ &  \eqref{xinvp-sau-xinnegvp}\\[1mm]
(2) & $\vdash  x\in \vp \sau x \in \neg\vp \to (\neg(x\in\vp) \to (x\in\neg\vp))$ &  
\eqref{Gamma-vptopsi-psitovp-iff-vpdndpsi} and
\eqref{vpsaupsi-negvptopsi} \\[1mm]
(3) & $\vdash \neg(x\in\vp) \to x\in\neg\vp$ &  \prule{modus-ponens}: (1), (2)\\[2mm]
\end{tabular}

\eqref{neg-xinnegvp-to-xinvp}\\[1mm]
\begin{tabular}{lll}
(1) & $\vdash \neg(x\in\neg\vp) \to x\in\neg\neg\vp$ &  \eqref{neg-xinvp-to-xinnegvp}\\[1mm]
(2) & $\vdash \neg\neg\vp \to \vp$ &  \eqref{negnegvp-dnd-vp} and \eqref{Gamma-vptopsi-psitovp-iff-vpdndpsi}\\[1mm]
(3) & $\vdash  x\in\neg\neg\vp \to x\in\vp$ &  \eqref{lemma-in-compat-to-vppsi}: (2)\\[1mm]
(4) & $\vdash \neg(x\in\neg\vp) \to x\in\vp$ &  \prule{syllogism}: (1), (3)\\[2mm]
\end{tabular}

\eqref{xinvpsaupsi-dnd-xinvpsauxinpsi}:\\[1mm]
\begin{tabular}{lll}
(1) & $\vdash\ceil{(x\si\vp)\sau(x\si\psi)}\dnd\ceil{x\si\vp}\sau\ceil{x\si\psi}$ & \eqref{lemma-ceil-compat-in-or}\\[1mm]
(2) & $\vdash x\si(\vp\sau\psi)\dnd(x\si\vp)\sau(x\si\psi)$ & \eqref{distrib-si-sau}\\[1mm]
(3) & $\vdash\ceil{x\si(\vp\sau\psi)}\dnd\ceil{(x\si\vp)\sau(x\si\psi)}$ & \eqref{lemma-iff-compat-in-floor}: (2)\\[1mm]
(4) & $\vdash x\in(\vp\sau\psi)\dnd x\in\vp \sau x\in\psi$ & \eqref{sim-trans}: (3), (1) and the definition of $\in$\\[2mm]
\end{tabular}

\eqref{xinvpsaupsi-eqdef-xinvpsauxinpsi}: Apply \eqref{xinvpsaupsi-dnd-xinvpsauxinpsi} and \eqref{Gamma-equal-iff-dnd}.
}

\begin{lemma}
\begin{align}
\vdash & \,(x\in\exists y\vp)\, \eqdef \, \exists y(x\in\vp) \quad \text{if $x\ne y$}. \label{mem-exists}
\end{align}
\end{lemma}
\solution{\,\\[1mm]
\begin{tabular}{lll}
(1) & $\vdash x\si\exists y\vp\dnd\exists y(x\si\vp)$ & \eqref{vp-si-existsxpsi-dnd-existsx-vp-si-psi},\quad 
as $y\notin FV(x)$\\[1mm]
(2) & $\vdash x\in\exists y\vp\dnd\ceil{\exists y(x\si\vp)}$ & \eqref{rule-iff-compat-in-ceil}: (1) and the definition of $\in$\\[1mm]
(3) & $\vdash\ceil{\exists y(x\si\vp)}\dnd\exists y\ceil{x\si\vp}$ & \eqref{existsxceilvp-dnd-ceilexistsxvp}\\[1mm]
(4) & $\vdash x\in\exists y\vp\dnd\exists y(x\in\vp)$ & \eqref{sim-trans}: (2), (3) and the definition of $\in$\\[1mm]
(5) & $\vdash x\in\exists y\vp \, \eqdef \, \exists y(x\in\vp)$ & \eqref{Gamma-equal-iff-dnd}: (4)
\end{tabular}

\, 
}


\mbox{}

We consider in the sequel that the following hold:
\begin{align*}
\prule{\singletons} & \vdash \, \neg(x\in \vp)\sau\neg (x\in \neg\vp), \\
\prule{existence} & \vdash \, \exists x x.	
\end{align*}

\begin{lemma}
\begin{align}
\vdash & \, \neg(x\in \vp) \sau (\neg x\sau \vp), \label{singleton-used-membership-elim}\\
\vdash & \, \neg(x\in \vp \si (x\si \neg\vp)), \label{singleton-used-membership-elim-2}\\
\vdash & \, x\in \vp \si x \to \vp. \label{xinvpsix-to-vp}
\end{align}
\end{lemma}
\solution{ \, \\[1mm]
\eqref{singleton-used-membership-elim}:\\[1mm]
\begin{tabular}{lll}
(1) & $\vdash \neg(x\in \vp)\sau\neg (x\in \neg\vp)$ & \prule{\singletons} \\[1mm]
(2) & $\vdash x\si \neg\vp \to (x\in \neg\vp)$ & \prule{\axvptoceilvp} and the definition of $\in$\\[1mm]
(3) & $\vdash \neg (x\in \neg\vp) \to \neg (x\si \neg\vp)$ & \eqref{rec-ax3-rule}: (2)\\[1mm]
(4) & $\vdash \neg (x\si \neg\vp)\to\neg x\sau\vp$ & \eqref{neg-psisinegvp-dnd-negpsisauvp}
and \eqref{Gamma-vptopsi-psitovp-iff-vpdndpsi}\\[1mm]
(5) & $\vdash \neg (x\in \neg\vp) \to \neg x \sau \vp$ & \prule{syllogism}: (3), (4) \\[1mm]
(6) & $\vdash \neg(x\in \vp) \sau \neg (x\in \neg\vp) \to \neg(x\in \vp) \sau (\neg x \sau \vp)$ & 
\prule{expansion}: (5)\\[1mm]
(7) & $\vdash \neg(x\in \vp) \sau (\neg x\sau \vp)$ & \prule{modus-ponens}: (1), (6) \\[2mm]
\end{tabular}

\eqref{singleton-used-membership-elim-2}:\\[1mm]
\begin{tabular}{lll}
(1) & $\vdash \neg(x\in \vp) \sau (\neg x\sau \vp)$ & \eqref{singleton-used-membership-elim} \\[1mm]
(2) & $\vdash  \neg (x \si \neg\vp) \dnd \neg x\sau \vp $ & \eqref{neg-psisinegvp-dnd-negpsisauvp}\\[1mm]
(3) & $\vdash \neg(x\in \vp) \sau \neg (x \si \neg\vp) \dnd  \neg(x\in \vp) \sau  (\neg x\sau \vp)$ & 
\eqref{sim-pairs-or-right}: (2)\\[1mm]
(4) & $\vdash \neg(x\in \vp) \sau \neg (x \si \neg\vp)$ & \eqref{vpdndpsi-imp-vp-iff-psi}:  (3), (2)\\[1mm]
(5) & $\vdash \neg(x\in \vp \si (x\si \neg\vp))$ & \eqref{vpdndpsi-imp-vp-iff-psi}: \eqref{de-morgan-or-dnd}, (4)\\[1mm]
\end{tabular}

\eqref{xinvpsix-to-vp}:\\[1mm]
\begin{tabular}{lll}
(1) & $\vdash \neg(x\in \vp) \sau (\neg x\sau \vp)$ & \eqref{singleton-used-membership-elim} \\[1mm]
(2) & $\vdash (\neg(x\in \vp) \sau \neg x) \sau \vp$ & \eqref{vpdndpsi-imp-vp-iff-psi}: \eqref{sau-assoc}, (1) \\[1mm]
(3) & $\vdash \neg(x\in \vp) \sau \neg x \dnd \neg(x\in \vp \si x)$ & \eqref{de-morgan-or-dnd}\\[1mm]
(4) & $\vdash (\neg(x\in \vp) \sau \neg x) \sau \vp \dnd \neg(x\in \vp \si x)\sau \vp $ & \eqref{sim-pairs-or-left}: (3)\\[1mm]
(5) & $\vdash \neg(x\in \vp \si x)\sau \vp$ & \eqref{vpdndpsi-imp-vp-iff-psi}:  (4), (2)\\[1mm]
(6) & $\vdash x\in \vp \si x \to \vp$ & \eqref{vpdndpsi-imp-vp-iff-psi}: \eqref{def-imp-or}, (5)\\[1mm]
\end{tabular}

\, 

}

\begin{lemma}
\begin{align}
\vdash & \, \neg(x\in \vp \si x\in \neg\vp), \label{singletons-equiv-1}\\
\vdash & \, (x\in \neg\vp) \to \neg(x\in \vp), \label{singletons-equiv-2}\\
\vdash & \, (x\in \neg\vp) \dnd \neg(x\in \vp), \label{singletons-equiv-2-dnd}\\
\vdash & \, (x\in\neg\vp)\, \eqdef\, \neg(x\in\vp),  \label{singletons-equiv-2-eqdef}\\
\vdash & \, (x\in \vp) \dnd \neg(x\in \neg\vp), \label{singletons-equiv-2-negvp-dnd}\\
\vdash & \, (x\in \vp)\, \eqdef\, \neg(x\in \neg\vp). \label{singletons-equiv-2-negvp-eqdef}
\end{align}
\end{lemma}
\solution{ \, \\[1mm]
\eqref{singletons-equiv-1}:\\[1mm]
\begin{tabular}{lll}
(1) & $\vdash \neg(x\in \vp)\sau\neg (x\in \neg\vp)$ & \prule{\singletons} \\[1mm]
(2) & $\vdash \neg(x\in \vp)\sau\neg (x\in \neg\vp) \to \neg(x\in \vp \si x\in \neg\vp)$ 
& \eqref{de-morgan-or-dnd} and \eqref{Gamma-vptopsi-psitovp-iff-vpdndpsi}\\[1mm]
(2) & $\vdash \neg(x\in \vp \si x\in \neg\vp)$ & \prule{modus-ponens}: (1), (2) \\[2mm]
\end{tabular}

\eqref{singletons-equiv-2}:\\[1mm]
\begin{tabular}{lll}
(1) & $\vdash\neg(x\in \vp) \sau \neg(x\in \neg \vp)$ & \prule{\singletons}\\[1mm]
(2) & $\vdash\neg(x\in \vp) \sau \neg(x\in \neg \vp) \to ((x\in \neg \vp)\to \neg(x\in \vp))$ & 
\eqref{negvp-sau-negpsi--dnd--vp-to-negpsi} and \eqref{Gamma-vptopsi-psitovp-iff-vpdndpsi}\\[1mm]
(3) & $\vdash (x\in \neg\vp) \to \neg(x\in \vp)$ & \prule{modus-ponens}: (1), (2)\\[2mm]
\end{tabular}

\eqref{singletons-equiv-2-dnd}:  Apply \eqref{neg-xinvp-to-xinnegvp}, \eqref{singletons-equiv-2} and 
\eqref{Gamma-vptopsi-psitovp-iff-vpdndpsi}.

\eqref{singletons-equiv-2-eqdef}: By \eqref{singletons-equiv-2-dnd} and \eqref{Gamma-equal-iff-dnd}.

\eqref{singletons-equiv-2-negvp-dnd}:\\[1mm]
\begin{tabular}{lll}
(1) & $\vdash x\in \neg\neg\vp \dnd \neg(x\in \neg\vp)$ & \eqref{singletons-equiv-2-dnd}\\[1mm]
(2) & $\vdash \vp \dnd \neg\neg\vp$ & \eqref{sim-symm}: \eqref{negnegvp-dnd-vp} \\[1mm]
(3) & $\vdash x\in \vp \dnd x\in \neg\neg\vp$ & \eqref{lemma-in-compat-dnd-vppsi}: (2)\\[1mm]
(4) & $\vdash x\in \vp \dnd \neg(x\in \neg\vp)$ & \eqref{sim-trans} : (3), (2) \\[2mm]
\end{tabular}

\eqref{singletons-equiv-2-negvp-eqdef}: By \eqref{singletons-equiv-2-negvp-dnd} and \eqref{Gamma-equal-iff-dnd}.
}

\begin{lemma}
\begin{align}
\vdash & \, x\in y \dnd  x\eqdef y, \label{xiny-dnd-xeqdefy}\\
\vdash & \, (x\in y) \, \eqdef \, (x\eqdef y). \label{xiny-eqdef-xeqdefy}
\end{align}
\end{lemma}
\solution{\,\\[1mm]
\eqref{xiny-dnd-xeqdefy}:\\[1mm]
\begin{tabular}{lll}
(1) & $\vdash  x\eqdef y  \dnd \neg(x \in \neg y) \si  \neg(y\in \neg x)$ & \eqref{xeqdefy-dnd-negxinnegy-si-negyinnegx}\\[1mm]
(2) & $\vdash x\in y \dnd \neg(x \in \neg y)$ & \eqref{singletons-equiv-2-negvp-dnd} \\[1mm]
(3) & $\vdash y\in x \dnd  \neg(y\in \neg x)$ & \eqref{singletons-equiv-2-negvp-dnd} \\[1mm]
(4) & $\vdash  x\in y \si y\in x \dnd  \neg(x \in \neg y) \si  \neg(y\in \neg x)$ & 
\eqref{sim-pairs-and}: (2), (3) \\[1mm]
(5) & $\vdash  x\in y \si y\in x \dnd x\eqdef y$ & \eqref {sim-trans-2}: (4), (1) \\[1mm]
(6) & $\vdash  x\in y \dnd y\in x $ & \eqref{xiny-dnd-yinx}\\[1mm]
(7) & $\vdash x\in y \si  x\in y \dnd x\in y\si y\in x$ & \eqref{sim-pairs-and-right}: (6)\\[1mm]
(8) & $\vdash x\in y  \dnd x\in y \si  x\in y$ & \eqref{sim-symm}: \eqref{si-idemp}\\[1mm]
(9) & $\vdash x\in y  \dnd x\in y\si y\in x$ & \eqref{sim-trans}: (8), (7)\\[1mm] 
(9) & $\vdash x\in y \dnd x\eqdef y$ & \eqref {sim-trans}: (9), (5) \\[1mm]
\end{tabular}

\eqref{xiny-eqdef-xeqdefy}: Apply \eqref{xiny-dnd-xeqdefy} and \eqref{Gamma-equal-iff-dnd}.
}

\begin{lemma}
\bea
\Gamma\vdash\forall x(x\in\vp) & \text{implies} & \Gamma\vdash x \to\vp,
\label{forallxxinvp-xtovp}\\
\Gamma\vdash\vp & \text{iff} &  \Gamma\vdash\forall x(x\in\vp) \quad \text{if~}x\notin FV(\vp).  
\label{mem-intro-elim}
\eea
\end{lemma}
\solution{\,\\[1mm]
\eqref{forallxxinvp-xtovp}:\\[1mm]
\begin{tabular}{lll}
(1) & $\Gamma\vdash\forall x(x\in\vp)$ & (Assumption)\\[1mm]
(2) & $\Gamma\vdash\forall x(x\in\vp)\to x\in\vp$ & \eqref{forall-quant-x-M} \\[1mm]
(3) & $\Gamma\vdash x\in\vp$ & \prule{modus-ponens}: (1), (2)\\[1mm]
(4) & $\Gamma\vdash \neg(x\in\vp)\sau(\neg x\sau\vp)$ & \eqref{singleton-used-membership-elim}\\[1mm]
(5) & $\Gamma\vdash \neg(x\in\vp)\sau(\neg x\sau\vp) \to (x\in\vp\to(\neg x\sau\vp))$ & \eqref{def-imp-or}
and \eqref{Gamma-vptopsi-psitovp-iff-vpdndpsi}\\[1mm]
(6) & $\Gamma\vdash x\in\vp \to \neg x\sau\vp$ & \prule{modus-ponens}: (4), (5)\\[1mm]
(7) & $\Gamma\vdash\neg x\sau\vp$ & \prule{modus-ponens}: (3), (6)\\[1mm]
(8) & $\Gamma\vdash\neg x\sau\vp\to(x\to\vp)$ & \eqref{def-imp-or} and \eqref{Gamma-vptopsi-psitovp-iff-vpdndpsi}\\[1mm]
(9) & $\Gamma\vdash x\to\vp$ & \prule{modus-ponens}: (7), (8)\\[2mm]
\end{tabular}

\eqref{mem-intro-elim}:\\[1mm]
$(\Ra)$: By \eqref{mem-intro}.

$(\La)$: \\[1mm]
\begin{tabular}{lll}
(1) & $\Gamma\vdash\forall x(x\in\vp)$ & (Assumption) \\[1mm]
(2) & $\Gamma\vdash x\to\vp$ & \eqref{forallxxinvp-xtovp}: (1)\\[1mm]
(3) & $\Gamma\vdash\exists x x\to\vp$ & \eqref{exists-quant-rule}: (2), as $x\notin FV(\vp)$ \\[1mm]
(4) & $\Gamma\vdash\exists x x$ & \prule{existence}\\[1mm]
(5) & $\Gamma\vdash\vp$ & \prule{modus-ponens}: (4), (3)
\end{tabular}

\, 
}

\begin{lemma}
\begin{align}
\vdash & \,  x\in(\vp\si\psi) \dnd x\in\vp \si x\in\psi, \label{xinvpsipsi-dnd-xinvp-si-xinpsi} \\
\vdash & \,  (x\in(\vp\si\psi))\eqdef(x\in\vp)\si(x\in\psi). \label{xinvpsipsi-eqdef-xinvp-si-xinpsi}
\end{align}
\end{lemma}
\solution{\,\\[1mm]
\eqref{xinvpsipsi-dnd-xinvp-si-xinpsi}:\,\\[1mm]
\begin{tabular}{lll}
(1) & $\vdash x\in(\vp\si\psi) \dnd x\in \neg(\neg\vp\sau\neg\psi)$ & \eqref{lemma-in-compat-dnd-vppsi}: \eqref{vpsipsi-demorgan}\\[1mm]
(2) & $\vdash x\in \neg(\neg\vp\sau\neg\psi) \dnd \neg(x\in(\neg\vp\sau\neg\psi))$ & \eqref{singletons-equiv-2-dnd}\\[1mm]
(3) & $\vdash x\in(\vp\si\psi) \dnd \neg(x\in(\neg\vp\sau\neg\psi))$ & \eqref{sim-trans}: (1), (2)\\[1mm]
(4) & $\vdash x\in(\neg\vp\sau\neg\psi) \dnd x\in\neg\vp \sau x\in\neg\psi$ & \eqref{xinvpsaupsi-dnd-xinvpsauxinpsi}\\[1mm]
(5) & $\vdash \neg(x\in(\neg\vp\sau\neg\psi)) \dnd  \neg(x\in\neg\vp \sau x\in\neg\psi)$ & 
\eqref{Gamma-dnd-cong-neg}: (4)\\[1mm]
(6) & $\vdash x\in(\vp\si\psi) \dnd \neg(x\in\neg\vp \sau x\in\neg\psi)$ & 
\eqref{sim-trans}: (3), (5)\\[1mm]
(7) & $\vdash \neg(x\in\neg\vp \sau x\in\neg\psi) \dnd   \neg(x\in\neg\vp) \si \neg(x\in\neg\psi)$ & 
\eqref{sim-symm}: \eqref{de-morgan-and-dnd}\\[1mm]
(8) & $\vdash x\in(\vp\si\psi) \dnd \neg(x\in\neg\vp) \si \neg(x\in\neg\psi)$ & \eqref{sim-trans}: (6), (7)\\[1mm]
(9) & $\vdash \neg(x\in\neg\vp) \si \neg(x\in\neg\psi) \dnd x\in\vp \si x\in\psi$ & 
\eqref{sim-pairs-and}: \eqref{singletons-equiv-2-negvp-dnd}\\[1mm]
(10) & $\vdash x\in(\vp\si\psi) \dnd x\in\vp \si x\in\psi$ & \eqref{sim-trans}: (8), (9)\\[2mm]
\end{tabular}

\eqref{xinvpsipsi-eqdef-xinvp-si-xinpsi}: Apply \eqref{xinvpsipsi-dnd-xinvp-si-xinpsi} and \eqref{Gamma-equal-iff-dnd}.
}

\section{$\cG^c$ is weaker than $\appMLnewGc$}\label{section-Gc-weaker-MGc}

In the sequel, we use Proposition \ref{absps1-weakerps-absps2} to prove that $\cG^c \weakerps \appMLnewGc$.

\bthm\label{Gc-weaker-MGc}
$\cG^c \weakerps \appMLnewGc$.
\ethm
\solution{
We prove first that $\vdash_{\appMLnewGc} \varphi$ for every axiom $\varphi$  of $\cG^c$:

\prule{contraction}: common.

\prule{weakening}: common.

\prule{ex falso quodlibet}: common.

\prule{lem}: common.

\prule{axiom-not-1}: common.

\prule{axiom-not-2}: common.

\prule{$\exists$-Quantifier}:  By Proposition \ref{subfxyvp-existsxvp-xfreey}.

\prule{$\forall$-Quantifier}: By Proposition \ref{forallxvp-subfxyvp-xfreey}.

\prule{Propagation$_\bot$}: By \eqref{propagation-bot-l}, \eqref{propagation-bot-r}.

\prule{Propagation$_\vee$}: common.

\prule{Propagation$_\exists$}: common. \\

Let $\Gamma$ be a set of $\LAML$-patterns. We prove now that $\amlthmgen{\Gamma}{\cG^c}$ is closed to 
the deduction rules of $\appMLnewGc$:

\prule{modus ponens}: common.

\prule{exportation}: common.

\prule{importation}: common.

\prule{expansion}: common.

\prule{$\exists$-quantifier rule}: By \eqref{exists-quant-rule}.

\prule{$\forall$-quantifier rule}: By \eqref{forall-quant-rule}.

\prule{framing-left}, \prule{framing-right}: common.
}

\section{Soundness of $\appMLnewGc$}\label{MG-sound}

We refer to \cite{Leu25} for all the semantic definitions and properties that we use in this section.

\bthm\label{soundness-AML-MG}
Let $\Gamma\cup\{\vp\}$ be a set of patterns. Then
\bce 
$\Gamma\vdash \vp$ implies $\Gamma\gsc\vp$.
\ece
\ethm
\solution{Let us denote $\Theta=\{\vp\in Pattern \mid \Gamma\gsc\vp\}$. We prove that 
$Thm(\Gamma)\se \Theta$ by induction on $\Gamma$-theorems.

\subsection*{Axioms and Notations:}

We prove that if $\vp$ is an axiom or a notation, then $\vp$ is valid. Hence, one 
can apply \cite[Remark 3.32(ii)]{Leu25} to get that $\Gamma\gsc\vp$, that is $\vp\in \Theta$.

\subsubsection*{Propositional:}

Assume that $\vp$  is a propositional axiom or notation. One can easily see that $\vp$ is a tautology, 
By \cite[Proposition 3.66]{Leu25}, $\vp$ is valid. 

\subsubsection*{First-order:}

\prule{\Monku}: Apply \cite[Proposition 3.85]{Leu25}. 

\prule{\Monkd}: Apply \cite[Proposition 3.84]{Leu25} and the fact that  if 
$x$ does not occur in $\vp$, then $x\notin FV (\vp)$.

\prule{\Monkt}: Apply \cite[Proposition 6.8]{Leu25}.

\prule{axiom-exists-1}, \prule{axiom-exists-2}: Obviously, by \cite[Section 3.1]{Leu25}.

\subsubsection*{Application:}

\prule{Propagation$_\vee$}: Apply  \cite[Proposition 3.90]{Leu25}.

\prule{Propagation$_\exists$}:  Apply \cite[Proposition 3.90]{Leu25} and the fact that  if 
$x$ does not occur in $\psi$, then $x\notin FV (\psi)$.

\prule{\propagationdef}: Apply \cite[Proposition 6.9]{Leu25}. 

\subsection*{Definedness}

\prule{\axdef}: Apply \cite[Proposition 6.6]{Leu25}. 

\prule{\axvptoceilvp}: Apply \cite[Proposition 6.7]{Leu25}. 

\prule{\axceilbot}: Apply \cite[Proposition 6.10]{Leu25}. 

\subsection*{Rules:}

\subsubsection*{Propositional:}

\prule{modus ponens}, \prule{syllogism}, \prule{exportation}, \prule{importation}, \prule{expansion}: Apply \cite[Pro-position 3.88]{Leu25}.

\subsubsection*{First-order:}

\prule{gen}: Apply \cite[Proposition 3.89]{Leu25}.

\subsubsection*{Application:}

\prule{framing-left}, \prule{framing-right}: Apply \cite[Proposition 3.97]{Leu25}.
}

\appendix 
\addcontentsline{toc}{chapter}{Appendices}

\chapter{Proof system $\cG^c$}\label{appML-Gc}

\section*{PROPOSITIONAL:}

\subsection*{AXIOMS}

\bt{ll}
\prule{contraction} & $\vp\vee\vp\to \vp$  \qquad $\vp \to \vp\wedge\vp$ \\[2mm]
\prule{weakening} &  $\vp\to \vp\vee\psi$  \qquad  $\vp\wedge\psi \to \vp$ \\[2mm]
\prule{permutation} &  $\vp\vee\psi \to \psi\vee\vp$ \qquad $\vp\wedge\psi \to \psi\wedge\vp$ \\[2mm]
\prule{ex falso quodlibet} & $\bot\to \vp$ \\[2mm]
\prule{lem} & $\vp\sau\neg\vp$
\et

\subsection*{DEDUCTION RULES}

\bt{ll}
\prule{modus ponens} &
$
\begin{prftree}
{\varphi}{\varphi \to \psi}
{\psi}
\end{prftree}
$
\\[2mm]
\prule{syllogism} &
$
\begin{prftree}
{\varphi\to \psi}{\psi \to \chi}
{\vp\to\chi}
\end{prftree}
$
\\[2mm]
\prule{exportation} &
$
\begin{prftree}
{\varphi\si \psi \to \chi}
{\varphi\to (\psi \to \chi)}
\end{prftree}
$
\\[2mm]
\prule{importation} &
$
\begin{prftree}
{\varphi\to (\psi \to \chi)}
{\varphi\si \psi \to \chi}
\end{prftree}
$
\\[2mm]
\prule{expansion} &
$
\begin{prftree}
{\varphi\to\psi}
{\chi\sau\varphi \to \chi\sau\psi}
\end{prftree}
$
\et

\subsection*{NOTATIONS}

\bt{ll}
\prule{axiom-not-1} & $\neg\vp\to(\vp\to\bot)$\\[2mm]
\prule{axiom-not-2} & $(\vp\to\bot)\to\neg\vp$
\et

\section*{FIRST-ORDER:}

\subsection*{AXIOMS}

\bt{ll}
\prule{$\exists$-Quantifier} & $\subfx{y}{\vp} \to \exists x\vp$\quad if $x$ is free 
for $y$ in $\vp$\\[2mm]
\prule{$\forall$-Quantifier} & $\forall x\vp \to \subfx{y}{\vp}$\quad if $x$ is free 
for $y$ in $\vp$
\et 

\subsection*{DEDUCTION RULES}

\bt{ll}
\prule{$\exists$-quantifier rule} &
$$
\begin{prftree}[r]{\quad if $x\not\in FV(\psi)$}
{\vp \to \psi}
{\exists x \vp \to \psi}
\end{prftree}
$$
\\[2mm]
\prule{$\forall$-quantifier rule} &
$$
\begin{prftree}[r]{\quad if $x \not\in FV(\vp)$}
{\vp \to \psi}
{\vp \to \forall x\psi}
\end{prftree}
$$
\et

\section*{APPLICATION:}

\subsection*{AXIOMS}

\bt{lll}
\prule{Propagation$_\bot$} &
$\appln{\vp}{\bot} \to \bot$ & 
$\appln{\bot}{\vp} \to \bot$ \\[4mm]
\et

\bt{lll}
\prule{Propagation$_\vee$} & 
$\appln{(\vp \sau\psi)}{\chi} \to \appln{\vp}{\chi}\sau \appln{\psi}{\chi}$ & 
$\appln{\chi}{(\vp \sau\psi)} \to \appln{\chi}{\vp}\sau \appln{\chi}{\psi}$
\\[4mm]
\et

\bt{lll}
\prule{Propagation$_\exists$} &
$\appln{(\exists x\vp)}{\psi} \to \exists x(\appln{\vp}{\psi})$ &  
$\appln{\psi}{(\exists x\vp)} \to \exists x(\appln{\psi}{\vp})$  \quad  if $x$ does not occur in $\psi$
\et

\subsection*{DEDUCTION RULES}

\bt{ll}
\prule{framing-left} &
$
\begin{prftree}
{\vp \to \psi}
{\appln{\vp}{\chi}\to\appln{\psi}{\chi}}
\end{prftree}
$
\\[2mm]
\prule{framing-right} &
$
\begin{prftree}
{\vp \to \psi}
{\appln{\chi}{\vp}\to\appln{\chi}{\psi}}
\end{prftree}
$
\et

\newpage
\chapter{Proof system  $\appMLnow$}\label{appML-now}

\section*{PROPOSITIONAL:}

\subsection*{AXIOMS}

\bt{ll}
\prule{axiom-1} & $\vp\to(\psi\to\vp)$ \\[2mm]
\prule{axiom-2} & $(\vp\to(\psi\to\chi))\to((\vp\to\psi)\to(\vp\to\chi))$ \\[2mm]
\prule{axiom-3} & $\neg \neg \vp \to \vp$
\et

\subsection*{DEDUCTION RULES}

\bt{ll}
\prule{modus ponens} &
$\ds \frac{\vp \quad \vp \to \psi}{\psi}$ 
\et

\subsection*{NOTATIONS}

\bt{ll}
\prule{axiom-not-1} & $\neg\vp\to(\vp\to\bot)$\\[2mm]
\prule{axiom-not-2} & $(\vp\to\bot)\to\neg\vp$\\[2mm]
\prule{axiom-or-1} & $\vp\sau\psi\to (\neg\vp\to\psi)$\\[2mm]
\prule{axiom-or-2} & $(\neg\vp\to\psi)\to\vp\sau\psi$\\[2mm]
\prule{axiom-and-1} & $\vp\si\psi\to\neg(\neg\vp\sau\neg\psi)$\\[2mm]
\prule{axiom-and-2} & $\neg(\neg\vp\sau\neg\psi)\to\vp\si\psi$
\et

\section*{FIRST-ORDER:}

\subsection*{AXIOMS}

\bt{ll}
\prule{$\exists$-Quantifier} & $\subfx{y}{\vp} \to \exists x\varphi$ \quad if $x$ is free 
for $y$ in $\vp$\\[2mm]
\et 

\subsection*{DEDUCTION RULES}

\bt{ll}
\prule{$\exists$-quantifier rule} &
$
\begin{prftree}[r]{\quad if $x \not\in FV(\psi)$}
{\vp \to \psi}
{\exists x \vp \to \psi}
\end{prftree}
$
\et 

\subsection*{NOTATIONS}

\bt{ll}
\prule{axiom-forall-1} & $\forall x\vp \to \neg\exists x\neg\vp$\\[2mm]
\prule{axiom-forall-2} & $\neg\exists x\neg\vp \to\forall x\vp$\\[2mm]
\et

\section*{APPLICATION:}

\subsection*{AXIOMS}

\bt{lll}
\prule{Propagation$_\bot$} &
$\appln{\vp}{\bot} \to \bot$ & 
$\appln{\bot}{\vp} \to \bot$ \\[4mm]
\prule{Propagation$_\vee$} & 
$\appln{(\vp \sau\psi)}{\chi} \to \appln{\vp}{\chi}\sau \appln{\psi}{\chi}$ & 
$\appln{\chi}{(\vp \sau\psi)} \to \appln{\chi}{\vp}\sau \appln{\chi}{\psi}$
\\[4mm]
\prule{Propagation$^*_\exists$} &
$\appln{(\exists x\vp)}{\psi} \to \exists x(\appln{\vp}{\psi})$ &  
$\appln{\psi}{(\exists x\vp)} \to \exists x(\appln{\psi}{\vp})$  \quad  if $x\notin FV(\psi)$
\et

\subsection*{DEDUCTION RULES}

\bt{ll}
\prule{framing-left} &
$
\begin{prftree}
{\vp \to \psi}
{\appln{\vp}{\chi}\to\appln{\psi}{\chi}}
\end{prftree}
$
\\[2mm]
\prule{framing-right} &
$
\begin{prftree}
{\vp \to \psi}
{\appln{\chi}{\vp}\to\appln{\chi}{\psi}}
\end{prftree}
$
\et

\newpage
\chapter{Proof system $\appMLnewGc$} \label{AML-proof-system-new-Gc}

\section*{PROPOSITIONAL:}

\subsection*{AXIOMS}

\bt{ll}
\prule{contraction} & $\vp\vee\vp\to \vp$  \qquad $\vp \to \vp\wedge\vp$ \\[2mm]
\prule{weakening} &  $\vp\to \vp\vee\psi$ \qquad  $\vp\wedge\psi \to \vp$ \\[2mm]
\prule{permutation} &  $\vp\vee\psi \to \psi\vee\vp$ \qquad $\vp\wedge\psi \to \psi\wedge\vp$ \\[2mm]
\prule{ex falso quodlibet} & $\bot\to \vp$ \\[2mm]
\prule{lem} & $\vp\sau\neg\vp$
\et

\subsection*{DEDUCTION RULES}

\bt{ll}
\prule{modus ponens} &
$
\begin{prftree}
{\varphi}{\varphi \to \psi}
{\psi}
\end{prftree}
$
\\[2mm]
\prule{syllogism} &
$
\begin{prftree}
{\varphi\to \psi}{\psi \to \chi}
{\vp\to\chi}
\end{prftree}
$
\\[2mm]
\prule{exportation} &
$
\begin{prftree}
{\varphi\si \psi \to \chi}
{\varphi\to (\psi \to \chi)}
\end{prftree}
$
\\[2mm]
\prule{importation} &
$
\begin{prftree}
{\varphi\to (\psi \to \chi)}
{\varphi\si \psi \to \chi}
\end{prftree}
$
\\[2mm]
\prule{expansion} &
$
\begin{prftree}
{\varphi\to\psi}
{\chi\sau\varphi \to \chi\sau\psi}
\end{prftree}
$
\et

\subsection*{NOTATIONS}

\bt{ll}
\prule{axiom-not-1} & $\neg\vp\to(\vp\to\bot)$\\[2mm]
\prule{axiom-not-2} & $(\vp\to\bot)\to\neg\vp$
\et

\section*{FIRST-ORDER:}

\subsection*{AXIOMS}

\bt{ll}
\prule{\Monku} & $\forall x(\vp \to \psi) \to  (\forall x\vp \to \forall x\psi)$\\[2mm]
\prule{\Monkd} & $\vp \to \forall x\vp$\quad if $x$ does not occur in $\vp$\\[2mm]
\prule{\Monkt} & $\exists x (x \eqdef y)$\quad if $y$ is a variable distinct from $x$\\[2mm]
\et 

\subsection*{DEDUCTION RULES}

\bt{ll}
\prule{gen} &
$$
\begin{prftree}[r]{}
{\vp }
{\forall x \vp}
\end{prftree}
$$
\et

\subsection*{NOTATIONS}

\bt{ll}
\prule{axiom-exists-1} & $\exists x\vp \to \neg \forall x\neg\vp$\\[2mm]
\prule{axiom-exists-2} & $\neg\forall x\neg\vp \to \exists x\vp$
\et

\section*{APPLICATION:}

\subsection*{AXIOMS}

\bt{lll}
\prule{Propagation$_\vee$} & 
$\appln{(\vp \sau\psi)}{\chi} \to \appln{\vp}{\chi}\sau \appln{\psi}{\chi}$ & 
$\appln{\chi}{(\vp \sau\psi)} \to \appln{\chi}{\vp}\sau \appln{\chi}{\psi}$
\\[4mm]
\et

\bt{lll}
\prule{Propagation$_\exists$} &
$\appln{(\exists x\vp)}{\psi} \to \exists x(\appln{\vp}{\psi})$ &  
$\appln{\psi}{(\exists x\vp)} \to \exists x(\appln{\psi}{\vp})$  \quad  if $x$ does not occur in $\psi$
\\[4mm]
\prule{\propagationdef} & $\appln{\ceil{\vp}}{\psi} \to \ceil{\vp}$ & $\appln{\psi}{\ceil{\vp}} \to \ceil{\vp}$
\et

\subsection*{DEDUCTION RULES}

\bt{ll}
\prule{framing-left} &
$
\begin{prftree}
{\vp \to \psi}
{\appln{\vp}{\chi}\to\appln{\psi}{\chi}}
\end{prftree}
$

\\[2mm]\prule{framing-right} &
$
\begin{prftree}
{\vp \to \psi}
{\appln{\chi}{\vp}\to\appln{\chi}{\psi}}
\end{prftree}
$
\et

\section*{DEFINEDNESS:}

\bt{ll}
\prule{\axdef} & $\ceil{x}$\\[2mm]
\prule{\axvptoceilvp} & $\vp\to\ceil{\vp}$\\[2mm]
\prule{\axceilbot} & $\ceil{\bot} \to \bot$
\et

\newpage

\newpage 

\chapter{Results from \cite{Leu25a}}

Let $\eqrepl$ be a congruence. 
 
\bthm\label{replacement-thm-contexts-abstractML}$\,$\\
For any context  $\fcontext$ and any patterns $\vp,\psi$,
\bce
$\eqrepla{\vp}{\psi}$ implies $\eqrepla{\fcontext[\vp]}{\fcontext[\psi]}$.
\ece
\ethm
\solution{See \cite[Theorem 4.1]{Leu25a}.}

\bthm\label{replacement-thm-more-occurence-abstractML}$\,$\\
Let $\vp,\psi, \chi,\theta$ be patterns such that $\vp$ is a subpattern of $\chi$ and $\theta$ is obtained by 
from $\chi$ by replacing one or more occurrences of $\vp$ with $\psi$. Then 
\bce
$\eqrepla{\vp}{\psi}$ implies $\eqrepla{\chi}{\theta}$.
\ece
\ethm 
\solution{See \cite[Theorem 4.3]{Leu25a}.}

\bthm\label{thm-vp-cong-subbxybvp-abstractML}
Assume that  $\eqrepl$ is reflexive and transitive and  that the following holds: \\[1mm]
\bt{ll}
(ASSUMPTION) & For any pattern $\vp$, distinct variables $x$, $z$ such that  \\
& $x$ does not occur bound in $\vp$ and $z$ does not occur 
in $\vp$, \\[1mm]
& $\eqrepla{Qx\vp}{Qz\subf{x}{z}{\vp}}$ \quad for all $Q\in\FolQ$.
\et

Then for any pattern $\vp$ and any variables $x$, $y$ such that $y$ does not occur in $\vp$,
\begin{align*}
\eqrepla{\vp}{\subb{x}{y}{\vp}}. 
\end{align*}
\ethm
\solution{See \cite[Theorem 8.1]{Leu25a}.}

\subsection{Free and bound substitution, $x$ free for $y$}

\blem\label{subf-lemma-useful-subfxy-3-abstractML}
Let $\vp$ be a pattern and $x$, $y$ be  distinct element variables such that $x$ does not occur bound in $\vp$.

Then $x$ does not occur in $\subf{x}{y}{\vp}$.
\elem
\solution{See \cite[Lemma 6.9(vi)]{Leu25a}.}

\blem\label{subf-lemma-useful-subfxy-1-abstractML}
Let $\vp$ be a pattern and $x,y\in \evar$ such that $y$ does not occur bound in $\vp$.

Then $y$ does not occur bound in $\subf{x}{y}{\vp}$.
\elem
\solution{See \cite[Lemma 6.9(ii)]{Leu25a}.}

\blem\label{subfzy-subfxzvp=subfxyvp-abstractML}
Let $\vp$ be a pattern and $x$, $y$, $z$ be element variables such that $z$ does not occur in $\vp$.

Then $\subf{z}{y}{\subf{x}{z}{\vp}}=\subf{x}{y}{\vp}$.
\elem
\solution{See \cite[Lemma 6.11]{Leu25a}.}
\blem\label{subf-lemma-useful-subfxy-2-abstractML}
Let $\vp$ be a pattern and $x$, $y$ be element variables such that $y$ does not occur in $\vp$.

Then $\subf{y}{x}{\subf{x}{y}{\vp}}=\vp$.
\elem
\solution{See \cite[Lemma 6.12]{Leu25a}.}

\blem\label{subf-lemma-useful-x-noccur-bound-subbxy-abstractML}
Let $\vp$ be a pattern and $x,y\in \evar$ be distinct.

Then $x$ does not occur bound in $\subb{x}{y}{\vp}$.
\elem
\solution{See \cite[Lemma 7.7(i)]{Leu25a}.}

\blem\label{subb-lemma-useful-x-noccur-free-vp-noccur-subbxyvp-abstractML}
Let $\vp$ be a pattern and $x,y\in \evar$ be distinct. Assume that $x\notin FV(\vp)$.

Then $x$ does not occur  in $\subb{x}{y}{\vp}$.
\elem
\solution{See \cite[Lemma 7.7(iii)]{Leu25a}.}

\blem\label{x-noccur-subfxysubbxzvp-abstractML}
Let $\vp$ be a pattern and $x$, $y$, $z$ be element  variables such that $x\ne y,z$. Then $x$ does not occur in 
$\subf{x}{y}{\subb{x}{z}{\vp}}$.
\elem
\solution{See \cite[Lemma 7.8]{Leu25a}.}

\blem\label{subb-xyz-2-abstractML}
Let $\vp$ be a pattern and $x$, $y$, $z$ be distinct element variables such that $z$ does not occur in $\vp$. Then 
$$ \subbf{z}{x}{\subf{x}{y}{\subbf{x}{z}{\vp}}} = \subf{x}{y}{\vp}.
$$
\elem
\solution{See \cite[Lemma 7.10]{Leu25a}.}

\blem\label{subf-lemma-useful-x-freefor-x-abstractML}
Let $x\in \evar$ and $\vp$  be a pattern. Then $x$ is free for $x$ in $\vp$. 
\elem
\solution{See \cite[Lemma 12.7(i)]{Leu25a}.}

\blem\label{lemma-useful-y-noccur-vp-imp-y-freefor-x-abstractML}
Let $x,y\in \evar$ and $\vp$  be a pattern such that $y$ does not occur in $\vp$.
Then $x$ is free for $y$ in $\vp$. 
\elem
\solution{See \cite[Lemma 12.7(v)]{Leu25a}.}

\blem\label{subf-lemma-useful-xfreeyvp-xfreeysubbxzvp}
Let $\vp$ be a pattern and $x$, $y$, $z$ be element variables such that  $x$ is free for $y$ in $\vp$.

Then $x$ is free for $y$ in $\subb{x}{z}{\vp}$. 
\elem
\solution{See \cite[Lemma 12.11]{Leu25a}.}

\blem\label{subf-us-xyz-1-abstractML}
Let $\vp$ be a pattern and $x$, $y$, $z$ be  variables such that $z$ does not occur in $\vp$. Assume that $x$ is free for $y$ in $\vp$.

Then $z$ is free for $y$ in $\subf{x}{z}{\vp}$.
\elem
\solution{See \cite[Lemma 12.12]{Leu25a}.}

\subsection{Comparison of proof systems}

Let $\absps_1=(Axm_1, \dedrules_1)$,  $\absps_2=(Axm_2, \dedrules_2)$ be two proof systems for $\appML$.

\bprop\label{absps1-weakerps-absps2}
Assume that 
\be
\item $\vdash_{\absps_2} \varphi$ for every axiom $\varphi\in Axm_1$,
\item For every set $\Gamma$ of $\LAML$-patterns, $\amlthmgen{\Gamma}{\absps_2}$ is closed to 
$\dedrules_1$.
\ee
Then $\absps_1 \weakerps \absps_2$.
\eprop
\solution{See \cite[Proposition 13.12]{Leu25a}.
}

\bprop\label{absps1-equivps-absps2}
The following are equivalent:
\be
\item $\absps_1 \equivps \absps_2$,
\item $\absps_1 \weakerps \absps_2$ and $\absps_2 \weakerps \absps_1$,
\ee
\eprop
\solution{See \cite[Proposition 13.14]{Leu25a}.}

\end{document}